%% file: manuscript.tex
\documentclass[]{aa}
\usepackage{graphicx}
\usepackage{txfonts}
\usepackage[dvipsnames]{xcolor}
\usepackage{natbib,twoopt}
\usepackage{dcolumn}
\usepackage{booktabs,textcomp}
\usepackage[breaklinks=true]{hyperref}
\usepackage[breaklinks=true]{hyperref}
    \hypersetup{
        colorlinks=true,
        linkcolor=magenta,
        filecolor=violet,      
        urlcolor=red,
        citecolor=blue 
    }

\usepackage[normalem]{ulem}
\usepackage{multirow}
\usepackage{pdflscape}
\usepackage{lscape}
\usepackage{longtable}
\usepackage{soul}
\bibpunct{(}{)}{;}{a}{}{,}

\begin{document}

\newcommand{\csixty}{C$_{60}$\xspace}
\newcommand{\cseventy}{C$_{70}$\xspace}
\newcommand{\csixtyplus}{C$_{60}^{+}$\xspace}
\newcommand{\jan}[1]{\textcolor{purple}{{JC: #1}}}
\newcommand{\nick}[1]{\textcolor{orange}{{NC: #1}}}
\newcommand{\amin}[1]{\textcolor{green}{{Amin: #1}}}
\newcommand{\jonathan}[1]{\textcolor{blue}{{JSm: #1}}}

\title{The EDIBLES survey}
\subtitle{VI. Searching for time variations of interstellar absorption features\thanks{Tables \ref{tab-targets}, \ref{tab:DIBresults}, and \ref{tab:ionresults} also Figs.~\ref{plt-dib-var1}--\ref{plt-dib-var122}, and Figs~\ref{plt-atoms1}--\ref{plt-atoms11} are also available in an electronic form at the https://cdsarc.cds.unistra.fr/cgi-bin/qcat?J/A+A/}}

\author{Amin~Farhang\inst{1,2}
\and
Jonathan~Smoker\inst{3,4}
\and
Nick L.J. Cox\inst{5}
\and
Jan Cami\inst{1,6,7}
\and
Harold Linnartz\inst{8}
\and
Jacco Th. van Loon\inst{9}
\and
Martin A.~Cordiner\inst{10,11}
\and
Peter J.~Sarre\inst{12}
\and
Habib G. Khosroshahi\inst{2}
\and
Pascale Ehrenfreund\inst{8,13}
\and
Bernard H. Foing\inst{14}
\and
Lex Kaper\inst{15}
\and
Mike Laverick\inst{16}
}

\institute{
% 1
Department of Physics and Astronomy, The University of Western Ontario, London, ON N6A 3K7, Canada\\  \email{a.farhang@ipm.ir}
%2
\and
School of Astronomy, Institute for Research in Fundamental Sciences, 19395-5531 Tehran, Iran
%3
\and
European Southern Observatory, Alonso de Cordova 3107, Vitacura, Santiago, Chile
%4
\and
UK Astronomy Technology Centre, Royal Observatory, Blackford Hill, Edinburgh EH9 3HJ, UK
%5
\and
Centre d’Etudes et de Recherche de Grasse, ACRI-ST, Av. Nicolas Copernic, Grasse 06130, France
%6
\and
Institute for Earth and Space Exploration, The University of Western Ontario, London ON N6A 3K7, Canada
%7
\and
SETI Institute, 339 Bernardo Ave, Suite 200, Mountain View, CA 94043, USA
%8
\and
Laboratory for Astrophysics, Leiden Observatory, Leiden
University, PO Box 9513, 2300 RA Leiden, The Netherlands
%9
\and
Lennard-Jones Laboratories, Keele University, ST5 5BG, UK
%10
\and
Astrochemistry Laboratory, NASA Goddard Space Flight Center,
Code 691, 8800 Greenbelt Road, Greenbelt, MD 20771, USA
%11
\and
Department of Physics, The Catholic University of America,
Washington, DC 20064, USA
%12
\and
School of Chemistry, The University of Nottingham, University Park, Nottingham NG7 2RD, UK
%13
\and
Space Policy Institute, George Washington University, Washington DC, USA
%14
\and
LUNEX EMMESI, Leiden Observatory 2300 RA Leiden, The Netherlands
%15
\and
Anton Pannekoek Institute for Astronomy, University of Amsterdam, 1090 GE Amsterdam, The Netherlands
%16
\and
Instituut voor Sterrenkunde, KU Leuven, Celestijnenlaan 200D, bus 2401, 3001 Leuven, Belgium
}

\date{Accepted Aug 2023}

%%%%%%%%%%%%%%%%%%%%%%%%%%%%%%%%
%%%%%%%%%%%%%%%%%%%%%%%%%%%%%%%%
\abstract
%Context heading (optional)
{Interstellar absorption observed toward stellar targets changes slowly over long timescales, mainly due to the proper motion of the background target relative to the intervening clouds, such that over time, different parts of the intervening cloud are probed. On longer timescales, the slowly changing physical and chemical conditions in the cloud can also cause variation. Detecting such time variations thus provides an opportunity to study cloud structure. }
%Aims heading (mandatory)
{We searched for systematic variations in the absorption profiles of the diffuse interstellar bands (DIBs) and interstellar atomic and molecular lines by comparing the high-quality data set from the recent ESO diffuse interstellar bands large exploration survey (EDIBLES) to older archival observations, bridging typical timescales of $\sim$10 years with a maximum timescale of 22 years. }
% Methods heading (mandatory)
{For 64 EDIBLES targets, we found adequate archival observations. We selected 31 strong DIBs, seven atomic lines, and five molecular lines to focus our search on. We carefully considered various systematic effects and used a robust Bayesian quantitative test to establish which of these absorption features could display significant variations.}
% Results heading
{While systematic effects greatly complicate our search, we find evidence for variations in the profiles of the $\lambda\lambda$4727 and 5780 DIBs in a few sightlines. Toward HD~167264, we find a new \ion{Ca}{i} cloud component that appears and becomes stronger after 2008. The same sightline furthermore displays marginal, but systematic changes in the column densities of the atomic lines originating from the main cloud component in the sightline. Similar variations are seen toward HD~147933.}
% Conclusions heading (optional)
{Our high-quality spectroscopic observations in combination with archival data show that it is possible to probe interstellar time variations on time scales of typically a decade. Despite the fact that systematic uncertainties as well as the generally somewhat lower quality of older data complicate matters, we can conclude that time variations can be made visible, both in atomic lines and DIB profiles for a few targets, but that generally, these features are stable along many lines of sight. We present this study as an archival baseline for future comparisons, bridging longer periods.}

\keywords{ISM: lines and bands -- ISM: dust, extinction -- ISM: clouds -- Line: profiles}

   \titlerunning{The EDIBLES Survey. VI.}
   \authorrunning{A. Farhang et al.}
   
\maketitle

%%%%%%%%%%%%%%%%%%%%%%%%%%%%%%%%%
%%%%%%%%%%%%%%%%%%%%%%%%%%%%%%%%%
%%%%%%%%%%%%%%%%%%%%%%%%%%%%%%%%%

\section{Introduction}
Interstellar clouds reveal their presence in stellar spectra through numerous absorption lines that are observed from the UV to the near-infrared (NIR). These include atomic lines (from various ionization stages), molecular lines and the diffuse interstellar bands (DIBs), a set of hundreds of interstellar absorption features that were first observed a century ago (see \citealp[]{Herbig1995} and \citealp[]{Sarre06} for reviews; \citealp{2014IAUS..297.....C} for an overview of the field; \citealp{2019ApJ...878..151F} for a recent survey and \citealp{Heger22} and \citealp{2013RSPSA.46920604M} for historical context). In contrast to atomic lines and spectral features with an origin in simple molecules in the interstellar medium (ISM), the carriers of the vast majority of the more than 500 interstellar DIBs have remained unidentified, with one notable exception to date: C$_{60}^+$ was recently identified as the carrier of several DIBs in the NIR (\citealp{2015Natur.523..322C}, \citealp{2017ApJ...846..168S}, \citealp{2019ApJ...875L..28C}, and \citealp{LINNARTZ2020111243}) confirming an earlier suggestion by \citet{1994Natur.369..296F}. This identification was not a surprise. Indeed, there is a large body of evidence that suggests that the carriers of most of the DIBs are carbonaceous molecules that are abundant and widespread, and that can survive the harsh conditions in the ISM. The most promising carrier candidates are therefore carbon chains, polycyclic aromatic hydrocarbons (PAHs), fullerenes, and their derivatives (\citealp{Maier2004}, \citealp{2014IAUS..297..364S} and \citealp{Omont16}). They constitute an important part of the organic inventory of the Universe, and thus identifying the DIB carriers is a top priority in the field of astrochemistry \citep{2009ARA&A..47..427H}.

To narrow down the number of possible DIB carrier species, observational studies aim to provide constraints on the physical properties and the nature of the DIB carriers. This is often done by comparing the properties of the DIBs to the physical conditions in different sightlines, for instance as inferred from other, known constituents (e.g. \ion{H}{i}, H$_2$, and other atoms and molecules). DIBs have been mapped on large (galactic) scales (e.g. \citealt{2011A&A...533A.129V,Kos2014a,Zasowski15,2015MNRAS.454.4013B} and \citealp{Lan15}) as well as on more localized scales (\citealt{2009MNRAS.399..195V,2013A&A...550A.108V,Farhang2015a,2016A&A...585A..12B} and \citealp{2017A&A...605L..10E}), and these studies all reveal clear differences in the DIB carrier distributions that can indeed be attributed to variations in local physical conditions, on relatively large (parsec) scales. Other studies probed the small-scale structure of the ISM using DIBs by comparing sightlines that are nearly in the same direction and at the same distance. In particular, \citet{2013ApJ...764L..10C} studied $\rho$~Oph A, B, C, and DE and found changes of five to nine percent in the equivalent widths (EWs) of the $\lambda$5780 and $\lambda$5797 DIBs between the A and B components that are physically separated by only $\sim$344~au (0.002~pc). In some cases, the relative variations in EW were larger in the DIBs than in the atomic lines towards the same targets. Similar results were reported by \citet{2013A&A...550A.108V} when comparing the \ion{Na}{i} line and 5780 and 5797~\AA\ DIBs in FLAMES spectra towards the Tarantula nebula. On Galactic scales of 0.04~pc, they found that the 5797~DIB profile showed variations of 71 percent compared to only seven percent for the \ion{Na}{i} D lines. On larger scales (20~pc) in the Large Magellanic Cloud (LMC), the same DIB showed variations of 77 percent compared to 25 percent for the \ion{Na}{i} line. The measured variations for the 4428~\AA\ and 5797~\AA\ DIBs are larger than for the 5780~\AA\ DIB. While such analyses are greatly complicated by saturation effects in the strongest \ion{Na}{i} components, as well as continuum uncertainties and stellar line blends especially for the 4428 DIB, these results may suggest that there is more structure (i.e. inner-cloud column density variation) in the colder gas traced by the 5797~\AA\ DIB carrier than found in warmer gas probed by \ion{Na}{i} or the 5780~\AA\ DIB carrier (\citealp{2003ApJ...587..278W,2011A&A...533A.129V} and \citealp{2013A&A...550A.108V}). If that is indeed the case, some DIBs may be more sensitive than atomic lines to certain variations in the physical conditions inside interstellar clouds.

When comparing the DIB properties in different targets, it is often assumed that the conditions in the sightline are changing so slowly that any variation is imperceptible over timescales of years. However, variability in atomic transitions of neutral (e.g. \ion{Ca}{i}, \ion{K}{i}, \ion{Na}{i}) and singly ionized interstellar species (\ion{Ca}{ii}) has been detected towards several diffuse ISM sightlines using multi-epoch spectra taken over several years (see e.g. \citealp{1991ApJ...378..586H, 1997ApJ...478..648B,2000MNRAS.312L..43P, 2001ApJ...551L.175W, 2003ApJ...591L.123L, 2003ApJ...598L..23S, 2007ASPC..365...40L, 2013MNRAS.429..939S, 2003MNRAS.346..119S, 2011MNRAS.414...59S} and \citealp{2015MNRAS.451.1396M} and also the reviews by \citealp{2003Ap&SS.285..661C} and \citealp{2018ARA&A..56..489S}). Atomic and molecular variations were also reported toward HD~34078, HD~219188, and HD~73882 (\citealp{2003A&A...401..215R, 2007ASPC..365...86W} and \citealp{2013PASP..125.1329G}). Temporal variations have also been reported as a consequence of supernova evolution (\citealp{1994ApJ...436..144F} and \citealp{2014ApJ...782L...5M}). Perhaps the most relevant to this work is the study by \citet{2015MNRAS.451.1396M} who used spectra with a resolving power $R=\lambda/\delta\lambda = \sim$80\,000--140\,000 with epochs separated by 6--20 years. They found that of their 104 sightlines (each typically containing several different absorption components), one percent showed evidence for time variation in the \ion{Ca}{i} absorption, two percent in \ion{Ca}{ii} H and K and four percent in \ion{Na}{i} D lines. No evidence for any variation was found in the \ion{Ti}{ii}, \ion{Fe}{i}, CN, CH$^+$, or \ion{K}{i} 4044~\AA~line. Perhaps surprisingly, there were no clear differences in the physical conditions between those sightlines that did and those that did not show variations, although uncertainties are large.

There are two likely explanations in the case signals change over time.  In the first case, the cloud itself does not change, but the target star in the background has moved significantly (due to proper motion) relative to the cloud so that the sightline toward the star now effectively probes a different part of the cloud. These are then essentially ``small-scale variations'', similar to the ones studied for specific clouds by, for instance, (\citealp{2013ApJ...764L..10C} and \citealp{2013A&A...550A.108V}). In the second case, the sightline remains the same, but now the cloud itself exhibits temporal differences in physical conditions  \citep{2015MNRAS.451.1396M}. For instance, if the impinging radiation field is variable, this could affect the ionization stages of different species that could affect also DIBs (\citealp{Cami97}, \citealp{2013A&A...559A.131B}, \citealp{2015ApJS..216...33F} and \citealp{2019NatAs...3..922F}). Whatever the true origin of any variations, they could offer insightful constraints on the DIB carriers if we can tie them to specific physical parameters. 

In this work, we present the results of a study of temporal variations and stability in atomic and molecular lines and the DIBs. Our aim is to search for time variability or stability in interstellar spectra of targets whose proper motions, of the order of several milli-arcseconds per year, cause them to probe slightly different parts of the intervening interstellar clouds. For example, for a target or a cloud at a distance of 100~pc and with a proper motion of 10~mas\,yr$^{-1}$ this equates to probing the ISM on scales of about one au\,yr$^{-1}$. In a similar project \citep{2023A&A...672A.181S} we searched for time-variability in the near-infrared DIBs at 1318~nm towards 16 targets at a spectral resolving power of $R$=50\,000 with baselines of 6--12 months and at $R$=8\,000 for two objects with a baseline of nine years. We found no variations for the 1318~nm data and only tentative variation in the C$_{60}^{+}$ lines for one of the latter two objects. Finally, non-detection of time variability in the 5780, 5797, and 6613~\AA\, DIBs towards $\zeta$~Oph was recently reported by \cite{Cox2020}.

Section~\ref{sec-dataused} describes the data used in this study which is a combination of EDIBLES and archival spectra. In Sect.~\ref{sec-method}, we describe the methods to assess spectral variations over time, and the sources of uncertainty involved. We present our results in Sect.~\ref{sec-results} and discuss these in Sect.~\ref{sec-discussion}.

%%%%%%%%%%%%%%%%%%%%%%%%%%%%%%%%%
%%%%%%%%%%%%%%%%%%%%%%%%%%%%%%%%%
\section{Observations, data reduction and sample selection} \label{sec-dataused}

\subsection{Observational data and target selection}
The main data for this paper are the spectra obtained as part of the ESO Diffuse Interstellar Band Large Exploration Survey \citep[EDIBLES;][]{2017A&A...606A..76C}, a Large Program that used VLT/UVES to obtain a unique DIB data set: we observed 123 DIB targets to obtain spectra at a high spectral resolving power ($R \sim 72\,000$ for the blue arm and $R \sim 107\,000$ for the red arm), with a very high signal-to-noise ratio (S/N $\sim$500-1000 per target) and covering a large spectral range (305--1042~nm). The targets are furthermore chosen to represent very different physical conditions in the sightlines, and our sample thus probes a wide range of interstellar environment parameters including interstellar reddening $E(B-V)$ (0--1~mag), visual extinction $A_V$ (0--4.5~mag), total-to-selective extinction ratio $R_V$ (2--6), and molecular hydrogen fraction $f_{\rm H_2}$ (0.0--0.8). The program is described in \citet{2017A&A...606A..76C}, and several first results have been published (\citealp{2018A&A...614A..28L, 2018A&A...616A.143E, 2019A&A...622A..31B} and \citealp{2022A&A...662A..24M}; see also \citealp{2018Msngr.171...31C}). 

While we observed some EDIBLES targets multiple times (months or years apart), they offer only a limited baseline in time to search for variability. We therefore also searched for additional high-resolution spectra of EDIBLES targets previously acquired with FEROS \citep{1999Msngr..95....8K}, HARPS \citep{2003Msngr.114...20M}, UVES \citep{2000SPIE.4008..534D}, ESPaDOnS \citep{2003ASPC..307...41D}, HERMES \citep{2011A&A...526A..69R}, HDS \citep{2002PASJ...54..855N}, and HIRES \citep{1994SPIE.2198..362V}. These instruments offer a spectral resolution within a factor of $\sim$2 of the EDIBLES spectra, thus offering the best opportunity for a detailed comparison of spectra taken at different epochs. Reduced spectra were retrieved from the ESO archive\footnote{\url{https://archive.eso.org}}, the CFHT archive\footnote{\url{http://www.cadc-ccda.hia-iha.nrc-cnrc.gc.ca/}}, and the Mercator archive\footnote{\url{ http://www.mercator.iac.es/instruments/hermes/}}.

The EDIBLES data set contains 123 different targets that represent different conditions in the ISM. From this sample, we removed the targets where DIBs are very weak or absent, or for which we only have observations with some UVES dichroic settings at this point but not all, thus only offering a partial spectrum. Furthermore, we also need good-quality archival spectra for comparison, and we only found archival spectra for 75 of the initial EDIBLES targets. After retrieving the archival data, we carried out a quality assessment. We found that several archival spectra are of too low quality (with a S/N$\le$100 or displaying significant artifacts such as fringes) to be useful for our purposes and thus we discarded them. After this quality assessment, we were left with 65 targets for which we have complete EDIBLES spectra and good-quality archival spectra. However, one target (HD~93030) has a very low reddening and displays no DIBs; we therefore only used this target for the analysis of atomic and molecular lines. Consequently, for the remainder of this paper, we will use 64 EDIBLES targets. For just a few targets, we only have EDIBLES spectra taken just a few days or weeks apart; those are unlikely to show any time variation; we included them as a baseline and as an internal consistency check. Apart from those, the median timescale between the oldest archival observation and the most recent EDIBLES observation is nine years, with the longest timescale as long as 22 years. All targets and their key properties are listed in Table~\ref{tab-targets}.

We note that the spectral coverage is not identical for data from different epochs, and thus not all absorption lines or DIBs may be available for each of these targets. We shifted all spectra to the heliocentric rest frame. Finally, if needed, we smoothed the higher-resolution spectra using a Gaussian kernel to match the resolution of the archival observations (see Sect.~\ref{sec:resolution}).

\subsection{Spectral feature selection}

\begin{table}
\centering
\caption{DIBs selected for this study}
\begin{tabular}{lcrl}
\hline
\multicolumn{1}{c}{$\lambda_{\rm c}$} & \multicolumn{1}{c}{FWHM$^{a}$} & \multicolumn{1}{c}{EW} & Ref.\\
\multicolumn{1}{c}{[\AA]} & \multicolumn{1}{c}{[\AA]} & \multicolumn{1}{c}{[m\AA]} \\
\hline
4726.83$^{*}$ &    2.74 &  283.80 & 2\\
4762.62        &    2.50 &  126.50 & 1\\
4780.24       &    1.72 &   68.10 & 1\\
4963.98$^{*}$ &    0.72 &   26.40 & 1\\
4984.79$^{*}$ &    0.50 &   31.10 & 2\\
5176.04$^{*}$ &    0.62 &   35.60 & 2\\
5404.64       &    1.11 &   52.90 & 1\\
5418.87$^{*}$ &    0.76 &   49.1 & 2 \\
5494.16       &    0.69 &   31.20 & 1\\
5512.68$^{*}$ &    0.48 &   20.80 & 2\\
5545.11       &    0.86 &   28.10 & 1\\
5705.31       &    2.68 &  172.50 & 1\\
5772.66       &    1.23 &   40.40 & 1\\
5775.97       &    0.91 &   25.30 & 1\\
5780.61       &    2.14 &  779.30 & 1\\
5797.20       &    0.91 &  186.40 & 1\\
5849.88       &    0.93 &   67.80 & 1\\
6113.29       &    0.92 &   41.50 & 1\\
6196.09       &    0.66 &   90.40 & 1\\
6203.14       &    1.42 &  206.20 & 1\\
6269.93       &    1.32 &  256.40 & 1\\
6353.59       &    1.87 &   51.50 & 1\\
6376.21       &    0.94 &   63.70 & 1\\
6379.32       &    0.68 &  105.40 & 1\\
6425.78       &    0.70 &   25.80 & 1\\
6439.62       &    0.74 &   26.90 & 1\\
6445.41       &    0.84 &   58.10 & 1\\
6456.02       &    1.01 &   54.50 & 1\\
6597.43       &    0.69 &   26.70 & 1\\
6613.70       &    1.08 &  341.60 & 1\\
6660.73       &    0.67 &   59.70 & 1\\
\hline
\end{tabular}
  \tablebib{
    (1)~\citealt{Hobbs2009} (values for HD~183143); (2)~\citealt{2008ApJ...680.1256H} (values for HD~204827). 
    }
\tablefoot{
\tablefoottext{a}{Values for HD~183143 from \citet{Hobbs2009}.}\\
\tablefoottext{*}{C$_2$-DIBs}
}
\label{tab:DIBs}
\end{table}

Our starting point to select DIBs for this study is the catalog produced by \citet{Hobbs2009} listing the DIBs observed in HD~183143. However, we can only expect to reliably establish variations in fairly strong and well-defined spectral features. We thus first excluded DIBs longward of 6800~\AA\ where telluric contamination is generally quite severe, making it much harder to reliably extract the DIB profiles. For the remaining DIBs, it is clear that the broader a DIB is, the stronger it needs to be for us to be able to establish any such variation confidently. We therefore included:
\begin{itemize}
    \item \textit{narrow, moderately strong DIBs} with FWHM $\leq 1$~\AA\ and EW $\geq 25$~m\AA. From this list, we exclude the $\lambda$6699.36 DIB since it falls in a wavelength gap not covered by UVES. This then leaves 16 DIBs in this category.

    \item \textit{slightly broader, but stronger DIBs} with 1~\AA\ $\leq$ FWHM $\leq 2$~\AA\ with EW $\geq 40$~m\AA; this results in nine additional DIBs.
    
    \item \textit{Broader, strong and well-known DIBs:} $\lambda\lambda$4762.62, 5705.31, 5780.61, 6284.28 (but see note below)
\end{itemize}
\noindent
This list contains a few so-called C$_2$ DIBs \citep[as listed in][]{2008ApJ...680.1256H} -- DIBs that show a particularly good correlation with the column density of C$_2$, and that are believed to probe cloud interiors (\citealp[see e.g.][]{2003ApJ...584..339T} and \citealp{2018A&A...616A.143E}), and we added two more C$_2$-DIBs (at 4726.83~\AA\ and 5512.68~\AA) to our sample.  During the course of this study, we removed three DIBs from our sample since they are in wavelength ranges that are highly contaminated by telluric absorption lines; given the filler nature of the EDIBLES program, telluric contamination is particularly bad and we often could not obtain a good telluric correction. We thus chose to remove the $\lambda\lambda$5923.62, 6284.28, and 6520.75 DIBs from our study, and are thus left with 31 DIBs. All 31 selected DIBs and their properties are listed in Table ~\ref{tab:DIBs}. 

\begin{table}
    \caption{Rest wavelengths $\lambda_{c, {\rm air}}$ and oscillator strengths $f$ of atomic and molecular transitions included in this study. }
    \centering
    \begin{tabular}{lllll}
    \toprule
    species        & & \multicolumn{1}{c}{$\lambda_\mathrm{c,air}$} & $f$ & Ref \\
                   & &  \multicolumn{1}{c}{[\AA]}  \\
    \midrule
    \ion{Ca}{i}    & &  4226.728  & 1.77   & 1   \\
    \ion{Ca}{ii} K & &  3933.661  & 0.627  & 1   \\
    \ion{Ca}{ii} H & &  3968.470   & 0.312  & 1   \\
    \ion{K}{i}     & &  7698.964   & 0.3327  & 1   \\
    \ion{Na}{i}    & &  3302.369  & 0.009  & 1   \\
                   & &  3302.978  & 0.005  & 1   \\
                   & &  5889.951  & 0.6408 & 1   \\
                   & &  5895.924  & 0.3201 & 1   \\
    CH             & &  4300.313  & 0.005  & 2   \\
    CH$^{+}$       & &  3957.692  & 0.003  & 2   \\
                   & &  4232.548  & 0.006  & 2   \\
    $^{12}$CN (0,0) & R(1) &  3873.998  & 0.023  & 3   \\
                   & R(0) &  3874.608  & 0.034  & 3   \\
                   & P(1) & 3875.763  & 0.011  & 3   \\
    \ion{Fe}{i}    & &  3719.935  & 0.041  & 1   \\
    \ion{Ti}{ii}   & &  3383.759  & 0.358  & 1   \\
    \bottomrule
    \end{tabular}
    \\
    \tablebib{
    (1)~\citet{2003ApJS..149..205M}; (2)~\citet{2011AN....332..167W}; (3)~\citet{1989ApJ...343L...1M}
    }
    \label{tab-atoms}
\end{table}

In order to compare any changes in the DIBs to varying physical conditions, we also searched for variations in a number of atomic absorption lines and transitions originating from diatomic molecules that can act as diagnostic tools (see Table.~\ref{tab-atoms}). Of particular importance here are the \ion{Na}{i} UV lines that we used to determine the interstellar velocity of the different cloud components. We used the strongest absorption component in these lines as a reference to define an interstellar rest frame for each sightline. This rest frame was then used in our study of the DIBs (for which the rest wavelengths are not known). We note that for all atomic and molecular lines, we report velocities in the heliocentric reference frame.

%%%%%%%%%%%%%%%%%%%%%%%%%%%%%%%%%
%%%%%%%%%%%%%%%%%%%%%%%%%%%%%%%%%
\section{Measuring temporal variations} \label{sec-method}

For each of the 64 available targets, we compared all observed spectra from archival observations to the EDIBLES data to search for temporal variations. Our starting point is that we consider having found a real change in an interstellar absorption feature if the change is significant and if we can exclude any other logical origin for this change.

% %%%%%%%%%%%%%%%%%%%%%%%%%%%%%%%%%%%%%%%
% %%%%%%%%%% Uncertainties %%%%%%%%%%%%%%
% %%%%%%%%%%%%%%%%%%%%%%%%%%%%%%%%%%%%%%%
A straightforward way to search for such temporal variations is to divide the most recent spectrum by the oldest spectrum of the same target \footnote{We compared the residual and ratio spectra and found them quite similar for DIBs; however, more subtle variations show up for ratio spectra which is why we use those in this study.}. If the two spectra are identical, the resulting ratio spectrum should essentially be flat everywhere within the uncertainties. Any variations in an interstellar line (atom, molecule, or DIB) on the other hand will lead to deviations from a flat line. The most common change would be a change in the column density, which would lead to a deeper or shallower absorption feature in either of the two spectra. All other things being equal, this would show up in the ratio spectrum as an absorption or emission feature with otherwise the same characteristics (wavelength, width, profile shape) as the feature in the original observations. However, artifacts may show up that at first sight suggest temporal variations, but that is really due to systematic errors.

\subsection{The effect of spectral resolution}
\label{sec:resolution}
Perhaps the most obvious type of artifacts originates from the difference in spectral resolving power when comparing EDIBLES spectra to archival spectra. A lower resolution will broaden absorption lines and result in absorption- or emission-like features in the ratio spectrum. To minimize these effects, we used a Gaussian convolution to degrade the higher-resolution spectrum to match the lower spectral resolution. Our convolution kernel has a Full Width at Half Max (FWHM) value of 
\begin{equation}
{\rm FWHM} = \sqrt{(\lambda/R_{\rm low})^2 - (\lambda/R_{\rm high})^2}
\end{equation}
with $R_{\rm high}$ the resolving power of the highest resolution spectrum and $R_{\rm low}$ the lowest one. This convolution effectively degrades the high-resolution spectrum to match the low-resolution one. Using this degraded spectrum greatly reduces the impact of these artifacts in the ratio spectrum. 

\subsection{Other artifacts}
\label{sec:artefacts}
We considered several other sources of uncertainty as well but found them to be less important since they can generally be recognized as an artifact quite readily. Such effects include \textit{(i)} telluric correction residuals, \textit{(ii)} small wavelength shifts between the two spectra caused by imperfections in the wavelength calibration, and \textit{(iii)} variations in the telescope/instrument/flat-fielding response. Furthermore, some targets display complex stellar spectra whose variability can be misidentified as variations in interstellar absorption, including stellar emission lines that may contaminate interstellar absorption bands (see e.g. \citealt{2015MNRAS.451.1396M}). Any variations of interstellar lines and bands seen towards such targets in wavelength ranges where strong stellar lines exist thus need to be carefully checked for stellar contamination effects. For such targets, we compared their spectra to DIB-free sightlines with similar spectral types and luminosity classes.

\subsection{Searching for temporal variations in the DIBs}
\label{Sec:DIBvar_method}

Taking into account the above considerations, we took the following steps to conclude whether there is any temporal variation for a specific DIB:  

\begin{enumerate}
\item For each of our targets, we first selected the earliest available good-quality archival spectrum for our comparison with the latest EDIBLES spectrum. If the earliest available spectrum did not have good wavelength coverage or quality, we took the next available. This provides us with the longest possible baseline for our search and thus offers the best prospects to detect gradual, monotonic changes. In doing so, we may miss some variations that happen on shorter timescales and that would show up in intermediate observations. Note however that we only miss those cases where variations are non-monotonic and in fact cancel each other out such that the earliest archival spectrum and the EDIBLES observations are indistinguishable. 
\item Second, we ensured that the EDIBLES and archival spectra were degraded to the same spectral resolution (see Sect.~\ref{sec:resolution}). 
\item Then, we fitted the DIB in each observation with a simple Gaussian absorption line. The purpose here is not to obtain a good fit (since many DIB profiles are not Gaussian), but to have a simple measure for the central wavelength and the width of the DIB in each of the observations. We also use this fit to obtain an estimate of the equivalent width; this may result in differences between the EWs we list here and surveys where the EW has been obtained by direct integration. We expect that the Gaussian parameters in both observations should be very similar if there is no significant time variation. Significant changes in the EWs could thus be a first indication of time variation. We note, however, that the estimate of the FWHM and EW we obtain this way depends on the data quality, and the lower quality of many archival spectra in fact results in several false positives at this point.
\item Next, we inspect the ratio spectrum. If there are no DIB variations, this ratio spectrum should essentially be flat everywhere. After dividing the two spectra, we rescaled the ratio spectrum to an average value of 1. However, systematic errors in the positioning of the continuum will lead to constant ratio values that are slightly different from unity. We thus first calculate a "baseline" reduced $\chi^2$ value by assuming that the model for the ratio spectrum is constant. We will denote this initial value by $\chi_{\rm base}^2$. If a constant baseline results in a good fit ($\chi^2_{\rm base}\le1$; assuming our uncertainty estimates are realistic), there can be no significant change in the absorption feature. 
\item In case the baseline model does not properly reproduce the ratio spectrum, we consider that the most likely temporal change is an increase or decrease in column density (and hence absorption depth). Consequently, we would expect a feature in the ratio spectrum with similar characteristics (central wavelength \& width) as the absorption feature itself. Thus, we next fit a Gaussian to the ratio spectrum as well. However, this time we greatly constrain the Gaussian parameters: while the Gaussian can be in emission or absorption, the central wavelength is restricted to within $\pm \sigma/2$ of the DIB central wavelength (determined in step 2), and similarly the width should be within 50\% of the actual DIB width, established in step 2. This latter choice will then also fit residuals that are a bit too broad to be due to DIB variations but might help us recognize contamination by stellar lines that tend to be broader than the DIBs included in this study. Once a Gaussian has been fit to the ratio spectrum, we calculate the reduced chi-square also with this model and denote it by $\chi^2_{\rm Gauss}$.
\item The final test is to establish whether the Gaussian model is a significant improvement over the baseline model. To do so, we first calculate the Bayesian Information Criterion \citep[BIC; ][]{Schwarz_1978}, defined as: 
\begin{equation}
    {\rm BIC} = N\ln(\frac{\chi^2}{N})+\ln(N)m
    \label{ref:Eq:BIC}
\end{equation}
with $N$ the number of data points included in the fit and $m$ the number of parameters in the corresponding model. This criterion is similar to evaluating a $\chi^2$ statistic, but offers a slightly different measure of the quality of the fit, and includes a "penalty" for adding parameters. Indeed, one would expect the $\chi^2$ to decrease when adding parameters to a model, and the BIC evaluates whether the $\chi^2$ changes enough to warrant adding these new parameters. Using the BIC to test whether the additional parameters are significantly improving the fit is thus comparable to an $F$-test for additional parameters. When comparing models, we should then stick to the model with the lowest BIC statistic. However, a small change in the BIC value is not necessarily a significant change and one needs to consider scales of evidence \citep{article}. There exists some level of subjectivity when it comes to deciding what sort of changes should be considered interesting. For this paper, we will deem a change significant if the difference in BIC is larger than 3. Thus, we will conclude that the change to the absorption band is significant only if the Gaussian fit to the ratio spectrum has a BIC that is at least three lower than the baseline model.

\end{enumerate}

\begin{figure}
    \hspace*{-3mm}\includegraphics[width=0.98\columnwidth]{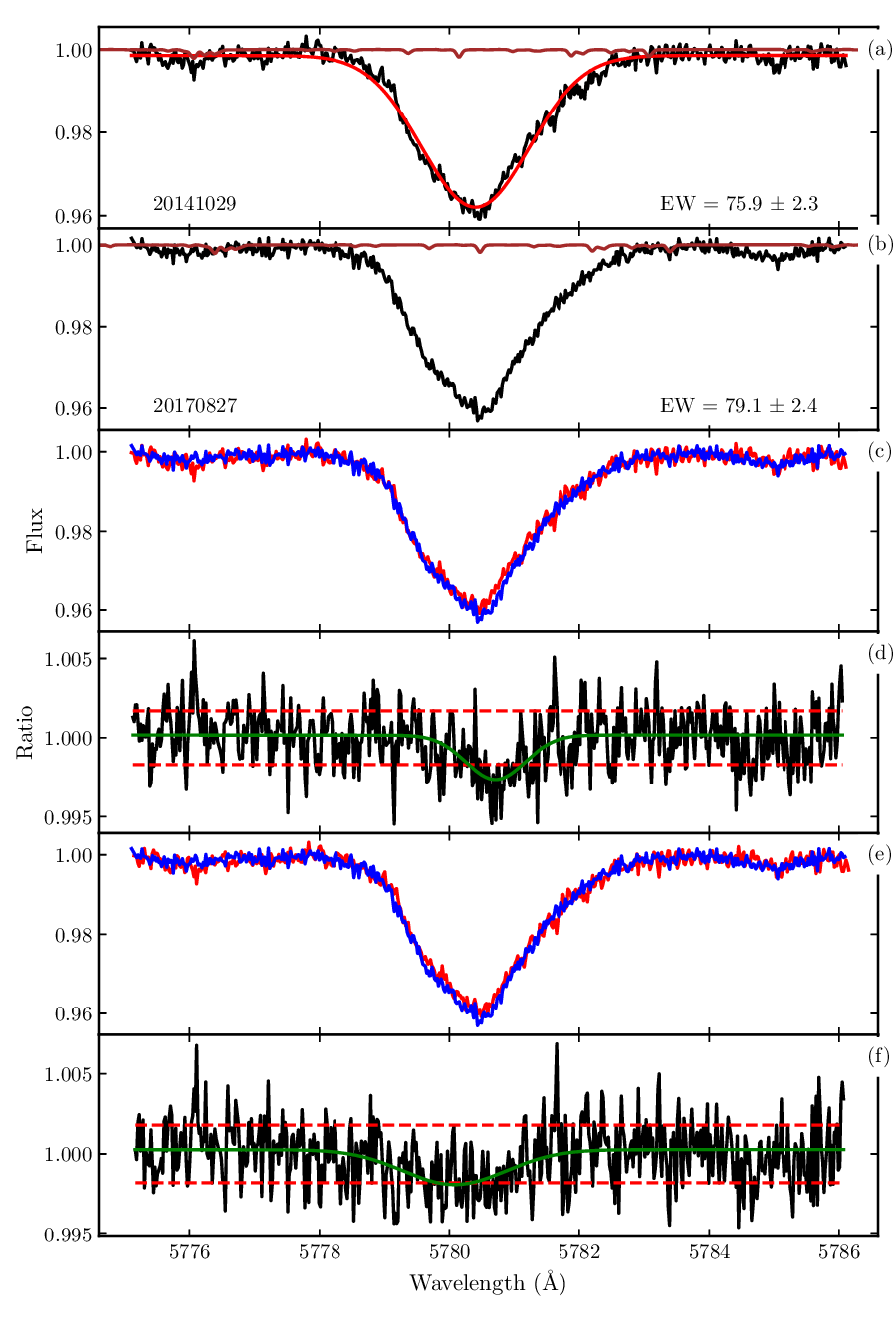}
	\caption{Variation in $\lambda$5780 DIB toward HD~23180. (a): EDIBLES "archival" observations on 2014-10-29 (black) and the best-fit Gaussian model (solid red line) and the atmospheric transmission spectrum (brown) showing weak telluric lines. (b) same as panel (a) but for the second epoch on 2017-08-27. (c) comparison of the first (solid red line) and second (solid blue line) epochs. We note that the profiles are very similar but not 100\% identical. (d) the ratio spectrum. The dashed red lines show the 1-$\sigma$ band around unity. The green solid line shows the best-fit Gaussian to the ratio spectrum. (e) same as (c), but now the 2014 archival spectrum has been shifted by 0.04~\AA. (f) the ratio spectrum of the two spectra in (e). We note that the apparent Gaussian feature from (d) has become much broader and is shifted to the blue, but has not disappeared. }
	\label{plt-method}
\end{figure}

Fig.~\ref{plt-method} shows an example of such a DIB comparison in practice, using the $\lambda$5780 DIB towards HD~23180. In this particular case, the comparison is fairly straightforward since both epochs are EDIBLES observations. Both observations (panels a, b, and c) look very similar, and indeed the Gaussian fit parameters are very close to one another. Both Gaussians have the same central wavelength of 5780.42~\AA, and absorption depths of 3.6\% and 3.8\% of the continuum with widths of 0.83~\AA\ and 0.82~\AA\ for the oldest versus the most recent observation, respectively. This then corresponds to equivalent widths of 75.9 $\pm$ 2.3 and 79.1 $\pm$ 2.4 respectively. We note that the 1-$\sigma$ confidence interval for both equivalent width measurements overlaps, and thus the equivalent width measurements are consistent with a constant value within uncertainties. 

Next, we look at the ratio spectrum (panel d in Fig.~\ref{plt-method}). The root-mean-square value of this ratio spectrum measured at the edges is $\sim$0.002 (indicated by the red dashed lines). We then compare this ratio spectrum to a baseline model with a constant value for this ratio. We find the best-fit constant to be slightly less than unity with a $\chi^2=860.2$ for $N=544$ data points, and thus a reduced $\chi_{\rm base}^2$ = 1.58 and ${\rm BIC}_{\rm base}=253.9$. It is clear from Fig.~\ref{plt-method} that the ratio spectrum shows an apparent absorption feature, and we could fit this with a Gaussian with an absorption depth of 0.3\% and a width of 0.46~\AA\ centered at 5780~\AA. These parameters are within the constraints imposed in step 4, and the resulting $\chi^2$ = 757.1 with now four parameters, so the reduced $\chi_{\rm Gauss}^2$ = 1.39, and  ${\rm BIC}_{\rm gauss}=203$. Since the BIC value for the Gaussian model is 50.9 units lower than for the straight line, our analysis concludes that there is a significant change in the $\lambda$5780 DIB. 

We further investigated promising cases like this in more detail. We note that there are a few more cases where the 5780 DIB exhibits similar subtle but (according to our method) significant profile variations (HD~37367 and HD~185859). We considered whether these could be due to chance superpositions of telluric lines in the two spectra, but as the transmission spectra in Fig.~\ref{plt-method} show, these telluric lines cannot be the origin of the residual absorption features. We also tested the role of wavelength calibration uncertainties. These can be as large as 0.04~\AA\ in this wavelength range. To test the effect of such errors, we shifted the archival spectrum by just 0.04~\AA\ and inspected the results. This shift makes the apparent Gaussian absorption feature all but disappear from the ratio spectrum (see panels e and f in Fig.~\ref{plt-method}) and instead, a more gradual change in the profile became apparent. This residual feature too can be fitted with a Gaussian profile as shown, but this time the absorption depth is shallower at 0.2\% and the width is broader at 0.67~\AA -- but still well within the constraints imposed in step 4. Once more we compare this to the baseline fit and find from the BIC that the model with the Gaussian is a significantly better fit than the baseline. So even when we allow for wavelength calibration errors, we are left with a significant time variation in our data. We thus have to conclude that in these cases, there are subtle but clear variations in the spectra of the DIBs. For many of the investigated DIBs, such subtle changes have not been found or were proven to be due to artifacts. Only for a small set of observations, changes as illustrated in Fig.~\ref{plt-method} actually show up. 

\subsection{Searching for temporal variations in interstellar atomic and molecular lines}

The absorption lines of interstellar atoms and molecules offer an opportunity to approach the search for time variations in a different way, by modeling each absorption line in more detail. Given the narrow nature of these lines, this only makes sense if both the archival and EDIBLES observations are high-resolution and high S/N UVES observations, and this is the case for 16 of our targets (see Table~\ref{tab-targets}).

For all lines, we first determined the local continuum by fitting a cubic spline to featureless regions of the spectra on either side of the absorption feature in question. The spectra were then normalized to this continuum. Next, we fitted a set of Voigt profiles to the absorption lines. The number of components for each of the sightlines was determined by the eye and can be different for different species. For lines originating from the same species (e.g. \ion{Na}{i}), we fixed the radial velocities and $b$-values. We also constrained the $b$ values to be the same between epochs. To do so, we first determined the $b$-value of each component from the spectrum with the highest resolving power and then modeled the lower resolution epoch with the same $b$-values. Figs.~\ref{plt-atoms1}--\ref{plt-atoms11} show the resulting best fits, and Table~\ref{tab:ionresults} lists the best-fit parameters for all lines.

For our DIB measurements, we fit the continuum (using a third-order Chebyshev function) and the DIB (using a Gaussian) at the same time. Uncertainties in the continuum can affect the Gaussian parameters and are likely the primary source of uncertainties in the derived parameters. To get an estimate of the magnitude of this effect, we then fitted the DIBs using a straight-line continuum and compared the DIB parameters. We used the difference between the two fits as an estimate of the systematic uncertainty, and added this to the statistical uncertainty on the parameters as returned by \texttt{lmfit} (\citealp{Levenberg44} and \citealp{Marquardt63}), derived from propagating the statistical uncertainties (noise) on the measurements.

For our atomic and molecular lines too, we expect that the most likely change over time is an increase or decrease in the column density of a species, and this time we have actual measurements for the column densities that include proper uncertainties. We inspected both the spectra (and profile fits) and plots of the column densities of each species over time to evaluate any changes. When comparing the line profiles at different epochs, many spectra show small or subtle changes in their line profiles. 
While we cannot exclude that some of these may be due to actual changes in the line profiles, in most cases the uncertainties on the column densities suggest that these variations are most likely due to differences in the resolution of the observations or continuum positioning uncertainties. There are some interesting exceptions though -- and those are discussed in Sect.~\ref{sec-results}. Once more, we stress that all velocities in this study are reported in the heliocentric reference frame.

% %%%%%%%%%%%%%%%%%%%%%%%%%%%%%%%%%%%%%%%
% %%%%%%%%%%%%% Results %%%%%%%%%%%%%%%%%
% %%%%%%%%%%%%%%%%%%%%%%%%%%%%%%%%%%%%%%%
\section{Results} \label{sec-results}

\subsection{The diffuse interstellar bands}
For all 64 sightlines, we first searched for possible variations in the DIB properties, following the method outlined in Section~\ref{Sec:DIBvar_method}. A comprehensive overview of our results is presented in Table~\ref{tab:DIBresults}, and we also created corresponding plots in Appendix~\ref{Sec:appendix_DIBs}. Our method results in a large number of cases where the ratio spectrum is significantly better reproduced by a constrained Gaussian than by a baseline model. However, this only indicates that there are differences between the spectra at the two epochs, but does not give any information about the source of these differences. To further assess these cases, we inspected the plots in detail. Often, a comparison of the same DIB in different sightlines can offer clues as to the origin of these changes. 

By and large, the most frequent occurrence is shallow, broad features in the ratio spectra, much broader than the width of the DIBs themselves. A visual inspection does not suggest the DIBs have changed between the two epochs. These features are then most likely induced by continuum positioning errors, and we have marked such cases with a \texttt{c} in Table~\ref{tab:DIBresults}. Good examples are for instance the $\lambda5513$ DIB in HD~23180 or the $\lambda4763$ DIB in HD~183143. In other cases, we recognize a blend of a stellar line with the DIB, and the target is a spectroscopic binary. In those cases, variations in the line profile are the consequence of the stellar line shifting in wavelength between the two epochs. We have indicated those cases with a $\star$. A good example of this effect is the $\lambda6440$ DIB in HD~147683. There are also a few cases where we believe the changes are due to differences in telluric contamination between the epochs (indicated with $\oplus$, for example, the $\lambda5419$ DIB in HD~41117), and finally, we also identified several instrumental artifacts. In particular, there were cases where we first noticed a significant variation between the two epochs, but after shifting the archival spectra by a fraction of a resolution element, the ratio spectra were flat. This is thus the consequence of inaccuracies in the wavelength calibration. A clear example is the $\lambda6379$ DIB toward HD~43384 which shows a clear feature in the ratio spectrum. However, this feature completely disappears if we apply a small wavelength shift to either of the two spectra. Another common artifact is a sudden jump in one of the spectra.
We denoted these artifacts as \texttt{a}. In the end, we find only two DIBs, the $\lambda\lambda4727$ and 5780~DIBs, that may show changes in their profiles between the two epochs that we cannot immediately explain by any of these causes.

\subsubsection{The $\lambda$4727 DIB} 
\begin{figure}[t!]
	\centering
	\includegraphics[width=0.98\hsize]{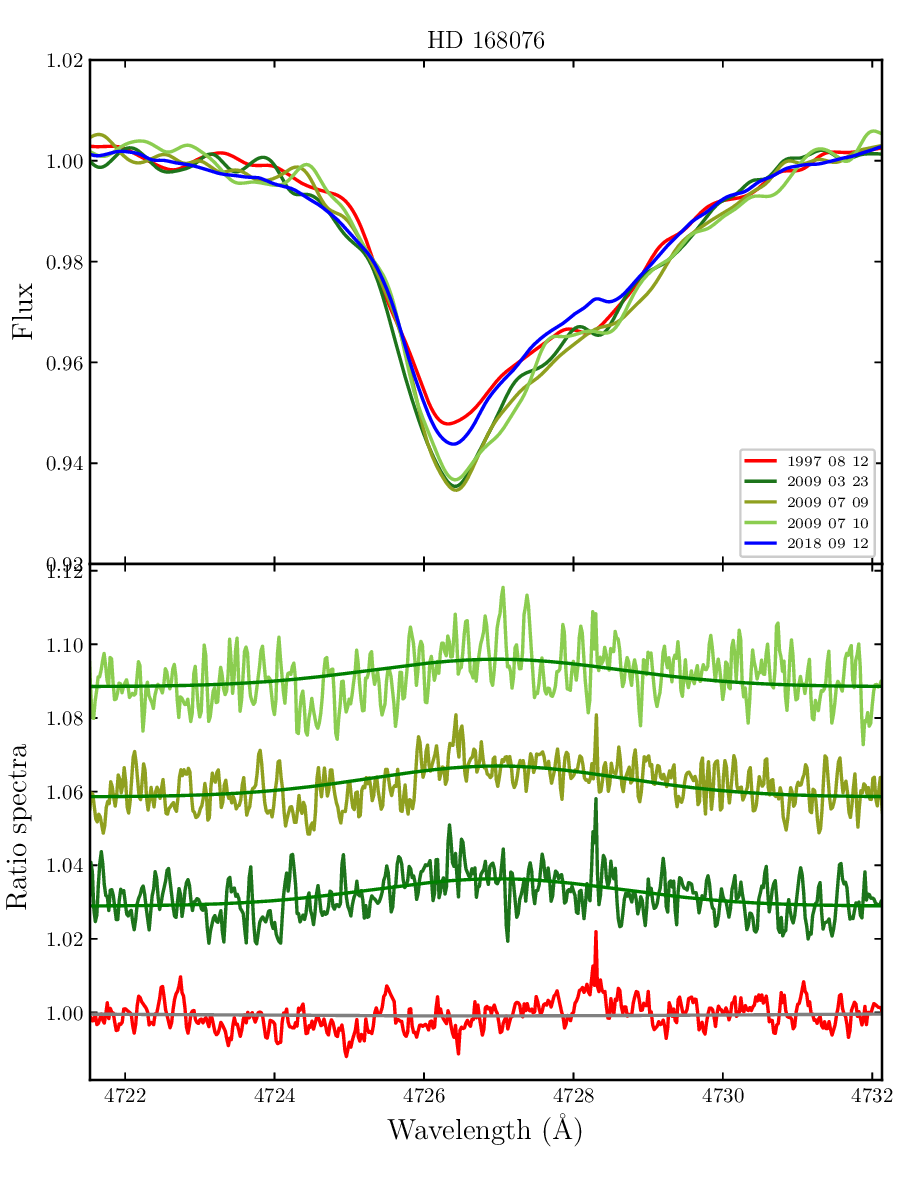}
	\caption{Variation in $\lambda$4727 DIB spectra towards HD~168076 used in this study. The EDIBLES spectrum is plotted in blue, and the earliest archival spectrum is in red; the intermediate observations are in green. The spectra are smoothed using a Gaussian convolution to minimize noise and better bring out the variations. The lower panel shows the ratio spectra and the fitted Gaussian model in each case. The top three ratio spectra (dividing the intermediate spectra by the EDIBLES spectrum) indicate significant variation, while the bottom ratio spectrum shows insignificant change when comparing the oldest archival to the EDIBLES observations.}
	\label{plt-hd168076}
\end{figure}

We possibly see changes in the profile of this DIB for three targets, but it is most significant in HD~168076 where we see variations on the order of 0.5--2\% of the absorption depth between the three 2009 archival observations and the EDIBLES 2018 observations (see Fig.~\ref{plt-hd168076}). 
The ratio spectrum shows a broad feature, of approximately the same width as the actual DIB, and the continuum levels on either side of the feature match up quite well. Two more objects (HD~24398, and HD~36861) show similar changes, but they are less pronounced or less significant ($\Delta BIC$ of 39 and 95 respectively; see Fig.~\ref{plt-dib-var1}). This residual feature in the ratio spectrum is too broad to be telluric in nature, and we can rule out a stellar origin for these differences as well since stellar lines are very weak in this range. Given these variations, we searched for more archival data and found a 1997 spectrum from HIRES. Surprisingly, the 1997 spectrum is more similar to the EDIBLES spectrum than the 2009 spectra, with perhaps the exception of the weak secondary absorption feature just redward of 4728~\AA.  We can rule out that these variations are due to, for instance, flatfield issues in the EDIBLES spectra since we also have several observations of the same DIB toward, for example, HD~170740, and for that target, we see no variations at all. Since all the 2009 observations are FEROS observations, this may point to an instrumental artifact in FEROS. It could also perhaps be an extreme case of continuum placement errors, but it is hard to tell.  Thus, even though the $\lambda$4727 DIB toward HD~168076 provides on first sight perhaps the most promising case for observing actual variations in the DIB strength, even in this case, it is hard to be confident that they are true changes in the sightline. Moreover, the magnitude of these variations is hard to understand in the context of physical changes in the sightline (see Discussion). 

In addition, we noticed that this DIB has an interesting, double-peaked profile that is clearly visible in many sightlines. This has been noticed before; \citet{2001MNRAS.323..293G} assigned the second profile of the $\lambda4727$  structure to \ion{C}{ii} stellar line contamination. \citet{2006A&A...448..221S} compared the DIB profile to synthetic stellar spectra containing lines of \ion{Ar}{ii} at 4726.85~\AA, \ion{C}{ii} at 4727.41~\AA, and \ion{Si}{iii} at 4730.52~\AA. Based on the depth of the absorption in the stellar model, they conclude that stellar contamination causes some minor broadening and increases the depth of the feature but does not affect the DIB peak positions. Consequently, they argued for an interstellar origin for the second absorption. \citet{2018A&A...616A.143E} similarly noticed the second absorption dip in the profile and pointed out that the shape of this second peak is independent of the stellar rotation rate. Instead, they found that its profile shape is perfectly correlated with the main DIB absorption.  They therefore also concluded that this absorption feature is indeed interstellar in nature and given the correlation with the main absorption band, part of one and the same DIB. 

We compared the line profile of this entire DIB (including the two absorption peaks) in more than 100 sightlines in the current work, and we find that the central wavelengths of the two peaks are constant in the interstellar reference frame, so that neither of the peaks can be a stellar feature. It is perhaps not clear whether these should be considered as two separate DIBs, but given that the ratio of the two features does not change much, we suspect that this is in fact one DIB with a rather large peak separation. 

\subsubsection{The $\lambda$5780 DIB}
\begin{figure}[ht!]
	\centering
	\includegraphics[width=0.98\hsize]{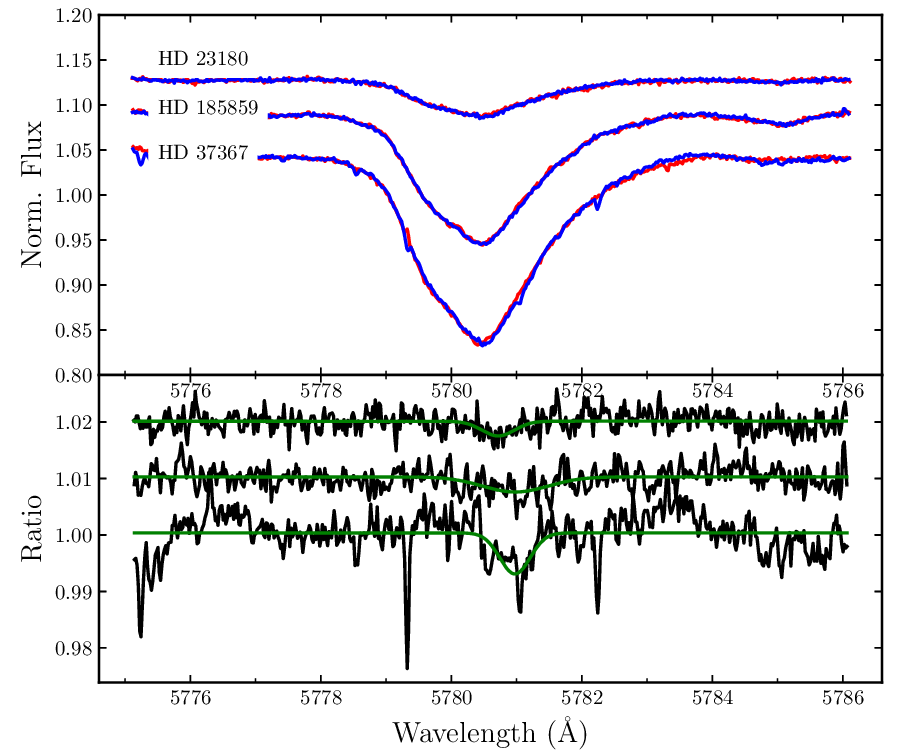}
	\caption{$\lambda$5780 profile toward HD~23180, HD~185859, and HD~37367 sightlines. The upper panel compares two epochs with red for the first observation and blue for the latest one. The lower panel shows the ratio spectra in the same order as the upper panel is sorting. The green line determines the Gaussian model to the ratio.}
	\label{plt-5780}
\end{figure}

As shown in Fig.~\ref{plt-method} and discussed in Sect.~\ref{Sec:DIBvar_method}, we find a small, but significant residual in the ratio spectrum for the $\lambda$5780 DIB toward HD~23180. Two more sightlines show a slight variation in the profile of the $\lambda$5780 DIB. We did several tests in the wavelength range covering the $\lambda$5780 DIB to rule out instrumental artifacts, wavelength calibration issues (considering wavelength shifts of up to 0.04~\AA), a telluric origin, and stellar contamination, but such effects cannot explain the observed discrepancies. Fig.~\ref{plt-5780} shows the spectra and the corresponding ratio spectra for these sightlines. For HD~23180, HD~185859, and HD~37367, the variations occur at roughly the same wavelength, just redward of the central peak (see Figs.~\ref{plt-dib-var54},~\ref{plt-dib-var55}, and~\ref{plt-dib-var59}). For all three targets, the changes represent a tiny increase in strength over time.

The very subtle variations we see in Fig.~\ref{plt-5780} are difficult to understand. They do not really appear to represent a change in the column density of the DIB carrier; they are too narrow for that. At the same time, there is no clear explanation for these residuals in terms of instrumental artifacts either. 

\subsection{Molecular and Atomic lines}
From a comparison to the literature results, it is clear that finding time variations is difficult even for narrow and strong atomic lines. Indeed, our data set includes four targets for which such variations have been reported before, and we first verified if we could reproduce some of these variations using our data set.\\

We have five targets in common with \citet{2015MNRAS.451.1396M}, and only one of those targets was reported to show variations: the \ion{Na}{i} D lines toward HD~113904 have a component at -60.5 km~s$^{-1}$ (LSR) that showed an increase of 10\% in intensity between 2002 and 2008 (corresponding to variations in the column density from $\log N=11.52 \pm 0.01$~cm$^{-2}$ to $\log N=11.56 \pm 0.01$~cm$^{-2}$). Our measurements of this component in archival UVES 2001 spectra yield $\log N=11.50 \pm 0.01$~cm$^{-2}$, in good agreement with their 2002 value; in the 2015 EDIBLES spectrum, we find $\log N=11.47\pm 0.01$~cm$^{-2}$ which is lower than their 2008 value. Perhaps this indicates a further change in this sightline.

\cite{2013PASP..125.1329G} studied spectra in the direction of HD~73882, and found significant variations in the EWs of the \ion{Ca}{i} line at 4227~\AA\ and the \ion{Fe}{i} line at 3860~\AA, and this over a period of only six years (2006--2012). For the \ion{Fe}{i} line component at 18-20 km s$^{-1}$, they reported changes in the EW from $1.68\pm0.15$~m\AA\ to $3.89\pm0.56$~m\AA, corresponding to changes in the column density from $\log~N=11.77\pm0.04$ to $\log~N=12.13\pm0.06$. We compared our 2017 EDIBLES spectrum of this source to the 2014 observations for all atomic lines listed in Table.~\ref{tab-atoms}. The region around the \ion{Ca}{i} line is rather noisy and does not result in good fits. We measured the stronger \ion{Fe}{1} line at 3719.9~\AA\ and found that the column density (and thus equivalent width) at best marginally decreases over these three years from $\log N=11.66\pm0.1$~cm$^{-2}$ to $\log N=11.54\pm0.04$~cm$^{-2}$ at heliocentric $\sim$18 km s$^{-1}$ (see Table~\ref{tab:ionresults}) -- i.e. a variation opposite to the one reported in the previous study. We note that the column density derived from the EDIBLES data matches their 2006 measurements better than their 2012 measurements. With our data set and approach, we can thus not reliably confirm these variations nor exclude them. For the other species we measured in this sightline, we find that the column densities are typically the same to within uncertainties as those listed by \citet{2013PASP..125.1329G}.

A third source for which time variation has been reported is HD~36486 ($\delta$ Ori A), for which \citet{2000MNRAS.312L..43P} studied ultra-high resolution observations. They found that an interstellar component at heliocentric $\varv=21.3$~km~s$^{-1}$ shows variation in the \ion{Na}{i} D1 lines. They suggest that the component consistently increased in strength from 1966 to 1994 and after that declined in intensity by 1999. At the same time, the corresponding lines of \ion{Ca}{ii} remained roughly constant.
The spectra used in the Price study had an ultra-high resolution, and the lower resolution of our EDIBLES observations prohibits resolving the $\varv=21.3$~km~s$^{-1}$ component. Nevertheless, we modeled the \ion{Na}{i} D1 line in the 2014 EDIBLES spectrum. Given our lower resolution, components six and seven from \citet{2000MNRAS.312L..43P} only show up as a single component at $22.15$~km~s$^{-1}$; we found that $\log N=11.2~\pm1.9$~cm$^{-2}$ for this component which compares well to the total column density of $\log N=11.13$~cm$^{-2}$ reported by \citet{2000MNRAS.312L..43P}. Within the (rather large) uncertainties, our results are thus consistent with no further change in these components. We also fitted the \ion{Ca}{ii} K lines in the 2014 EDIBLES spectrum at heliocentric velocities of 15.7, 21.3, and 23.5~km~s$^{-1}$, yielding column densities of respectively $\log N =$ 11.1$\pm$0.7, 11.0$\pm$1.2, and 10.9$\pm$0.2~cm$^{-2}$. These values compare well to the 1994 values of 10.6$\pm$0.1, 11.1$\pm$0.1, and 10.6$\pm$0.2 from \citet{2000MNRAS.312L..43P}, thus once more proof for no substantial change in this component.

Finally, \cite{2000MNRAS.319L...1C} reported clear evidence for variation in the profiles of the \ion{Na}{i} and \ion{K}{i} lines toward HD~81188 between 1996 and 2000, an increase in the column density of 16\% and 40\% respectively. We compared the EDIBLES spectrum (recorded in 2017) with the available archival data from FEROS (observed in 2019), but unfortunately, the archival spectrum is of low quality, and the comparison did not yield any useful results. However, the EDIBLES spectrum for this target has a good quality, albeit with significant telluric contamination in the wavelength range of the \ion{Na}{i} D line, and we can thus compare the EW in our spectra to the earlier results. We first used {\sc MOLECFIT} \citep{2015A&A...576A..77S, 2015A&A...576A..78K} to perform the telluric correction. This however leaves large residuals leading to significant uncertainties in our measurements. We determined the EW of the \ion{K}{i} and \ion{Na}{i} by integration. \citet{2000MNRAS.319L...1C} report an EW of 4.4$\pm$0.4~m\AA \, for the \ion{K}{i} line; our measurements yield EW=3.6 $\pm$ 0.1~m\AA, so perhaps a slight decrease. For the \ion{Na}{i} line, they report 58.9$\pm$2.4~m\AA\ whereas we find 71$\pm$8~m\AA, again perhaps a slight increase. Given the uncertainties on the measurements, however, this is at best a marginal change. 

This exercise emphasizes the importance of good-quality archival spectra that are unfortunately not yet available for all targets and thus highlights the value of the EDIBLES data set to be used as archival observations in future studies. 

\bigskip
The results of our detailed Voigt modeling of the atomic and molecular lines are listed in Table ~\ref{tab:ionresults}, and the corresponding best-fit models are shown together with the observations in Figs.~\ref{plt-atoms1} -- \ref{plt-atoms11}. For each interstellar cloud component, we searched for variations in the column density that are significant compared to the measurement uncertainty. We found only one indisputable case for a very significant change. Indeed, toward HD~167264, a new \ion{Ca}{i} component at $\varv = 5.6$~km~s$^{-1}$ shows up very clearly in the 2016 EDIBLES spectrum, whereas it was absent in the 2001 archival spectrum. A closer inspection reveals perhaps a small absorption dip in the 2001 spectrum. We searched for additional archival spectra, and they reveal intermediate absorption depths, thus strengthening the case that this is indeed a real-time variation (see Fig.~\ref{plt-hd167264-CaI}). In addition to this new component, we noticed that for the main component in this target (at $\varv=-7.5$~km~s$^{-1}$), the neutral species (and molecules) all show a marginal decrease in their column densities (10--30 percent change for the neutral atoms; five to ten percent change for the molecules, but note the large uncertainties on the column densities) while the ionized species \ion{Ca}{ii} and \ion{Ti}{ii} show a marginal increase ($\sim$12\% change; see Fig.~\ref{plt-atom-vars}). The \ion{Na}{i} D lines are very saturated, and thus we also inspected the \ion{Na}{i} UV doublet, revealing first of all very significant changes to the line depths for the main component, but on closer inspection, these lines also show the newly appearing component at $\varv = 5.6$~km~s$^{-1}$, at the same strength in both the archival and EDIBLES spectra. So whereas this component greatly increases in strength in the \ion{Ca}{i} line, it has not changed noticeably in the \ion{Na}{i} lines. The \ion{K}{i} or CN or C$_2$ lines are not covered in our archival observations. With the exception of the new \ion{Ca}{i} component and the main lines in the \ion{Na}{i} UV lines, the changes in the column densities that we detect are not very significant by themselves (given the uncertainties). However, the systematic nature of these changes may indicate that this too is a real effect representing a gradual change in the sightline conditions. A similar systematic set of changes we find for HD~147933 (second row in Fig.~\ref{plt-atom-vars}). We will discuss this further below.

% %%%%%%%%%%%%%%%%%%%%%%%%%%%%%%%%%%%%%%%
% %%%%%%%%%%%%%%%%%%%%%%%%%%%%%%%%%%%%%%%
% %%%%%%%%%%%%%%%%%%%%%%%%%%%%%%%%%%%%%%%
\section{Discussion} \label{sec-discussion}
To the best of our knowledge, the current study using a rigorous approach to search for time variations in interstellar lines and DIBs is the most comprehensive such effort done to date, and given the superb EDIBLES data quality, also potentially the most sensitive survey of its kind. The 31 DIBs we have selected offer the best prospects for detecting temporal changes given their strength and narrow widths, and we have also included the spectral lines of nine different atomic or small molecular species to trace possible changes in the physical conditions in the sightlines. 

The main finding from our study is that there is essentially almost no noticeable time variation in the interstellar features for the time frames studied here, i.e. typical periods of the order of nine years. Only two sightlines show small but systematic variations in their atomic lines, with one of them, HD~167264, showing unambiguously that a new interstellar line appears and increases in strength over time. Two DIBs possibly exhibit small variations across a few targets. While these detections show that we can detect even small variations, they also point to various systematic uncertainties as the key reason for the low number of DIBs, atomic or molecular lines, and sightlines for which we actually do find significant, conclusive physical variations. The main confounding systematics are too low a spectral resolution (especially in the archival spectra) to separate different components in interstellar atomic lines in some cases; small uncertainties in the wavelength calibration and, or continuum placement, and, or flat fielding for the DIBs; and in just a few cases severe contamination by telluric lines for both DIBs and atomic and molecular lines. 

In spite of concerns about systematic uncertainties, we must conclude that, in general, interstellar sightlines do not change noticeably on the time scales studied here -- which for some targets are as long as 22 years. For all our targets, proper motions have been measured, and thus we can determine the physical distance (transverse to our sightline) that our target stars have moved behind the intervening clouds over the time scales between the epochs we consider here. Those distances (expressed in au) are listed in Table ~\ref{tab-targets} and are typically several tens of au; the largest value is 294 au for HD~183143. If the interstellar clouds in the sightline are physically close to the target stars, the lack of detectable variations then implies that the physical conditions and chemical composition of most interstellar clouds do not change much over distance scales of several tens of aus; if the clouds are much closer to us, this, of course, corresponds to smaller distances. However, there are some interesting exceptions in our data set, pointing to tiny scale structures in diffuse interstellar material.

\subsection{Variations in HD~167264}
\begin{figure}[t!]
	\centering
	\includegraphics[width=0.98\hsize]{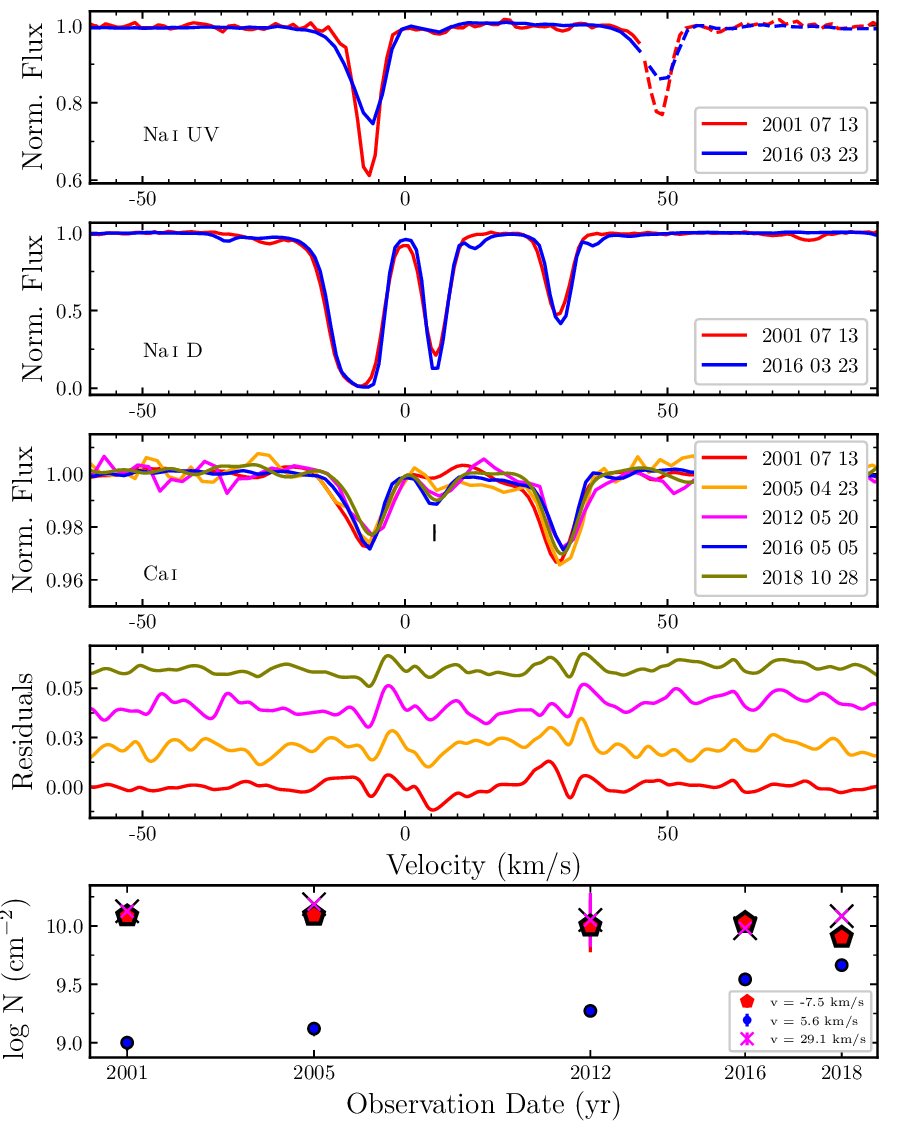}
	\caption{Archival and the EDIBLES spectra in the wavelength range of the \ion{Na}{i} lines (top two panels) and the \ion{Ca}{i} line (third panel) toward HD~167264. The fourth panel shows the residuals of the archival spectra after subtracting the 2016 EDIBLES spectrum in the range of the \ion{Ca}{i} lines. The bottom panel shows the corresponding column densities for the three different \ion{Ca}{i} components over time. Note the very clear increase in column density for the central component (at $\varv=5.6$~km~s$^{-1}$) corresponding to the feature indicated by the arrow in the \ion{Ca}{i} plot. The main component (at $\varv=-7.5~$km~s$^{-1}$) may show a small but systematic decrease in the column density over the same time period; no systematic changes are seen for the third component (at $\varv=29.1$~km~s$^{-1}$). We note that the apparent component at $\varv\approx 49$~km~s$^{-1}$ in the \ion{Na}{i} UV line is in fact the second line in the doublet and thus not a real component.}
	\label{plt-hd167264-CaI}
\end{figure}

\begin{figure*}[t!]
    \centering
    \includegraphics[width=0.85\hsize]{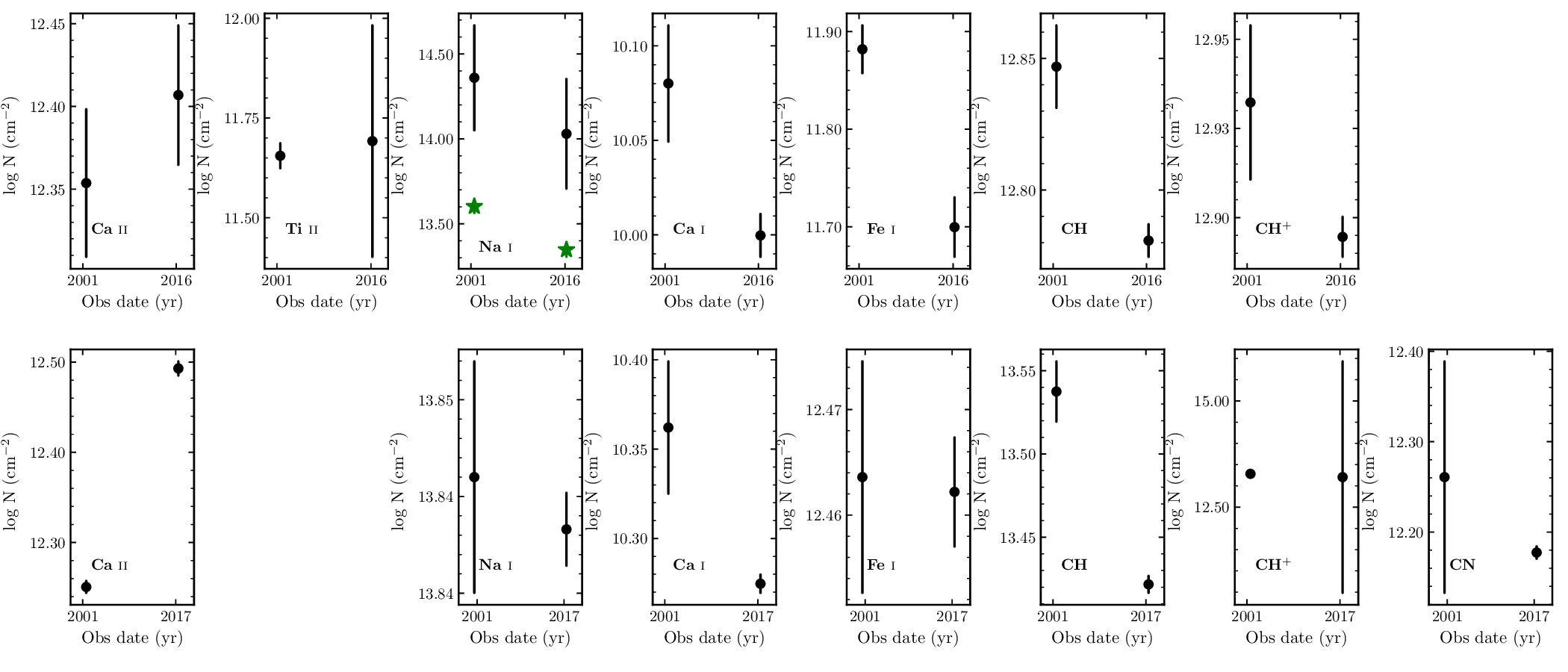} % width=1.01\columnwidth
    \caption{Column densities of various atoms and molecules as measured in the main component (at $\varv\sim-7.5$~km~s$^{-1}$) toward HD~167264 between the oldest available archival spectrum and the EDIBLES data. For Na, the black bar indicates the measurements and uncertainties using the \ion{Na}{i}~D lines; the green star represents those from the UV lines at 3300~\AA. The lower panel shows similar measurements for HD~147933. We note that we used the \ion{Na}{i} UV lines rather than the saturated \ion{Na}{i} D lines.}
    \label{plt-atom-vars}
\end{figure*}

We see very clear variations in the \ion{Ca}{i} line toward HD~167264. In particular, a new component shows up at $\varv=5.6$~km~s$^{-1}$ in our most recent spectra that was absent or at best very weakly present in archival spectra. Interestingly, \cite{2015MNRAS.451.1396M} investigated this sightline and reported no variability for this target for \ion{Na}{i}, \ion{Ca}{ii}, and \ion{Ca}{i}. We note that their study covered observations up to 2008. We checked all available spectra in the archive for different instruments for this sightline and found useful observations from 2001, 2005, 2008, 2012, and 2018 with FEROS, UVES, and ESPRESSO. These convincingly show that the \ion{Ca}{i} component is getting stronger \textit{after} 2008, with the feature very clearly visible from about 2012 (see Fig.~\ref{plt-hd167264-CaI}). Given the presence of this feature in the later observations, one may also consider that the absorption line is very weakly present also in the 2001 and 2005 observations, but their depth is at the noise level. Our Voigt model fits clearly show how the column density increased by at least a factor of five over the past 17 years (see lower panel in Fig.~\ref{plt-hd167264-CaI}). The very clear variations we see in the \ion{Ca}{i} line suggest that the sightline towards this object started crossing a region of enhanced \ion{Ca}{i} recently, with the most significant changes happening shortly after 2008. Interestingly, the \ion{Na}{i} D lines also exhibit a component at this velocity, that is already clearly present in the 2001 observations. Since these lines are saturated, we also included the \ion{Na}{i} lines in the UV (at 3300~\AA). There too, an interstellar component is clearly (though weakly) present in both archival and EDIBLES observations, and does not display any variation between the two epochs.  

It is interesting to note that also the main component (at $\varv=-7.5$~km~s$^{-1}$) of the \ion{Ca}{i} line exhibits much smaller, but systematic changes in the line strength as well: the column density appears to decrease by 0.08 dex over a period of 15 years (see Fig.~\ref{plt-atom-vars}).  The uncertainties on the column densities are derived from the Monte Carlo simulations, and appear perhaps small compared to the epoch-to-epoch variations of $\sim$0.1 dex that we see especially in the third component (at $\varv=29.1$~km~s$^{-1}$); however, the main component (at $\varv=-7.5$~km~s$^{-1}$) shows a systematic decrease over time in the column density that we do not see in this latter component. At the same time, it is interesting to look at the column densities of other species in this main $\varv=-7.5$~km~s$^{-1}$ component (top panel Fig.~\ref{plt-atom-vars}). All the neutral atoms and key molecular species show small, but systematic decreases in their column densities while the ionized species show small increases. The \ion{Na}{i} UV lines are perhaps the most extreme and show clearly that the column density decreases very significantly by 0.25 dex! This then suggests that the changes in the sightline probed by the main component reflect a change towards a more exposed part of the cloud. A similar change is also apparent for HD~147933 (discussed below). 

HD~167264 (15 Sgr) is located at a distance of 1140~pc \citep[with uncertainties on the parallax allowing a range of 958--1408~pc; ][]{2020yCat.1350....0G} and is moving with a proper motion of 2.2 mas~yr$^{-1}$ \citep{2020yCat.1350....0G}, which corresponds to a transverse velocity of 2.5 au~yr$^{-1}$ (or thus 11.8 km/s) at that distance. Given how clearly the feature shows up in our 2012 spectra, most of the change must have happened in only about 4 years, corresponding to a transverse distance of only 10~au. If the region probed by these variations is located close to the background star, we are thus probing clear variations at the scale of $\sim$10~au; if this region would be much closer to us, we are probing even smaller scales.

Clearly, the region probed by the variations in the \ion{Ca}{i} line represents the smallest scales, and it is interesting to compare this to some of the DIBs. In particular, \citet{1997A&A...318L..28E} compared the DIBs toward BD+63$^{\circ}$ 1964 with those toward HD~183143. While both targets represent comparable overall reddening, the DIB spectra are rather different. The sightline toward BD+63$^{\circ}$ 1964 was found to represent a more neutral environment, with a \ion{Ca}{i}/\ion{Ca}{ii} ratio that is higher toward BD+63$^{\circ}$ 1964 than toward HD~183143. At the same time, most of the narrow DIBs are much stronger toward BD+63$^{\circ}$ 1964 than toward HD~183143 and thus represent DIB carriers that reside more in the more neutral parts of interstellar clouds. The most prominent differences in EW between those two objects were measured for the $\lambda\lambda$5797, 5849, 6379, and 6614 DIBs. We had a closer look at these DIBs and compared the archival spectra to the EDIBLES spectra to see if there are any notable changes. While there are some small differences between the spectra for some DIBs, we believe those originate mostly from contamination, and thus we do not detect any changes in the DIBs whose carriers are likely to reside in the environment that produces the \ion{Ca}{i} absorption. This is somewhat surprising: the column density of the new \ion{Ca}{i} component is about 40\% that of the main component (see Table~\ref{tab:ionresults}), so if the DIB EWs would scale directly with the \ion{Ca}{i} column density, this should result in a significant change. Thus, a more neutral environment alone is not enough to activate these DIB carriers. 

\subsection{Variations in HD~147933}

The bottom panel in Fig.~\ref{plt-atom-vars} shows the column densities derived for the interstellar sightline toward HD~147933. While none of the variations are significant given the uncertainties on the measurements, we once more notice a systematic trend: all the neutral and molecular species show a slight decrease in column density, while the \ion{Ca}{ii} line shows an increase. Here too, we may be witnessing a subtle change in the sightline properties. 
 
It is interesting that variability on small spatial scales has been established already for the environment of this object. Indeed, HD~147933 is also known as $\rho$~Oph A and forms a double star system with HD~147934 ($\rho$~Oph B) whose separation on the sky corresponds to 344~au \citep{Cordiner13}. The interstellar cloud towards HD~147933 is characterized by kinetic temperatures of several tens of Kelvin in the molecular gas sampled by H$_2$ (46 K) and a molecular hydrogen fraction of 0.1 \citep{1977ApJ...216..291S}. \cite{2006FaDi..133..403C} used high signal-to-noise observation toward HD~147933 and HD~147934 and found minor differences in their CN/CH ratio, a small density contrast in the molecular gas toward the dark cloud and small variations in some of the DIBs. The derived densities toward components A and B are respectively 625 and 450 cm$^{-3}$ \citep{Cordiner13}, and for component A, we also know that the (total) hydrogen column density is $\log N=21.68$~cm$^{-2}$ \citep{2009ApJ...700.1299J}. This implies that for component A, we are probing a cloud with a length scale of $\sim$2.5~pc. Thus, the small scale variations probed between components A and B is only a tiny fraction of the overall length scale of the cloud, and the possible proper motion variations we probe here are even smaller (of the order of 65 au; see Table~\ref{tab-targets}).

\subsection{Variations in the $\lambda\lambda$4727 and 5780 DIBs}

The aim of this paper was to search for variations in the DIBs, and the only two DIBs for which we possibly see such variations are the $\lambda$4727 and $\lambda$5780 DIBs, and even then, we have to be cautious as we cannot rule out further systematics that could affect our results. Especially for the $\lambda$4727 DIB, instrumental issues appear to be the main reason for the observed variations. For the $\lambda$5780 DIB, the changes do not correspond to an increase in column density; if these changes are real, they would imply variations in the shape of the profile, which could then be a subtle expression of a change in physical conditions in the sightline. 
The changes appear to be subtle, and a good characterization of these changes may require a better quantitative description of the profiles of these DIBs.

\bigskip 
As we noted before, the lack of variations for most of our targets indicates that interstellar environments on average do not change much on distance scales of several tens of aus. The two cases noted above on the other hand correspond to sightlines where changes happen in only a few years, and thus for a transverse distance of only $\sim$10 au for HD~167264 and 65 au for HD~147933. The scales on which we see variations here correspond well with those traced by H{\sc i} as observed towards quasars, where clear variations are observed on scales as small as 10-25 au (\citealp{2005AJ....130..698B} and \citealp{2009AJ....137.4526L}; see also \citealp{2018ARA&A..56..489S}). The lack of variations in the DIB properties then suggests that the DIB carriers do not reside in these tiny scale structures. 

% %%%%%%%%%%%%%%%%%%%%%%%%%%%%%%%%%%%%%%%
% %%%%%%%%%%%%%%%%%%%%%%%%%%%%%%%%%%%%%%%
% %%%%%%%%%%%%%%%%%%%%%%%%%%%%%%%%%%%%%%%
\section{Summary and conclusions}
\label{sec-conclusions}

In this study, we compared the high-quality EDIBLES spectra for 64 sightlines to available archival observations to search for time variation in the profiles of 31 narrow and strong DIBs, and additionally also in nine atomic and molecular lines. We considered false positives due to various systematic uncertainties and used a mathematical formalism with a robust Bayesian approach to establish potential significant physical variations.

For the 31 DIBs we considered, we primarily searched for changes in the equivalent width with corresponding changes in the DIB profiles. We found that only two DIBs, those at $\lambda$4727 and $\lambda$5780, show possible significant variations in only a few sightlines, but even in these cases, some caution is needed since the changes are small and there may still be systematic effects that we underestimate or that we have not accounted for. For the $\lambda$4727 DIB, we see profile changes in three different sightlines. The most significant variation (a change of 0.5--2\% of the absorption depth between the two epochs) occurs for HD~168076, while HD~24398 and HD~36861 show less pronounced variations. Furthermore, we have several observations of HD 168076, and at least three of those show considerable deviations from the EDIBLES reference spectrum. We also confirm in our data set that this DIB has a double-peaked profile that appears to be intrinsic. For the $\lambda$5780 DIB, we found a small but significant residual in the ratio spectrum toward HD~23180, HD~185859, and HD~37367. The profile variation for these targets occurs at roughly the same wavelength and shows increasing strengths over time.

We fitted the atomic and molecular lines with a set of Voigt profiles, which allows for studying variations in the different line parameters more quantitatively. We found one incontrovertible case for a very significant change: toward HD~167264, a new \ion{Ca}{i} component shows up very prominently at 5.6 km s$^{-1}$ in the 2016 EDIBLES spectrum, whereas this component was absent in the 2001 archival spectrum and previous research indicated no variability until 2008. Additional archival spectra reveal that this component does indeed increase in strength over time. In addition, we noticed that for the main cloud component toward this target, the neutral species all show a marginal decrease in their column densities, in contrast, the ionized species show a marginal increase. These sightline changes are most likely induced by the proper motion of the background target and imply variations at scales of 10~au or smaller. Finally, we see a similar set of marginal, but systematic variations of the atomic lines toward HD~147933, a target for which small-scale structure variations (at the $\sim$344 au scale) have already been established. 

The fact that we can detect some variations in both DIBs and atomic and molecular lines shows that high-quality data like those in the EDIBLES data set have the potential to reveal these subtle changes. The archival baseline we present with this elaborate dataset allows us to trace subtle changes in the environmental cloud parameters and their implications on the life cycle of atomic and molecular interstellar species. Future work that can use the EDIBLES data as the archival reference will undoubtedly find more such variations crucial to understand the evolution of interstellar species.

\begin{acknowledgements}
This research has made use of the services of the ESO Science Archive Facility. Based on observations collected at the European Southern Observatory under ESO programme 193.C-0833 as well as ESO programmes 60.A-9036(A), 60.A-9700(G), 65.I-0498(A), 65.I-0526(A), 65.L-0165(A), 67.C-0281(A), 68.C-0149(A), 69.A-0529(A), 69.A-0613(A), 70.D-0191(A), 70.D-0607(A), 71.C-0513(C), 71.D-0168(A), 072.D-0196(A), 072.D-0410(A), 072.D-0524(A), 073.D-0234(A), 073.D-0291(A), 073.D-0609(A), 074.D-0021(A), 074.D-0300(A), 075.B-0190(A), 075.D-0061(A), 075.D-0103(A), 075.D-0369(A), 076.A-0860(A), 076.B-0055(A), 076.C-0431(A), 076.C-0503(A), 076.D-0294(A), 077.B-0348(A), 077.C-0547(A), 077.D-0146(A), 078.C-0403(A), 078.D-0080(A), 079.A-9008(A), 079.B-0856(A), 079.C-0597(A), 079.D-0564(A), 079.D-0564(C), 079.D-0567(A), 080.D-2006(A), 081.A-9006(A), 081.C-0475(A), 081.D-2002(A), 081.D-2008(A), 082.C-0271(A), 082.C-0390(A), 082.C-0427(C), 082.C-0446(B), 082.C-0566(A), 083.C-0503(A), 083.D-0040(A), 083.D-0475(A), 083.D-0589(A), 084.D-0067(A), 085.C-0799(A), 086.D-0236(A), 087.D-0264(F), 088.A-9003(A), 088.D-0424(D), 089.D-0189(A), 089.D-0730(A), 089.D-0975(A), 090.D-0153(A), 090.D-0358(A), 090.D-0600(A), 091.C-0713(A), 091.C-0851(A), 091.D-0061(A), 091.D-0221(A), 091.D-0622(A), 092.A-9018(A), 092.C-0173(A), 266.D-5655(A), 267.B-5698(A). We have also used observations obtained at the Canada-France-Hawaii Telescope (CFHT) which is operated by the National Research Council of Canada, the Institut National des Sciences de l´Univers of the Centre National de la R'echerche Scientifique of France, and the University of Hawaii. This research used the facilities of the Canadian Astronomy Data Centre operated by the National Research Council of Canada with the support of the Canadian Space Agency.
We have used observations made with the Mercator Telescope, operated on the island of La Palma by the Flemish Community, at the Spanish Observatorio del Roque de los Muchachos of the Instituto de Astrofísica de Canarias. In particular, we used observations obtained with the HERMES spectrograph, which is supported by the Research Foundation - Flanders (FWO), Belgium, the Research Council of KU Leuven, Belgium, the Fonds National de la Recherche Scientifique (F.R.S.-FNRS), Belgium, the Royal Observatory of Belgium, the Observatoire de Genève, Switzerland and the Thüringer Landessternwarte Tautenburg, Germany.\\
Peter J. Sarre thanks the Leverherhulme Trust for an Emeritus Fellowship.
A.~Farhang would like to thank Iranian National Observatory (INO) for providing computing facilities in support of the data reduction of the EDIBLES project. J.~Cami acknowledges support from an NSERC Discovery Grant and a SERB Accelerator Award from Western University.
\end{acknowledgements}

\bibliographystyle{aa}
\bibliography{dib.bib}

% %%%%%%%%%%%%%%%%%%%%%%%%%%%%%%%%%%%%%%%
% APPENDIX A: TARGETS
% %%%%%%%%%%%%%%%%%%%%%%%%%%%%%%%%%%%%%%%
\begin{appendix}

\onecolumn
\section{Targets}
Table~\ref{tab-targets} lists the properties of the sightlines studied in this paper. 
\input{table_target_list.tex}
\onecolumn

% %%%%%%%%%%%%%%%%%%%%%%%%%%%%%%%%%%%%%%%
% APPENDIX B: DIB MEASUREMENTS & PLOTS
% %%%%%%%%%%%%%%%%%%%%%%%%%%%%%%%%%%%%%%%
\section{DIB comparisons (online only)}
\label{Sec:appendix_DIBs}
Table~\ref{tab:DIBresults} lists the results of our comparison between the archival and EDIBLES observations for the 31 DIBs selected in this study. Figs.~\ref{plt-dib-var1}--\ref{plt-dib-var122} show the corresponding Figures. 
\input{table_dib_fit_results.tex}

\input{appendixfigures}

% %%%%%%%%%%%%%%%%%%%%%%%%%%%%%%%%%%%%%%%
% APPENDIX C: ATOMS & MOLECULES MEASUREMENTS AND PLOTS
% %%%%%%%%%%%%%%%%%%%%%%%%%%%%%%%%%%%%%%%
\onecolumn
\section{Atomic \& molecular line measurements (online only)}
\label{Sec:appendix_atoms}
\begin{figure*}[ht!]
	\centering
	\includegraphics[width=0.99\hsize]{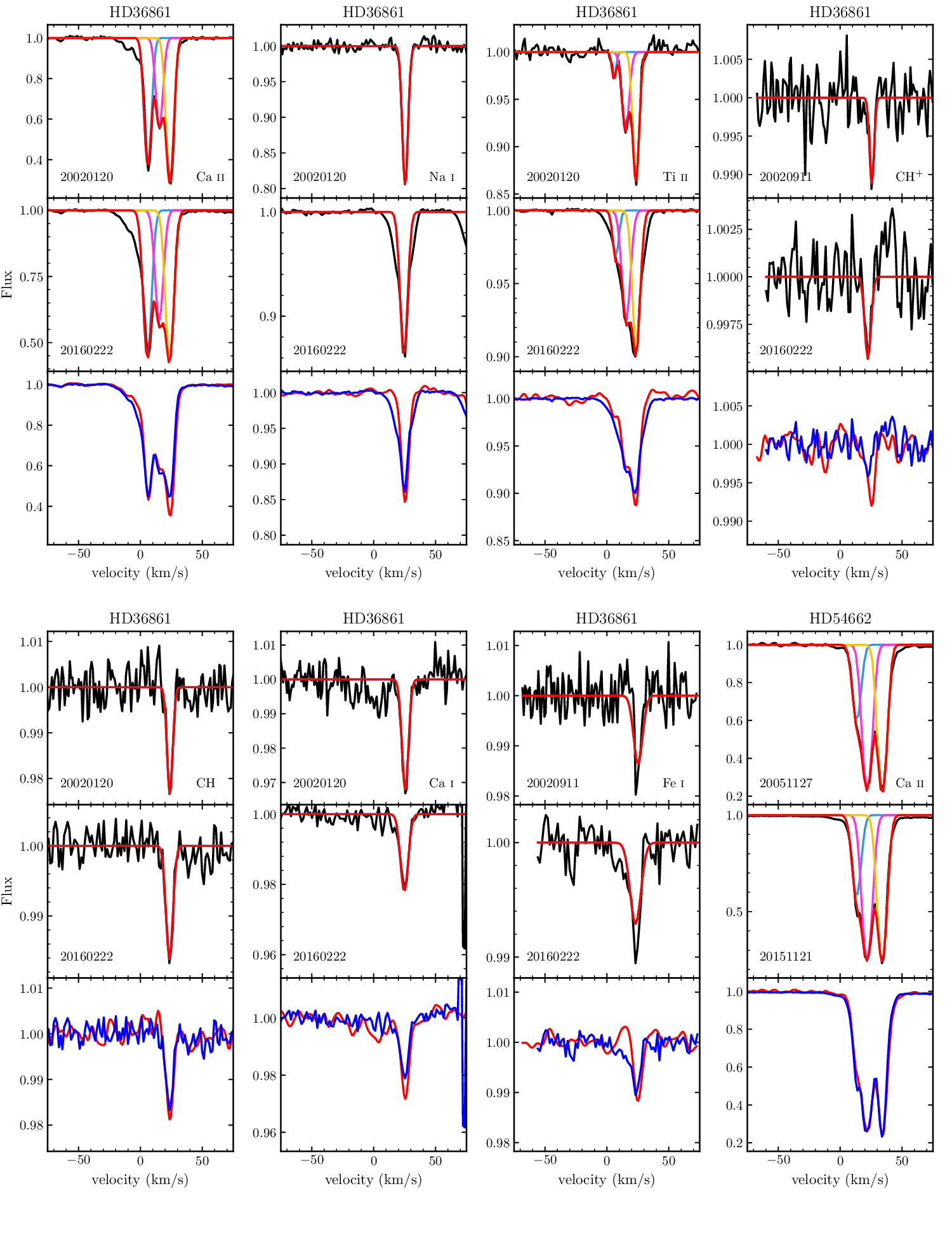}
	\caption{Atomic line profiles observed with UVES in two different epochs. The upper panels show the archival UVES spectra in black and our model in red, at the actual resolution. The middle panels are the same for the EDIBLES spectra. The lower panel is the comparison of two epochs. We note that in the lower panel, we have degraded the highest-resolution spectrum to match that of the lower-resolution spectrum. The blue spectra show the EDIBLES observations, and the red the archival data.}
	\label{plt-atoms1}
\end{figure*}

\begin{figure*}[ht!]
	\centering
	\includegraphics[width=0.99\hsize]{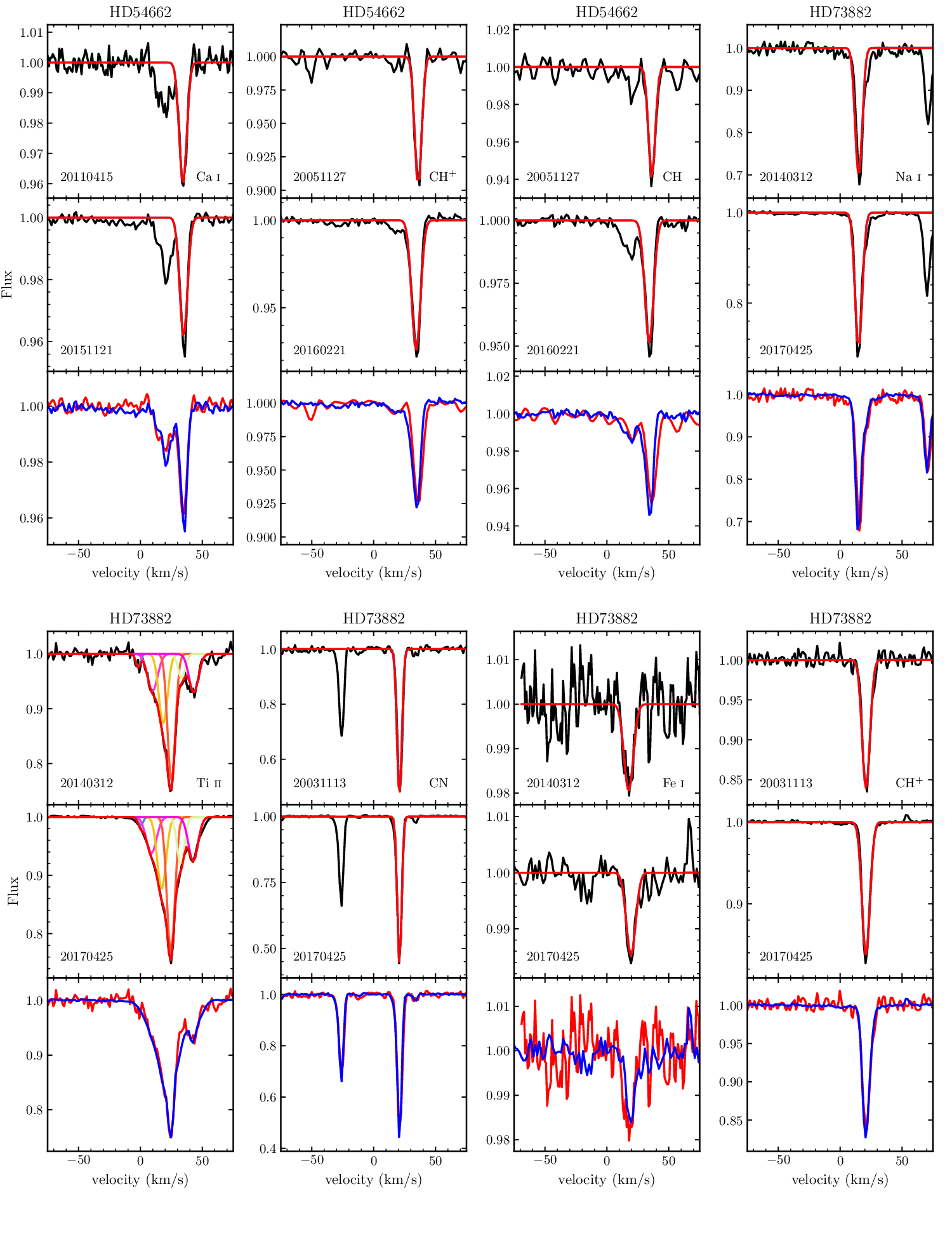}
	\caption{The same as Fig.~\ref{plt-atoms1}}
	\label{plt-atoms2}
\end{figure*}

\begin{figure*}[ht!]
	\centering
	\includegraphics[width=0.99\hsize]{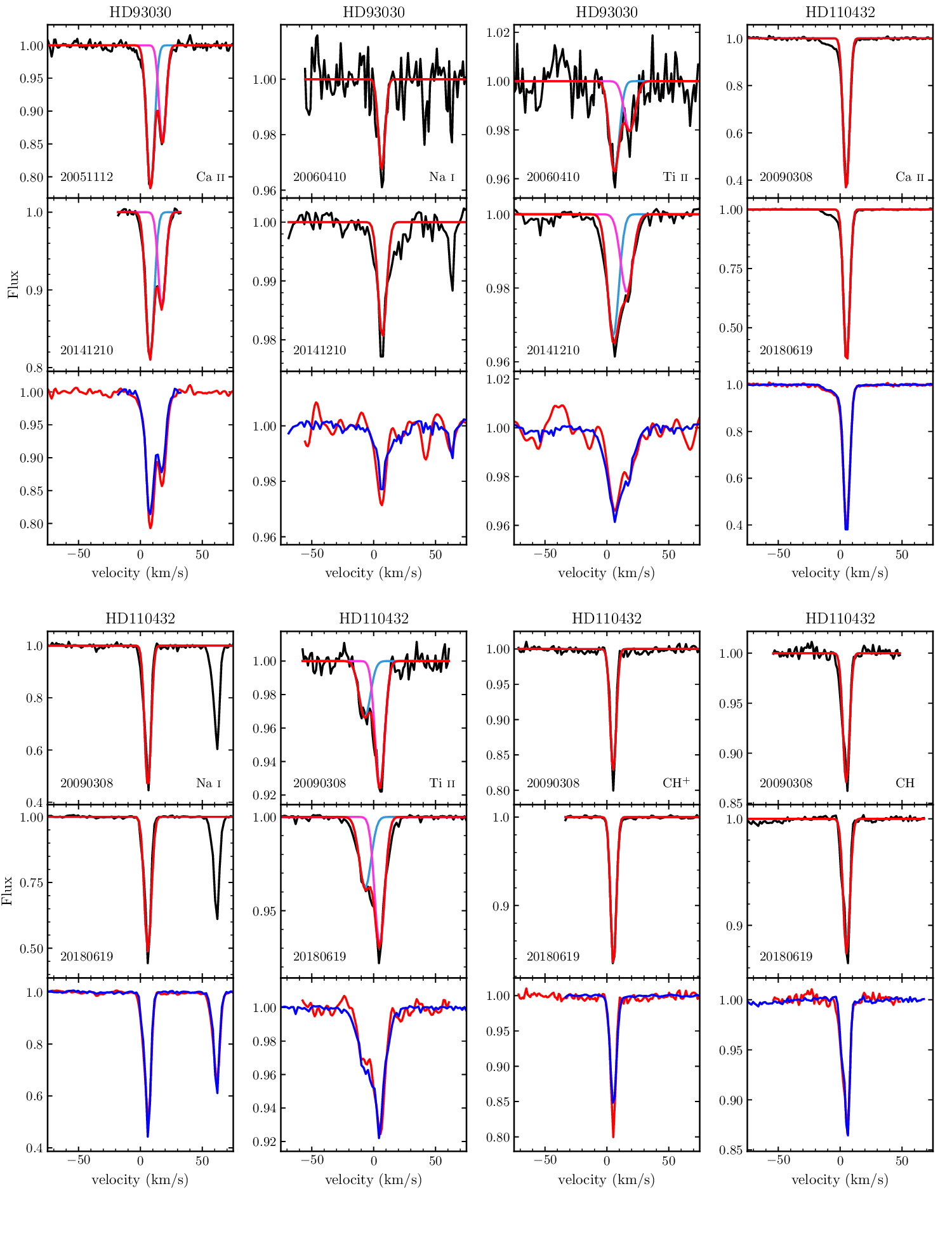}
	\caption{The same as Fig.~\ref{plt-atoms1}}
	\label{plt-atoms3}
\end{figure*}

\begin{figure*}[ht!]
	\centering
	\includegraphics[width=0.99\hsize]{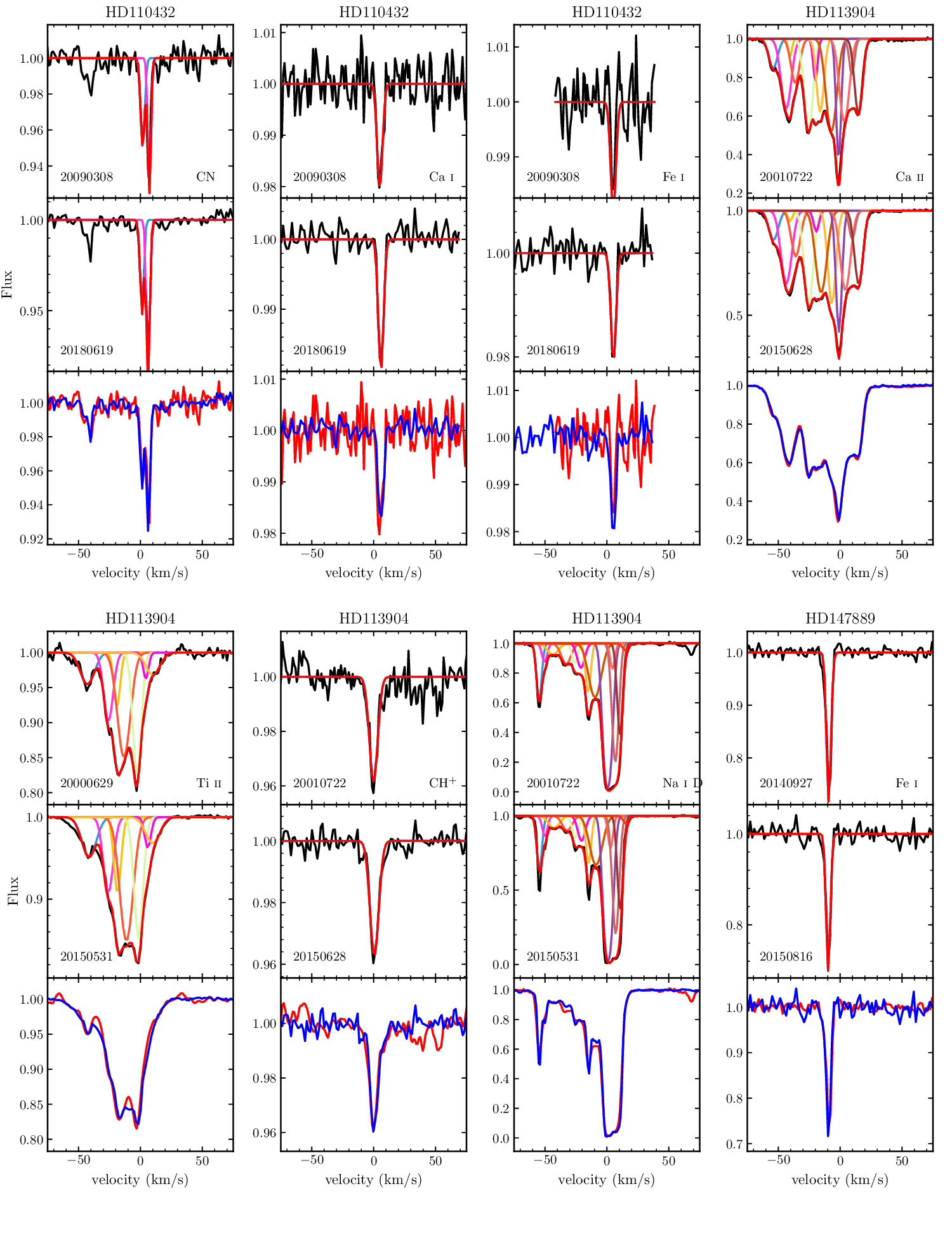}
	\caption{The same as Fig.~\ref{plt-atoms1}}
	\label{plt-atoms4}
\end{figure*}

\begin{figure*}[ht!]
	\centering
	\includegraphics[width=0.99\hsize]{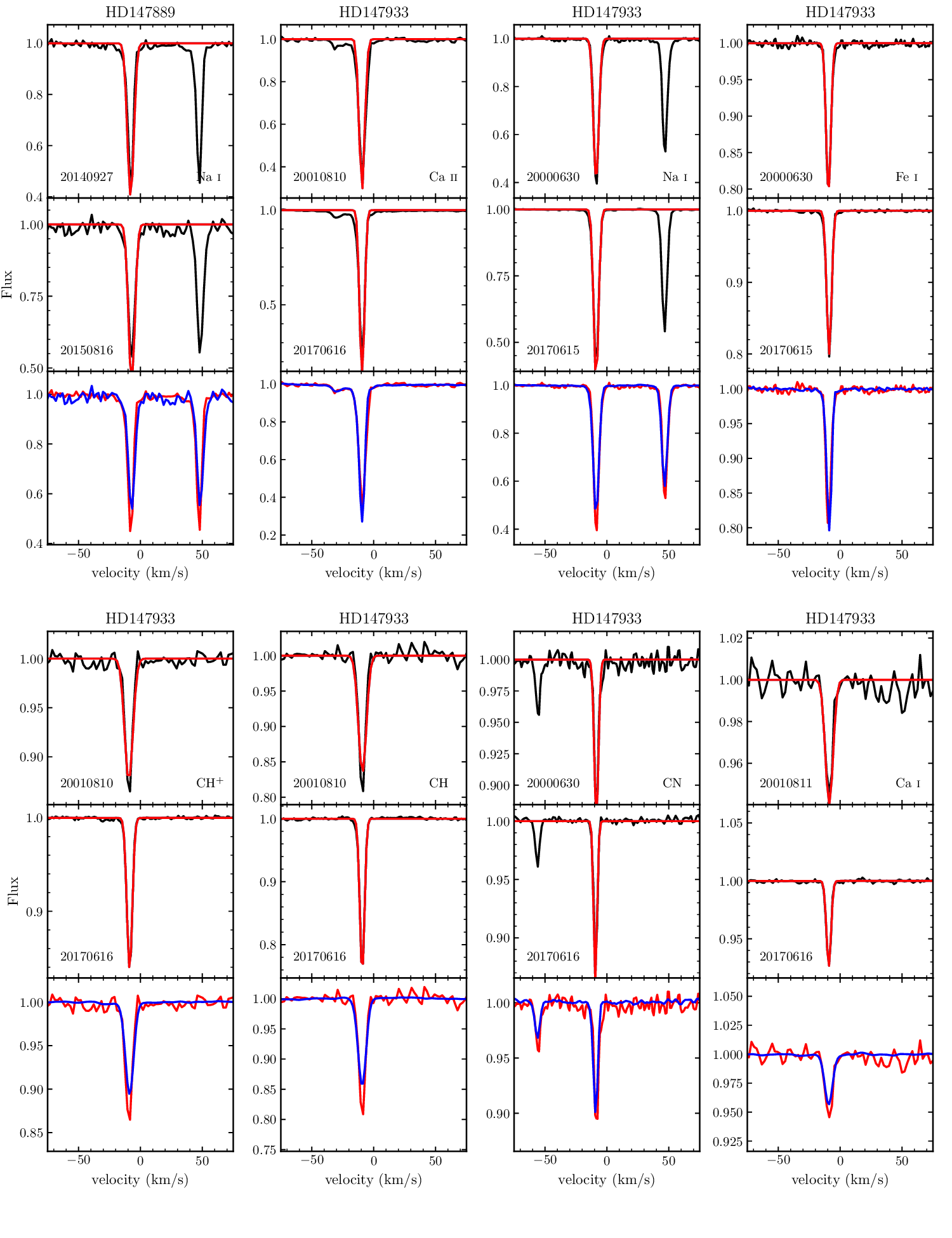}
	\caption{The same as Fig.~\ref{plt-atoms1}}
	\label{plt-atoms5}
\end{figure*}

\begin{figure*}[ht!]
	\centering
	\includegraphics[width=0.99\hsize]{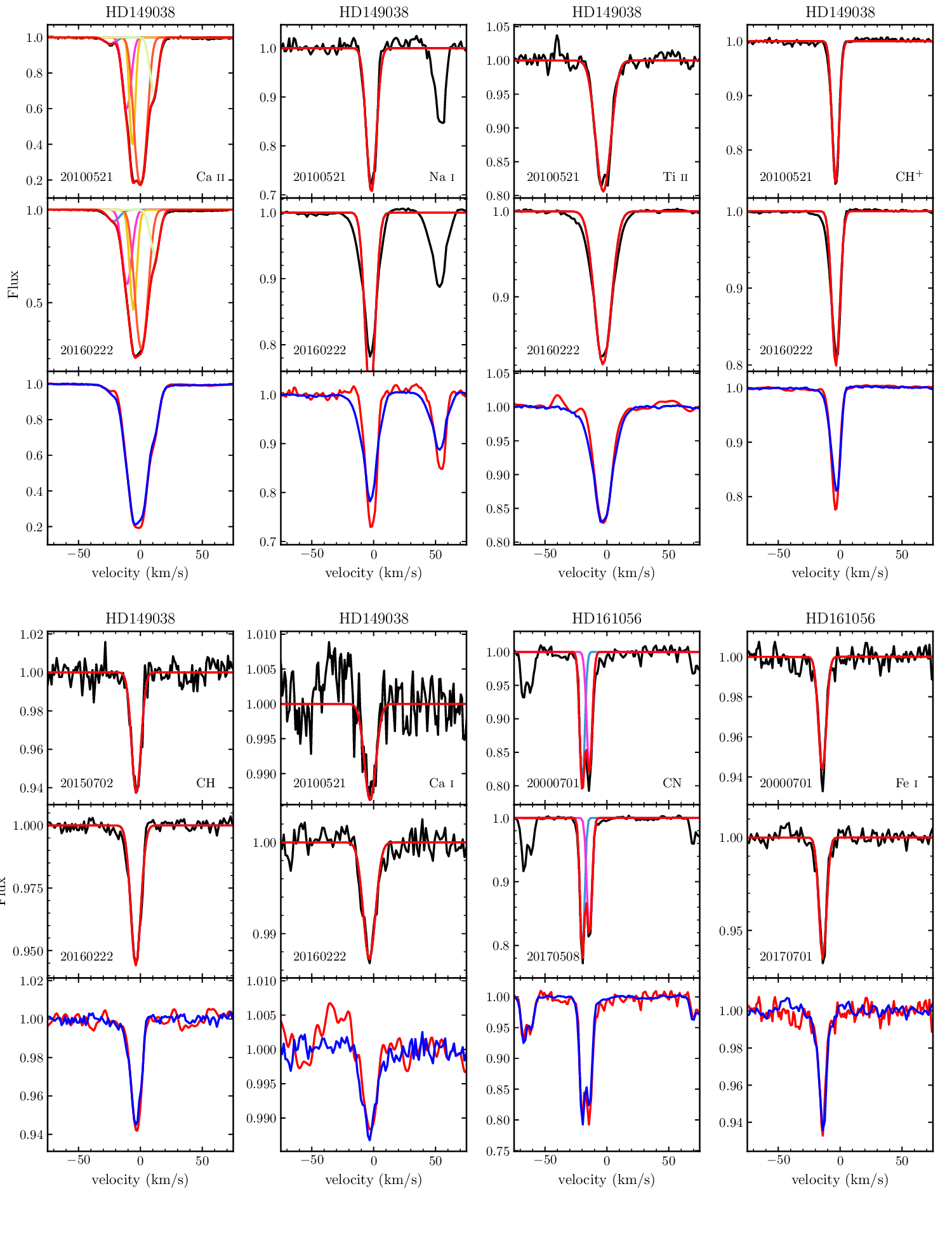}
	\caption{The same as Fig.~\ref{plt-atoms1}}
	\label{plt-atoms6}
\end{figure*}

\begin{figure*}[ht!]
	\centering
	\includegraphics[width=0.99\hsize]{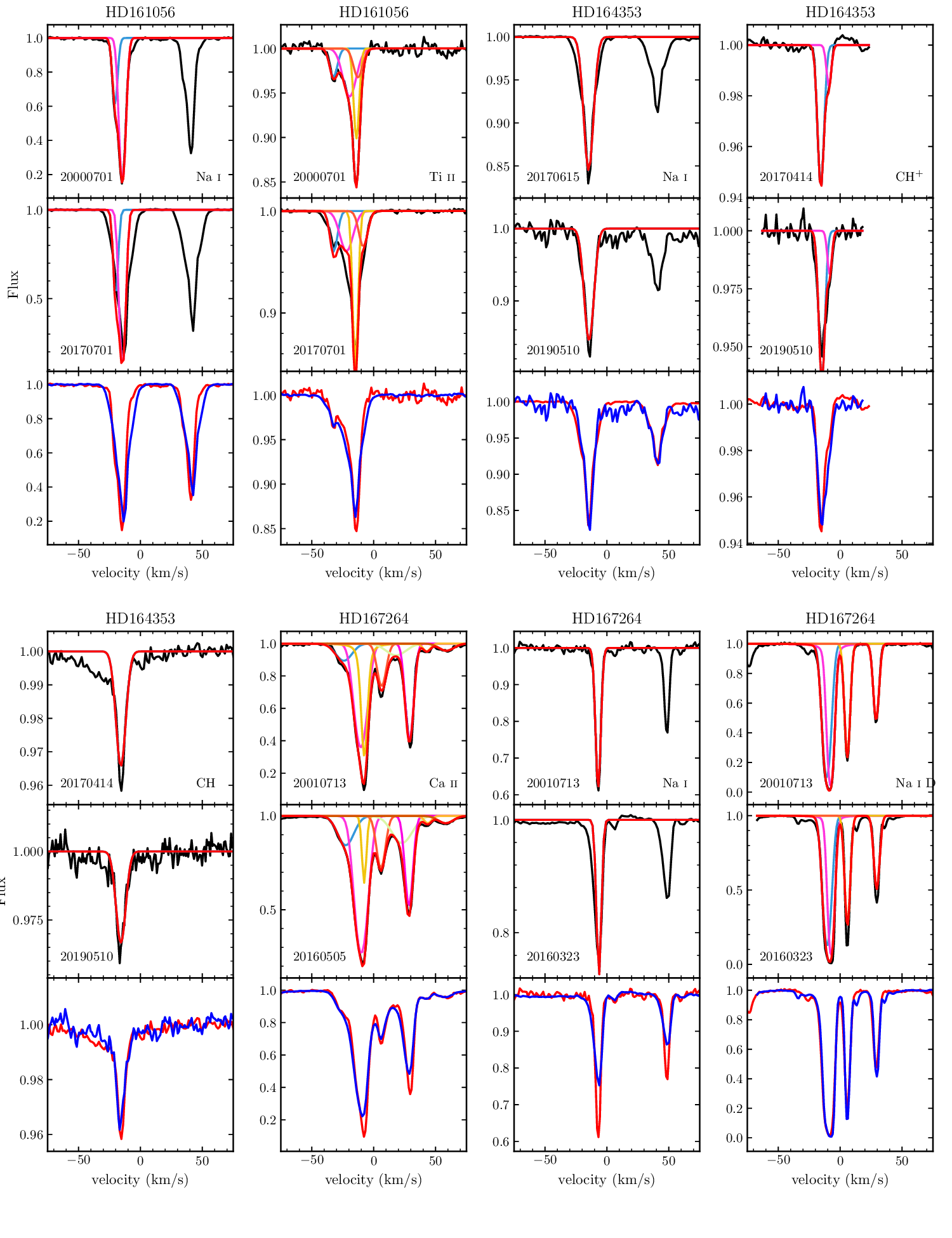}
	\caption{The same as Fig.~\ref{plt-atoms1}}
	\label{plt-atoms7}
\end{figure*}

\begin{figure*}[ht!]
	\centering
	\includegraphics[width=0.99\hsize]{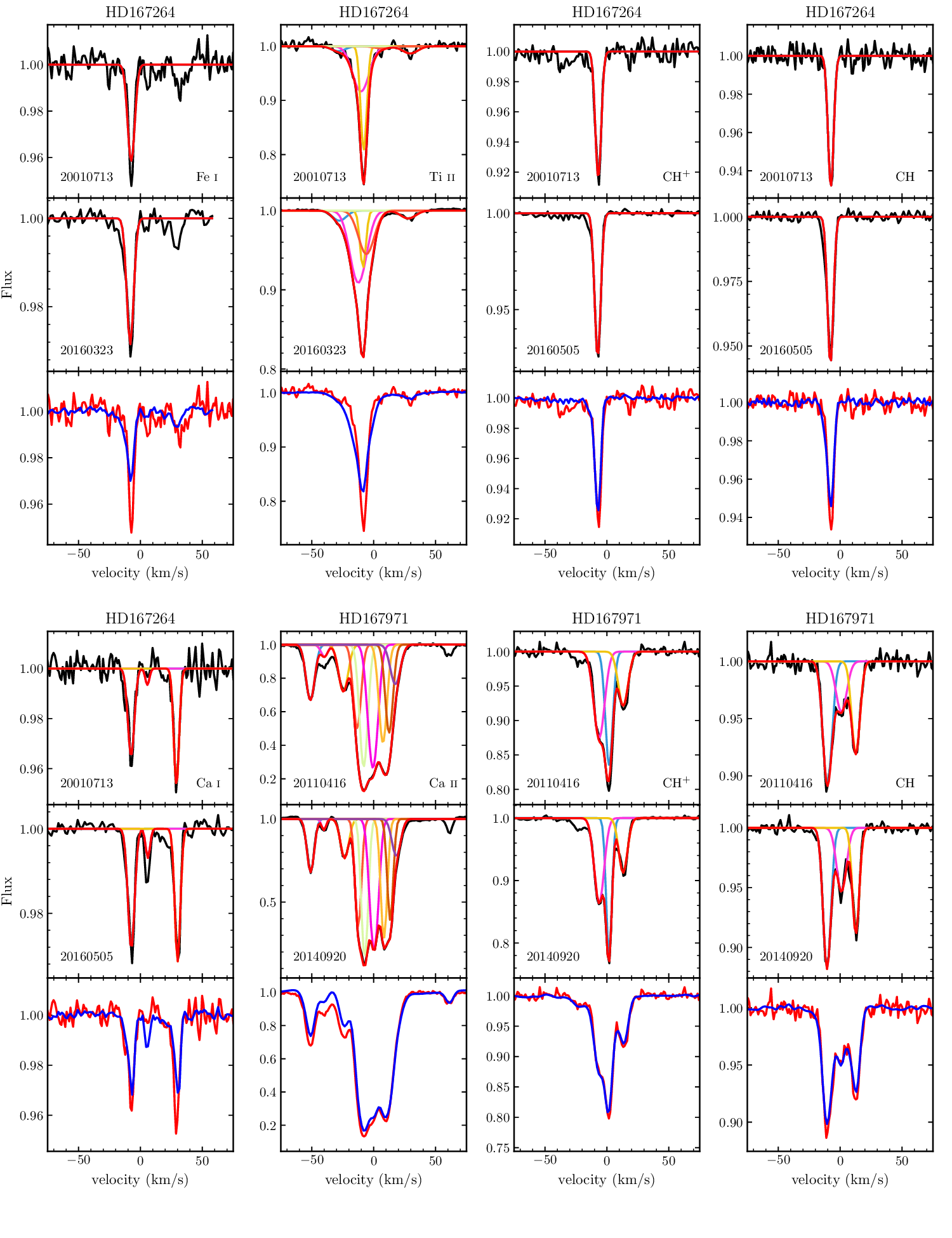}
	\caption{The same as Fig.~\ref{plt-atoms1}}
	\label{plt-atoms8}
\end{figure*}

\begin{figure*}[ht!]
	\centering
	\includegraphics[width=0.99\hsize]{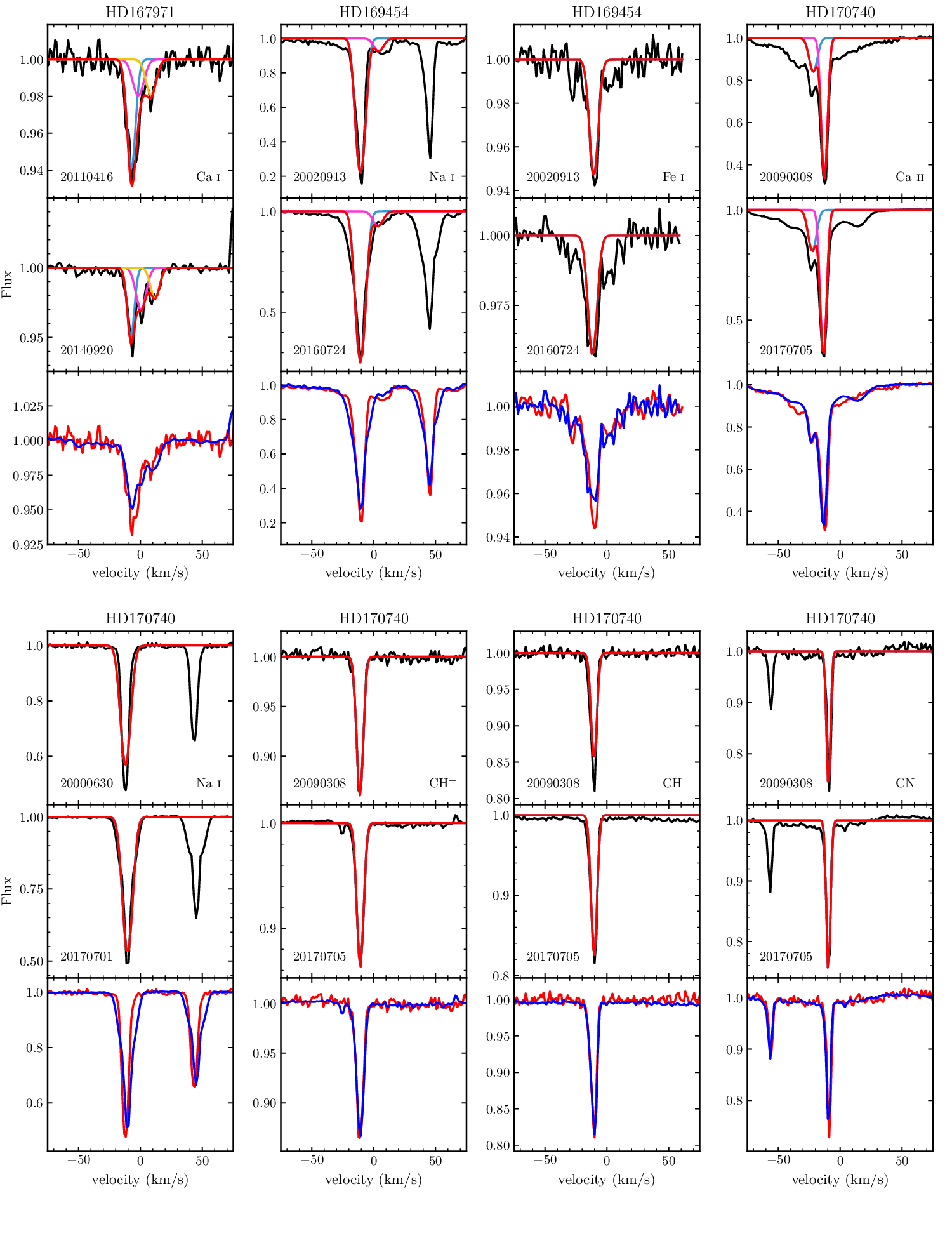}
	\caption{The same as Fig.~\ref{plt-atoms1}}
	\label{plt-atoms9}
\end{figure*}

\begin{figure*}[ht!]
	\centering
	\includegraphics[width=0.99\hsize]{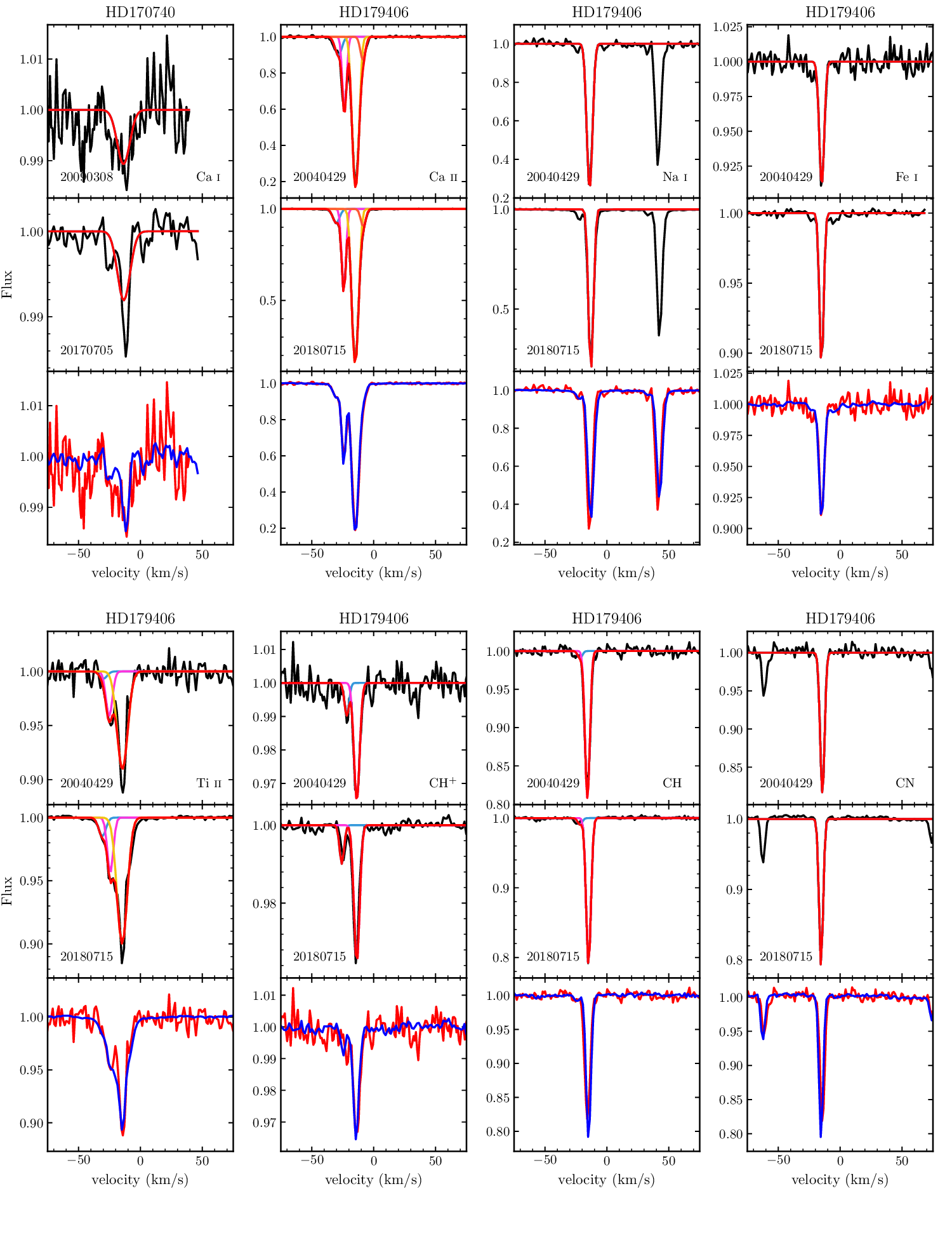}
	\caption{The same as Fig.~\ref{plt-atoms1}}
	\label{plt-atoms10}
\end{figure*}

\begin{figure*}[ht!]
	\centering
	\includegraphics[width=0.99\hsize]{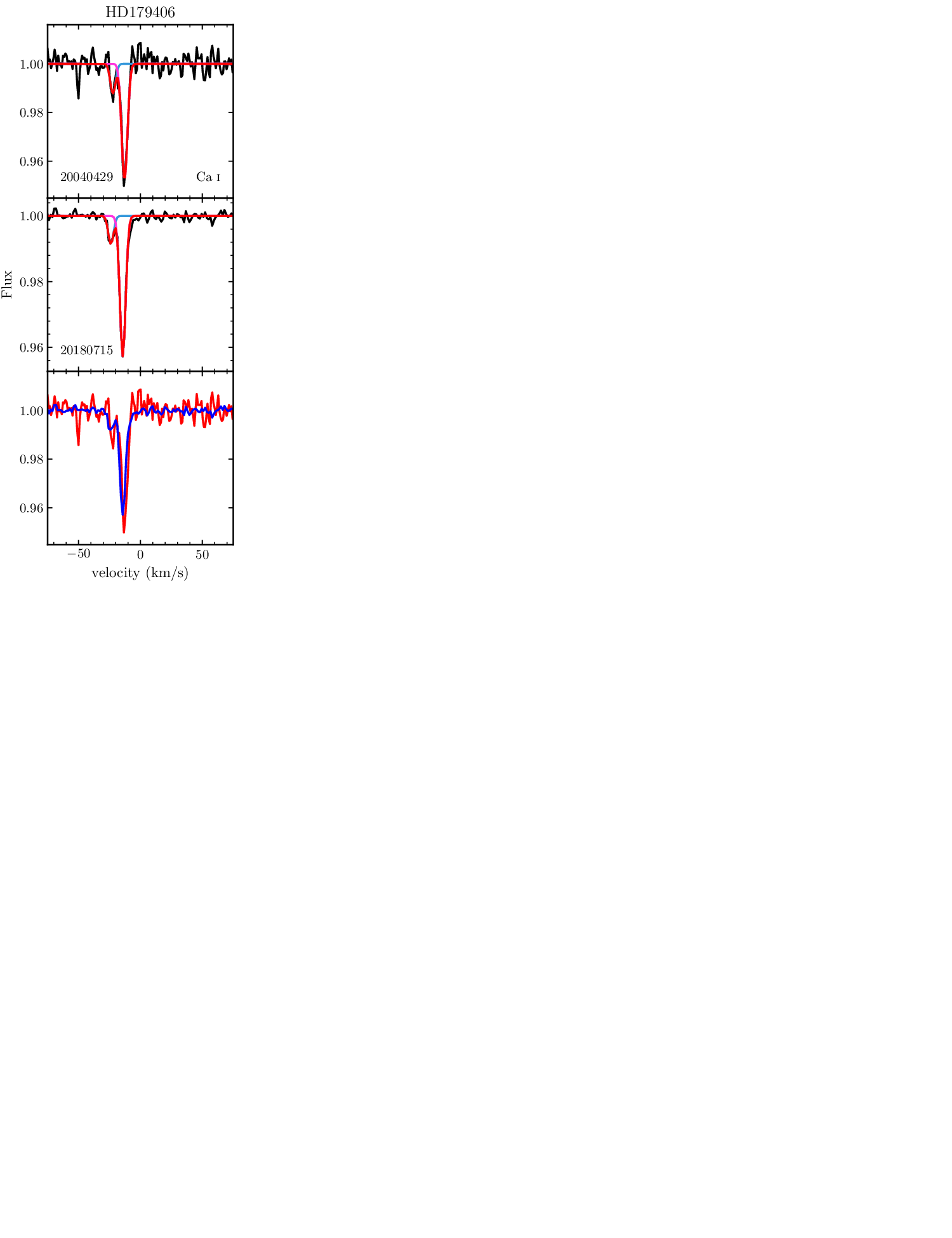}
	\caption{The same as Fig.~\ref{plt-atoms1}}
	\label{plt-atoms11}
\end{figure*}

Table~\ref{tab:ionresults} lists the best-fit parameters for the atomic and molecular lines that we modeled in the 16 targets for which we have UVES archival data in addition to the EDIBLES observations. Figs.\ref{plt-atoms1}--\ref{plt-atoms11} show the corresponding Figures. We note that all velocities are reported in the heliocentric rest frame.

\input{table_atomic_fit_results}

\end{appendix}
\end{document}

%% file: table_target_list.tex
\begin{landscape}

\begin{longtable}{lcccccccclcc}

\caption{The sample. Columns show HD number, E(B-V), interstellar cloud component radial velocity, rotational velocity, spectral type, RA \& Dec., Distance (pc), Proper motion (mas\,yr$^{-1}$), Observation Dates and Distance traveled (au).}\\

\toprule
\hline

Target  &  E(B-V)  & v$_\mathrm{ISM}$ & v\,sin$i$  & Spectral type  &  \multicolumn{2}{c}{Coordinate}  &  Distance  &  Proper motion  &  Obs date  & Instrument & Traversed \\
\cline{6-7}
~ &  (mag)  &  (km~s$^{-1}$) & (km~s$^{-1}$) & ~  &  ra (deg)  &  dec (deg)  & (pc)  &  (mas\,yr$^{-1}$)  &   ~  & ~  & distance (au) \\

    \hline
    \endfirsthead
    \caption{Continued.}\\
    \hline\hline

    Target  &  E(B-V)  &  v$_\mathrm{ISM}$ & v\,sin$i$  & Spectral type  &  \multicolumn{2}{c}{Coordinate}  &  Distance  &  Proper motion  &  Obs date  & Instrument & Traversed \\
    \cline{6-7}
    ~ &  (mag)  &  (km~s$^{-1}$) & (km~s$^{-1})$ & ~  &  ra (deg)  &  dec (deg)  & (pc)  &  (mas\,yr$^{-1}$)  &   ~  & ~  & distance (au) \\

    \hline
    \endhead
    \hline
    \endfoot

HD~22951 & 0.22 & 12.56 & 10$^{(1)}$ & B0.5V & 55.59 & 33.96 & 324 & 6.5 & 1997-08-12 & HIRES & ~ \\
~ & ~ & ~ & ~ & ~ & ~ & ~ & ~ & ~ & 1997-08-13 & HIRES &  0.0 \\
~ & ~ & ~ & ~ & ~ & ~ & ~ & ~ & ~ & 2016-08-29 & EDIBLES &  40.06 \\

HD~23016 & 0.05 & 18.0 & 291$^{(2)}$ & B8Ve & 55.58 & 19.7 & 159 & 13.76 & 2014-09-20 & EDIBLES & ~ \\
~ & ~ & ~ & ~ & ~ & ~ & ~ & ~ & ~ & 2014-09-23 & EDIBLES &  0.02 \\

HD~23180 & 0.26 & 12.85 & 100$^{(1)}$ & B1III & 56.08 & 32.29 & 344 & 13.26 & 2007-03-30 & HARPS & ~ \\
~ & ~ & ~ & ~ & ~ & ~ & ~ & ~ & ~ & 2009-03-01 & HARPS &  8.74 \\
~ & ~ & ~ & ~ & ~ & ~ & ~ & ~ & ~ & 2009-03-02 & HARPS &  8.74 \\
\end{longtable}
\tablebib{
(1)~\citet{2002ApJ...573..359A}; 
(2)~\citet{2007A&A...463..671R}; 
(3)~\citet{2020A&A...639A..81B}; 
(4)~\citet{2017A&A...597A..22S}; 
(5)~\citet{2010ApJ...722..605H}; 
(6)~\citet{2012AJ....144..130B}; 
(7)~\citet{2019AJ....157..196K}; 
(8)~\citet{1970CoKwa.189....0U}; 
(9)~\citet{1970CoAsi.239....1B}; 
(10)~\citet{2012A&A...542A.116A}; 
(11)~\citet{2018A&A...613A..65H}; 
(12)~\citet{2014A&A...562A.135S}
}\\
\textbf{($\dagger$)} This target shows no DIBs and was therefore only used to search for variations in the atomic and molecular lines.

\end{landscape}

%% file: table_dib_fit_results.tex
%\begin{landscape}
\onecolumn
\begin{longtable}{p{.5cm}rccccrrrrrc}
\caption{Results of our DIB measurements. For each sightline, we list only those DIBs from our sample that we can reliably detect. Note that the EW values are obtained from the Gaussian fits and not from direct integration. $\chi^{2}_{\rm base}$ is the reduced chi-square value for a model in which the ratio spectrum is equal to unity; $\chi^{2}_{\rm Gauss}$ the value for the best-fitting Gaussian model. The BIC values for baseline and Gaussian models are listed as well; see Sect.~\ref{Sec:DIBvar_method} for details. For sightlines where the BIC decreases by more than 3 units, we also list $|\Delta~{\rm BIC}|$. For those cases, the last column is our final assessment of what is the root cause for this change in BIC: continuum positioning errors are indicated by \texttt{c}; instrumental artifacts by \texttt{a}; blends with stellar lines by $\star$; telluric artifacts by $\oplus$ and a potential changes in the DIB properties by \texttt{DIB}. The $+$ and $-$ subscripts in those cases indicate whether the DIB has increased or decreased in strength over time in those cases. }
\label{tab:DIBresults}\\
\toprule

Target & \multicolumn{1}{c}{DIB} & Epoch 1 & Epoch 2 & \multicolumn{1}{c}{EW$_{1}$} & \multicolumn{1}{c}{EW$_{2}$} & \multicolumn{1}{c}{$\chi^2_{base}$} & \multicolumn{1}{c}{$\chi^2_{Gauss}$} & BIC$_{base}$ & BIC$_{Gauss}$ & $|\Delta {\rm BIC}|$ & Variation \\
~ & [\AA] & ~ & ~ & [m\AA] & [m\AA] & ~ & ~ & ~ & ~ & ~ & \\

\midrule
\endfirsthead
\caption{Continued.}\\
\toprule

Target & \multicolumn{1}{c}{DIB} & Epoch 1 & Epoch 2 & \multicolumn{1}{c}{EW$_{1}$} & \multicolumn{1}{c}{EW$_{2}$} & \multicolumn{1}{c}{$\chi^2_{base}$} & \multicolumn{1}{c}{$\chi^2_{Gauss}$} & BIC$_{base}$ & BIC$_{Gauss}$ & $|\Delta {\rm BIC}|$ & Variation \\
~ & [\AA] & ~ & ~ & [m\AA] & [m\AA] & ~ & ~ & ~ & ~ & ~ & \\

\midrule
\endhead

\multicolumn{2}{l}{HD~22951} & ~ \\
~ &  5176 &  1997-08-12 &  2016-08-29 &  11 $\pm$ 1 &  12 $\pm$ 1 &  1.61 &  1.51 &  258 &  239 &  19 & \texttt{c} \\
~ &  5781 &  1997-08-12 &  2016-08-29 &  91 $\pm$ 2 &  99 $\pm$ 1 &  1.07 &  0.94 &  44 &  -12 &  56 & \texttt{c} \\
~ &  6196 &  1997-08-13 &  2016-08-29 &  13 $\pm$ 1 &  14 $\pm$ 1 &  1.6 &  1.55 &  261 &  259 &  ~ & ~ \\
~ &  6270 &  1997-08-12 &  2016-08-29 &  13 $\pm$ 1 &  12 $\pm$ 1 &  0.6 &  0.59 &  -274 &  -268 &  ~ & ~ \\
~ &  6379 &  1997-08-12 &  2016-08-29 &  26 $\pm$ 1 &  21 $\pm$ 1 &  3.14 &  2.48 &  628 &  516 &  112 & \texttt{c} \\
~ &  6597 &  1997-08-13 &  2016-08-29 &  2 $\pm$ 1 &  2 $\pm$ 1 &  0.56 &  0.56 &  -307 &  -297 &  ~ & ~ \\
~ &  6614 &  1997-08-13 &  2016-08-29 &  36 $\pm$ 2 &  30 $\pm$ 3 &  1.04 &  1.0 &  28 &  23 &  6 & \texttt{c} \\
\bottomrule
\end{longtable}
\twocolumn

%\end{landscape}

%% file: appendixfigures.tex
\begin{figure*}[ht!]
	\centering
	\includegraphics[width=0.99\hsize]{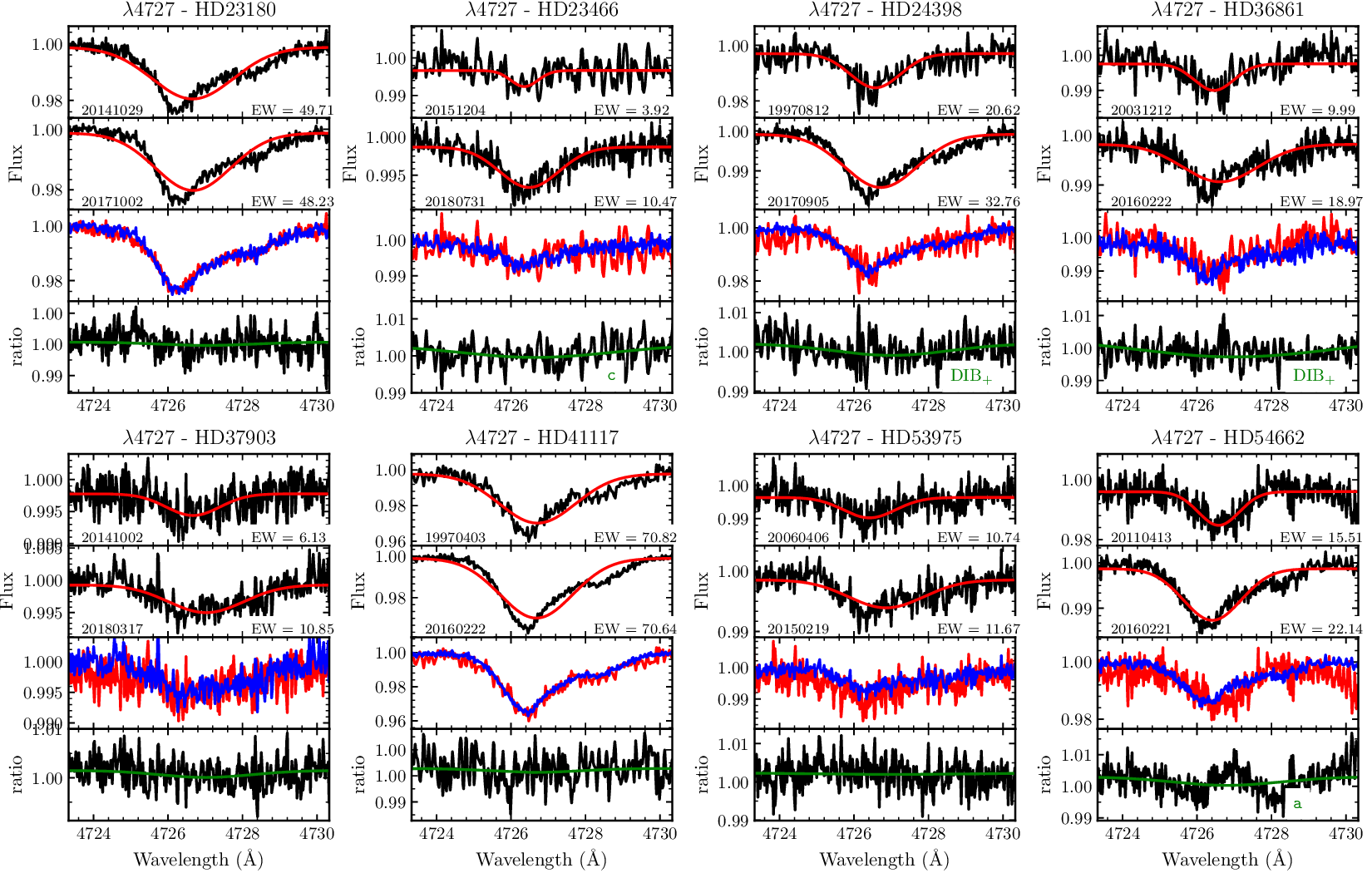}
	\caption{Comparison of the DIB profiles. The epochs of observation, target name, equivalent width (m$\AA$), and the DIB wavelength ($\AA$) are printed on the top panels. The upper panel is the first epoch. The second panel is the final observation, the third compares two observations, and the lower panel is the ratio spectrum. A Gaussian profile is fitted to both spectra without constraining the continuum or width (red line). Likewise, the Gaussian fitted to the ratio spectra is shown in the green line. The ratio spectra with significant variation are determined with the legend of the symbols in the lower panel (see Tab.~\ref{tab:DIBresults} for symbol descriptions). Note that the fluxes are normalized.}
	\label{plt-dib-var1}
\end{figure*}

\begin{figure*}[ht!]
	\centering
	\includegraphics[width=0.99\hsize]{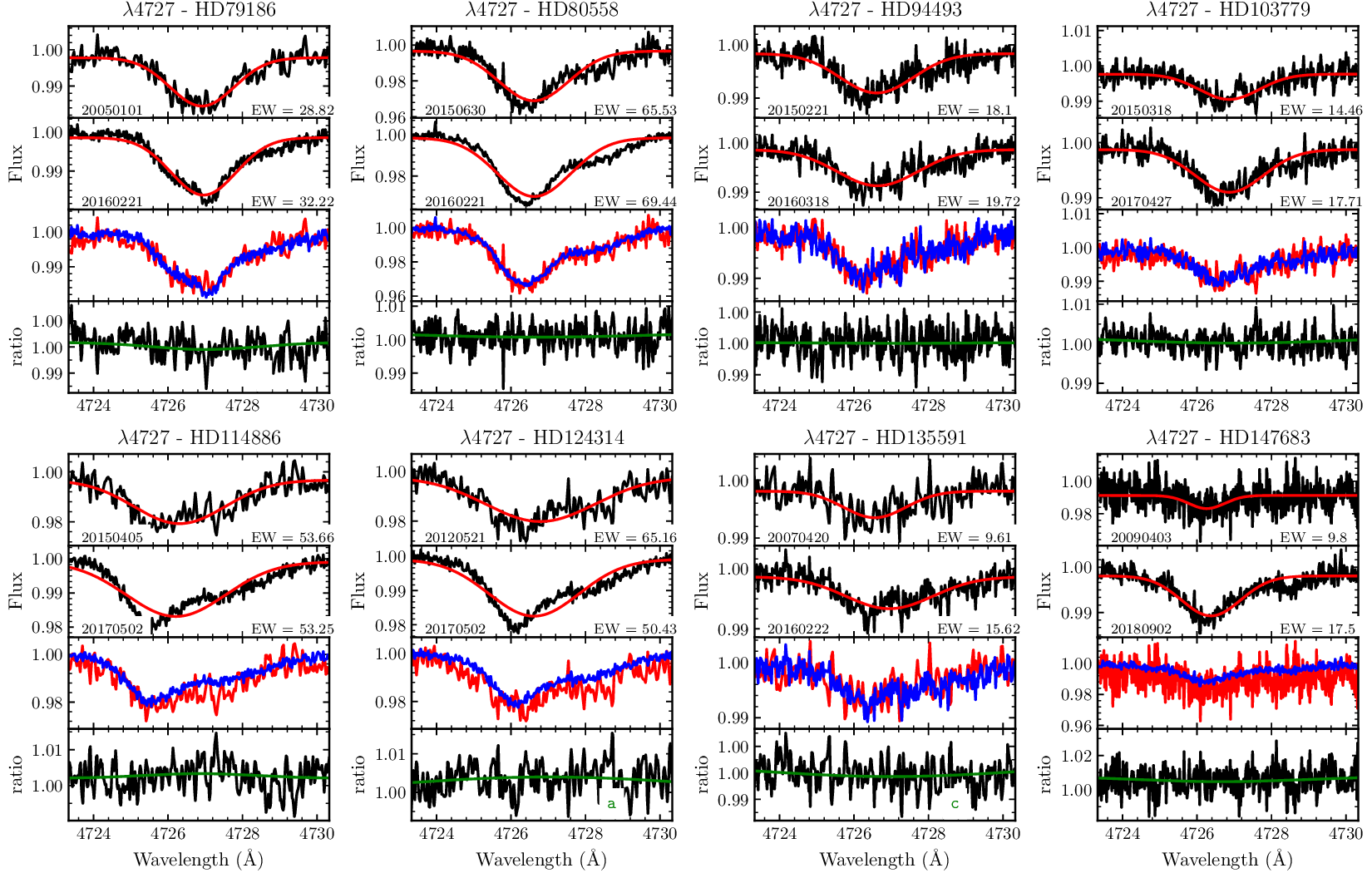}
	\caption{The same as Fig.~\ref{plt-dib-var1}}
	\label{plt-dib-var2}
\end{figure*}

\begin{figure*}[ht!]
	\centering
	\includegraphics[width=0.99\hsize]{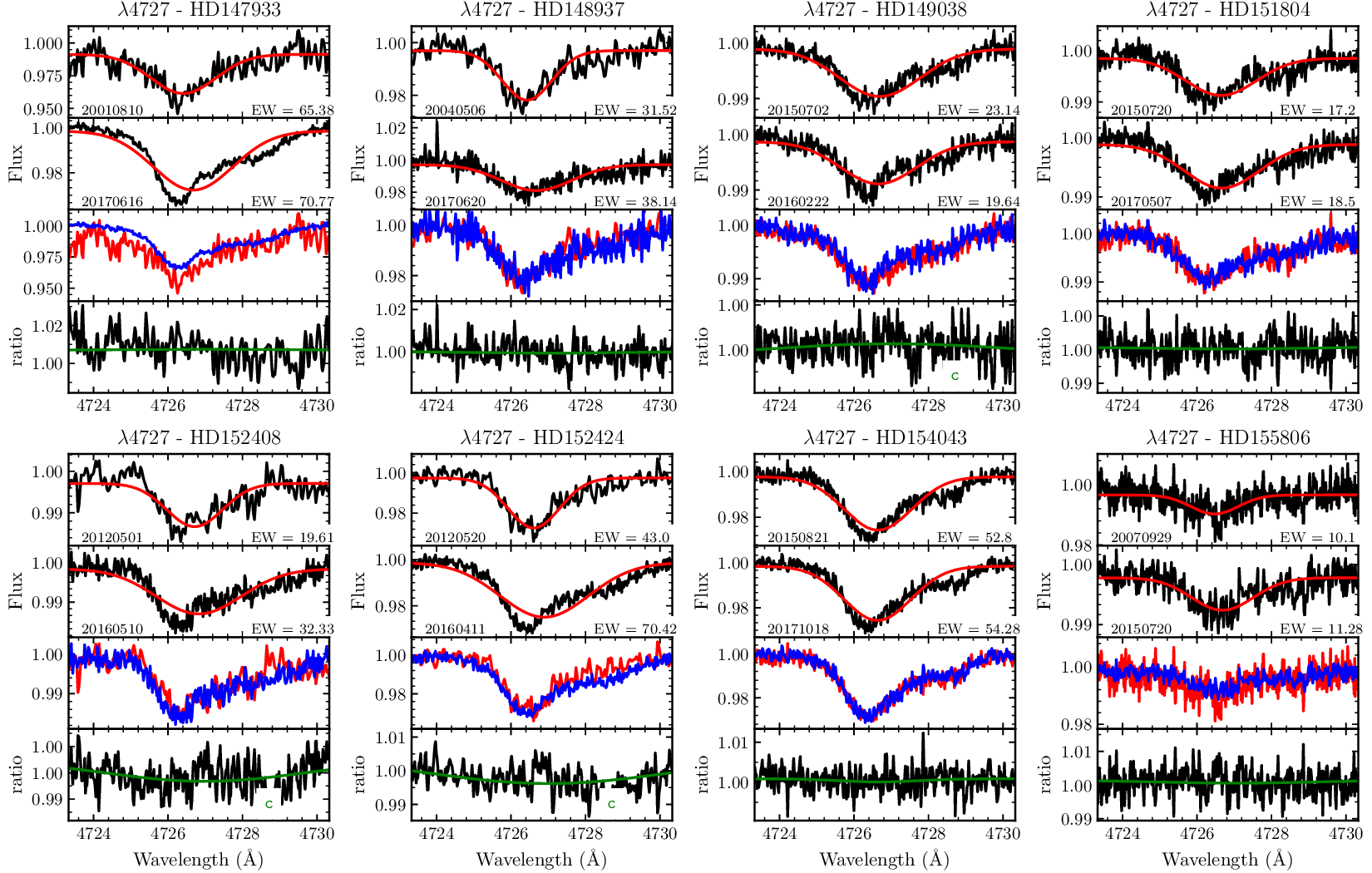}
	\caption{The same as Fig.~\ref{plt-dib-var1}}
	\label{plt-dib-var3}
\end{figure*}

\begin{figure*}[ht!]
	\centering
	\includegraphics[width=0.99\hsize]{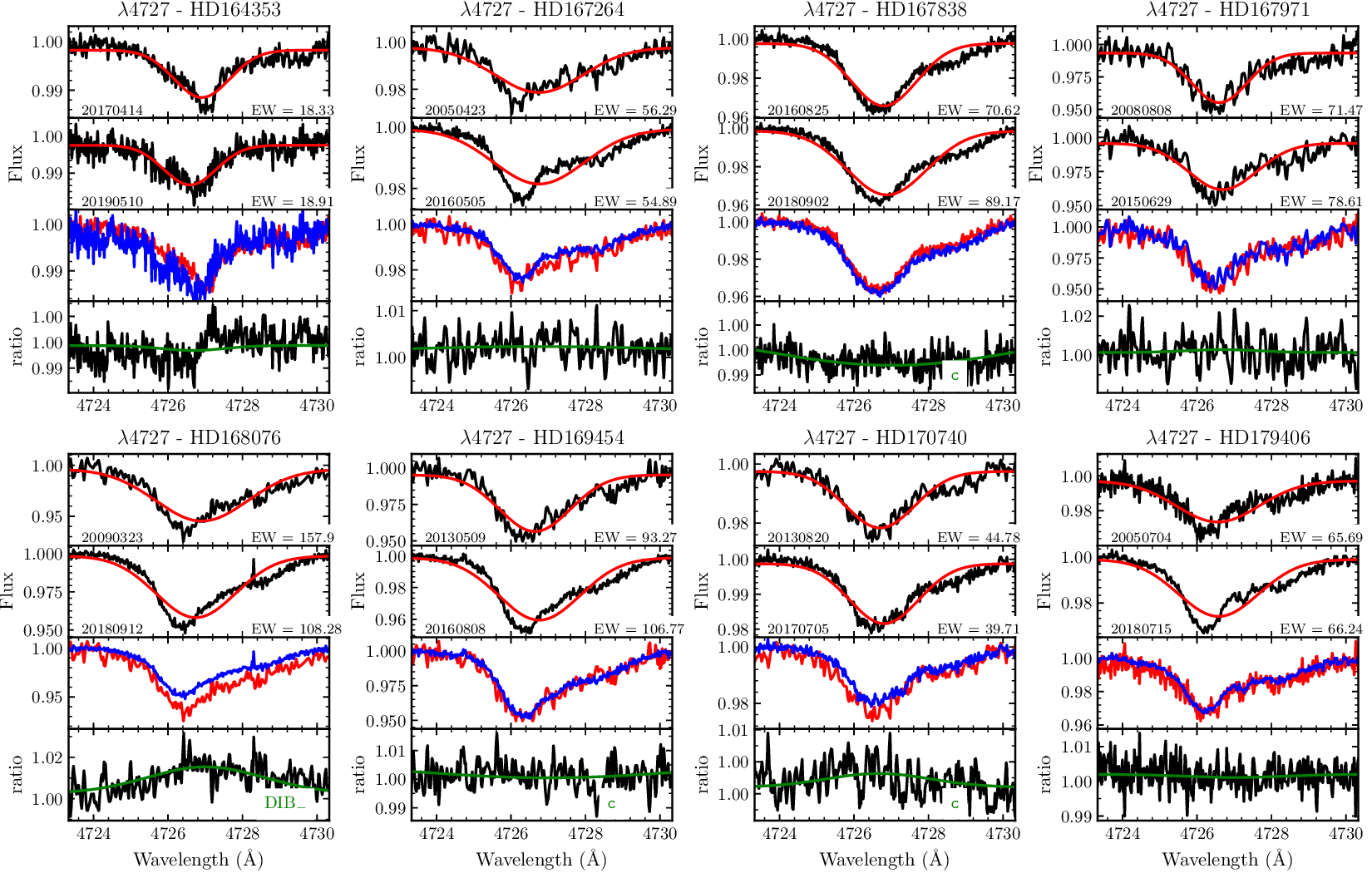}
	\caption{The same as Fig.~\ref{plt-dib-var1}}
	\label{plt-dib-var4}
\end{figure*}

\begin{figure*}[ht!]
	\centering
	\includegraphics[width=0.99\hsize]{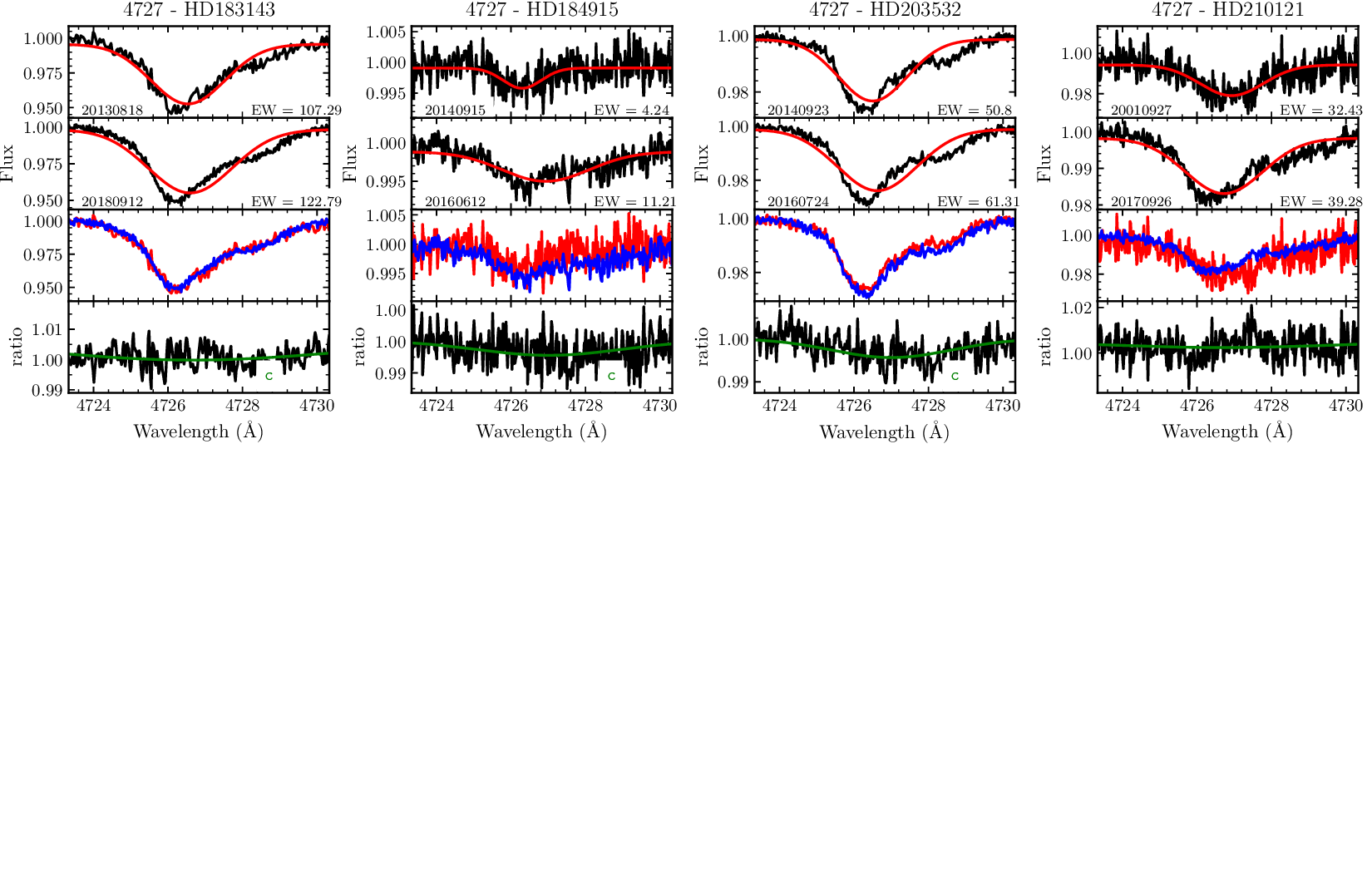}
	\caption{The same as Fig.~\ref{plt-dib-var1}}
	\label{plt-dib-var5}
\end{figure*}

%%%%%%%%%%%%%%%%%%
%%%%%%%%%%%%%%%%%%
\begin{figure*}[ht!]
	\centering
	\includegraphics[width=0.99\hsize]{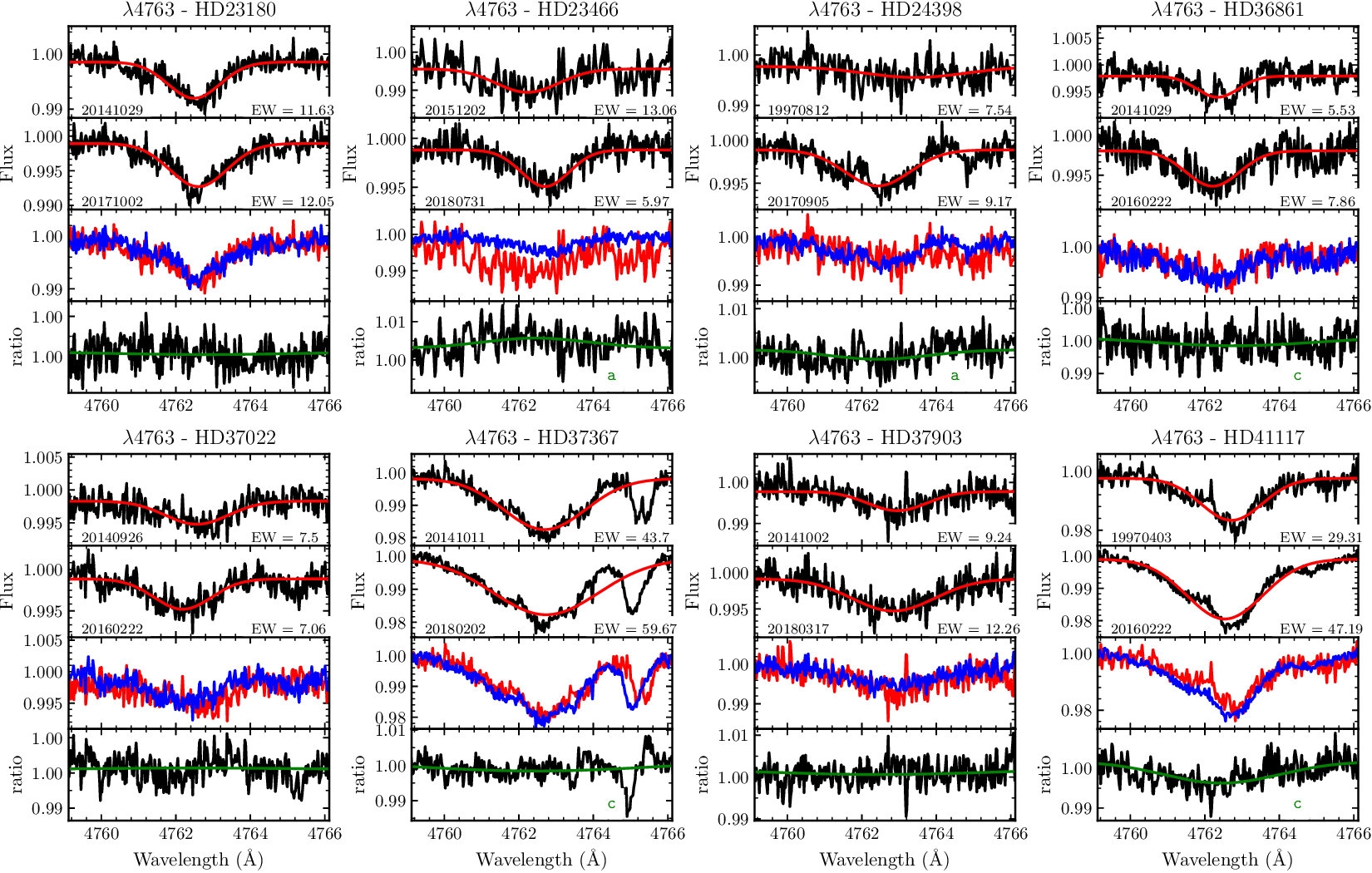}
	\caption{The same as Fig.~\ref{plt-dib-var1}}
	\label{plt-dib-var6}
\end{figure*}

\begin{figure*}[ht!]
	\centering
	\includegraphics[width=0.99\hsize]{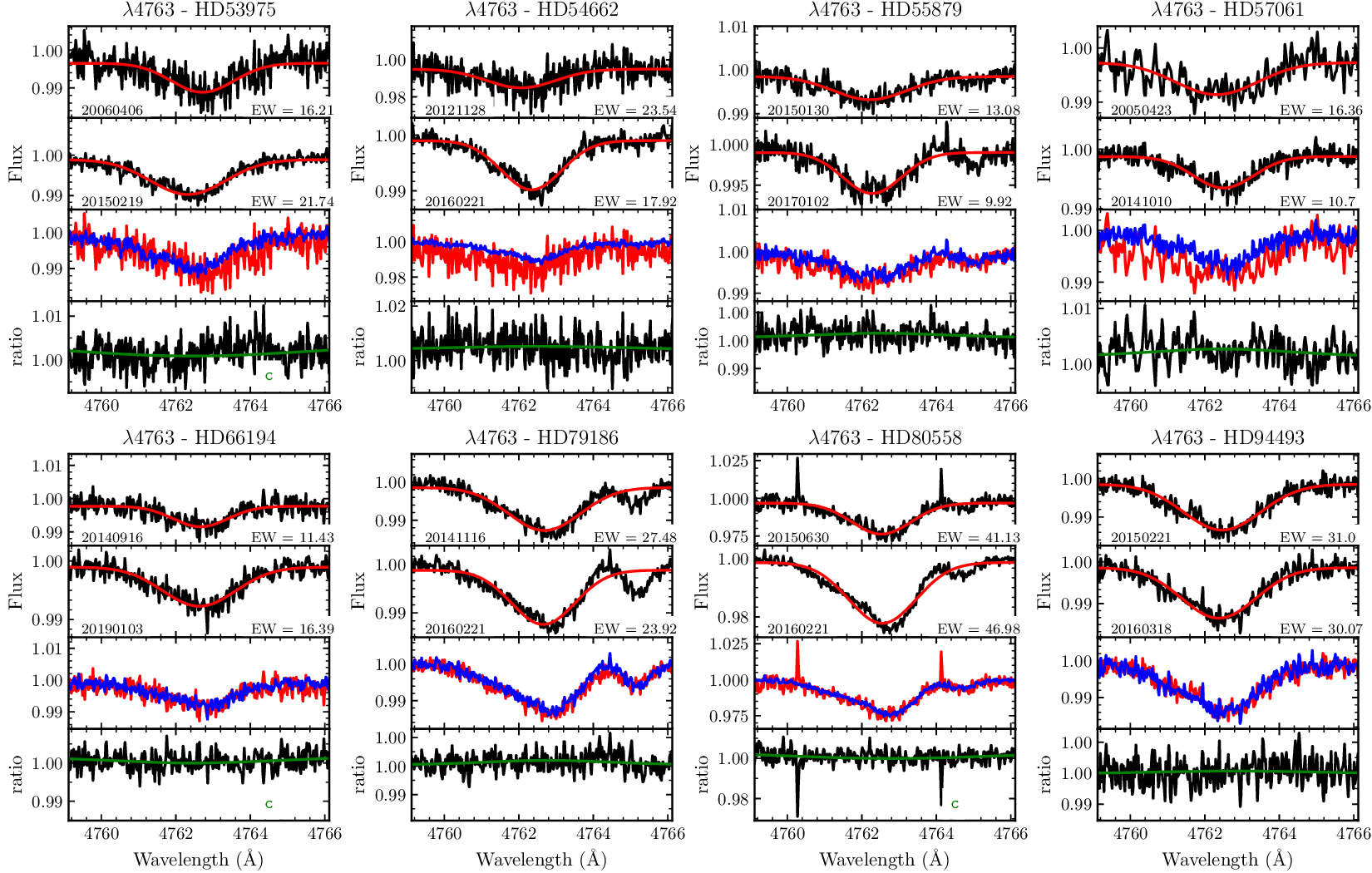}
	\caption{The same as Fig.~\ref{plt-dib-var1}}
	\label{plt-dib-var7}
\end{figure*}

\begin{figure*}[ht!]
	\centering
	\includegraphics[width=0.99\hsize]{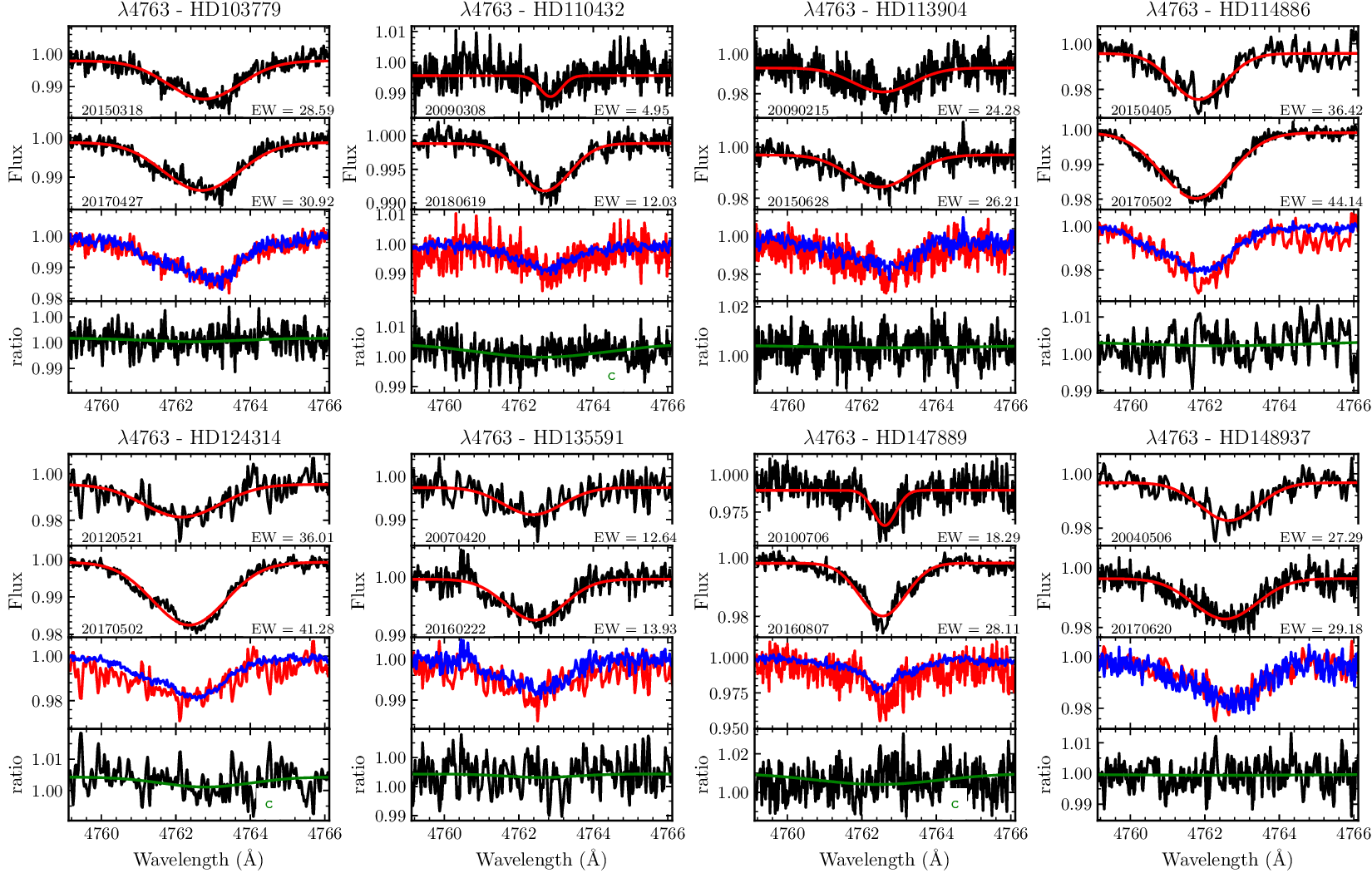}
	\caption{The same as Fig.~\ref{plt-dib-var1}}
	\label{plt-dib-var8}
\end{figure*}

\begin{figure*}[ht!]
	\centering
	\includegraphics[width=0.99\hsize]{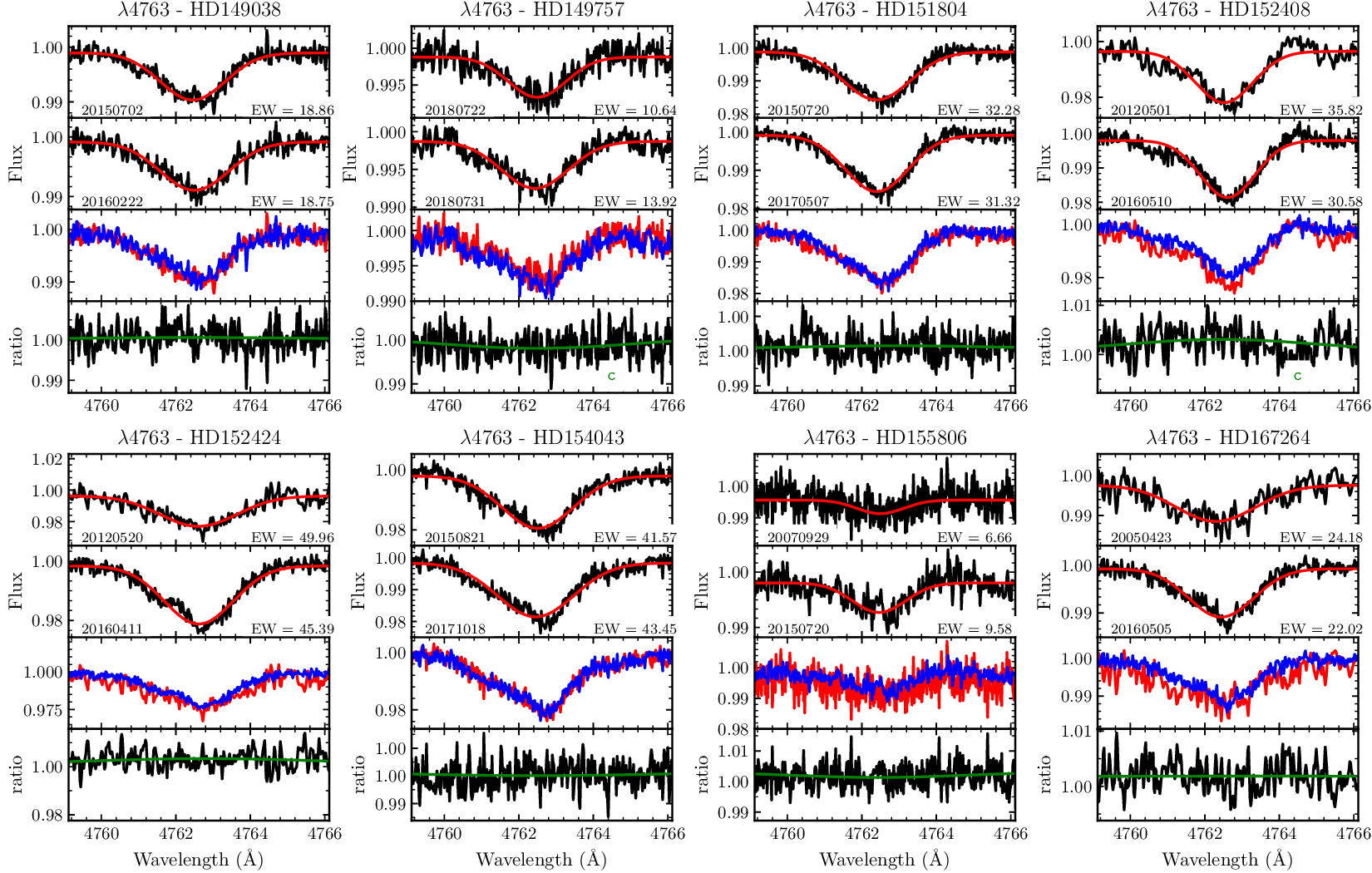}
	\caption{The same as Fig.~\ref{plt-dib-var1}}
	\label{plt-dib-var9}
\end{figure*}

\begin{figure*}[ht!]
	\centering
	\includegraphics[width=0.99\hsize]{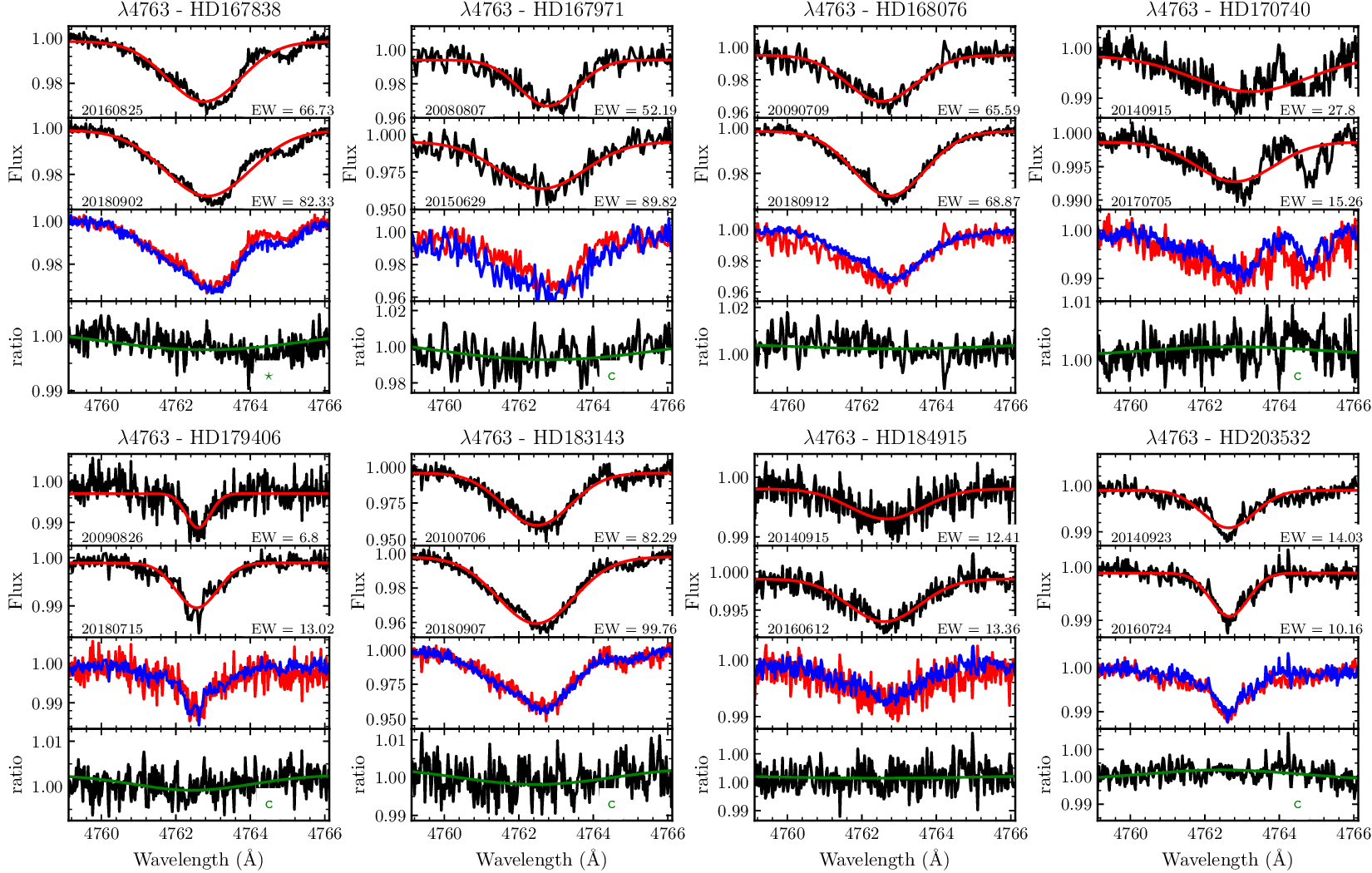}
	\caption{The same as Fig.~\ref{plt-dib-var1}}
	\label{plt-dib-var10}
\end{figure*}

\begin{figure*}[ht!]
	\centering
	\includegraphics[width=0.99\hsize]{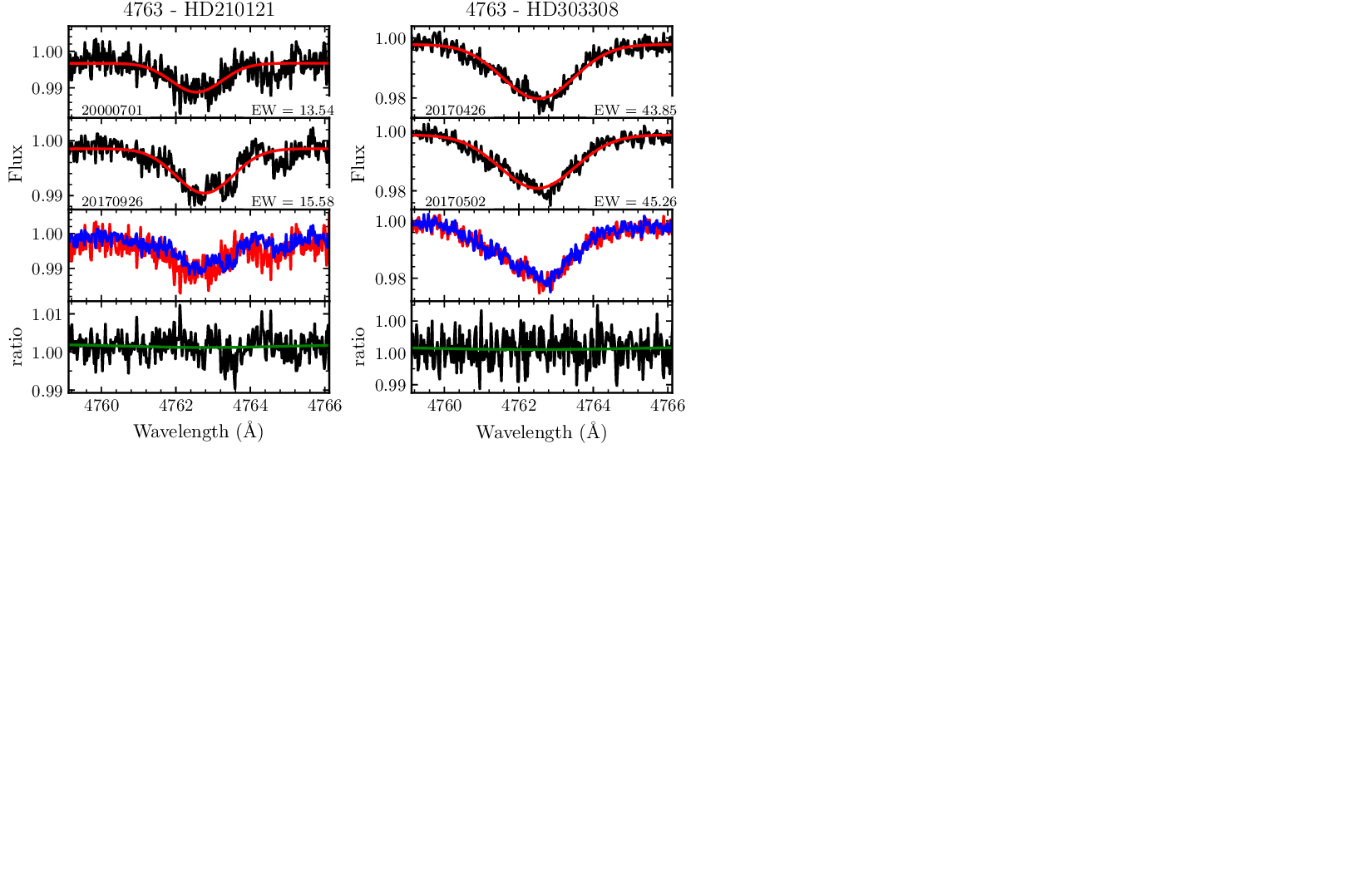}
	\caption{The same as Fig.~\ref{plt-dib-var1}}
	\label{plt-dib-var11}
\end{figure*}

%%%%%%%%%%%%%%%%%%
%%%%%%%%%%%%%%%%%%

\begin{figure*}[ht!]
    \centering
    \includegraphics[width=0.99\hsize]{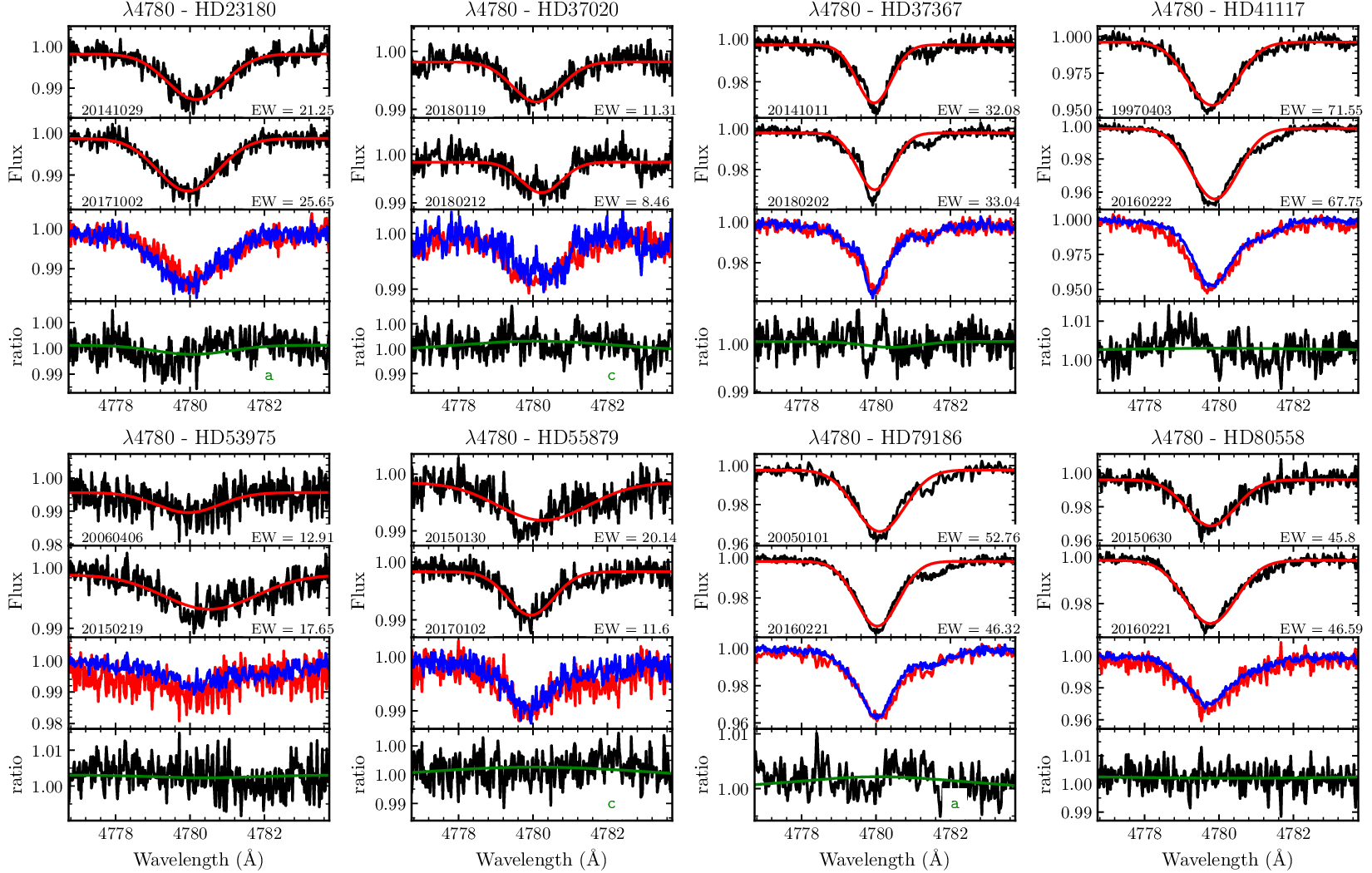}
    \caption{The same as \ref{plt-dib-var1}}
    \label{plt-dib-var12}
\end{figure*}

\begin{figure*}[ht!]
    \centering
    \includegraphics[width=0.99\hsize]{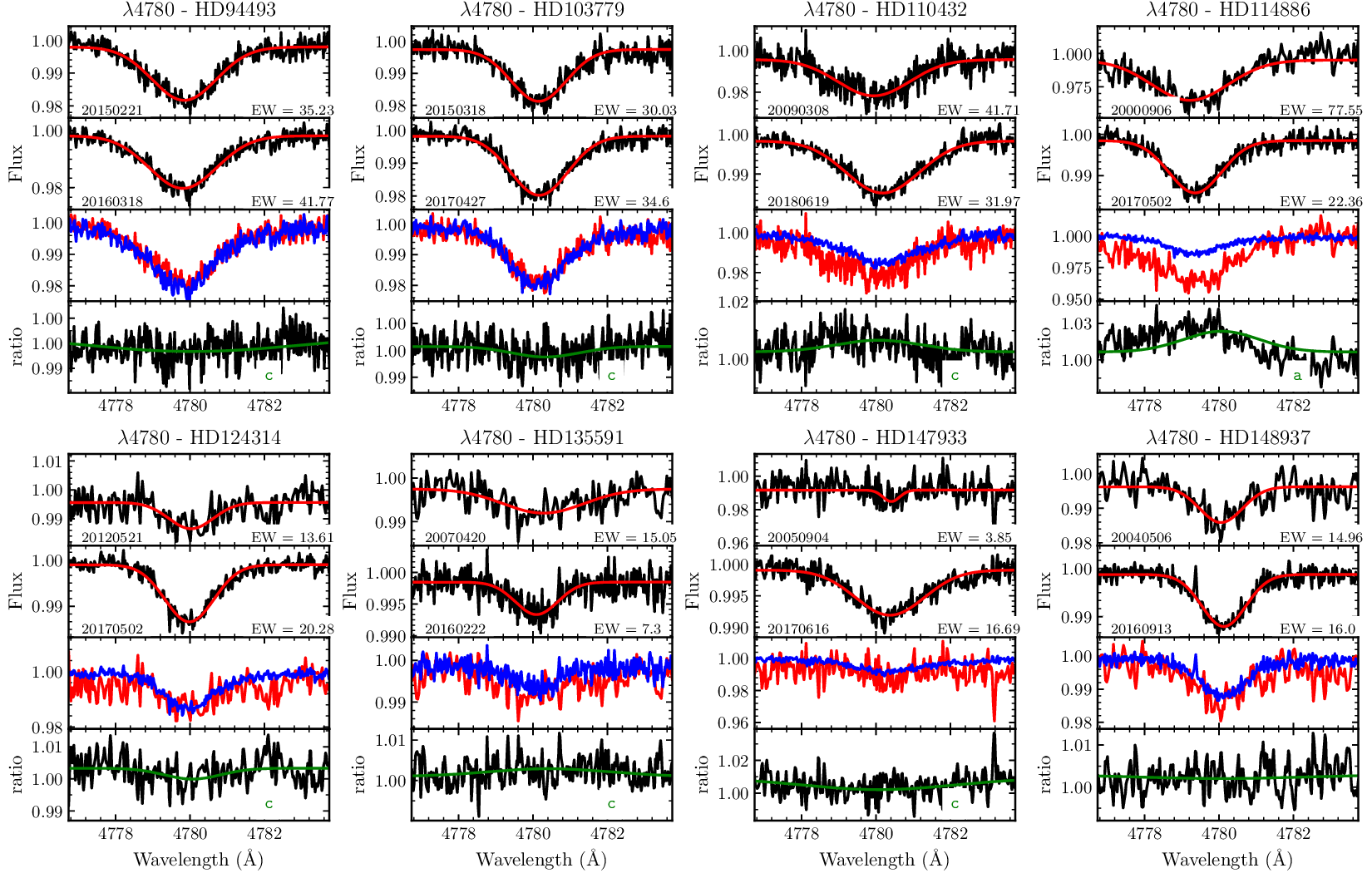}
    \caption{The same as \ref{plt-dib-var1}}
    \label{plt-dib-var13}
\end{figure*}

\begin{figure*}[ht!]
    \centering
    \includegraphics[width=0.99\hsize]{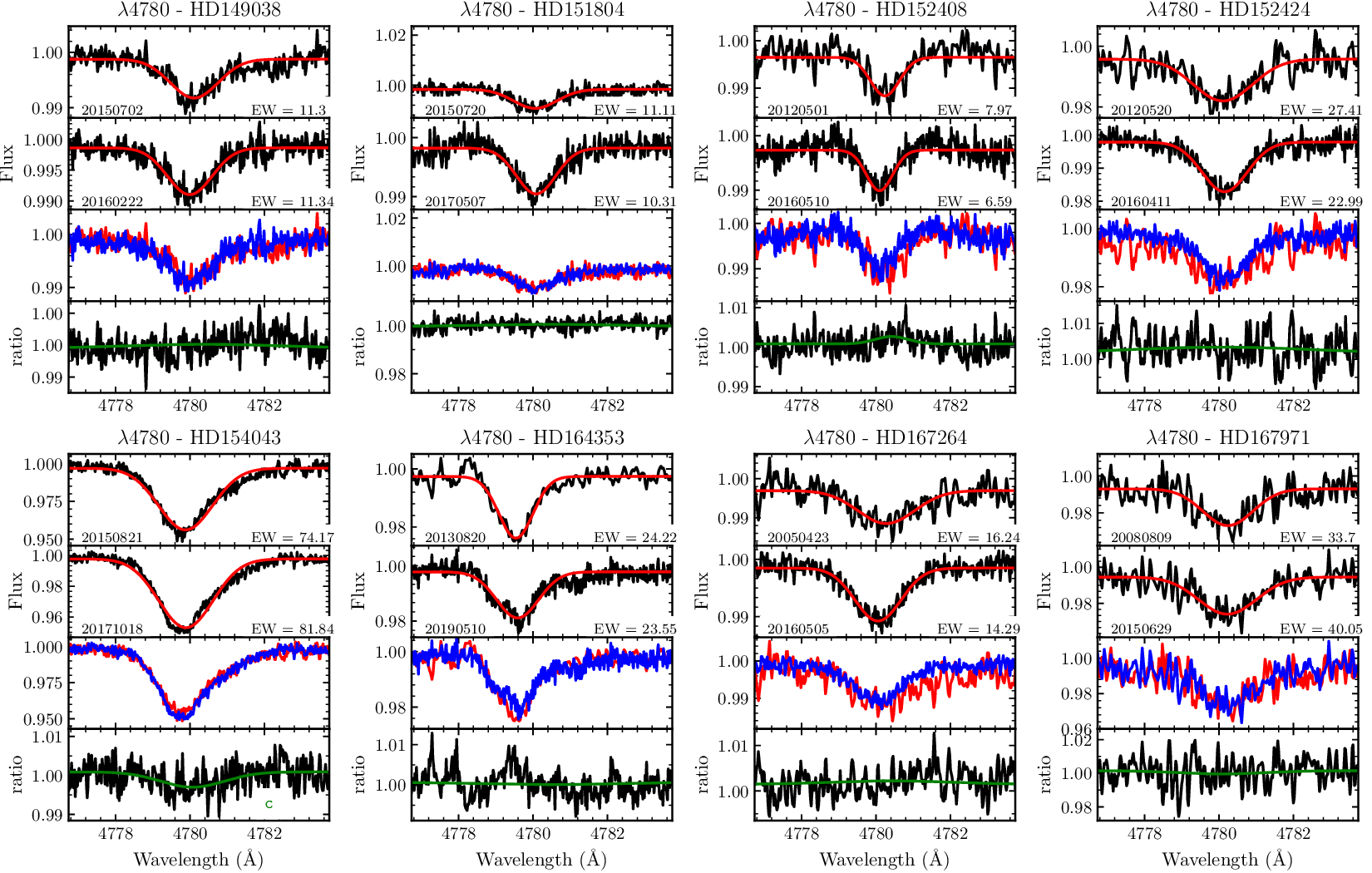}
    \caption{The same as \ref{plt-dib-var1}}
    \label{plt-dib-var14}
\end{figure*}

\begin{figure*}[ht!]
    \centering
    \includegraphics[width=0.99\hsize]{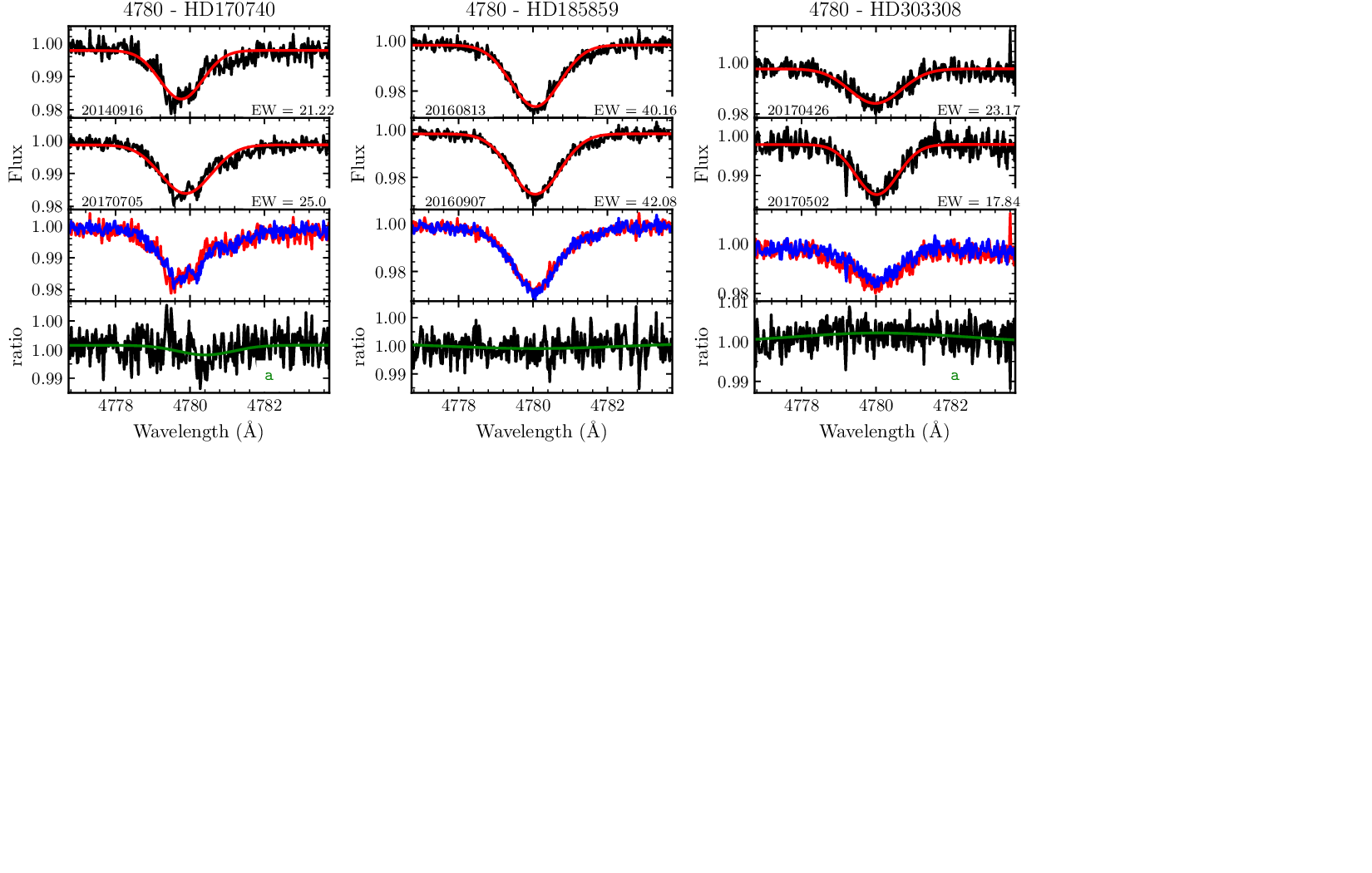}
    \caption{The same as \ref{plt-dib-var1}}
    \label{plt-dib-var15}
\end{figure*}

%%%%%%%%%%%%%%%%%%
%%%%%%%%%%%%%%%%%%

\begin{figure*}[ht!]
    \centering
    \includegraphics[width=0.99\hsize]{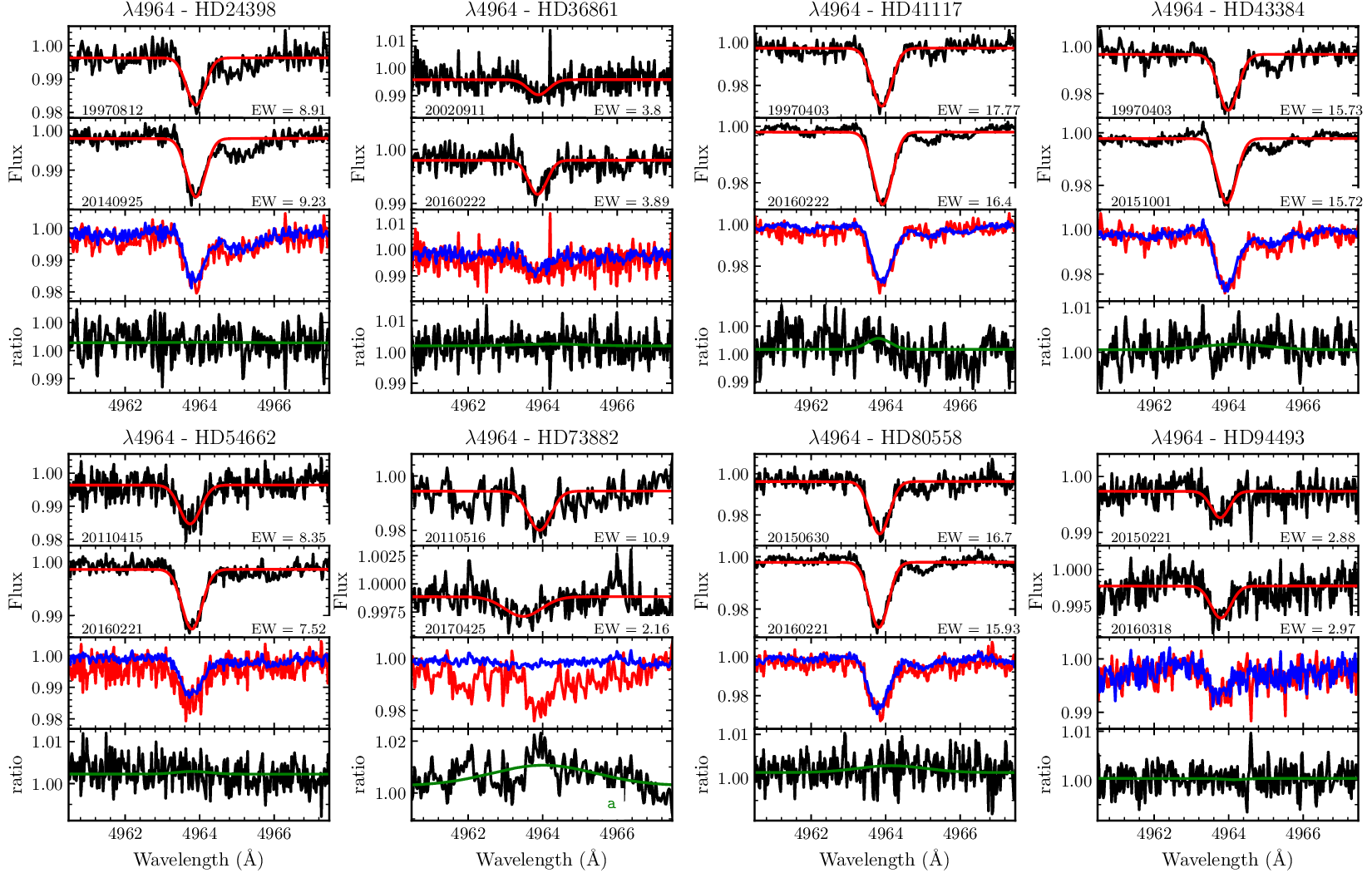}
    \caption{The same as \ref{plt-dib-var1}}
    \label{plt-dib-var16}
\end{figure*}

\begin{figure*}[ht!]
    \centering
    \includegraphics[width=0.99\hsize]{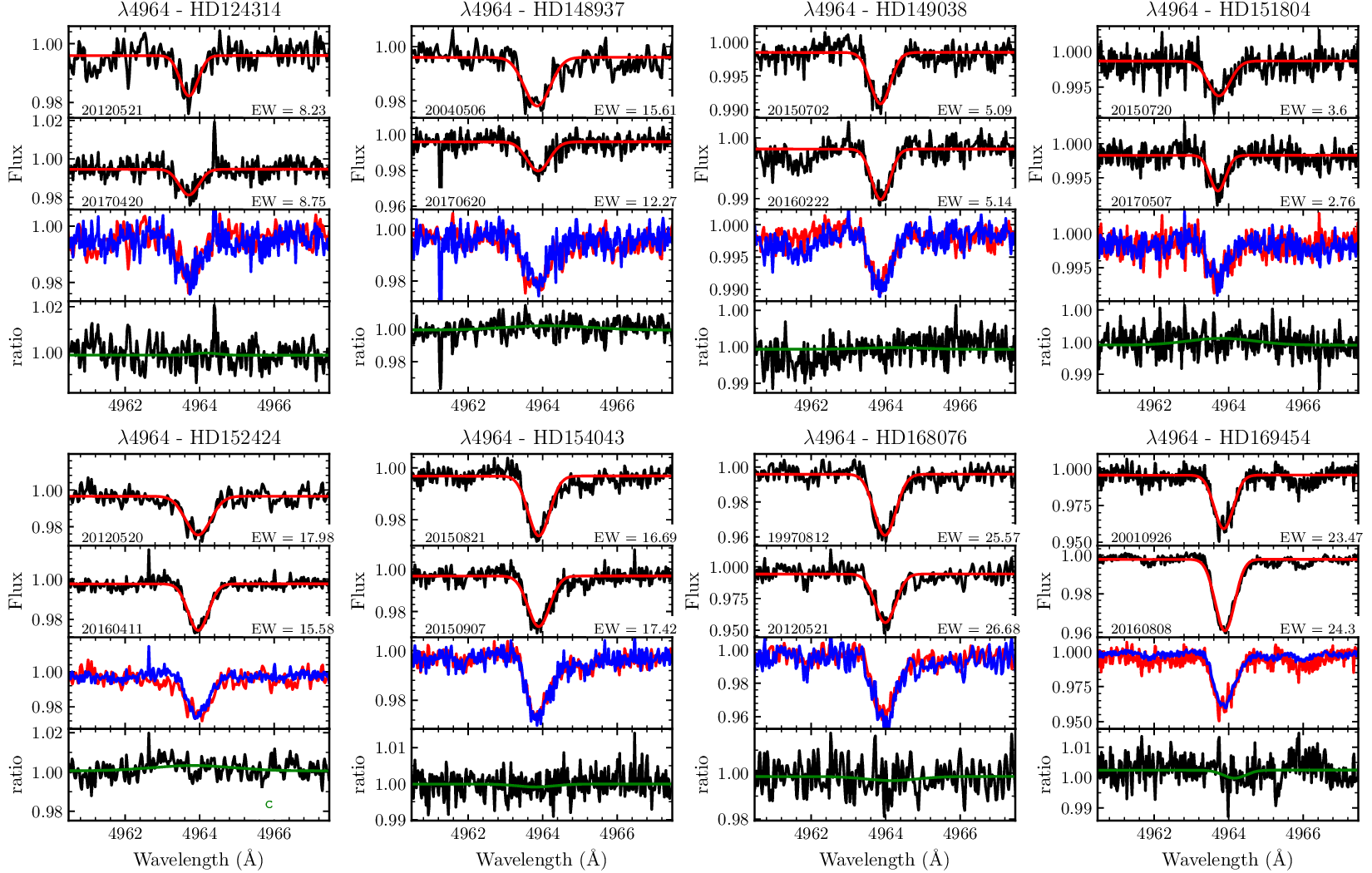}
    \caption{The same as \ref{plt-dib-var1}}
    \label{plt-dib-var17}
\end{figure*}

\begin{figure*}[ht!]
    \centering
    \includegraphics[width=0.99\hsize]{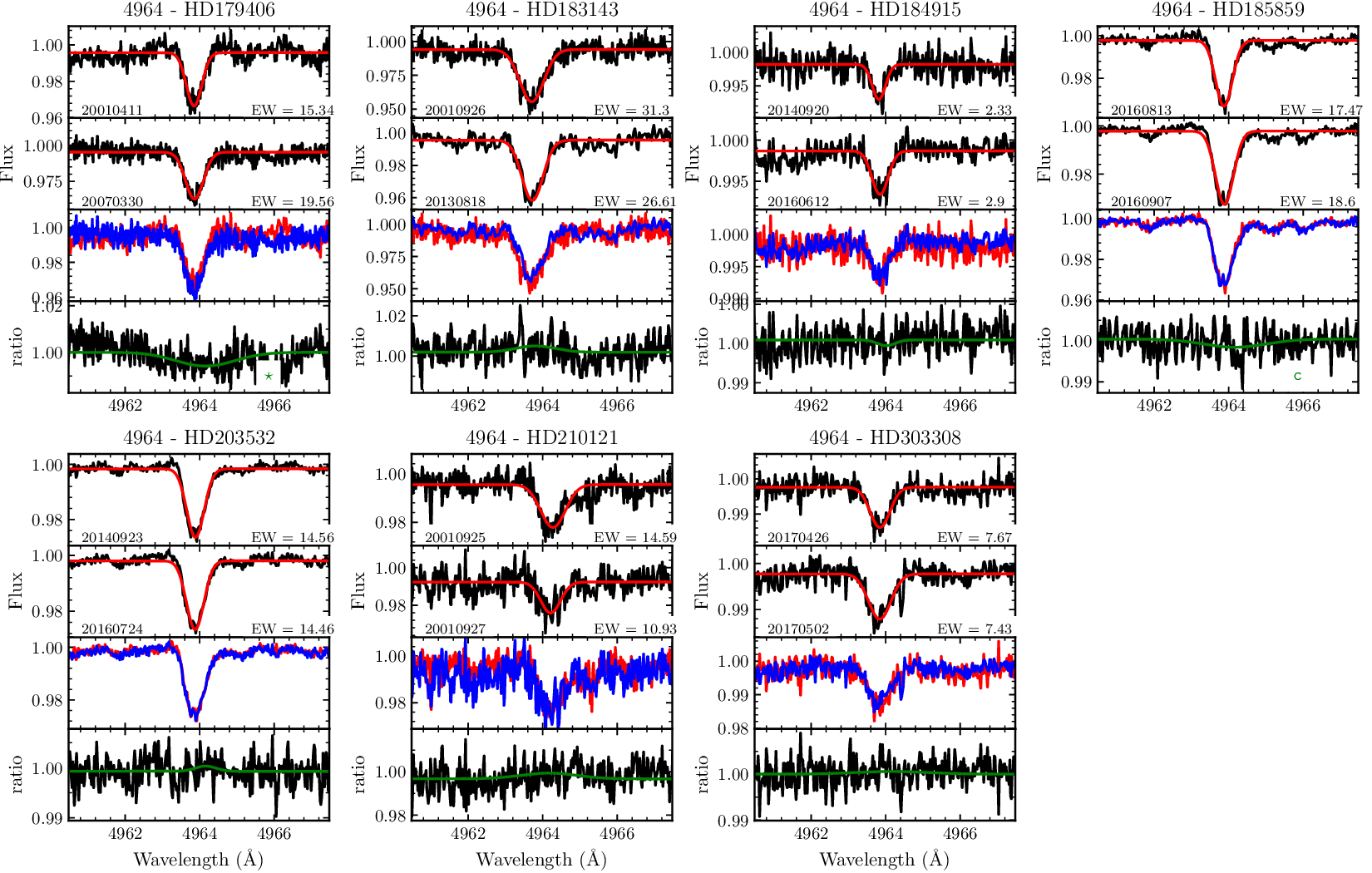}
    \caption{The same as \ref{plt-dib-var1}}
    \label{plt-dib-var18}
\end{figure*}

%%%%%%%%%%%%%%%%%%
%%%%%%%%%%%%%%%%%%

\begin{figure*}[ht!]
    \centering
    \includegraphics[width=0.99\hsize]{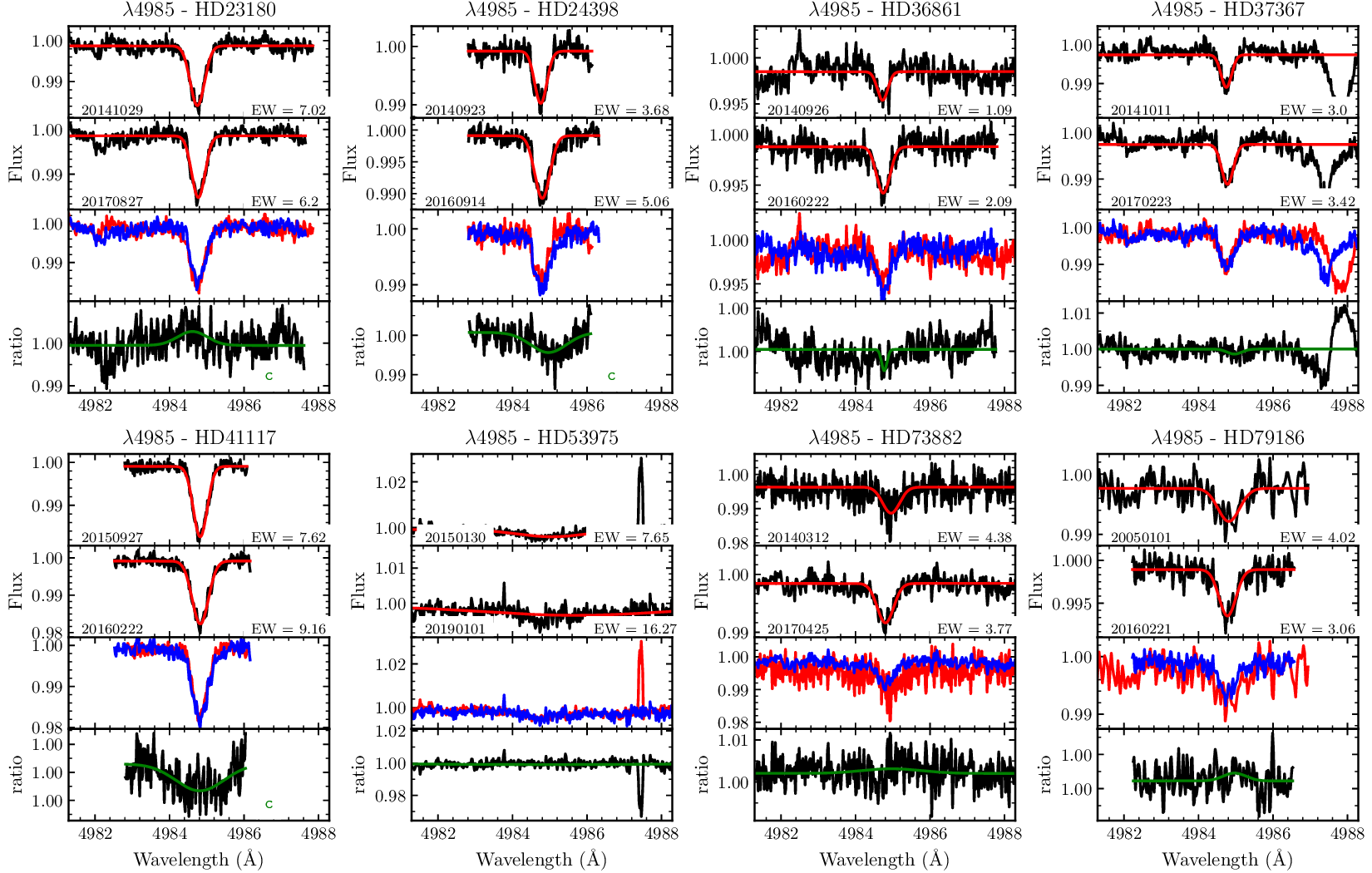}
    \caption{The same as \ref{plt-dib-var1}}
    \label{plt-dib-var19}
\end{figure*}

\begin{figure*}[ht!]
    \centering
    \includegraphics[width=0.99\hsize]{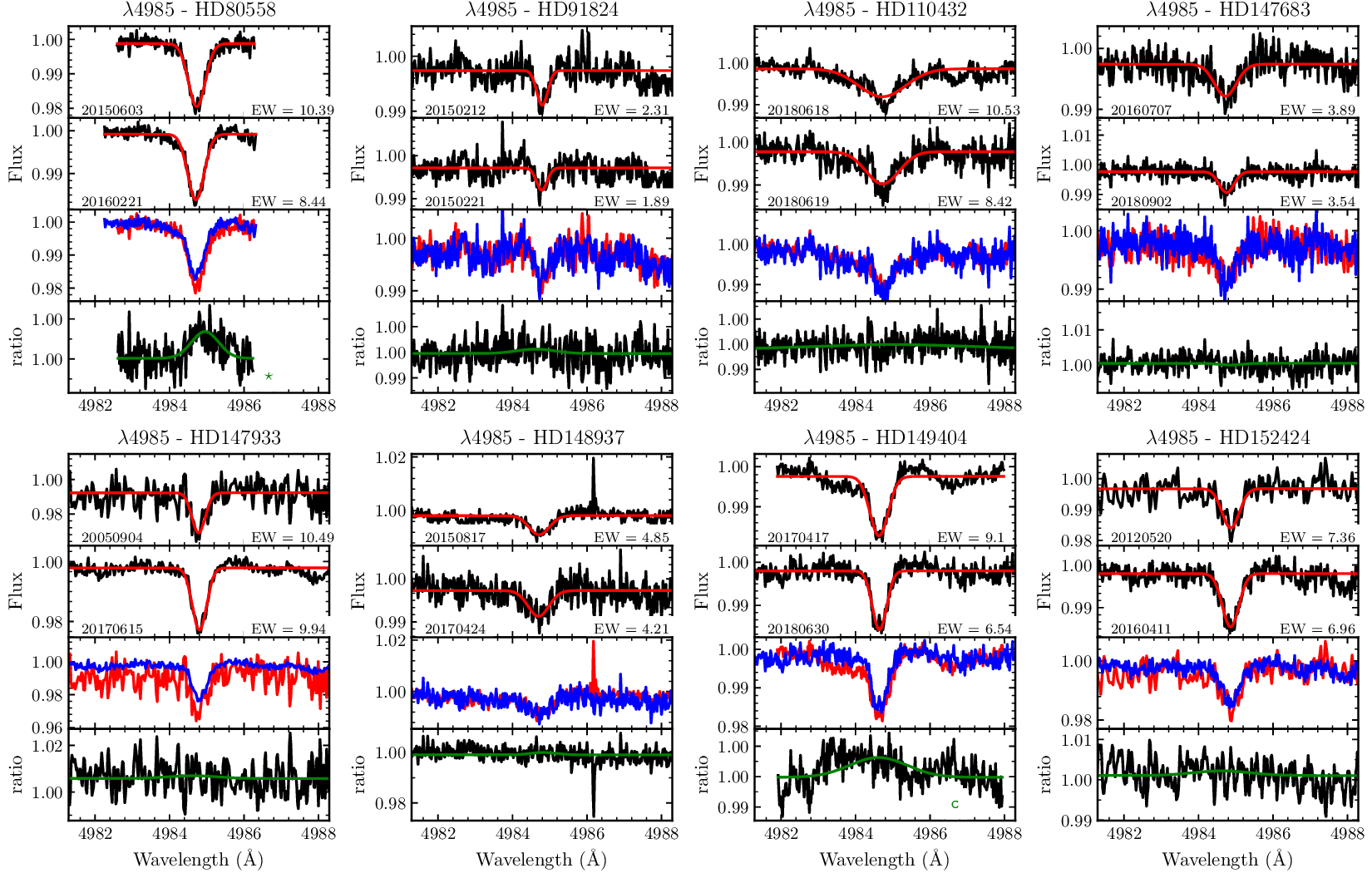}
    \caption{The same as \ref{plt-dib-var1}}
    \label{plt-dib-var20}
\end{figure*}

\begin{figure*}[ht!]
    \centering
    \includegraphics[width=0.99\hsize]{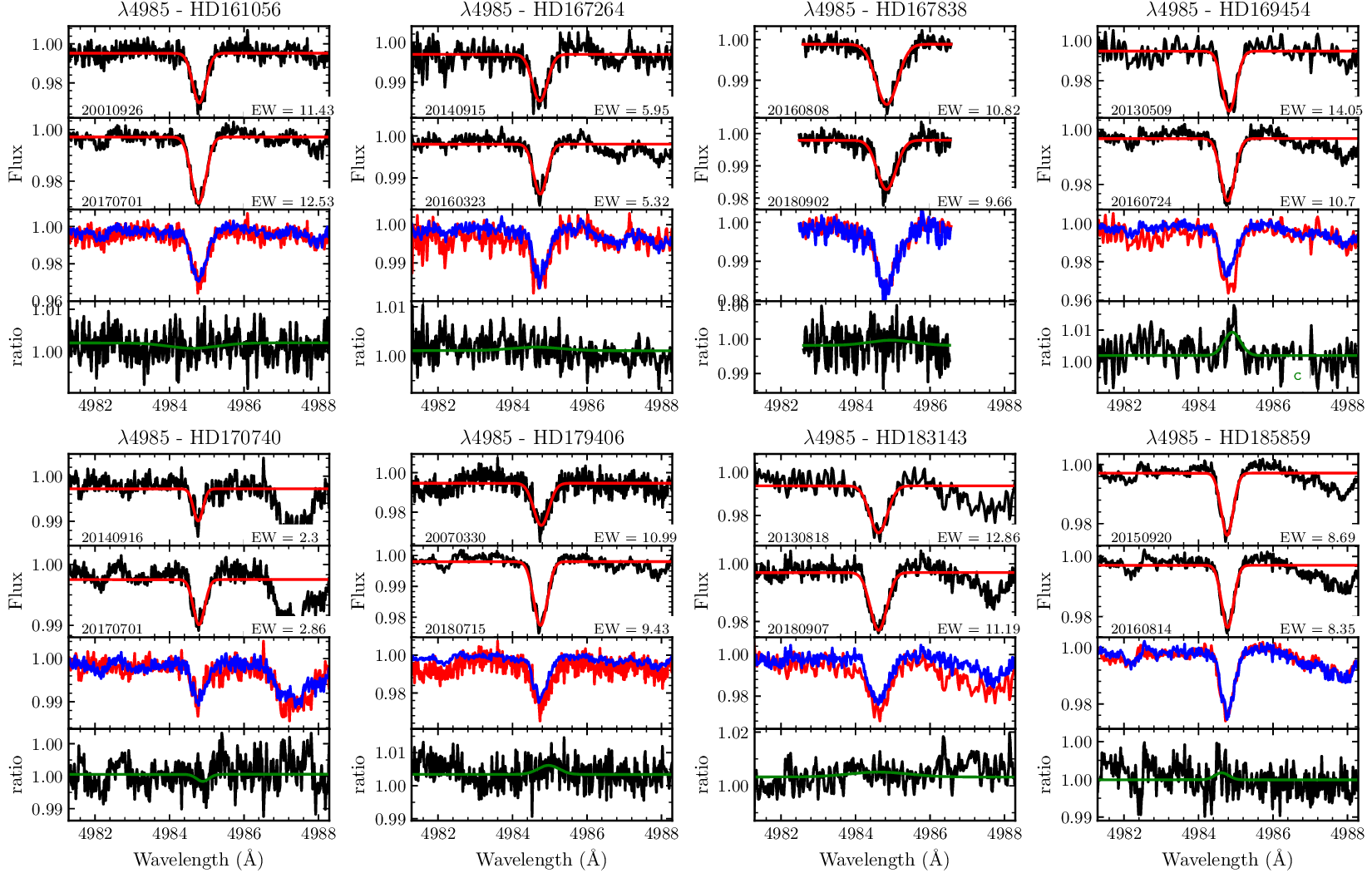}
    \caption{The same as \ref{plt-dib-var1}}
    \label{plt-dib-var21}
\end{figure*}

\begin{figure*}[ht!]
    \centering
    \includegraphics[width=0.99\hsize]{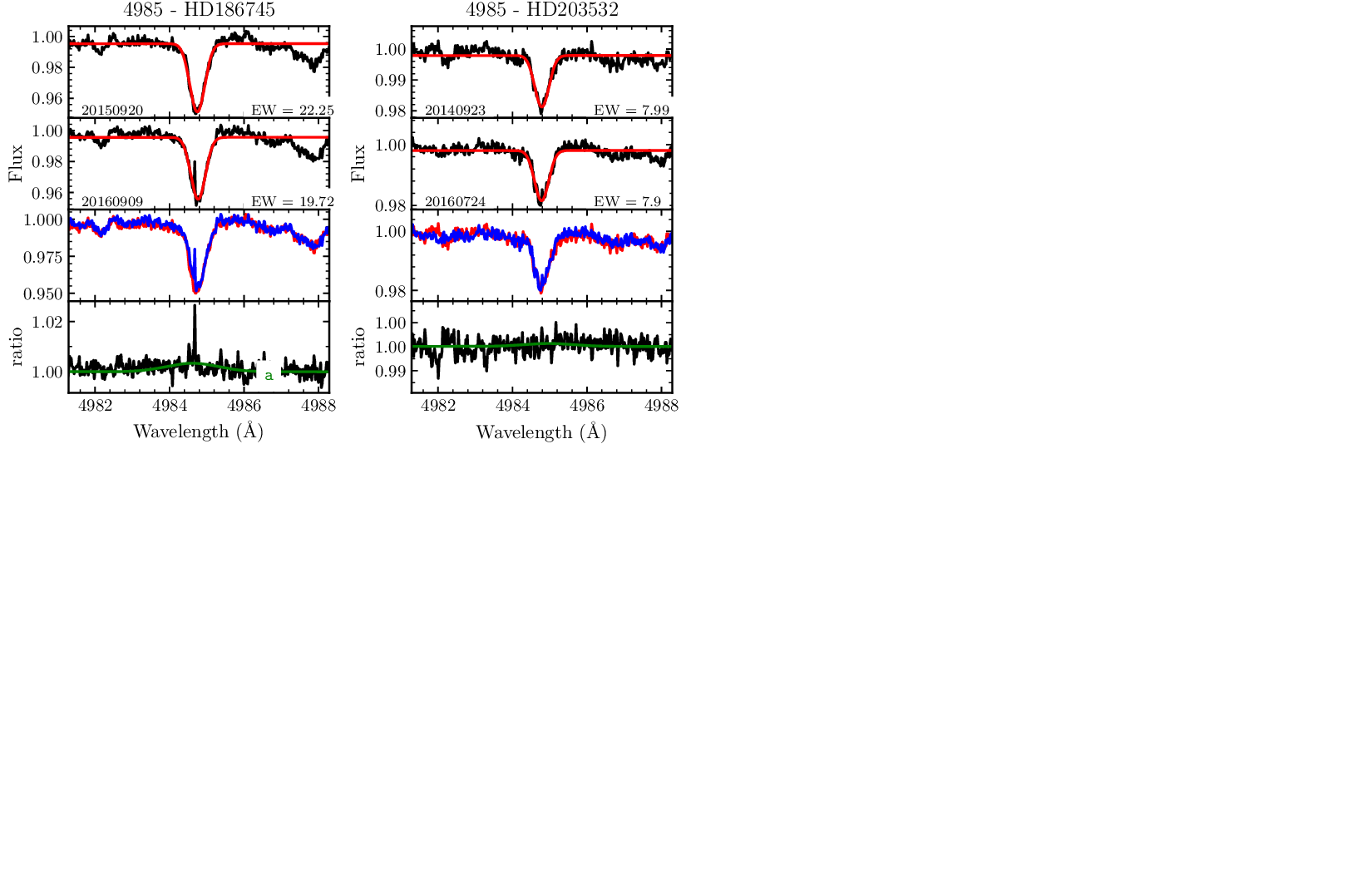}
    \caption{The same as \ref{plt-dib-var1}}
    \label{plt-dib-var22}
\end{figure*}

%%%%%%%%%%%%%%%%%%
%%%%%%%%%%%%%%%%%%
\begin{figure*}[ht!]
    \centering
    \includegraphics[width=0.99\hsize]{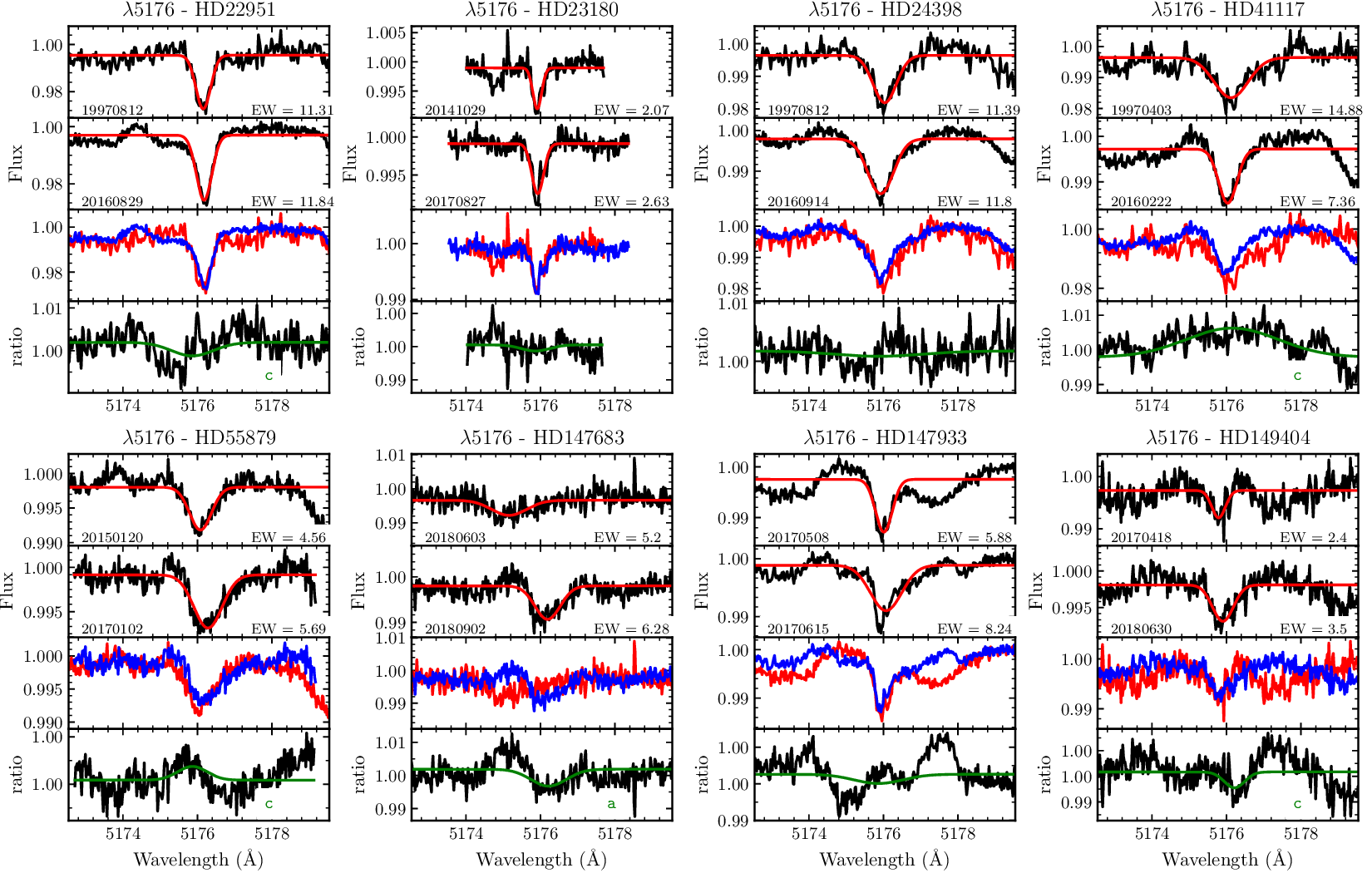}
    \caption{The same as \ref{plt-dib-var1}}
    \label{plt-dib-var23}
\end{figure*}

\begin{figure*}[ht!]
    \centering
    \includegraphics[width=0.99\hsize]{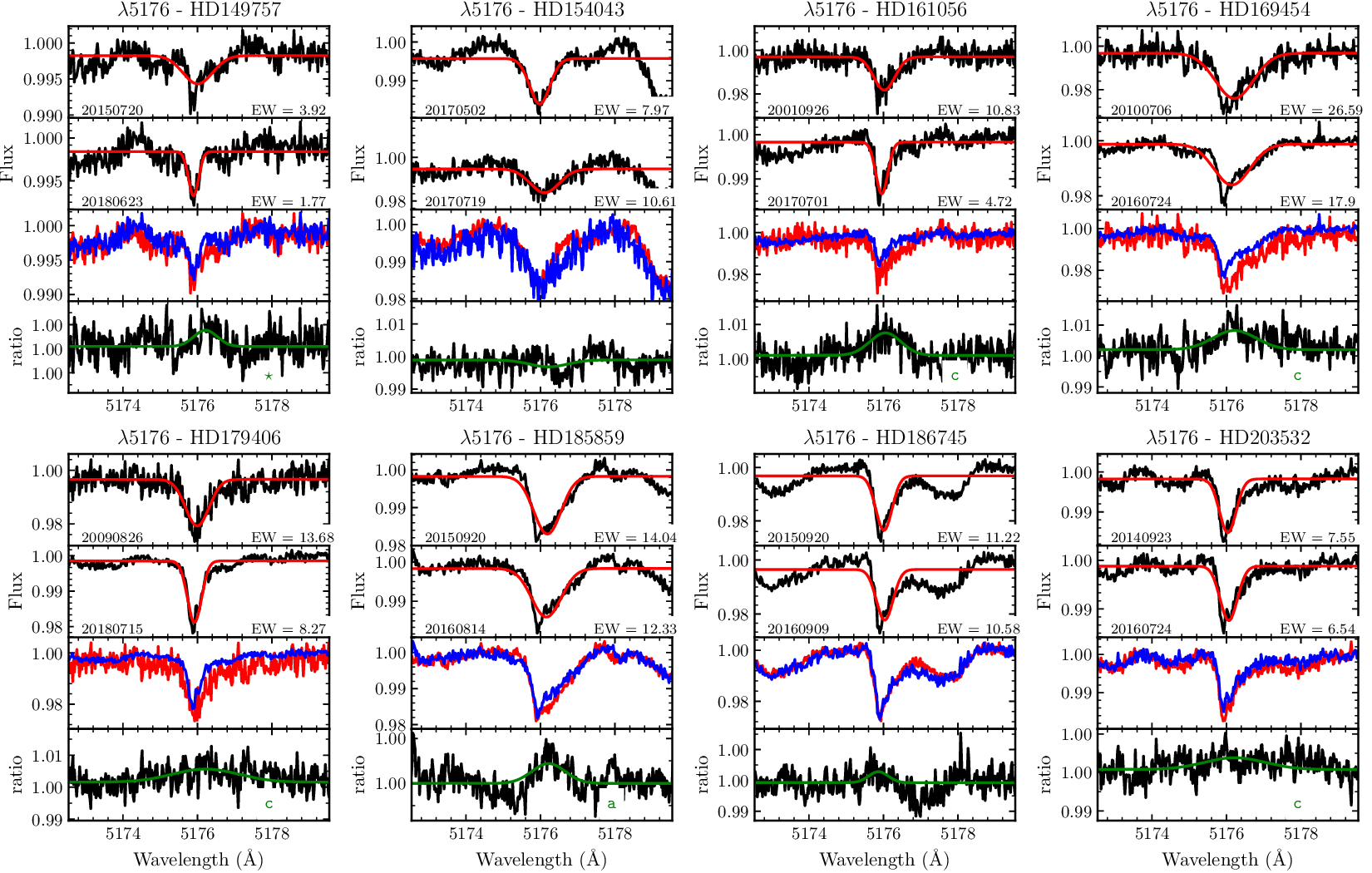}
    \caption{The same as \ref{plt-dib-var1}}
    \label{plt-dib-var24}
\end{figure*}

%%%%%%%%%%%%%%%%%%
%%%%%%%%%%%%%%%%%%
\begin{figure*}[ht!]
    \centering
    \includegraphics[width=0.99\hsize]{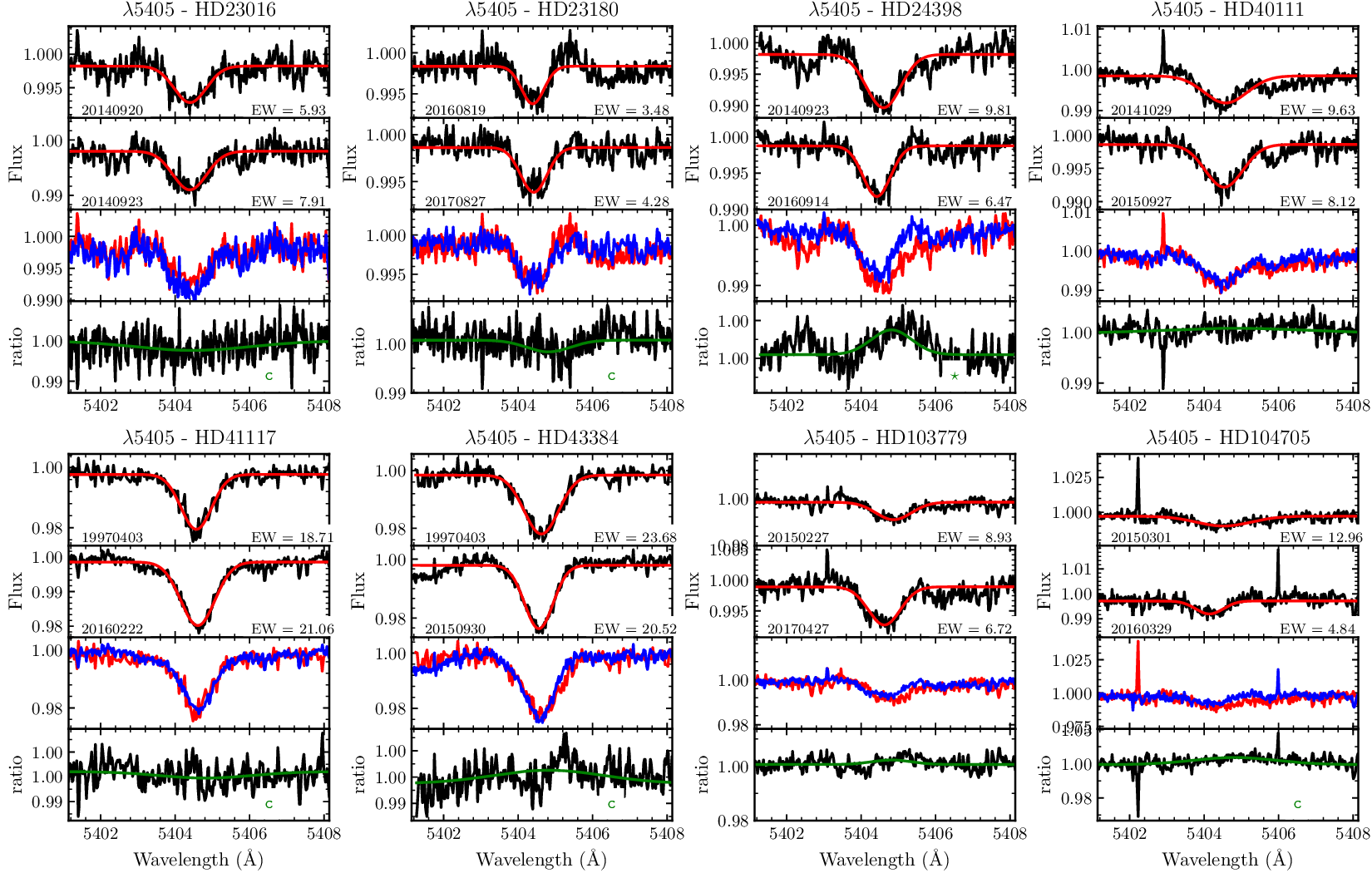}
    \caption{The same as \ref{plt-dib-var1}}
    \label{plt-dib-var25}
\end{figure*}

\begin{figure*}[ht!]
    \centering
    \includegraphics[width=0.99\hsize]{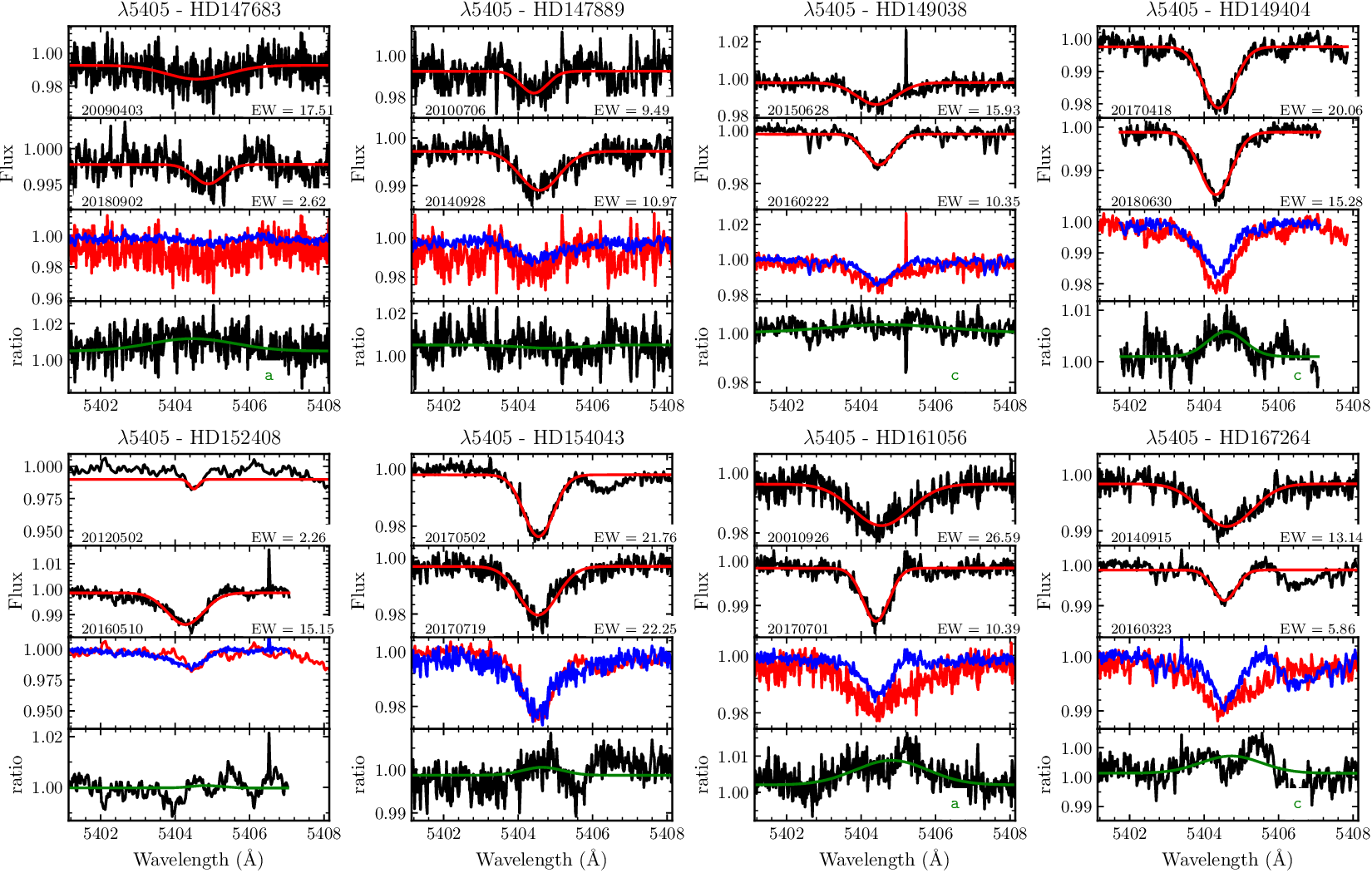}
    \caption{The same as \ref{plt-dib-var1}}
    \label{plt-dib-var26}
\end{figure*}

\begin{figure*}[ht!]
    \centering
    \includegraphics[width=0.99\hsize]{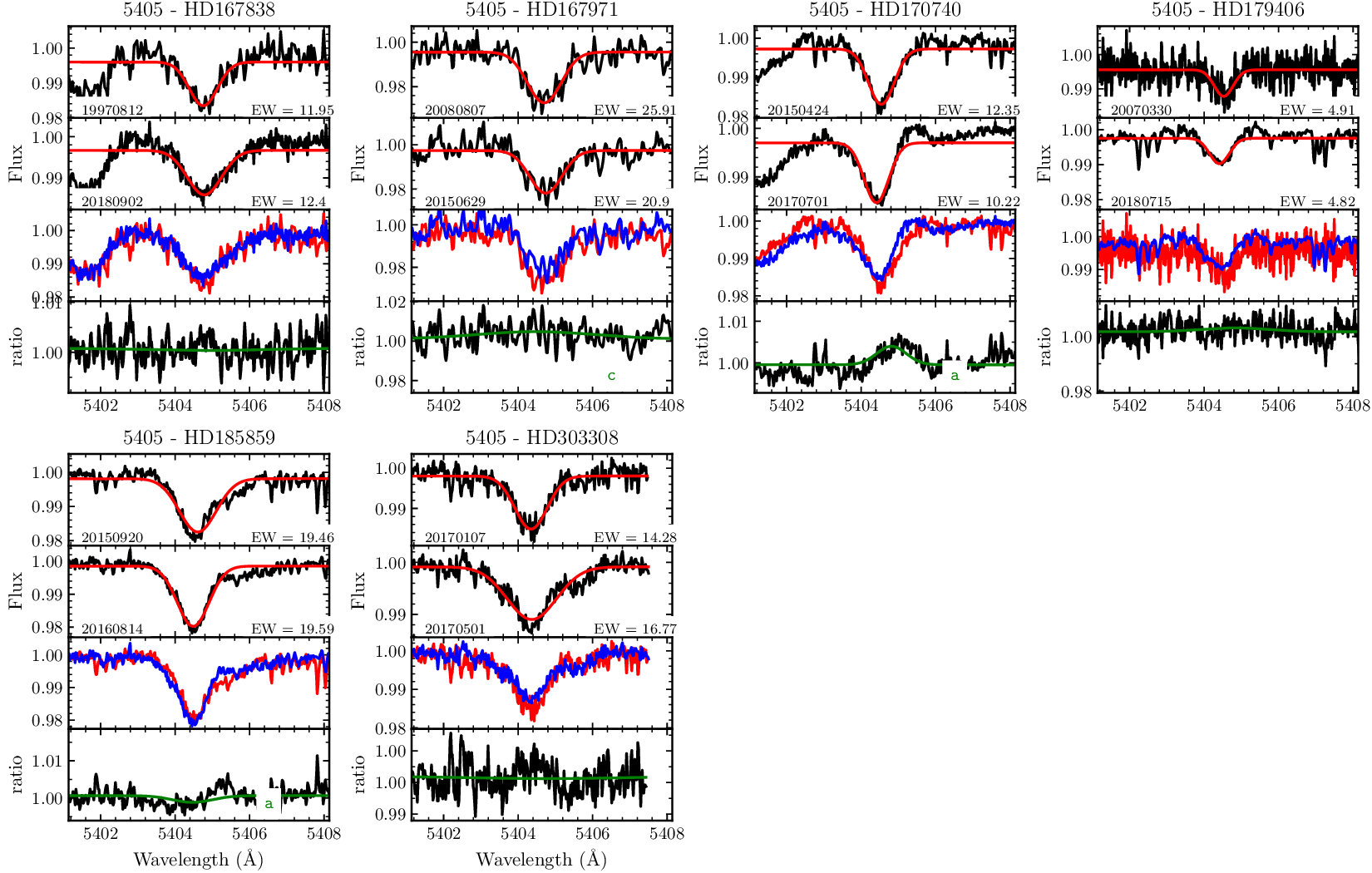}
    \caption{The same as \ref{plt-dib-var1}}
    \label{plt-dib-var27}
\end{figure*}

%%%%%%%%%%%%%%%%%%
%%%%%%%%%%%%%%%%%%
\begin{figure*}[ht!]
    \centering
    \includegraphics[width=0.99\hsize]{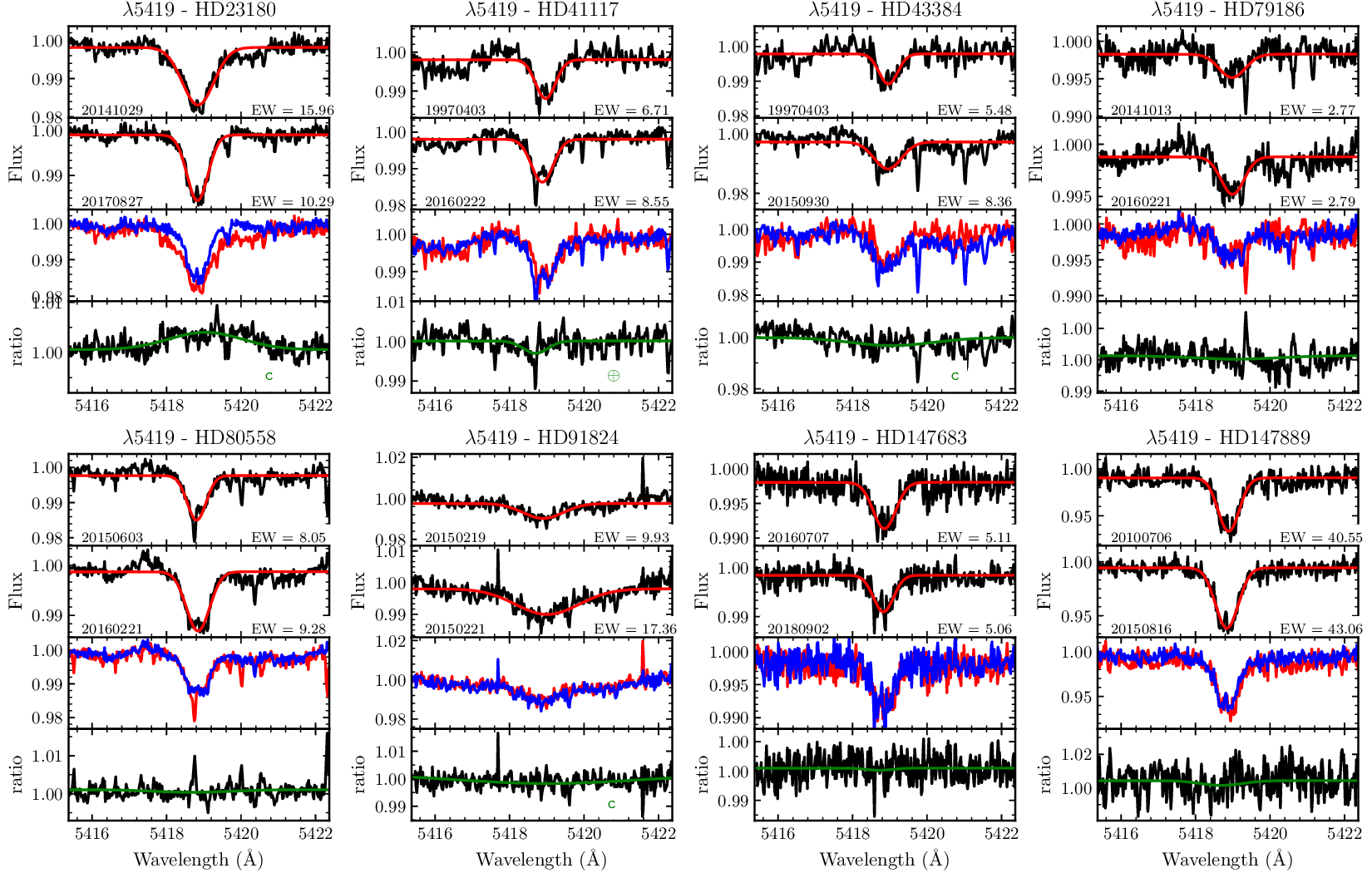}
    \caption{The same as \ref{plt-dib-var1}}
    \label{plt-dib-var28}
\end{figure*}

\begin{figure*}[ht!]
    \centering
    \includegraphics[width=0.99\hsize]{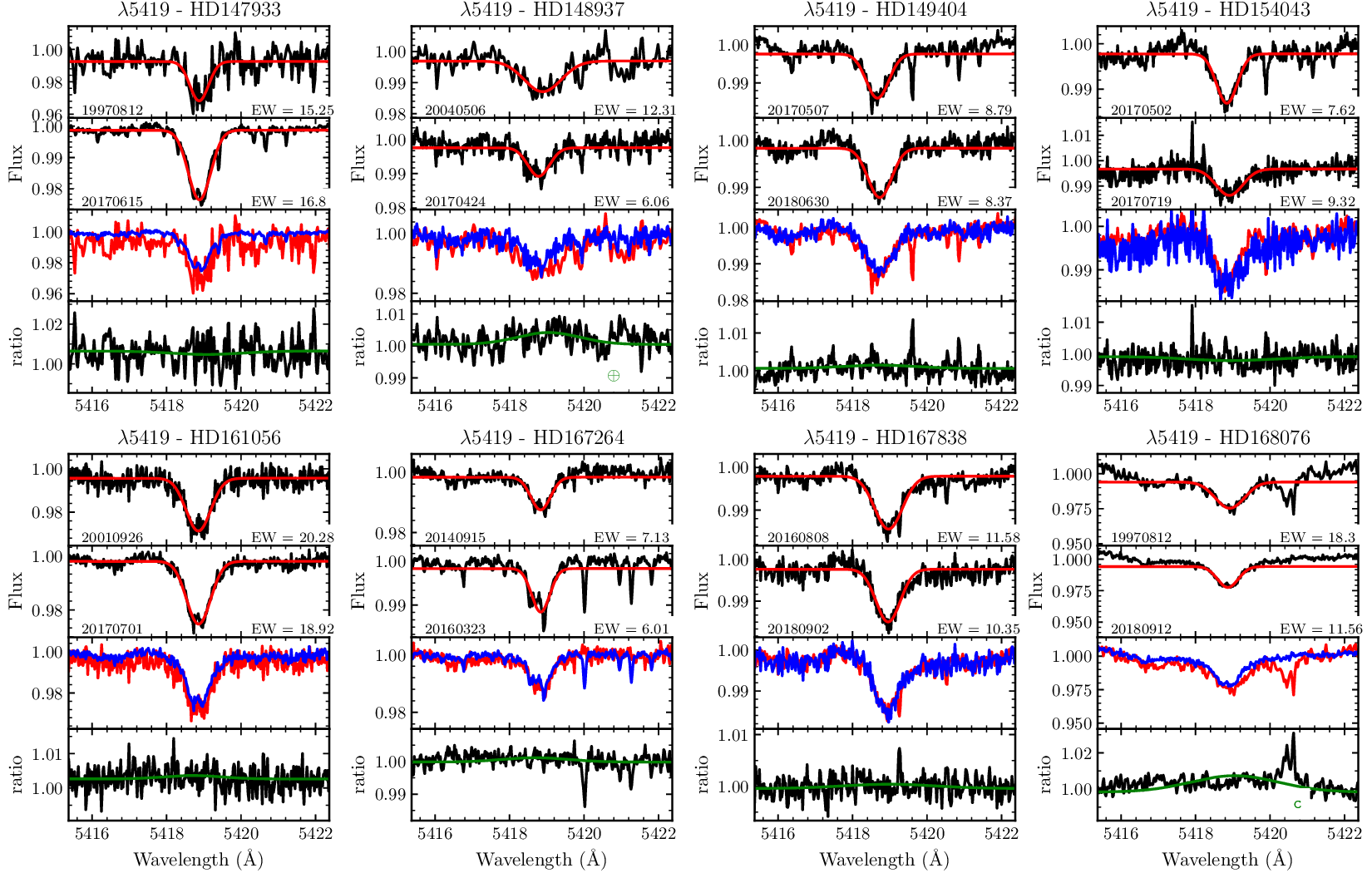}
    \caption{The same as \ref{plt-dib-var1}}
    \label{plt-dib-var29}
\end{figure*}

\begin{figure*}[ht!]
    \centering
    \includegraphics[width=0.99\hsize]{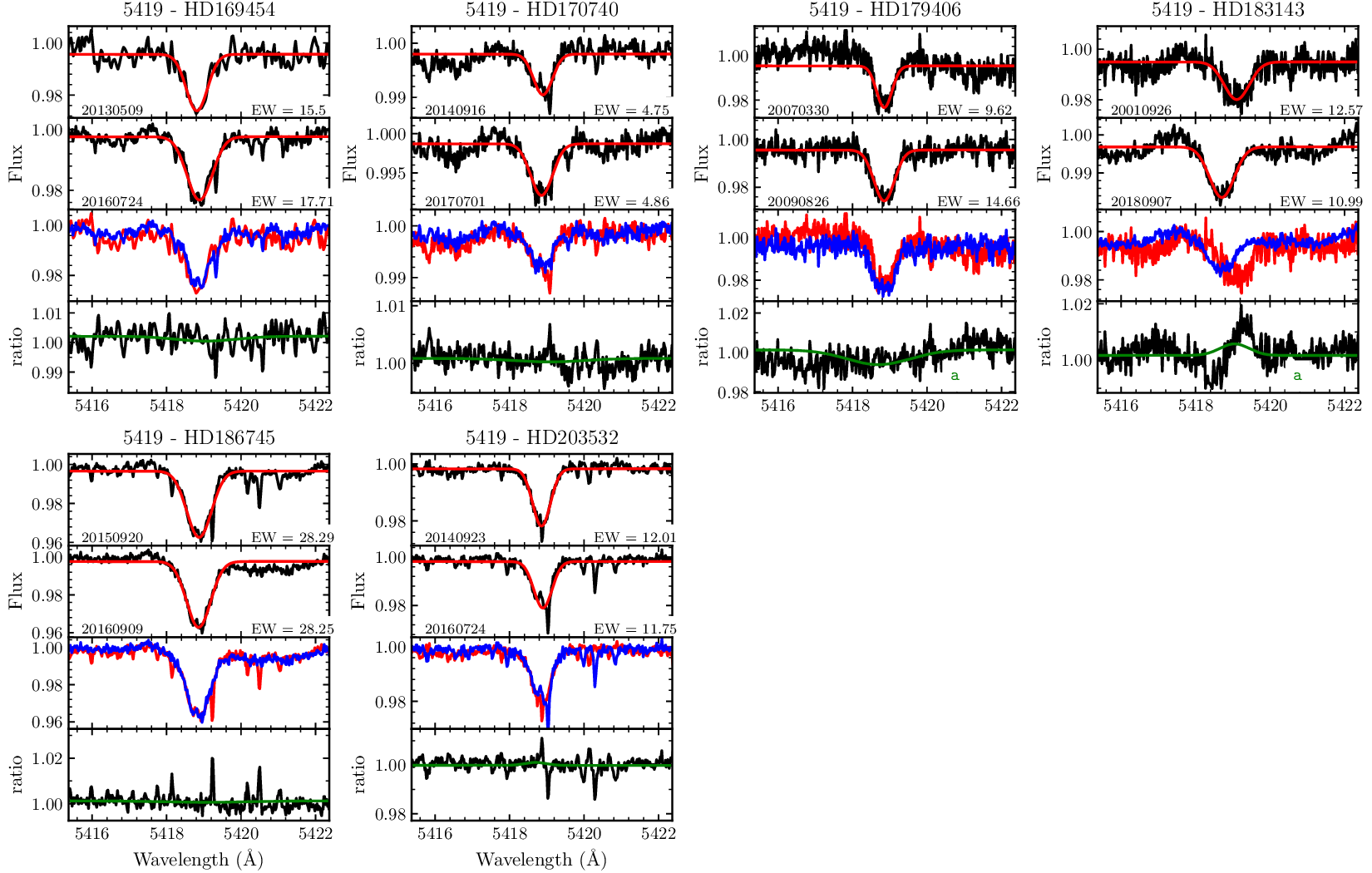}
    \caption{The same as \ref{plt-dib-var1}}
    \label{plt-dib-var30}
\end{figure*}

%%%%%%%%%%%%%%%%%%
%%%%%%%%%%%%%%%%%%
\begin{figure*}[ht!]
    \centering
    \includegraphics[width=0.99\hsize]{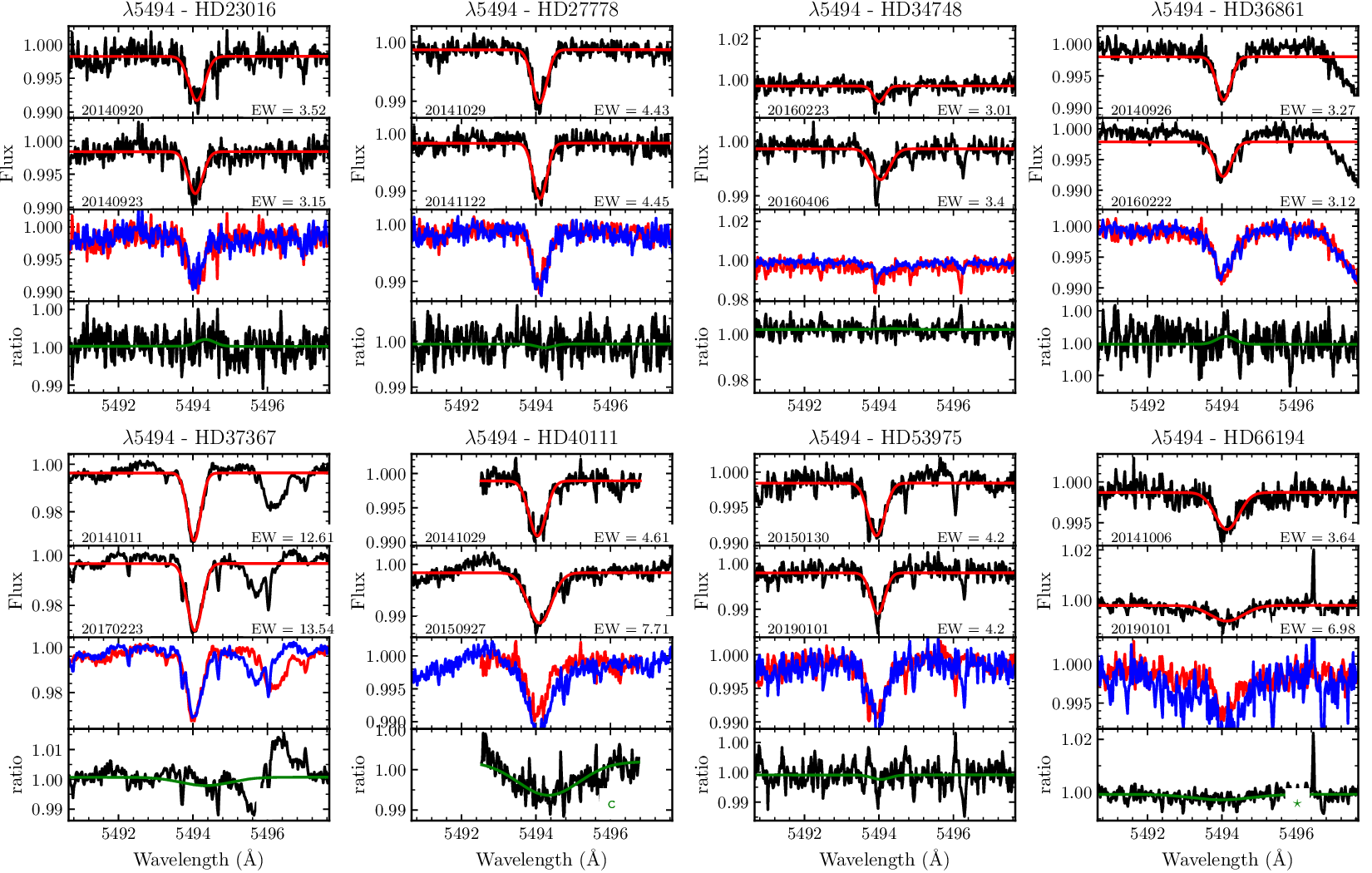}
    \caption{The same as \ref{plt-dib-var1}}
    \label{plt-dib-var31}
\end{figure*}

\begin{figure*}[ht!]
    \centering
    \includegraphics[width=0.99\hsize]{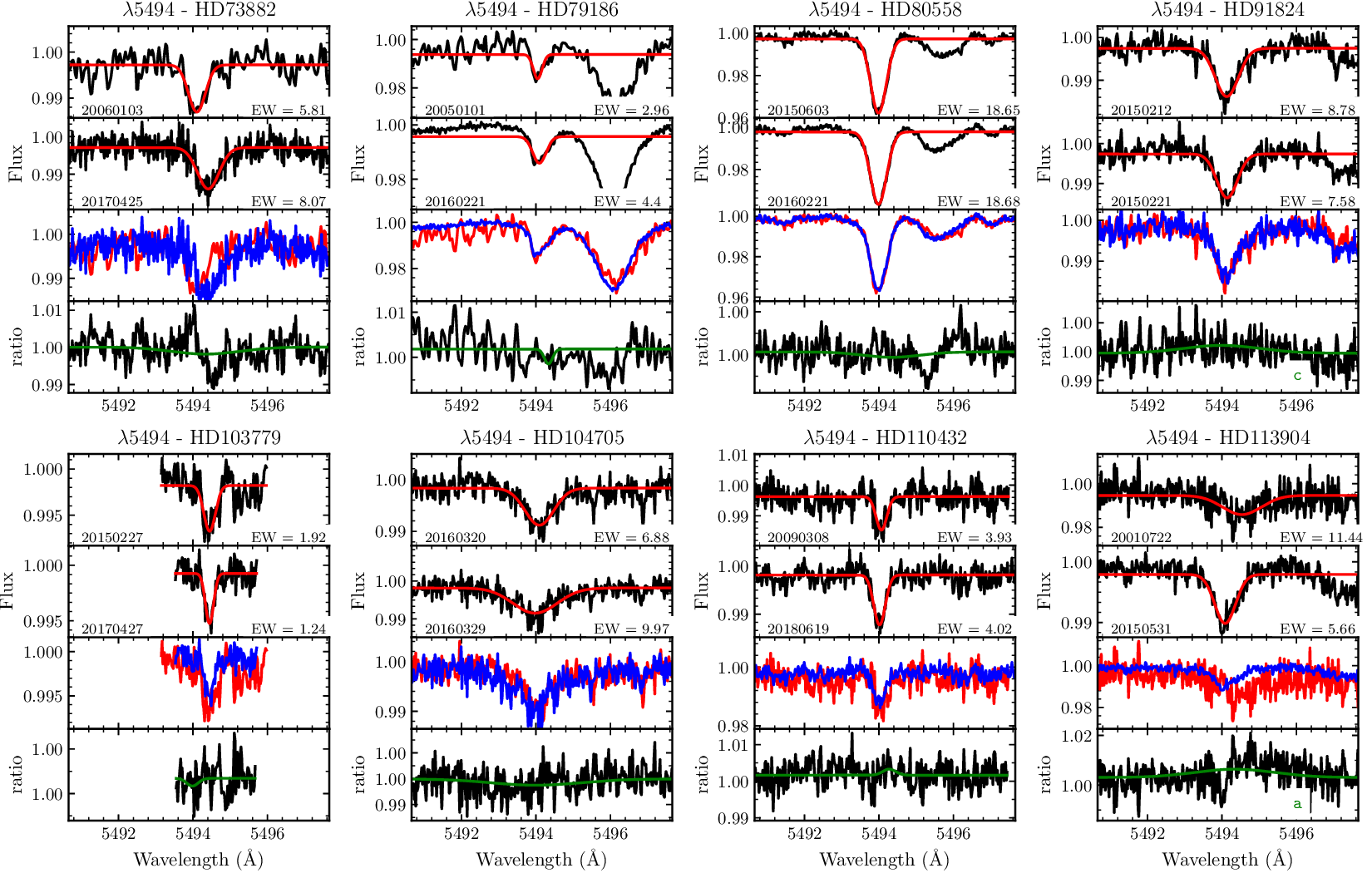}
    \caption{The same as \ref{plt-dib-var1}}
    \label{plt-dib-var32}
\end{figure*}

\begin{figure*}[ht!]
    \centering
    \includegraphics[width=0.99\hsize]{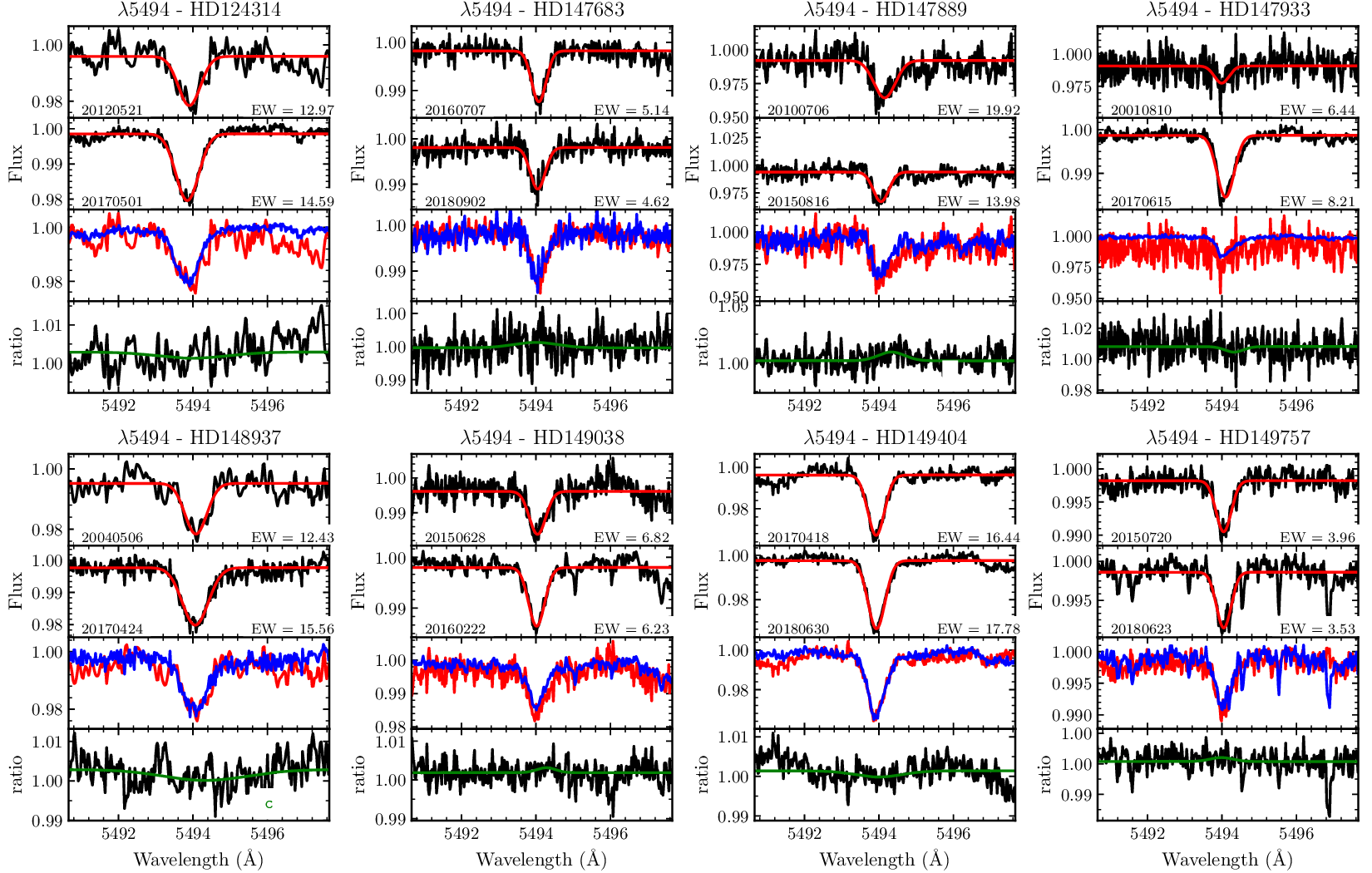}
    \caption{The same as \ref{plt-dib-var1}}
    \label{plt-dib-var33}
\end{figure*}

\begin{figure*}[ht!]
    \centering
    \includegraphics[width=0.99\hsize]{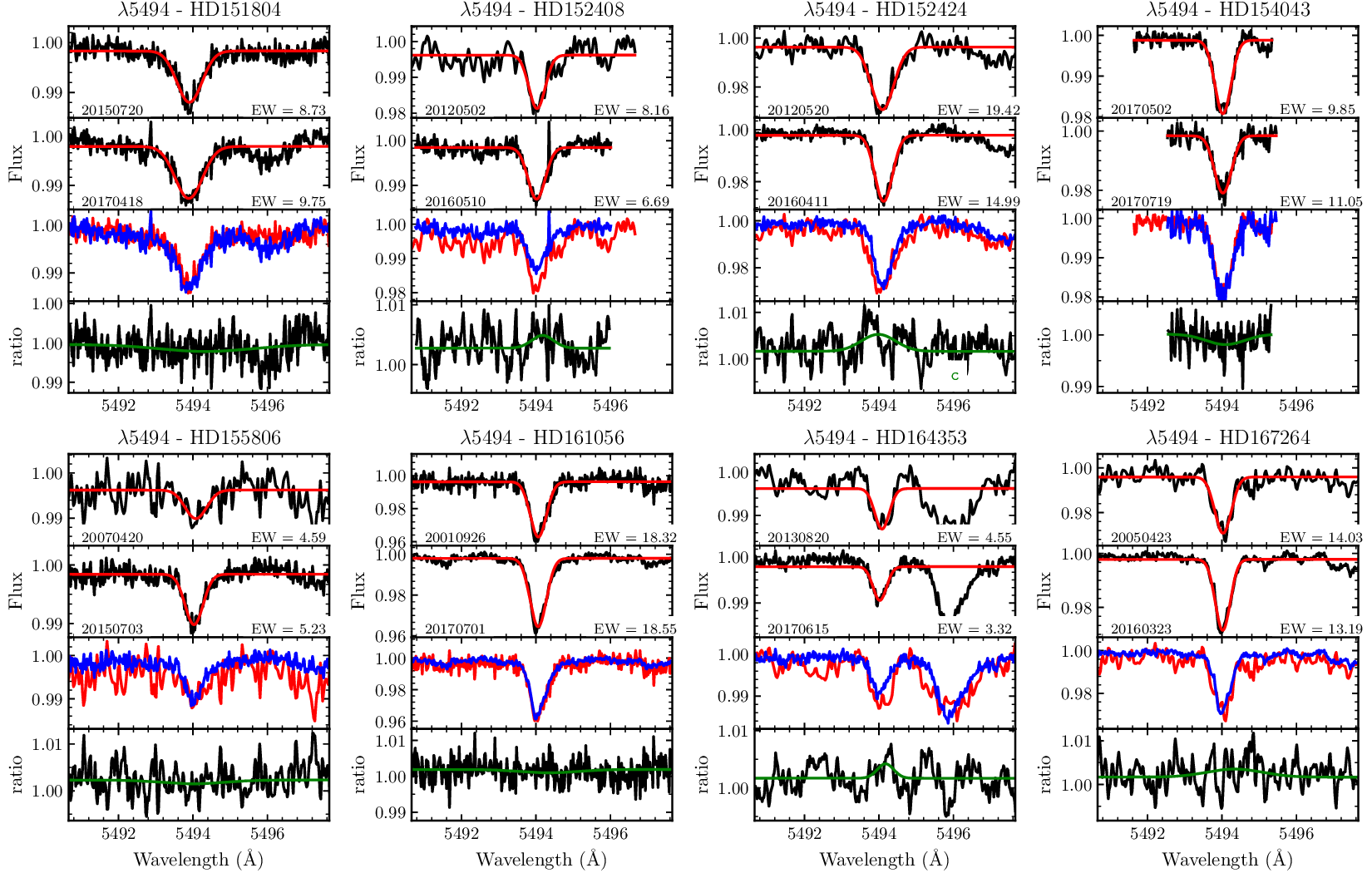}
    \caption{The same as \ref{plt-dib-var1}}
    \label{plt-dib-var34}
\end{figure*}

\begin{figure*}[ht!]
    \centering
    \includegraphics[width=0.99\hsize]{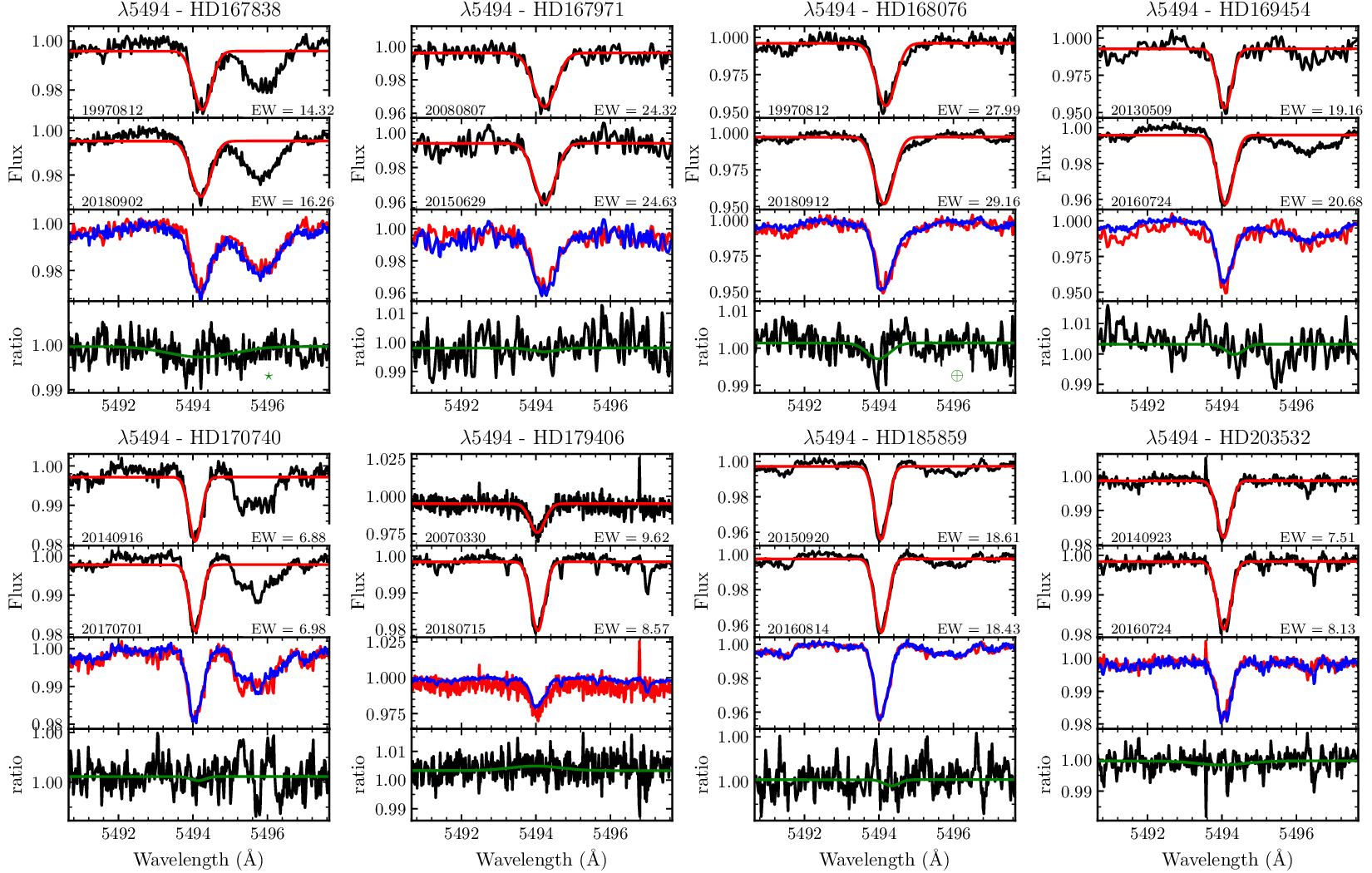}
    \caption{The same as \ref{plt-dib-var1}}
    \label{plt-dib-var35}
\end{figure*}

\begin{figure*}[ht!]
    \centering
    \includegraphics[width=0.99\hsize]{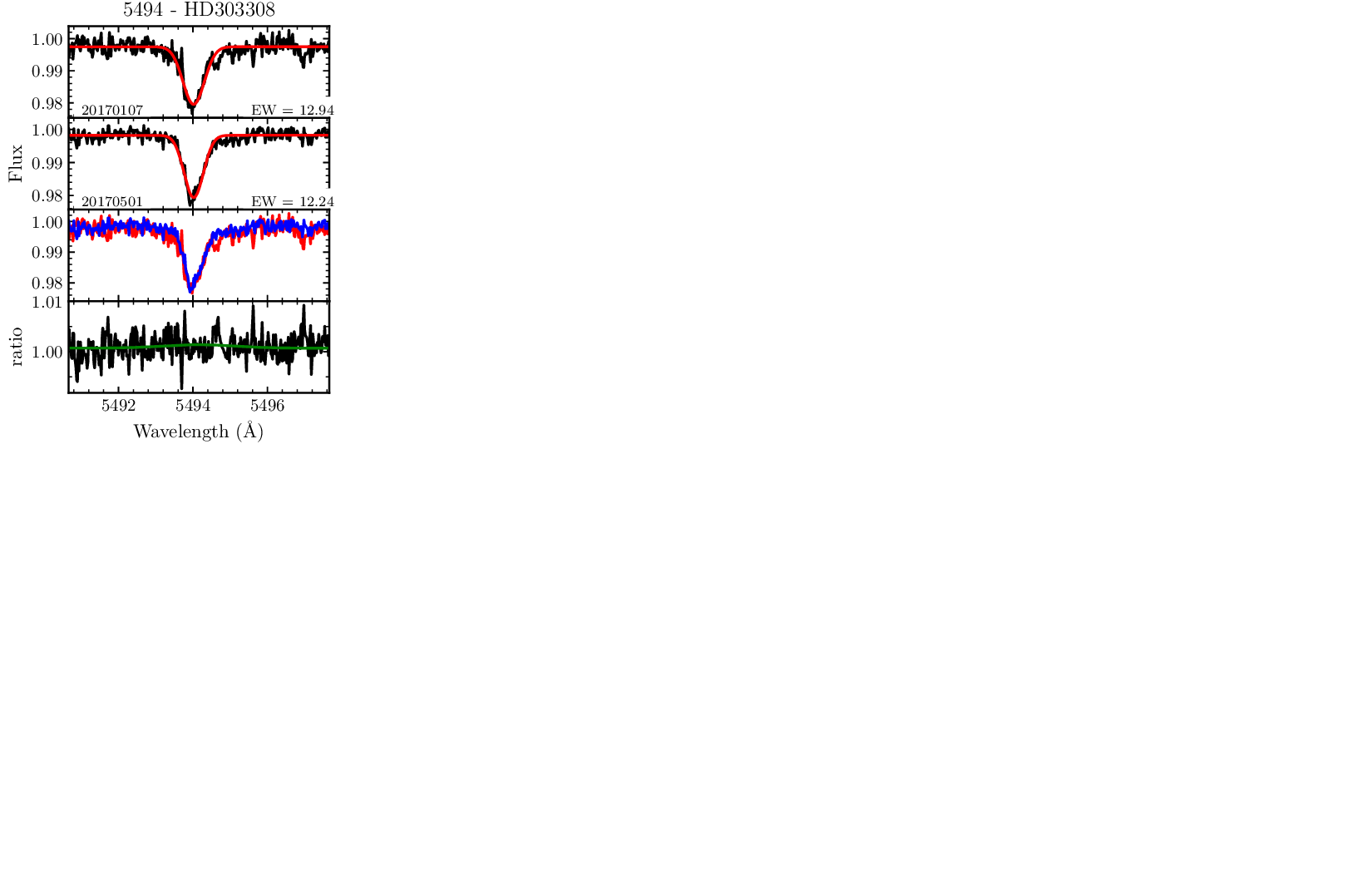}
    \caption{The same as \ref{plt-dib-var1}}
    \label{plt-dib-var36}
\end{figure*}

%%%%%%%%%%%%%%%%%%
%%%%%%%%%%%%%%%%%%
\begin{figure*}[ht!]
    \centering
    \includegraphics[width=0.99\hsize]{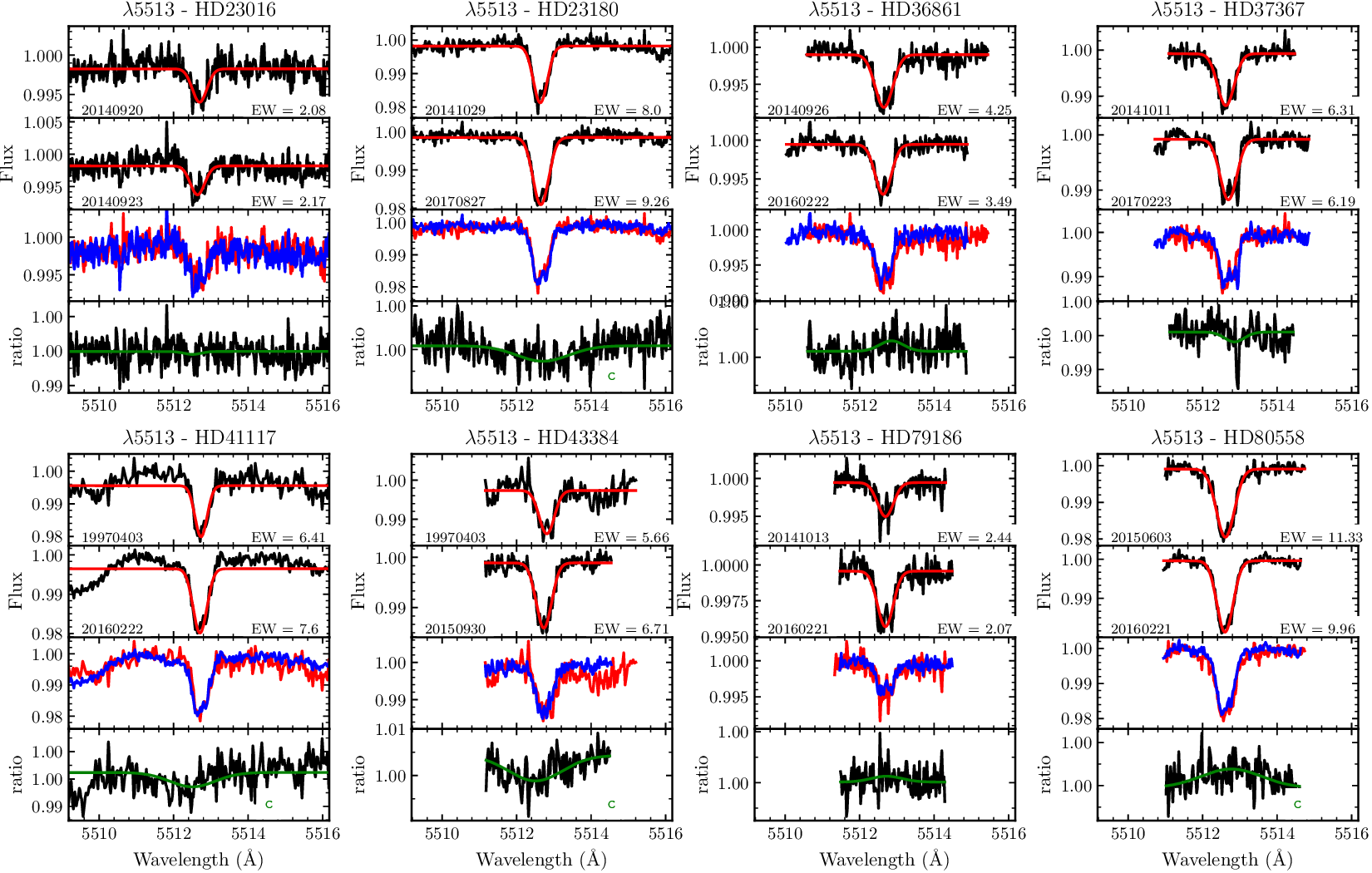}
    \caption{The same as \ref{plt-dib-var1}}
    \label{plt-dib-var37}
\end{figure*}

\begin{figure*}[ht!]
    \centering
    \includegraphics[width=0.99\hsize]{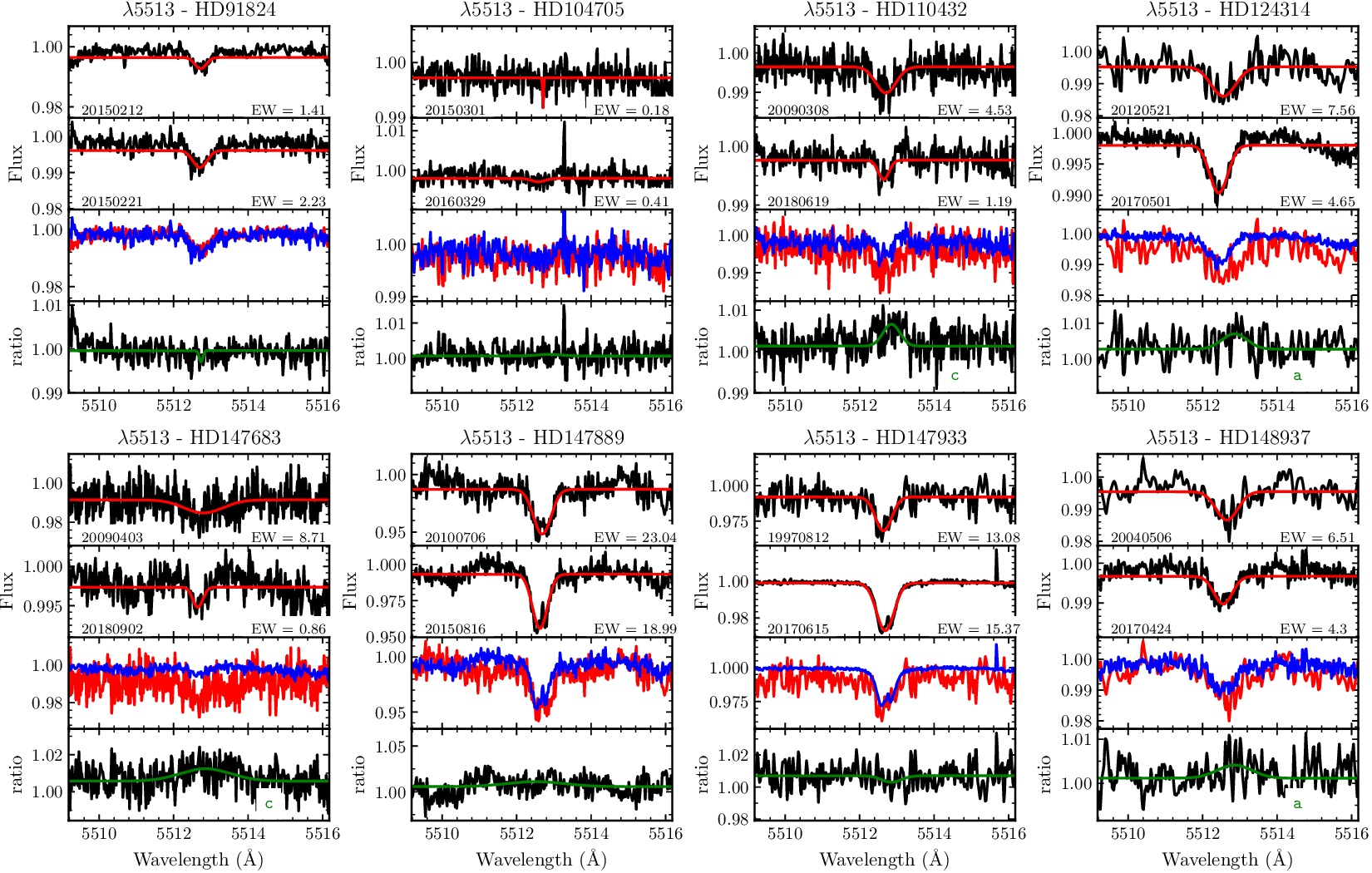}
    \caption{The same as \ref{plt-dib-var1}}
    \label{plt-dib-var38}
\end{figure*}

\begin{figure*}[ht!]
    \centering
    \includegraphics[width=0.99\hsize]{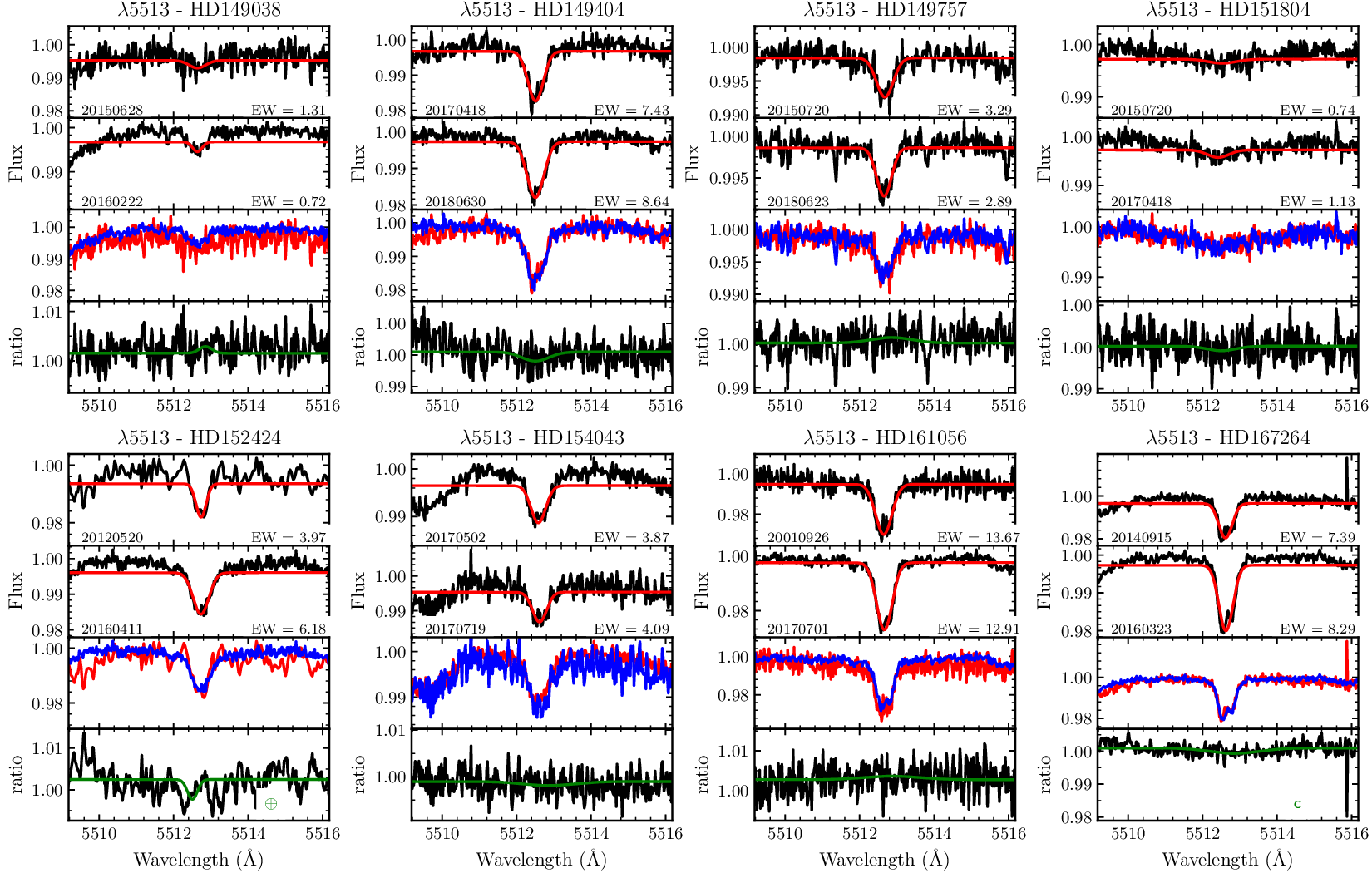}
    \caption{The same as \ref{plt-dib-var1}}
    \label{plt-dib-var39}
\end{figure*}

\begin{figure*}[ht!]
    \centering
    \includegraphics[width=0.99\hsize]{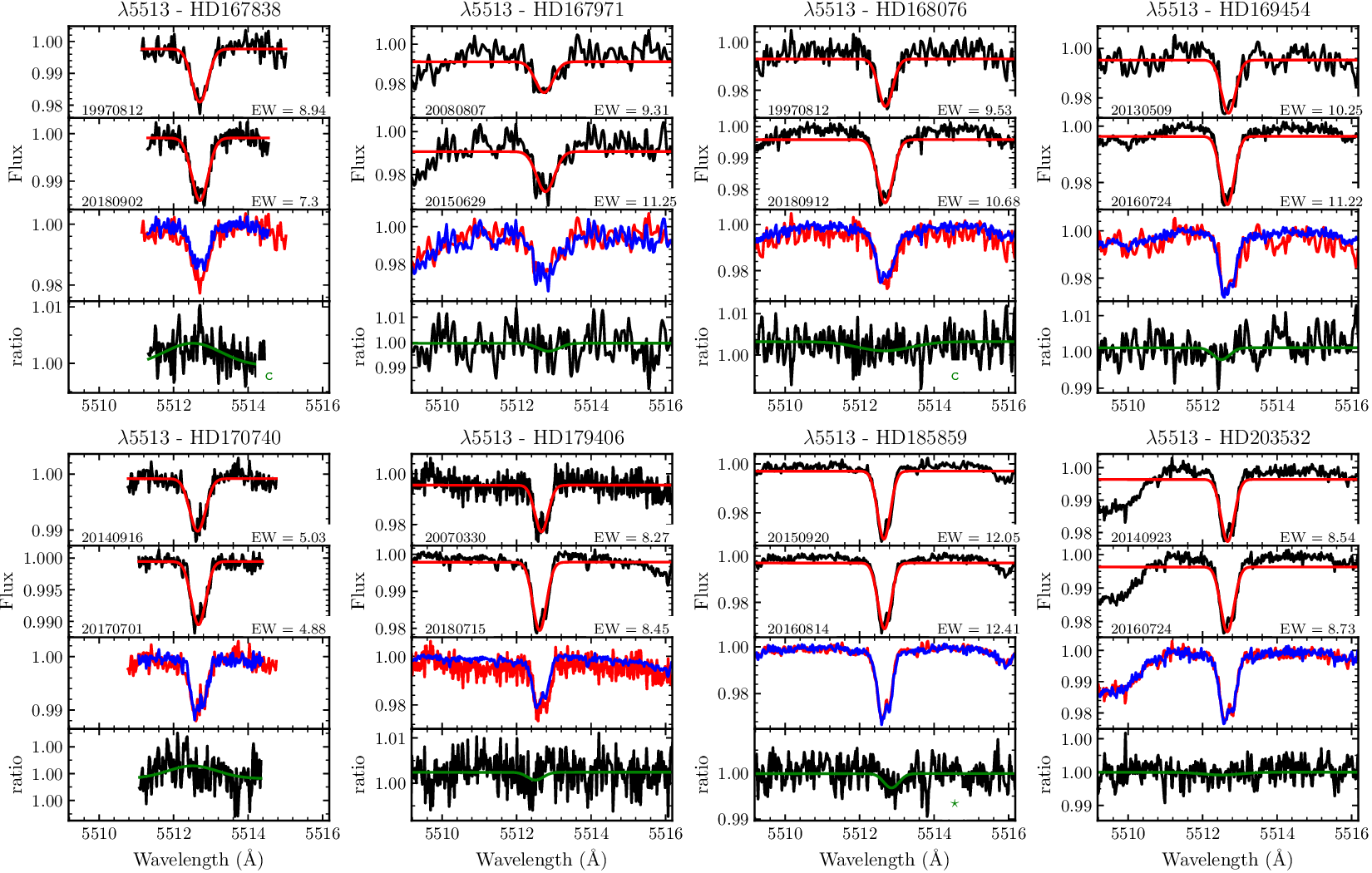}
    \caption{The same as \ref{plt-dib-var1}}
    \label{plt-dib-var40}
\end{figure*}

\begin{figure*}[ht!]
    \centering
    \includegraphics[width=0.99\hsize]{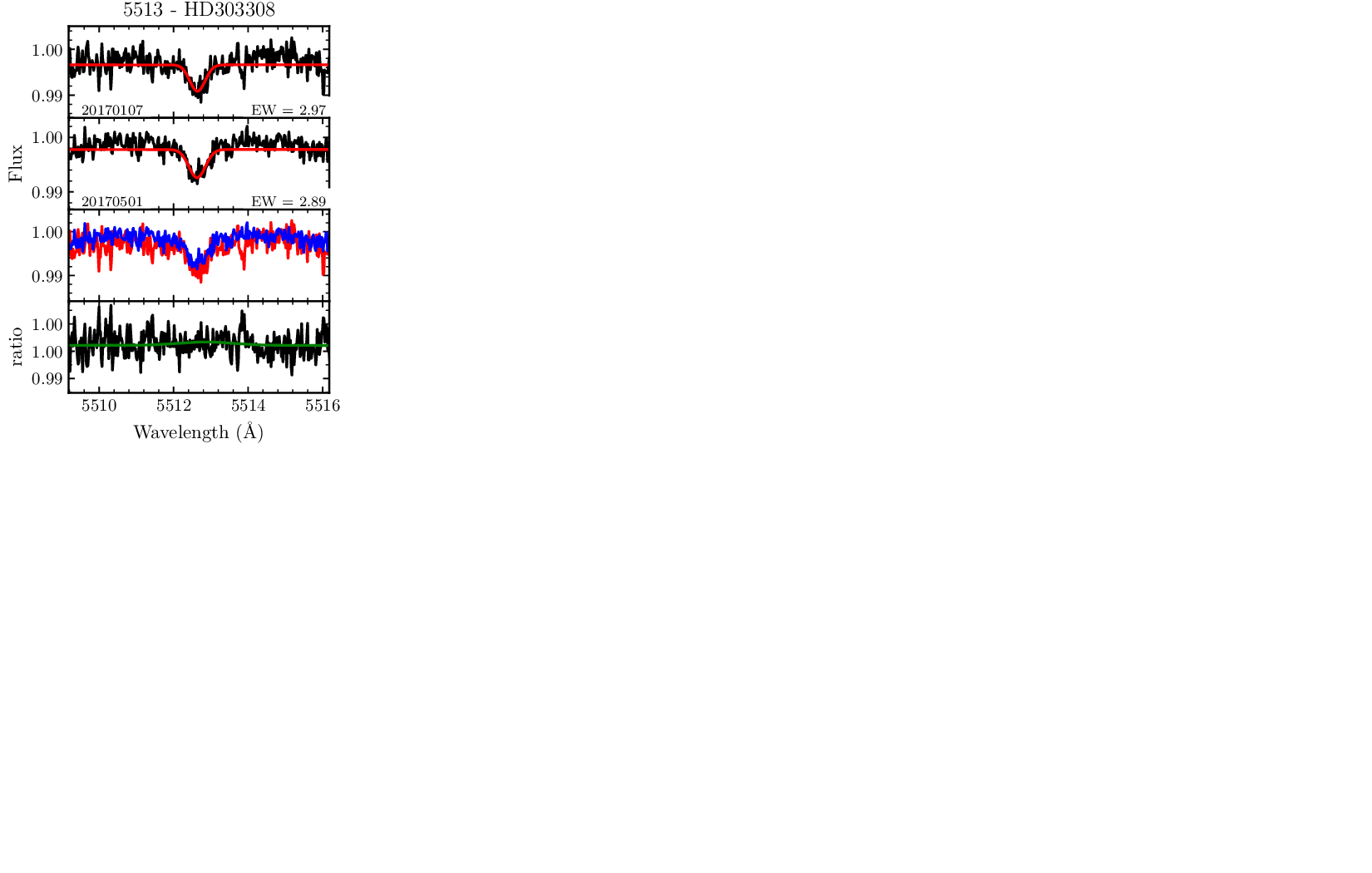}
    \caption{The same as \ref{plt-dib-var1}}
    \label{plt-dib-var41}
\end{figure*}

%%%%%%%%%%%%%%%%%%
%%%%%%%%%%%%%%%%%%
\begin{figure*}[ht!]
    \centering
    \includegraphics[width=0.99\hsize]{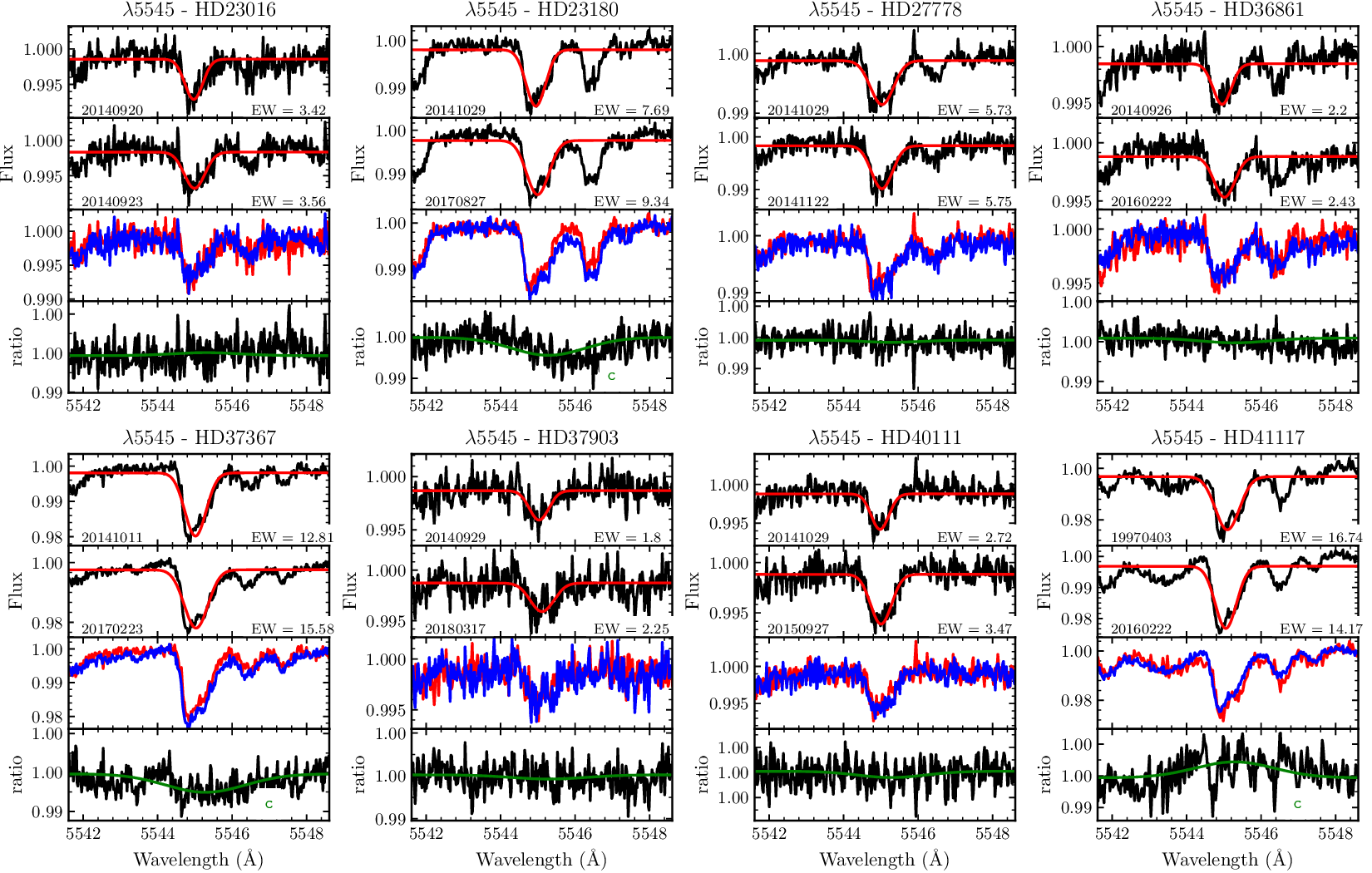}
    \caption{The same as \ref{plt-dib-var1}}
    \label{plt-dib-var42}
\end{figure*}

\begin{figure*}[ht!]
    \centering
    \includegraphics[width=0.99\hsize]{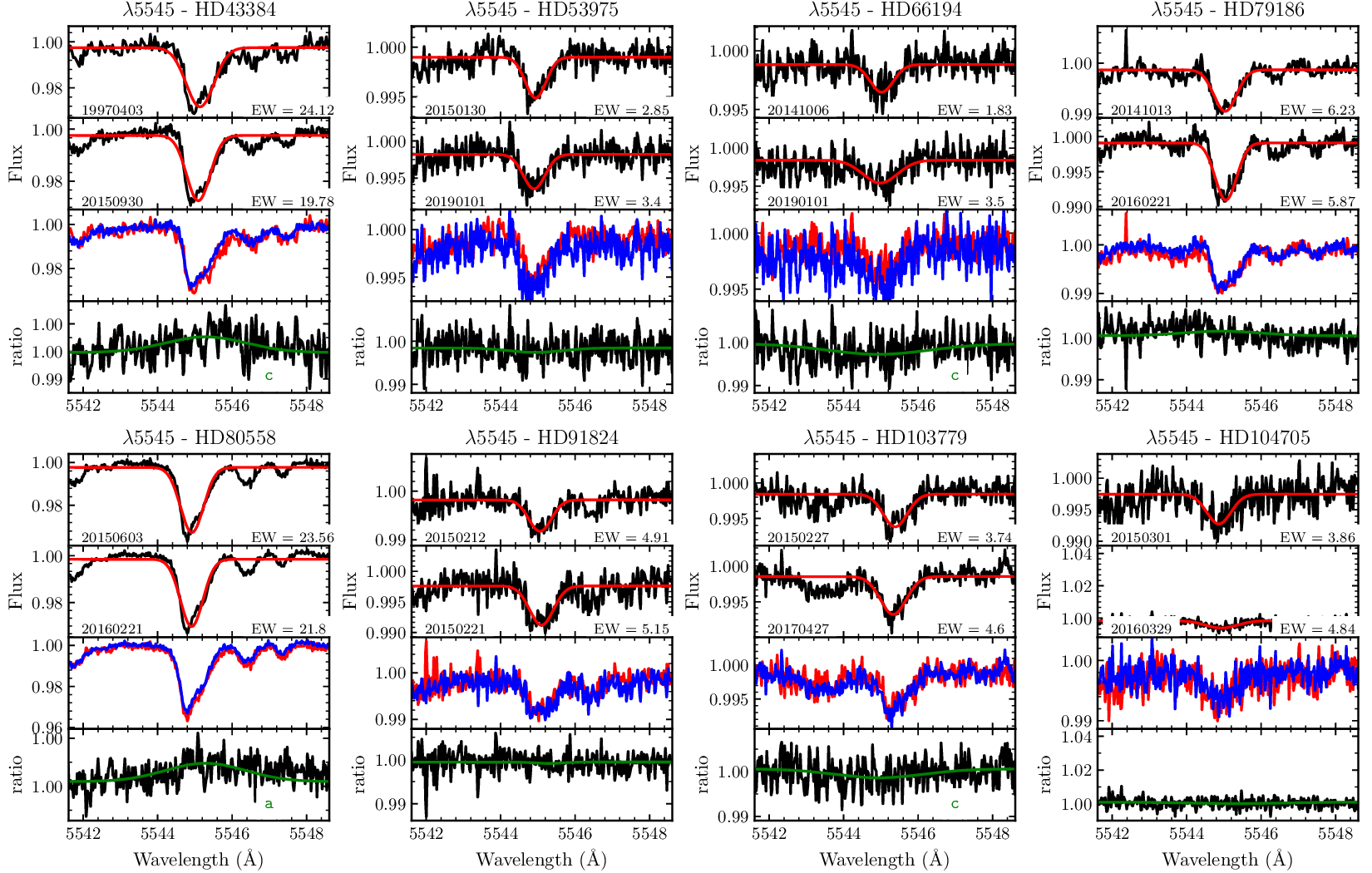}
    \caption{The same as \ref{plt-dib-var1}}
    \label{plt-dib-var43}
\end{figure*}

\begin{figure*}[ht!]
    \centering
    \includegraphics[width=0.99\hsize]{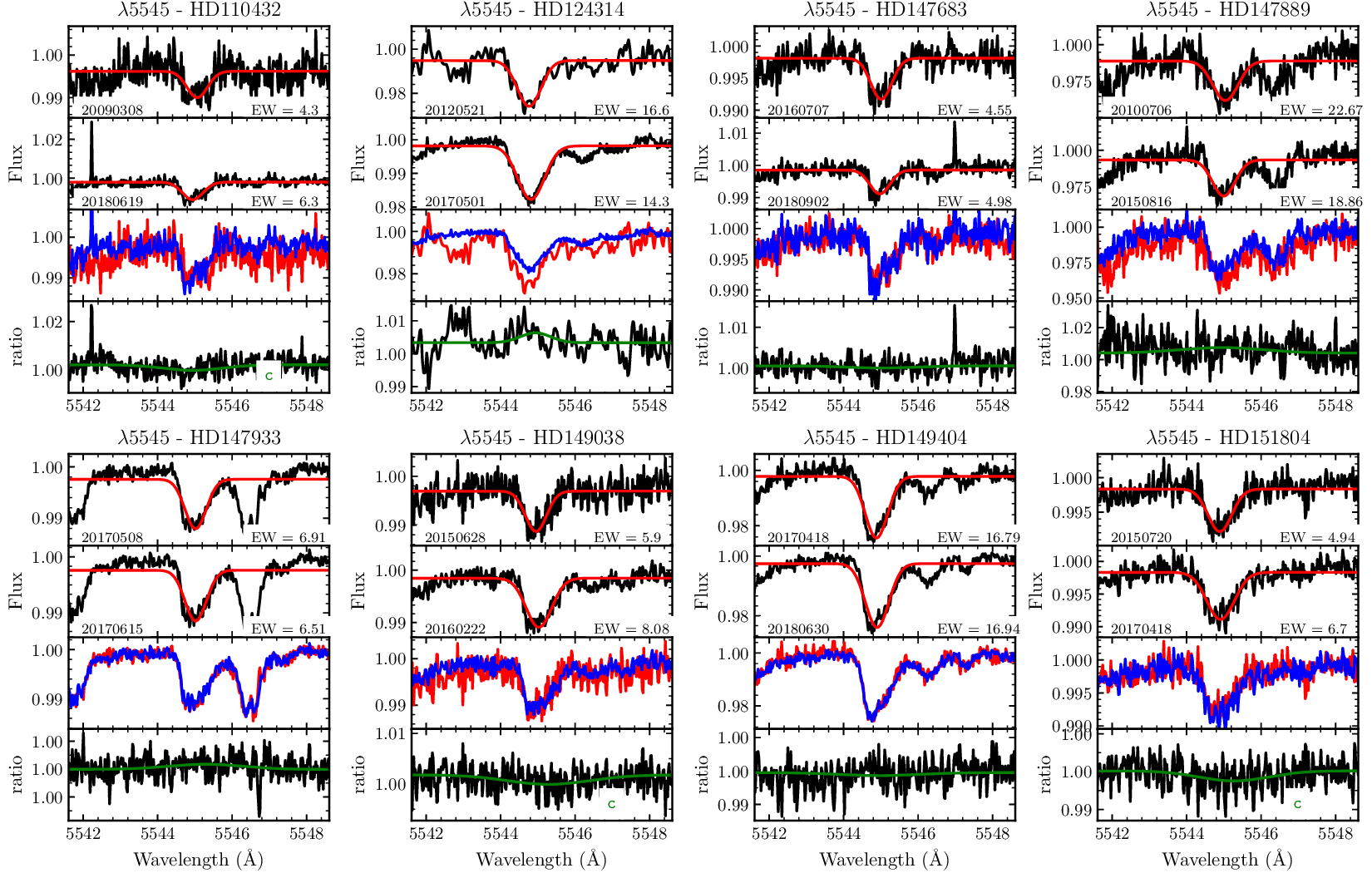}
    \caption{The same as \ref{plt-dib-var1}}
    \label{plt-dib-var44}
\end{figure*}

\begin{figure*}[ht!]
    \centering
    \includegraphics[width=0.99\hsize]{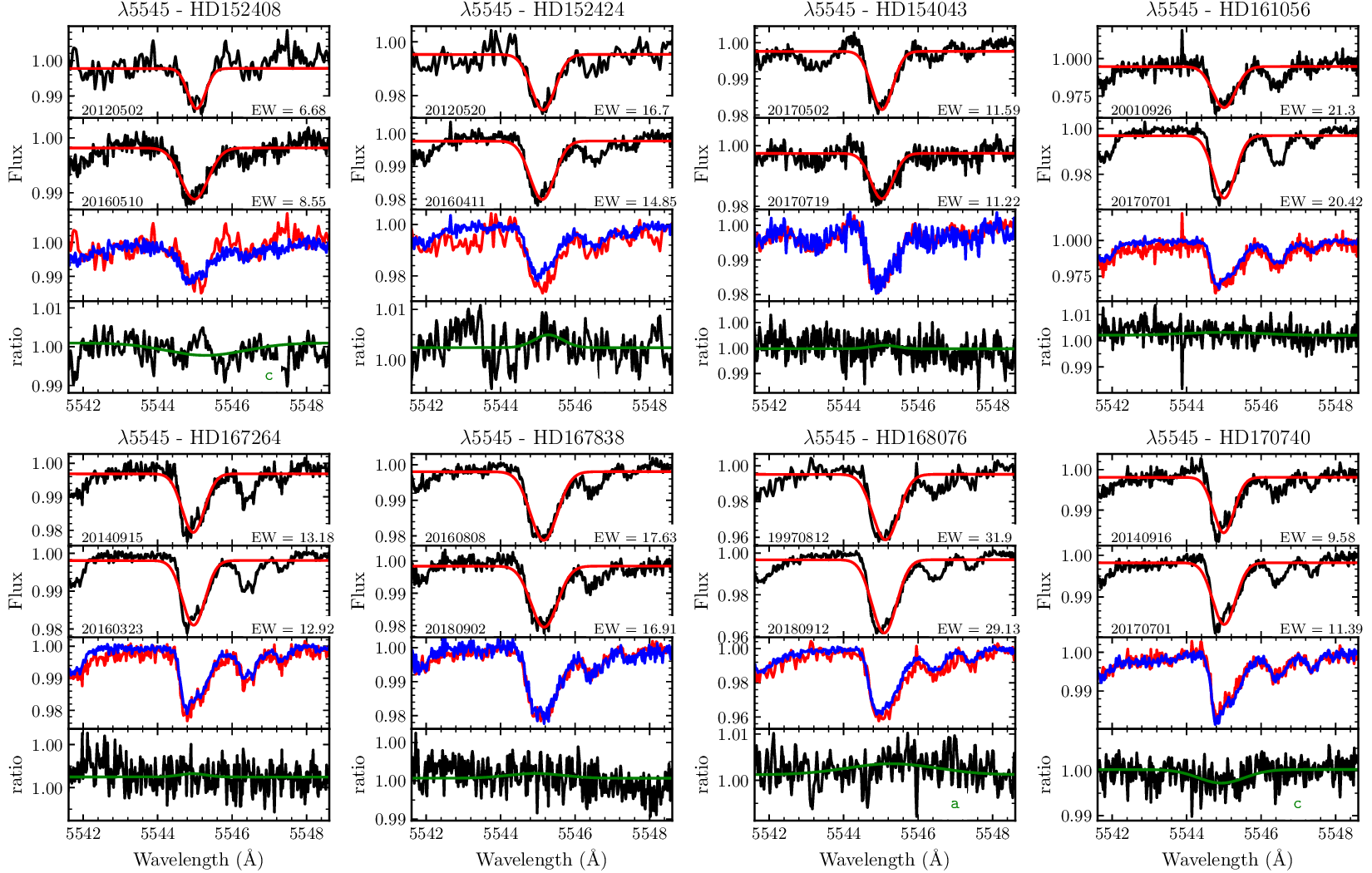}
    \caption{The same as \ref{plt-dib-var1}}
    \label{plt-dib-var45}
\end{figure*}

\begin{figure*}[ht!]
    \centering
    \includegraphics[width=0.99\hsize]{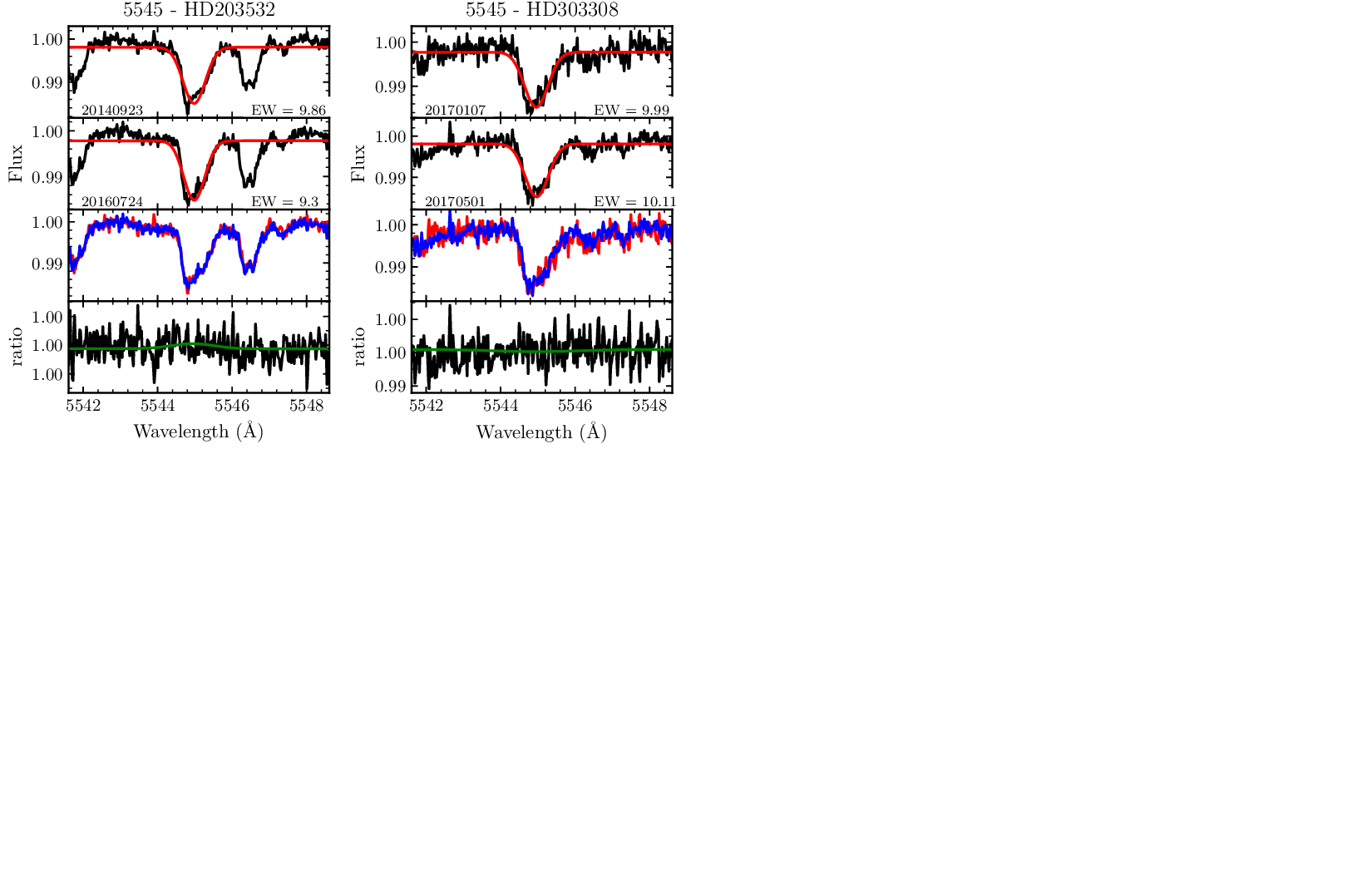}
    \caption{The same as \ref{plt-dib-var1}}
    \label{plt-dib-var46}
\end{figure*}

%%%%%%%%%%%%%%%%%%
%%%%%%%%%%%%%%%%%%
\begin{figure*}[ht!]
    \centering
    \includegraphics[width=0.99\hsize]{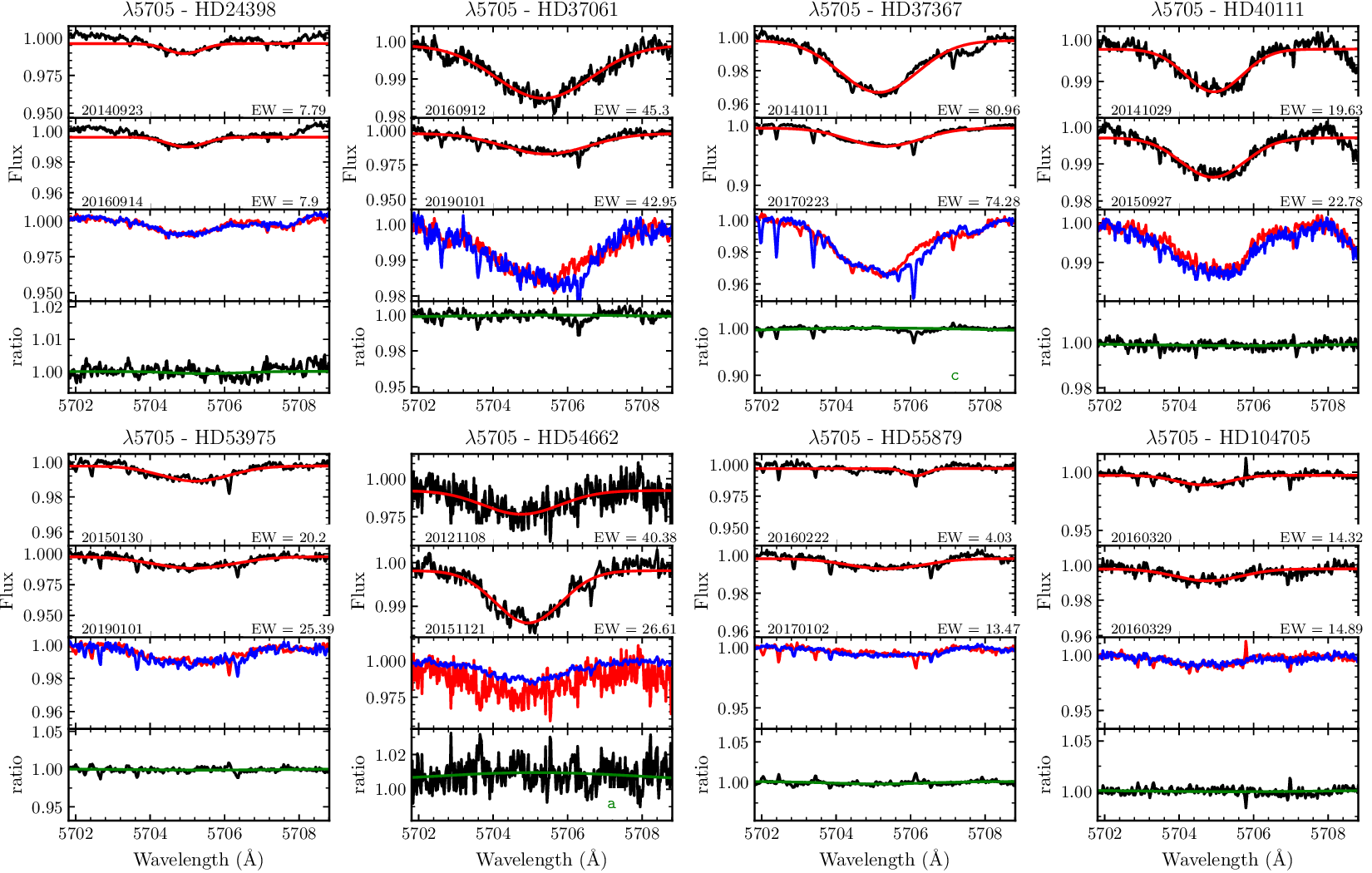}
    \caption{The same as \ref{plt-dib-var1}}
    \label{plt-dib-var47}
\end{figure*}

\begin{figure*}[ht!]
    \centering
    \includegraphics[width=0.99\hsize]{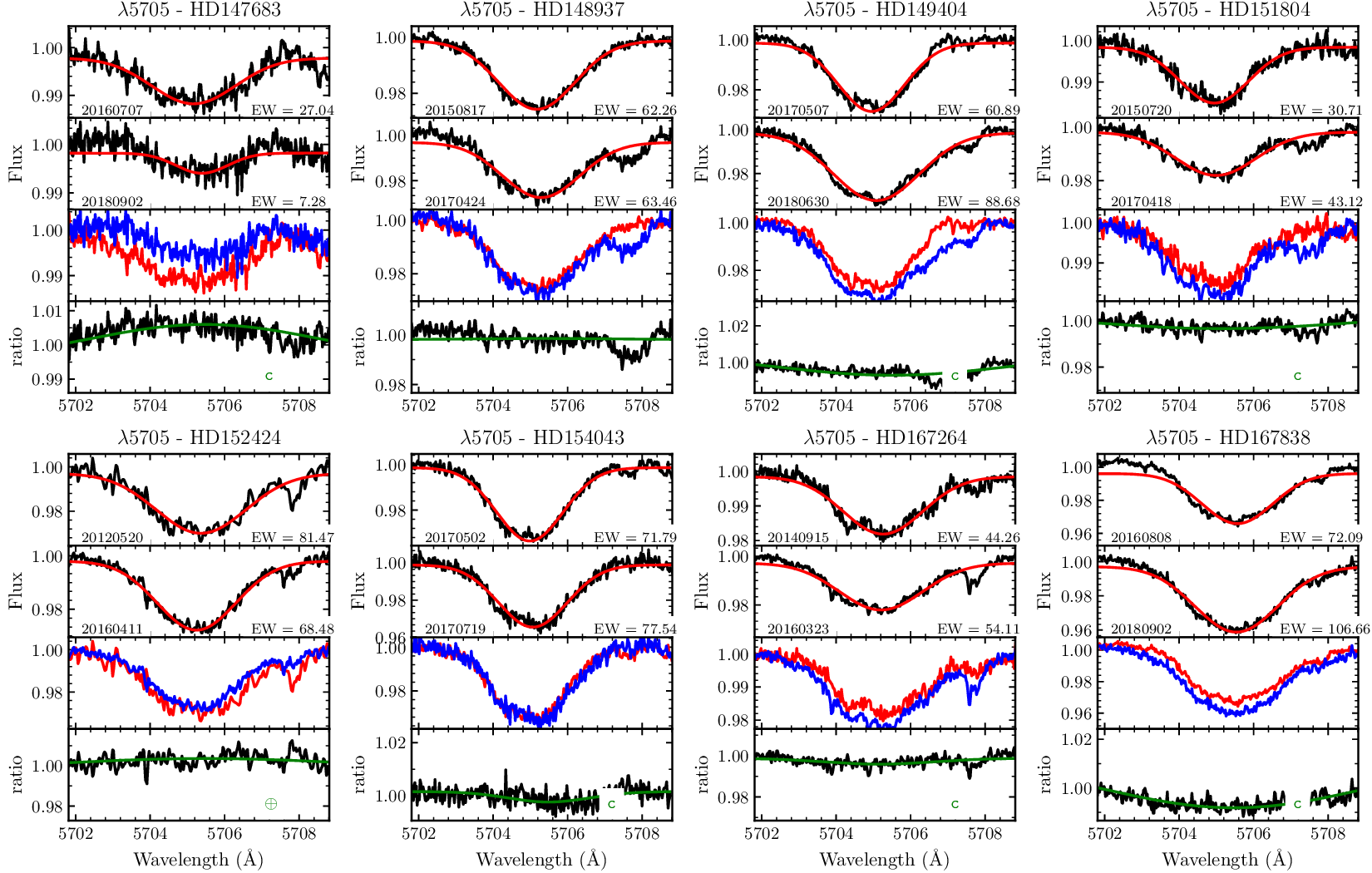}
    \caption{The same as \ref{plt-dib-var1}}
    \label{plt-dib-var48}
\end{figure*}

\begin{figure*}[ht!]
    \centering
    \includegraphics[width=0.99\hsize]{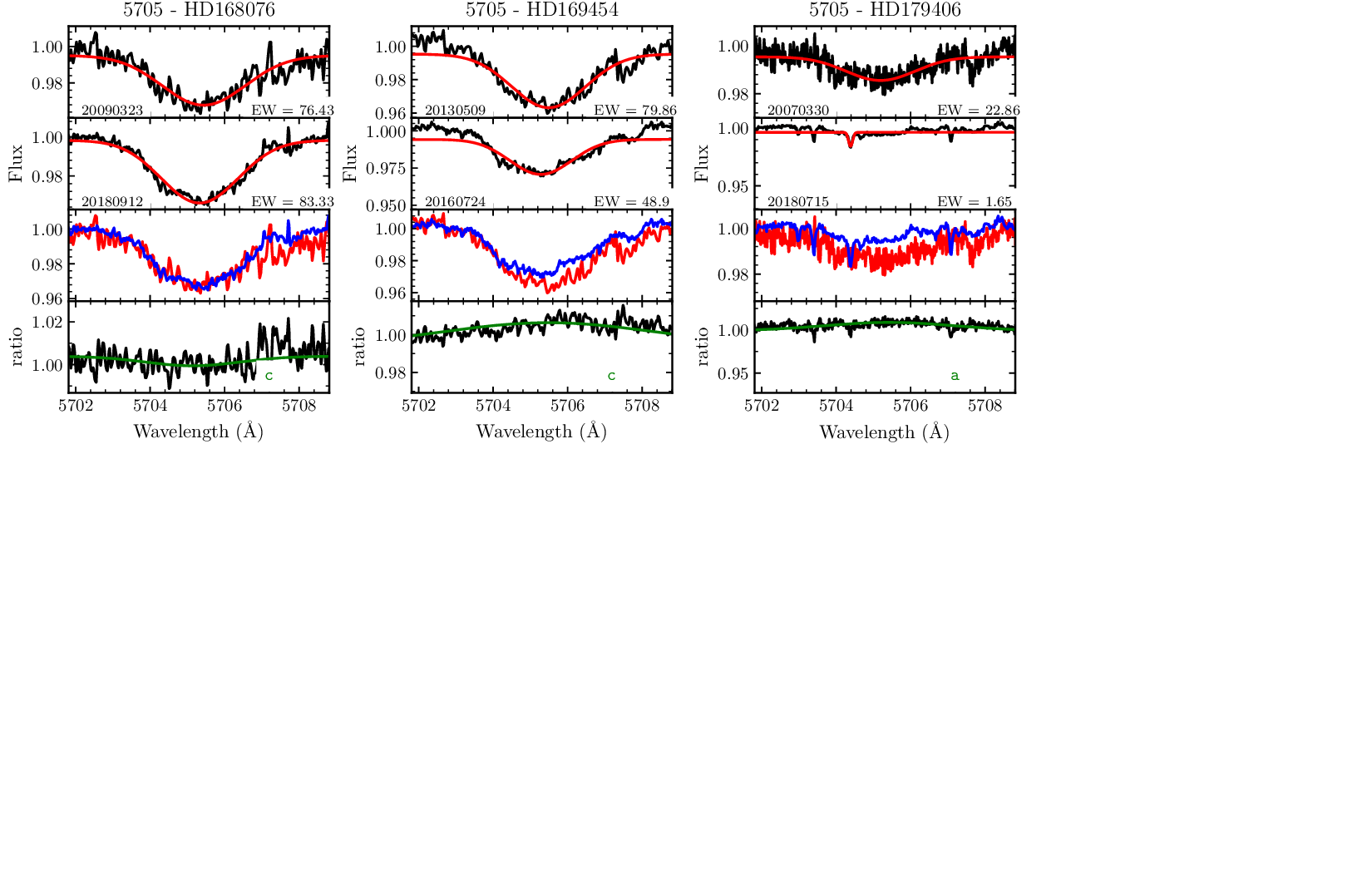}
    \caption{The same as \ref{plt-dib-var1}}
    \label{plt-dib-var49}
\end{figure*}

%%%%%%%%%%%%%%%%%%
%%%%%%%%%%%%%%%%%%
\begin{figure*}[ht!]
    \centering
    \includegraphics[width=0.99\hsize]{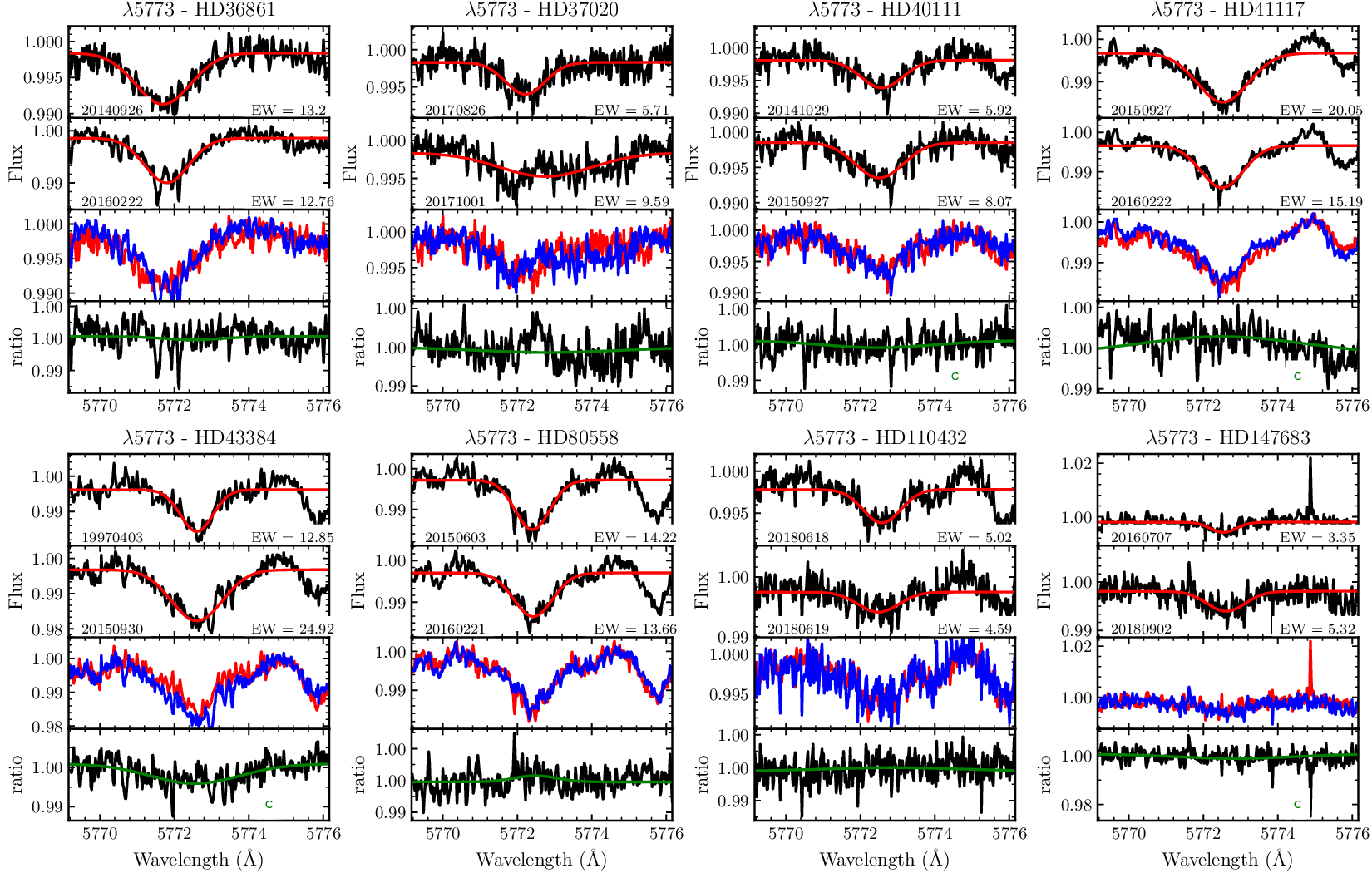}
    \caption{The same as \ref{plt-dib-var1}}
    \label{plt-dib-var50}
\end{figure*}

\begin{figure*}[ht!]
    \centering
    \includegraphics[width=0.99\hsize]{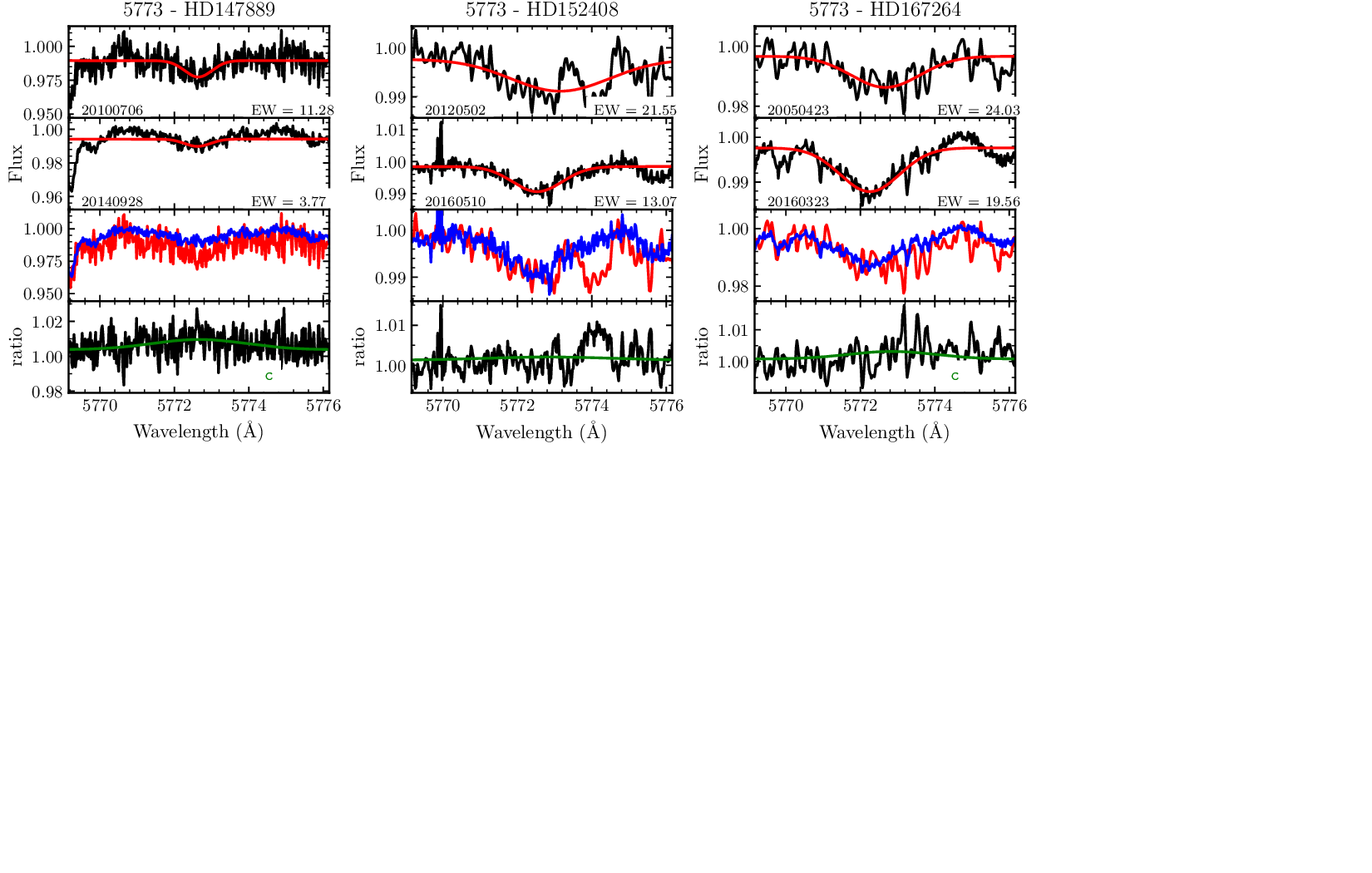}
    \caption{The same as \ref{plt-dib-var1}}
    \label{plt-dib-var51}
\end{figure*}

%%%%%%%%%%%%%%%%%%
%%%%%%%%%%%%%%%%%%
\begin{figure*}[ht!]
    \centering
    \includegraphics[width=0.99\hsize]{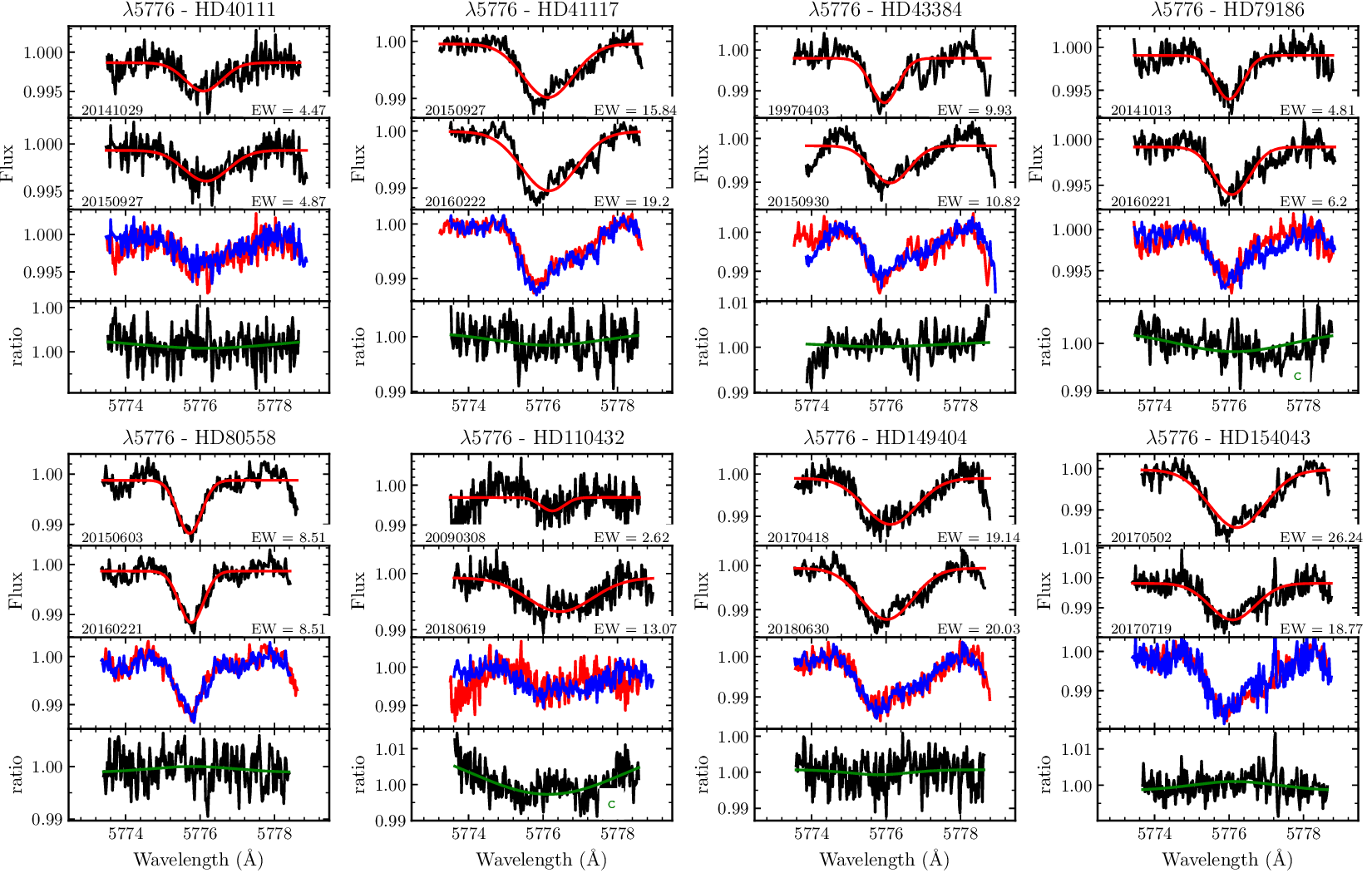}
    \caption{The same as \ref{plt-dib-var1}}
    \label{plt-dib-var52}
\end{figure*}

\clearpage
\begin{figure*}[ht!]
    \centering
    \includegraphics[width=0.99\hsize]{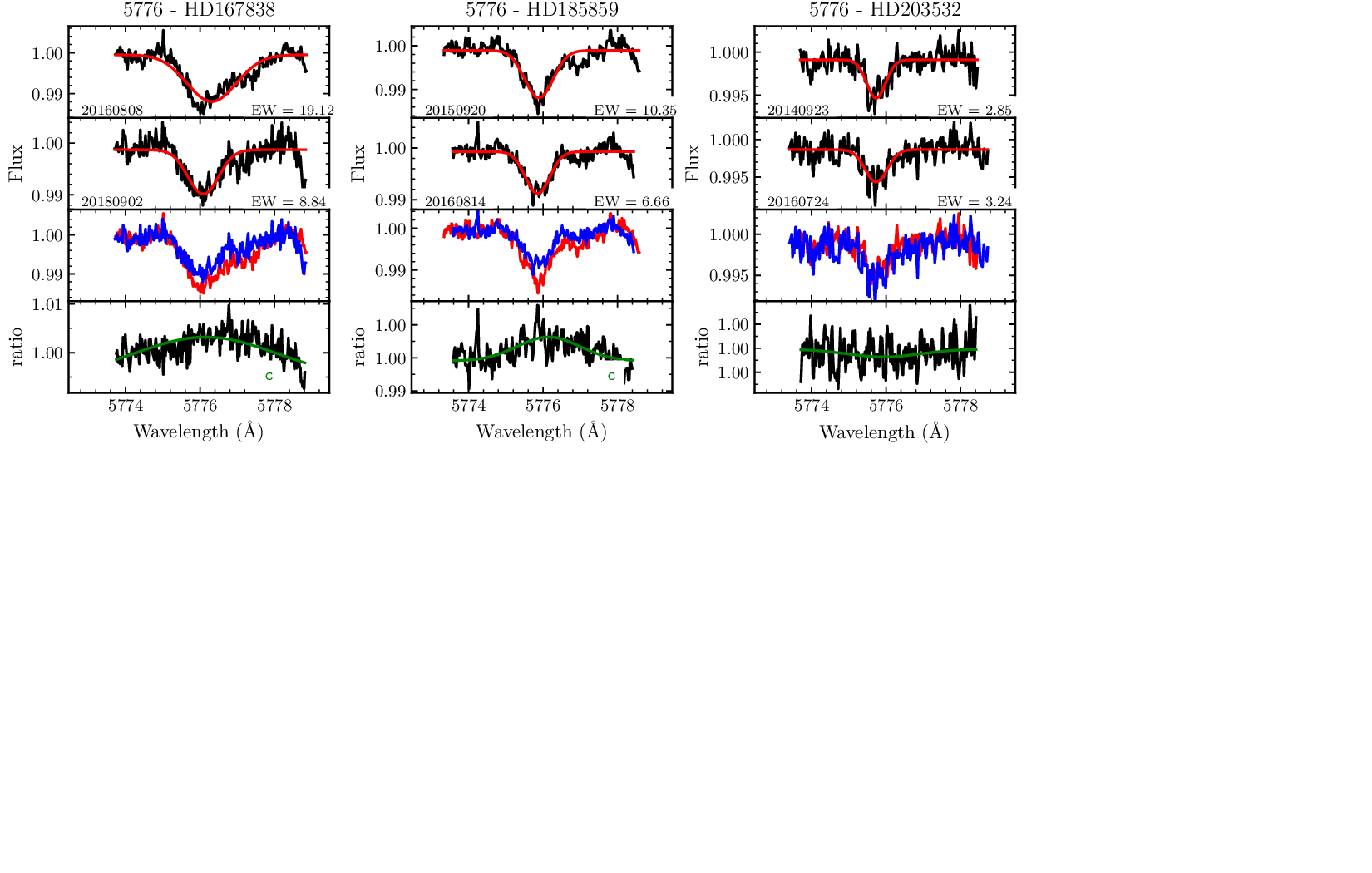}
    \caption{The same as \ref{plt-dib-var1}}
    \label{plt-dib-var53}
\end{figure*}

% %%%%%%%%%%%%%%%%%%
% %%%%%%%%%%%%%%%%%%
\clearpage
\begin{figure*}[ht!]
    \centering
    \includegraphics[width=0.99\hsize]{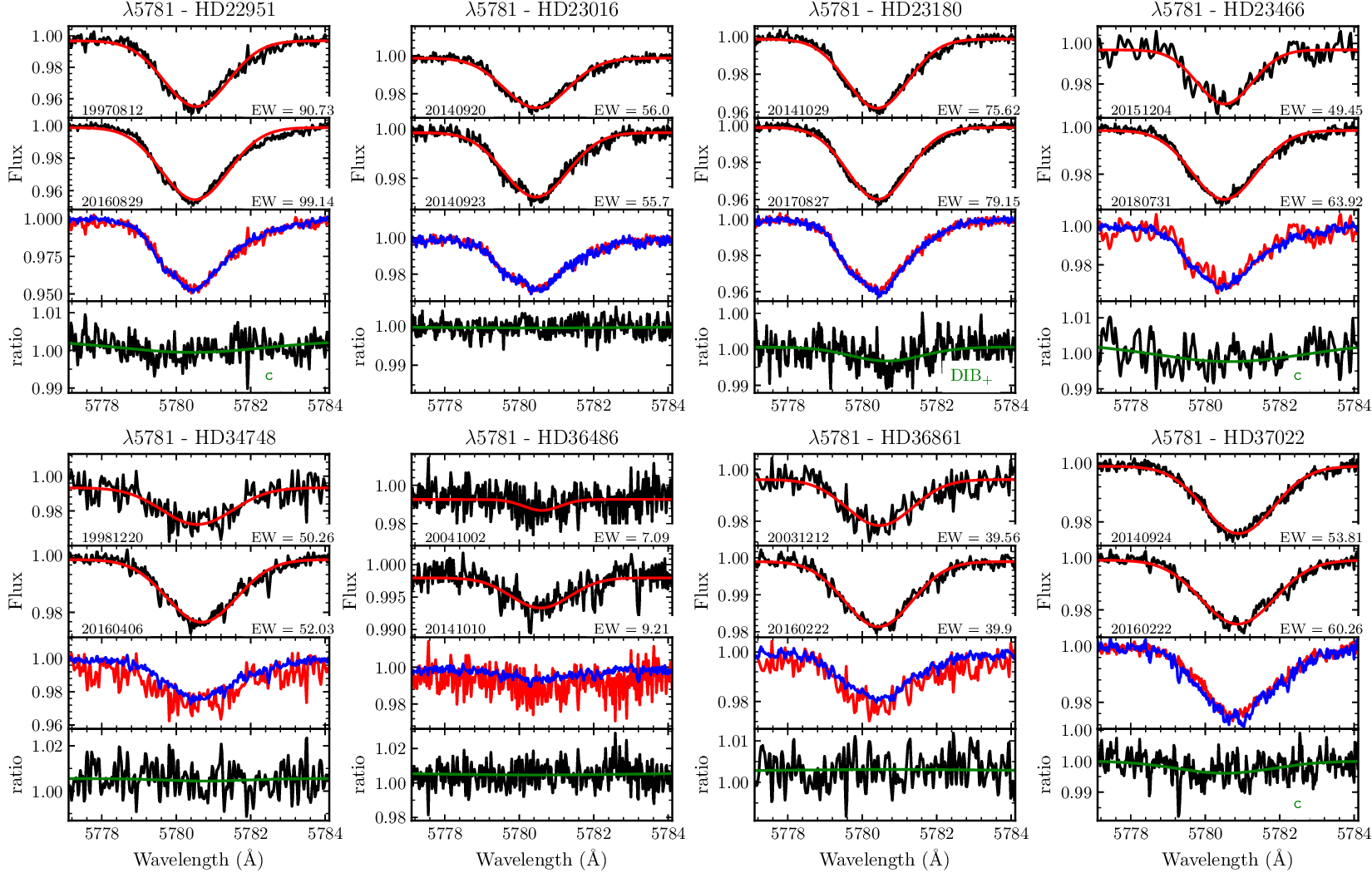}
    \caption{The same as \ref{plt-dib-var1}}
    \label{plt-dib-var54}
\end{figure*}

% \newpage
\begin{figure*}[ht!]
    \centering
    \includegraphics[width=0.99\hsize]{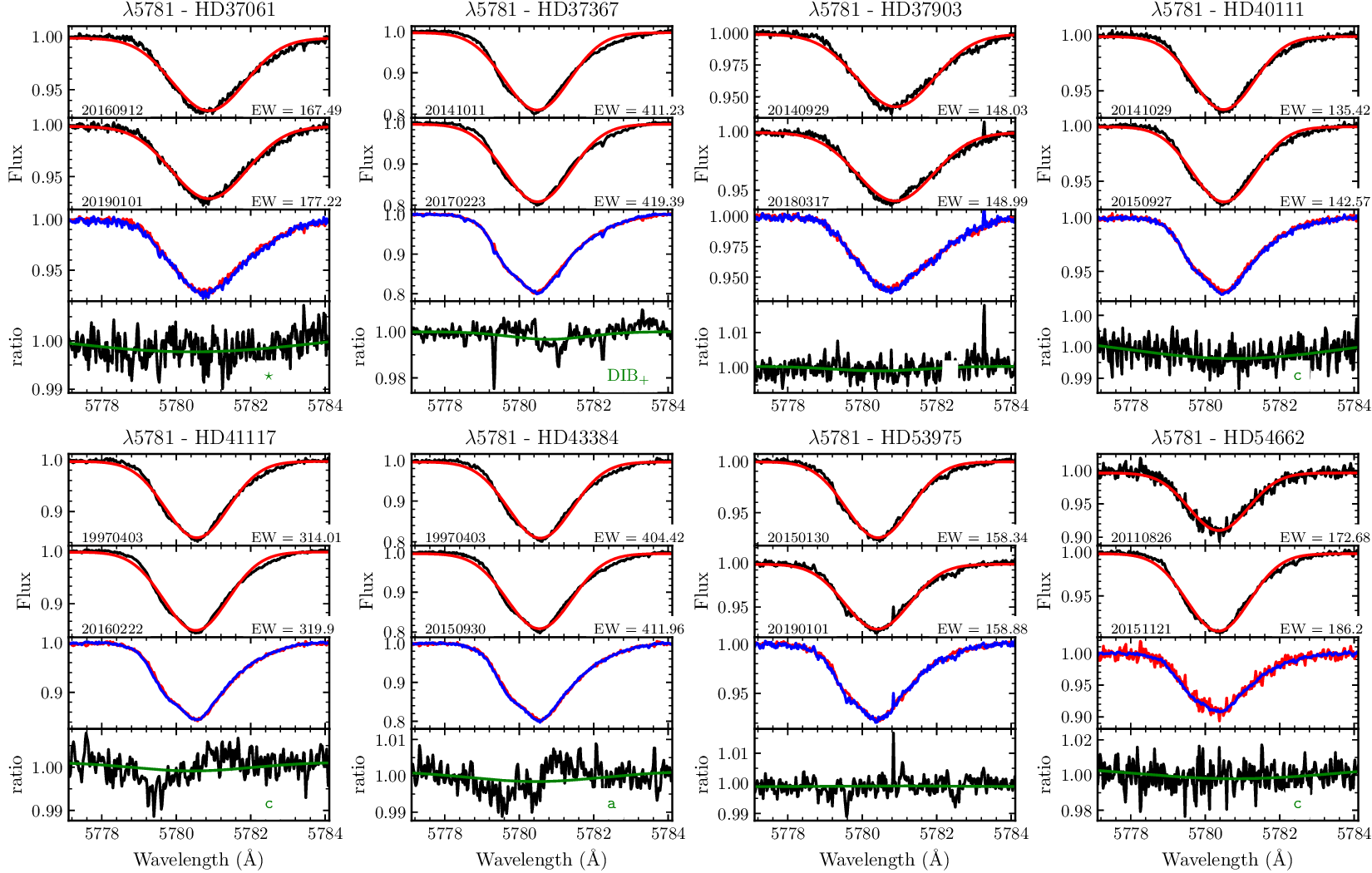}
    \caption{The same as \ref{plt-dib-var1}}
    \label{plt-dib-var55}
\end{figure*}

% \newpage
\begin{figure*}[ht!]
    \centering
    \includegraphics[width=0.99\hsize]{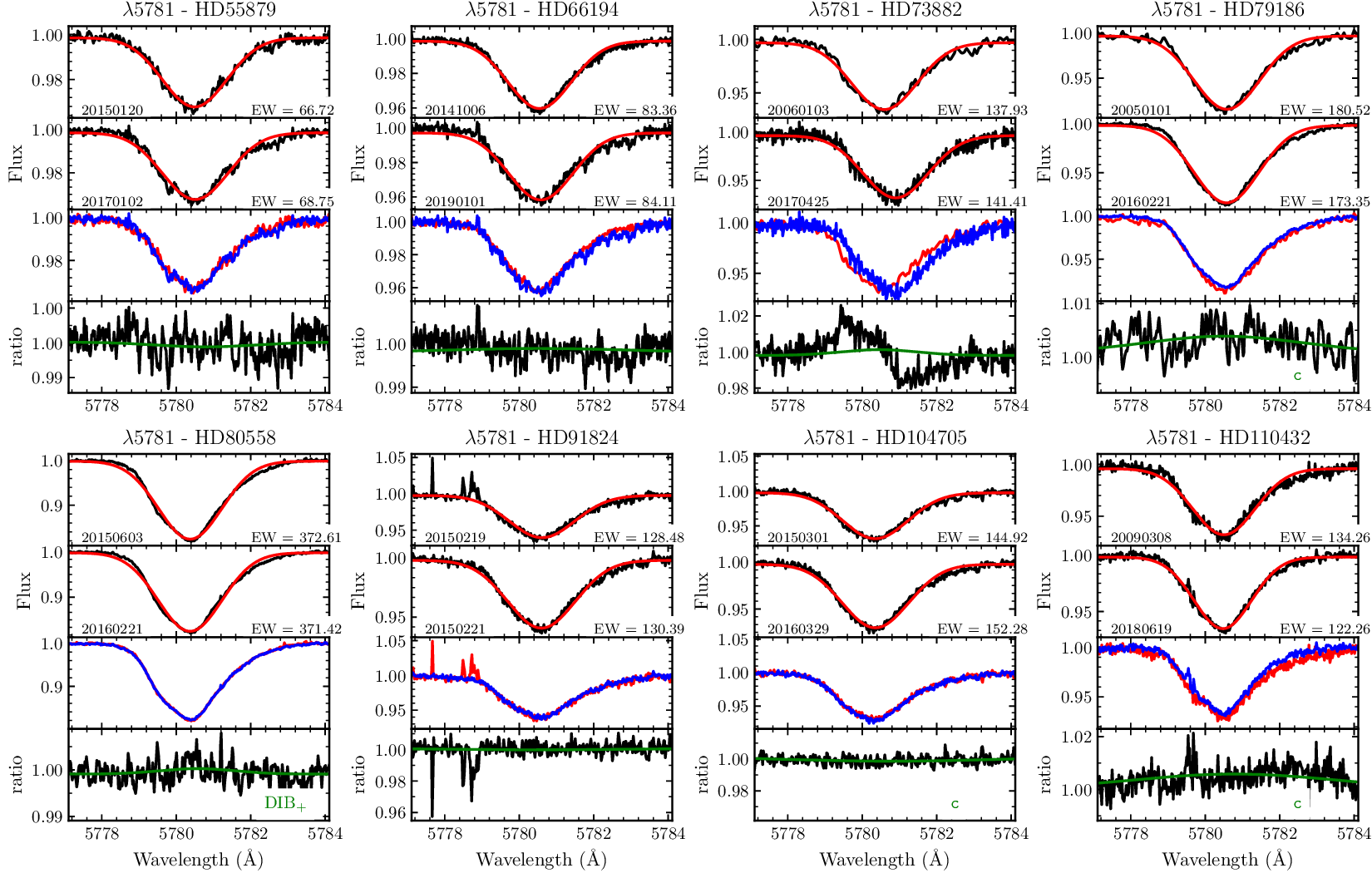}
    \caption{The same as \ref{plt-dib-var1}}
    \label{plt-dib-var56}
\end{figure*}

% \newpage
\begin{figure*}[ht!]
    \centering
    \includegraphics[width=0.99\hsize]{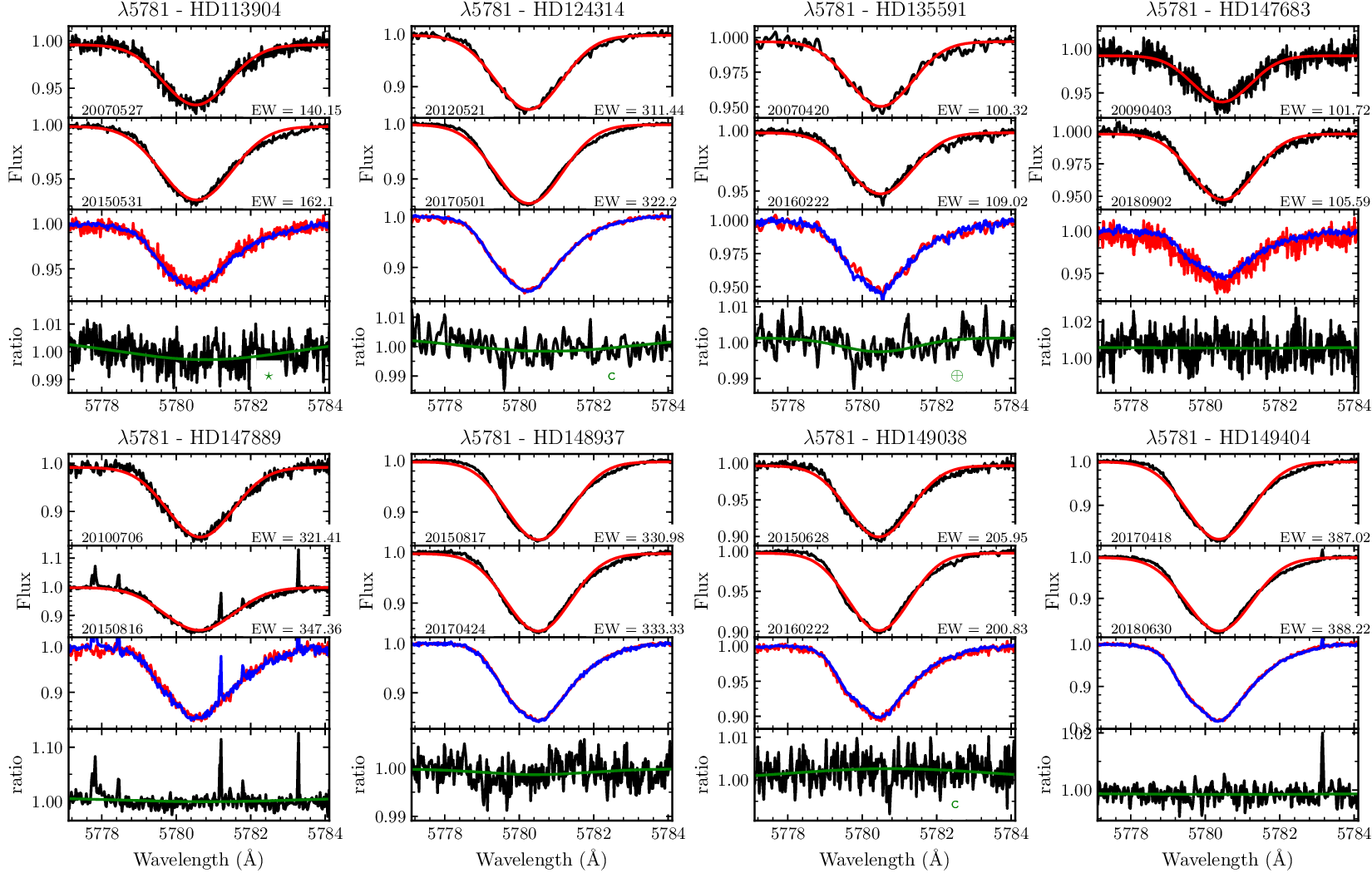}
    \caption{The same as \ref{plt-dib-var1}}
    \label{plt-dib-var57}
\end{figure*}

% \newpage
\begin{figure*}[ht!]
    \centering
    \includegraphics[width=0.99\hsize]{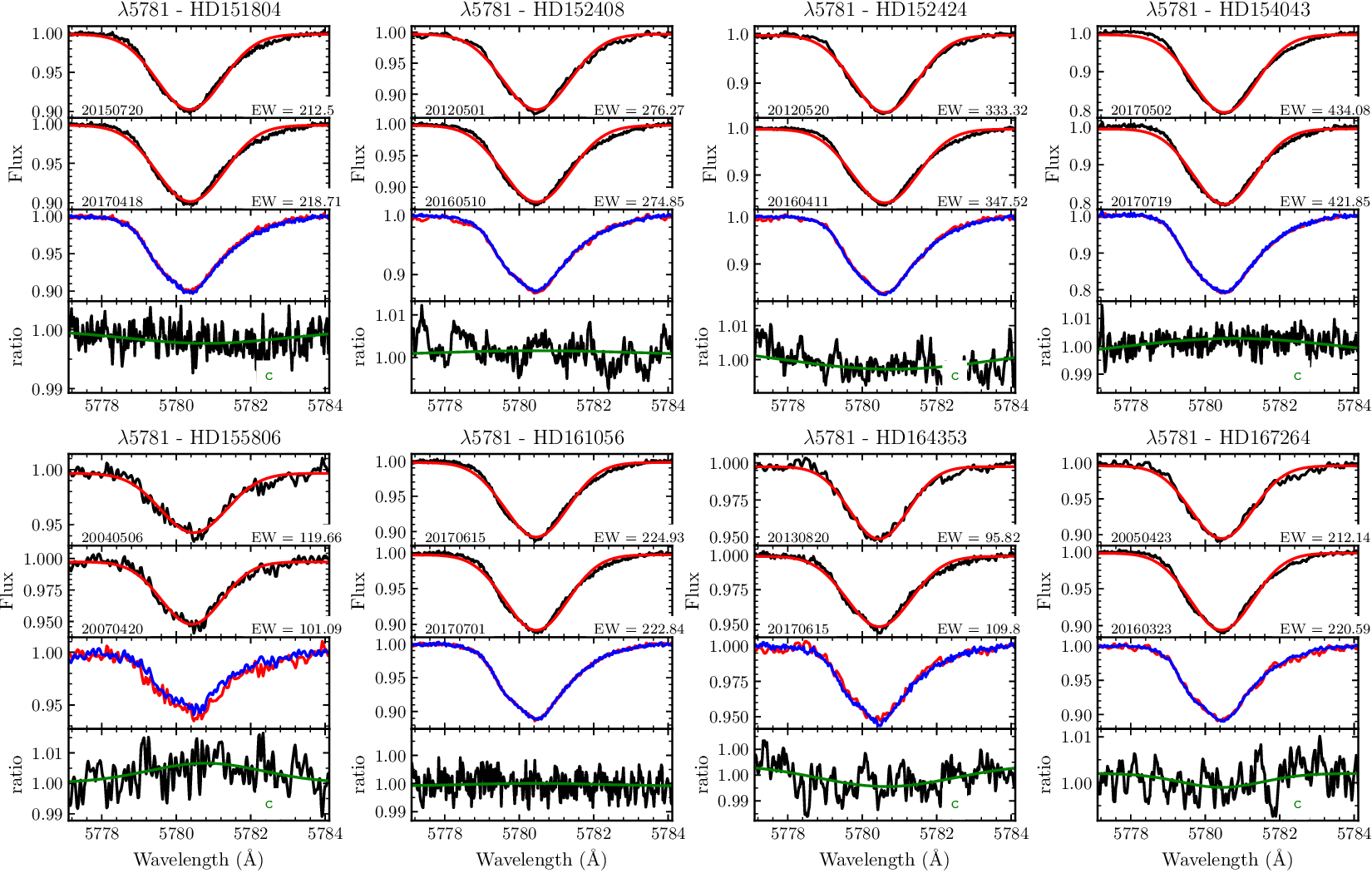}
    \caption{The same as \ref{plt-dib-var1}}
    \label{plt-dib-var58}
\end{figure*}

% \newpage
\begin{figure*}[ht!]
    \centering
    \includegraphics[width=0.99\hsize]{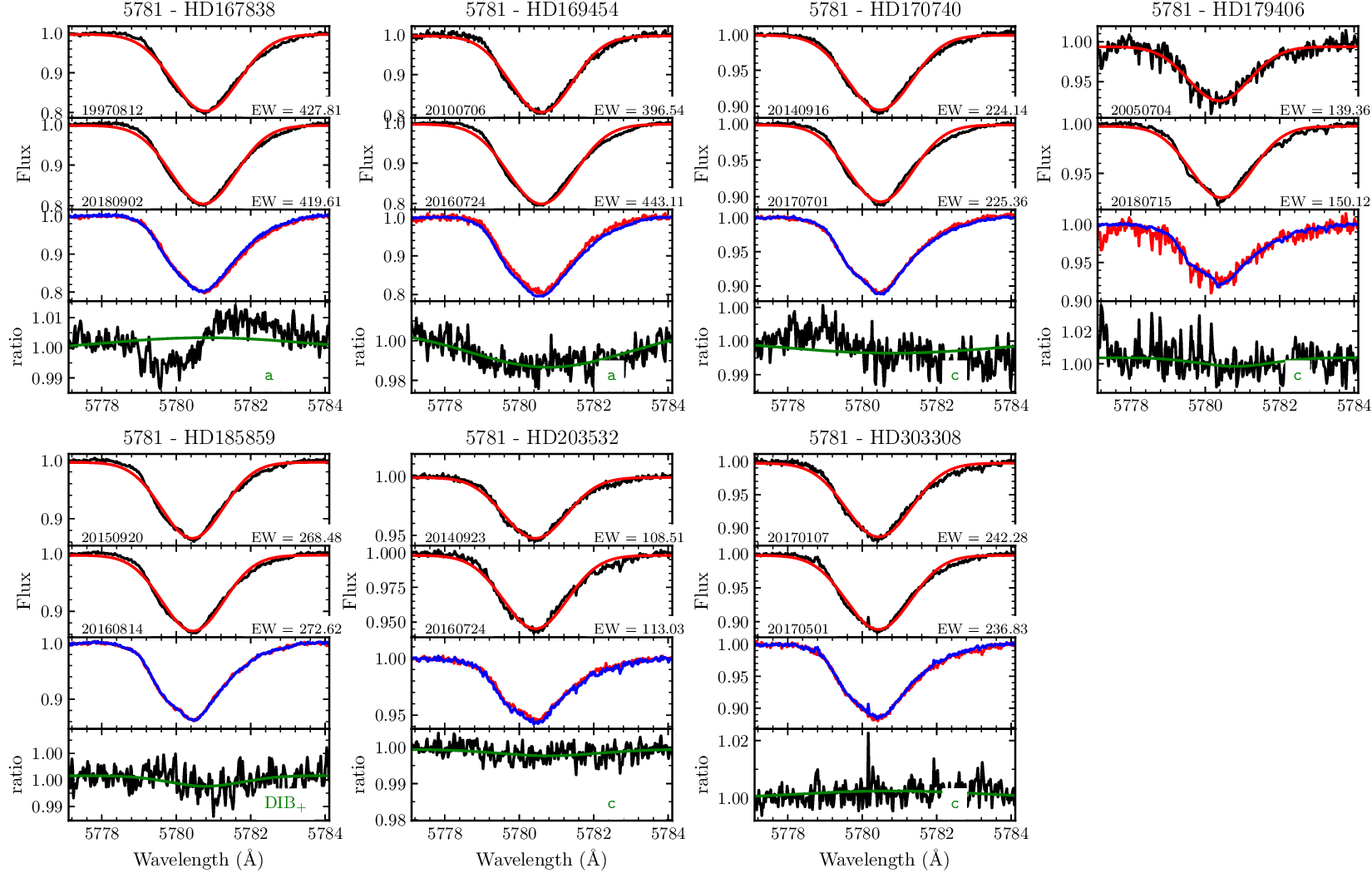}
    \caption{The same as \ref{plt-dib-var1}}
    \label{plt-dib-var59}
\end{figure*}

% %%%%%%%%%%%%%%%%%%
% %%%%%%%%%%%%%%%%%%
\begin{figure*}[ht!]
    \centering
    \includegraphics[width=0.99\hsize]{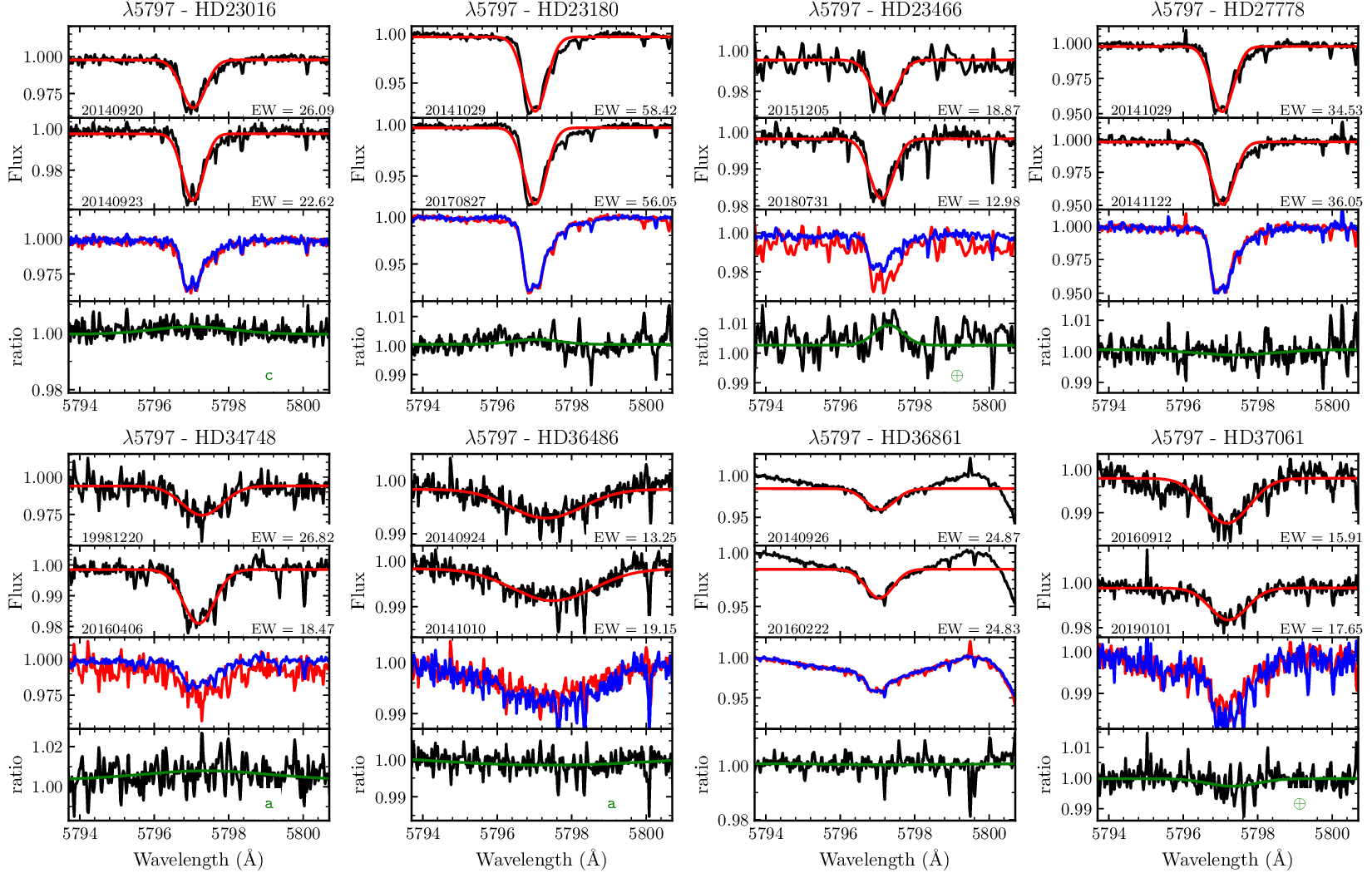}
    \caption{The same as \ref{plt-dib-var1}}
    \label{plt-dib-var60}
\end{figure*}

\begin{figure*}[ht!]
    \centering
    \includegraphics[width=0.99\hsize]{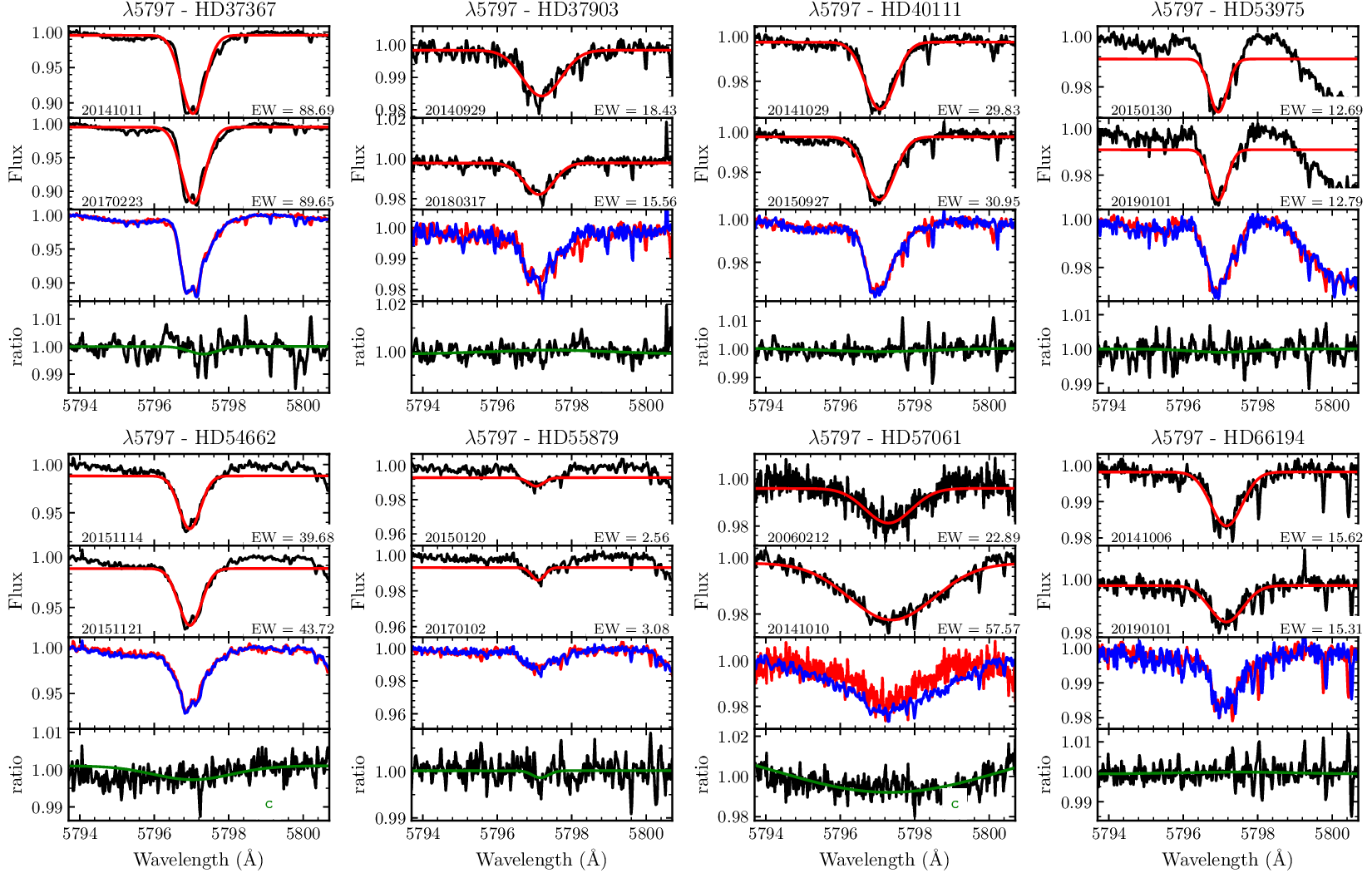}
    \caption{The same as \ref{plt-dib-var1}}
    \label{plt-dib-var61}
\end{figure*}

\begin{figure*}[ht!]
    \centering
    \includegraphics[width=0.99\hsize]{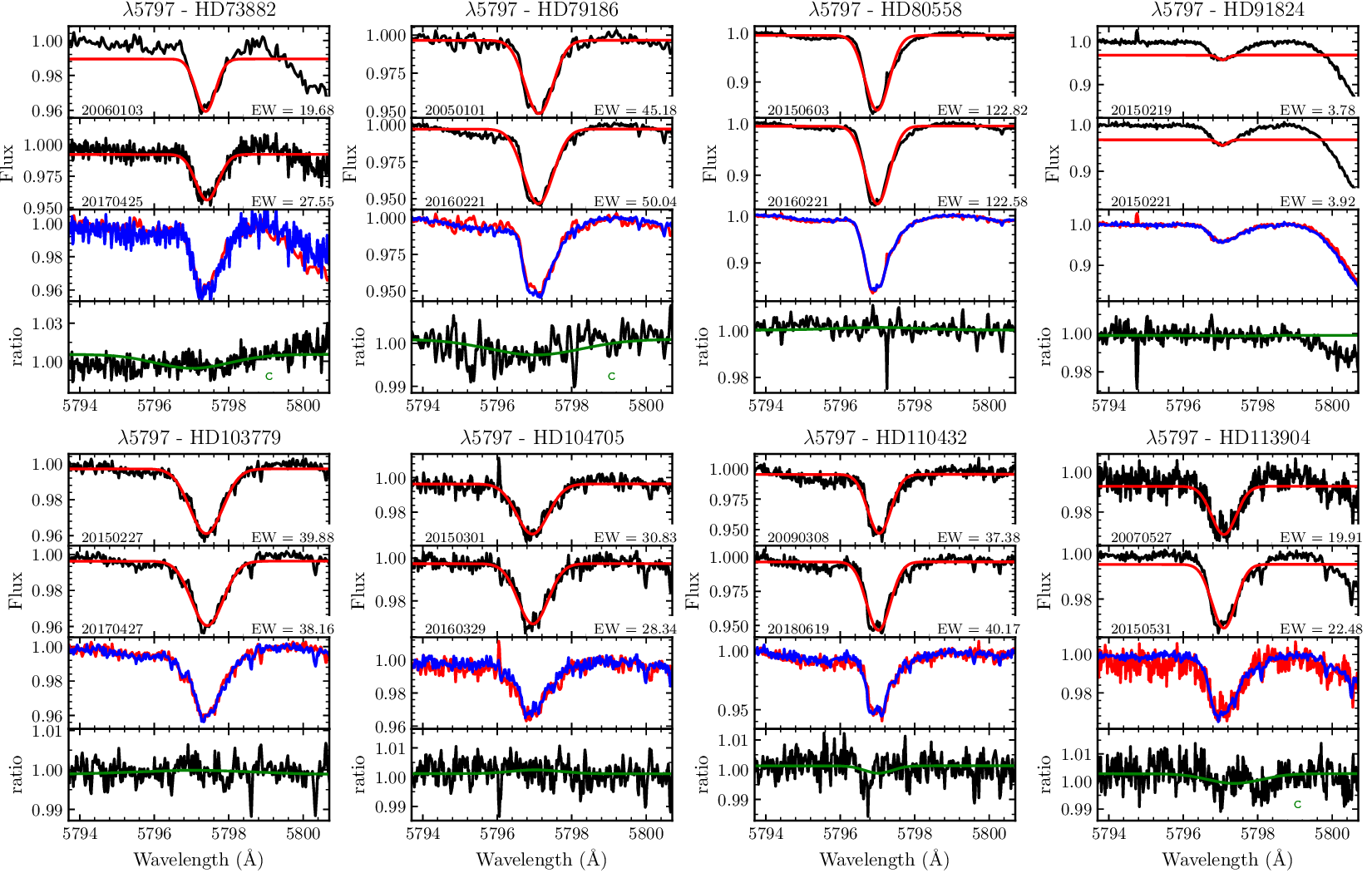}
    \caption{The same as \ref{plt-dib-var1}}
    \label{plt-dib-var62}
\end{figure*}

\begin{figure*}[ht!]
    \centering
    \includegraphics[width=0.99\hsize]{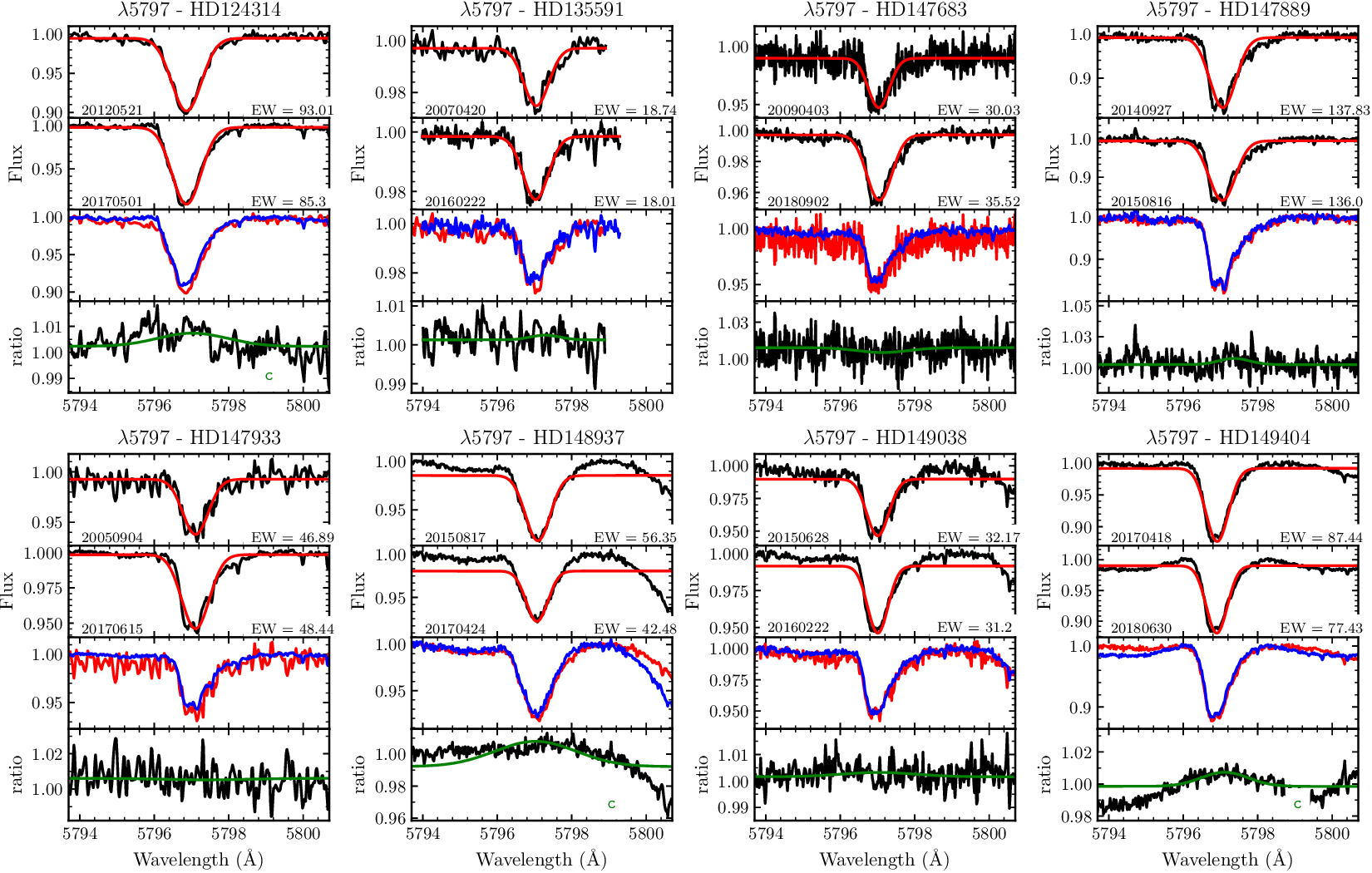}
    \caption{The same as \ref{plt-dib-var1}}
    \label{plt-dib-var63}
\end{figure*}

\begin{figure*}[ht!]
    \centering
    \includegraphics[width=0.99\hsize]{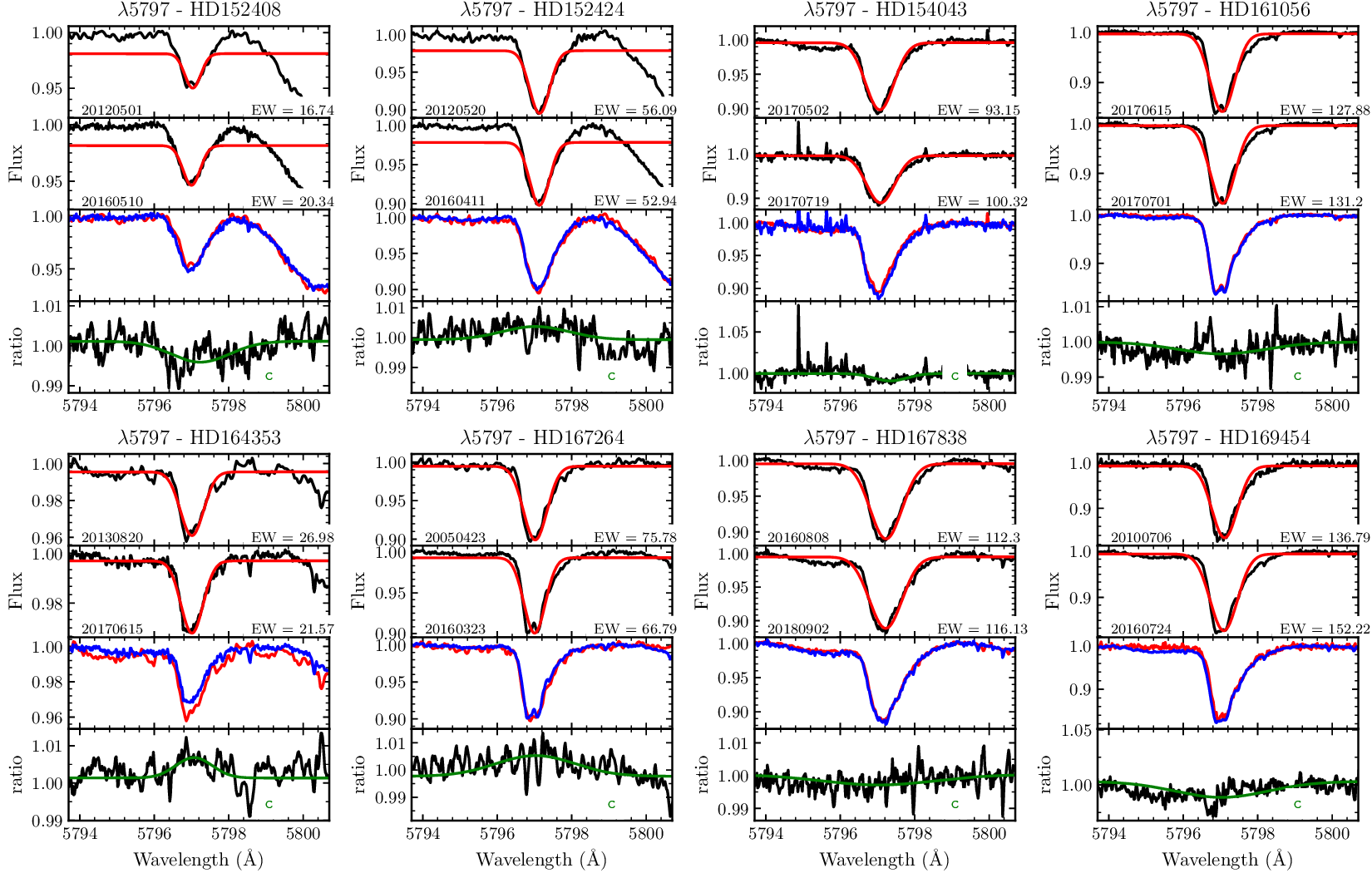}
    \caption{The same as \ref{plt-dib-var1}}
    \label{plt-dib-var64}
\end{figure*}

% %%%%%%%%%%%%%%%%%%
% %%%%%%%%%%%%%%%%%%
\begin{figure*}[ht!]
    \centering
    \includegraphics[width=0.99\hsize]{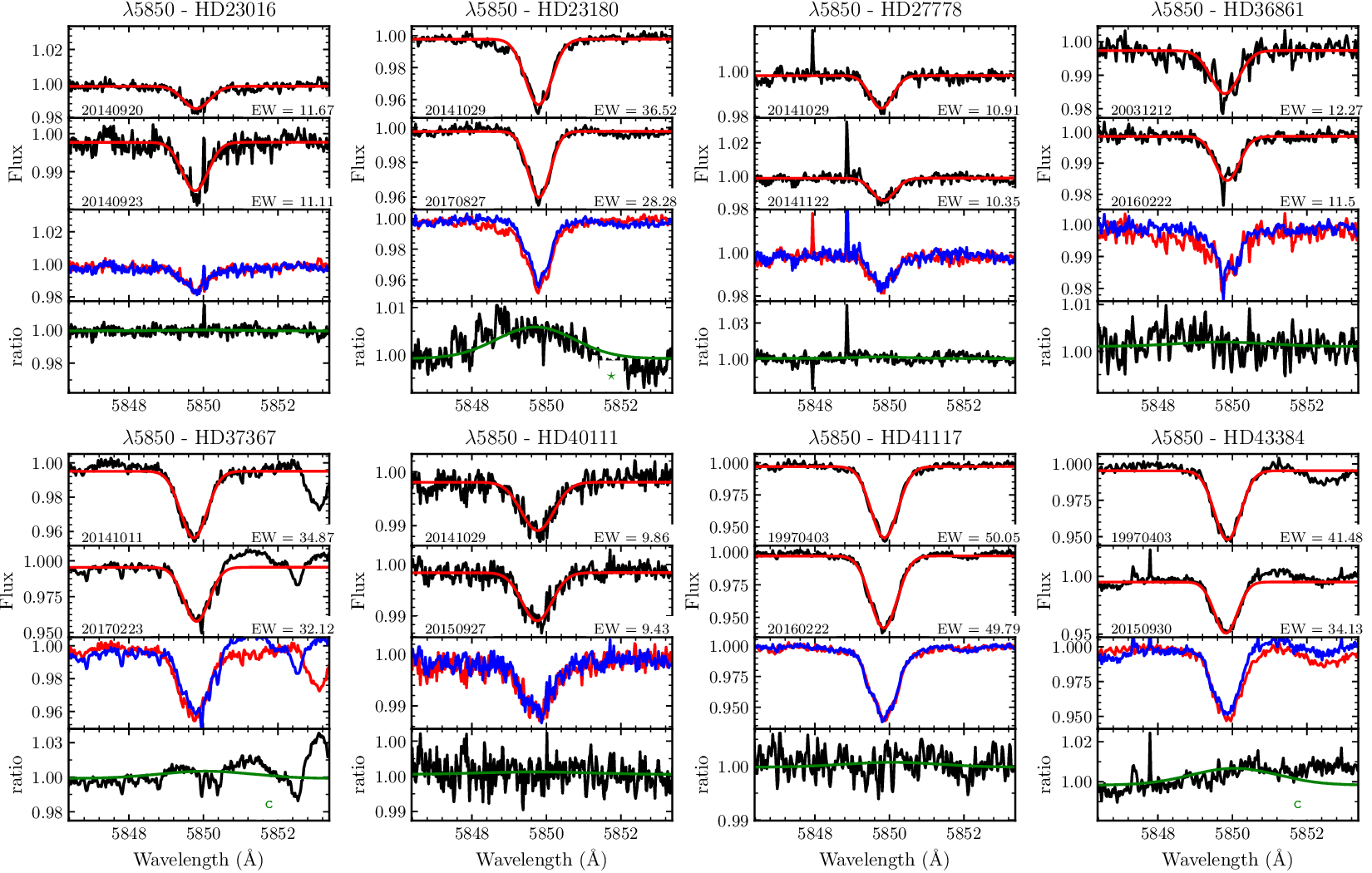}
    \caption{The same as \ref{plt-dib-var1}}
    \label{plt-dib-var65}
\end{figure*}

\begin{figure*}[ht!]
    \centering
    \includegraphics[width=0.99\hsize]{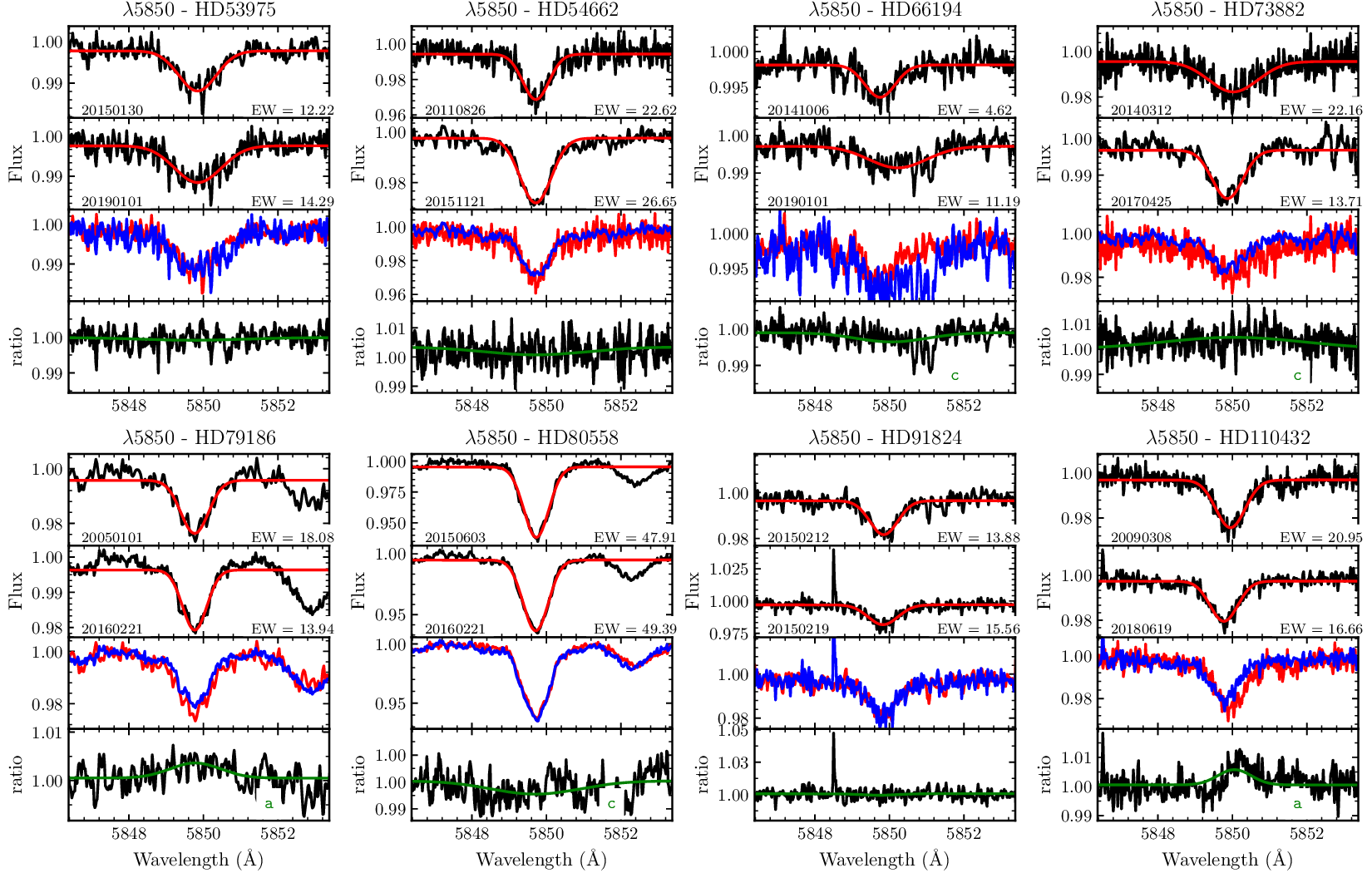}
    \caption{The same as \ref{plt-dib-var1}}
    \label{plt-dib-var66}
\end{figure*}

\begin{figure*}[ht!]
    \centering
    \includegraphics[width=0.99\hsize]{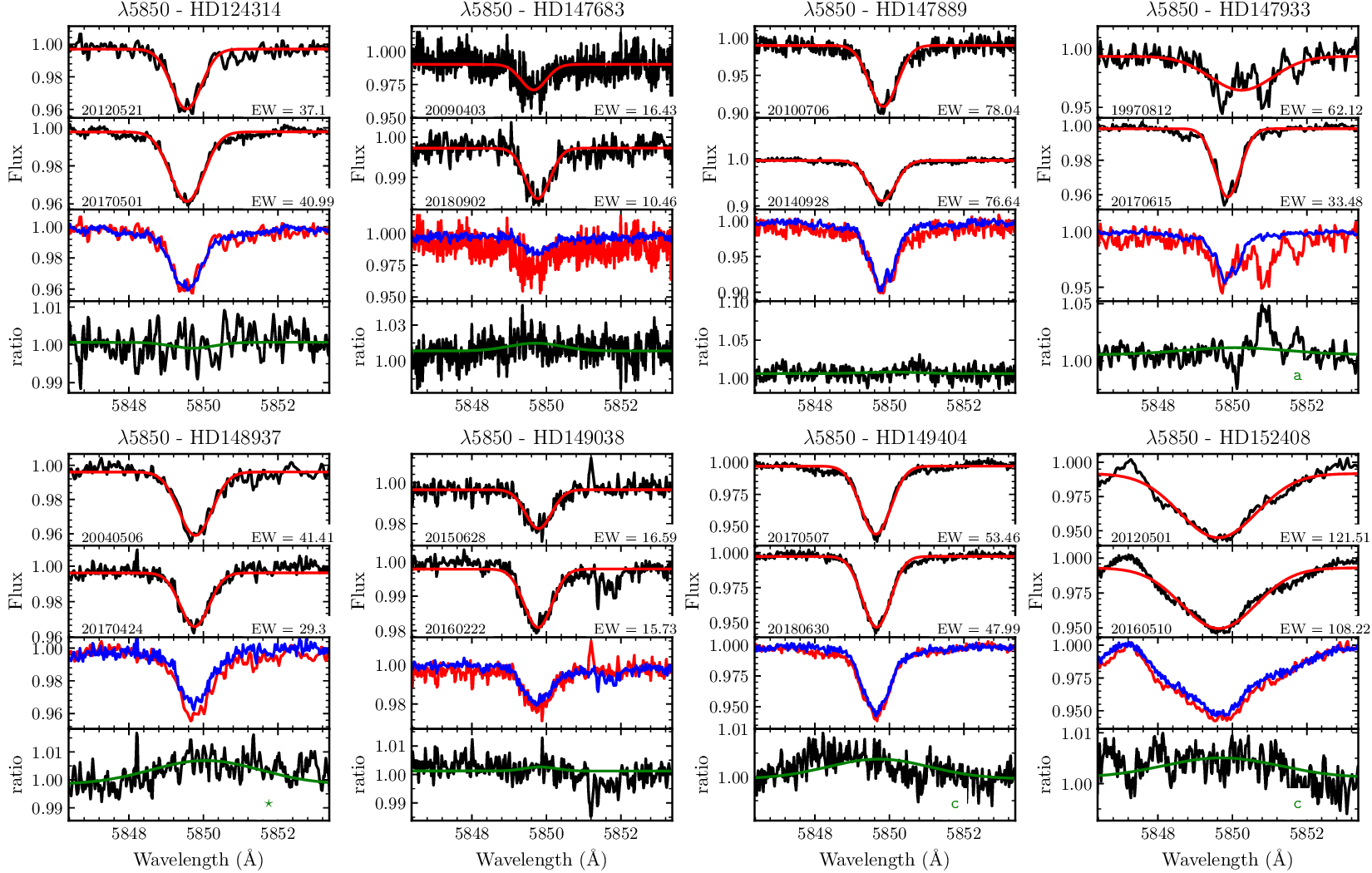}
    \caption{The same as \ref{plt-dib-var1}}
    \label{plt-dib-var67}
\end{figure*}

\begin{figure*}[ht!]
    \centering
    \includegraphics[width=0.99\hsize]{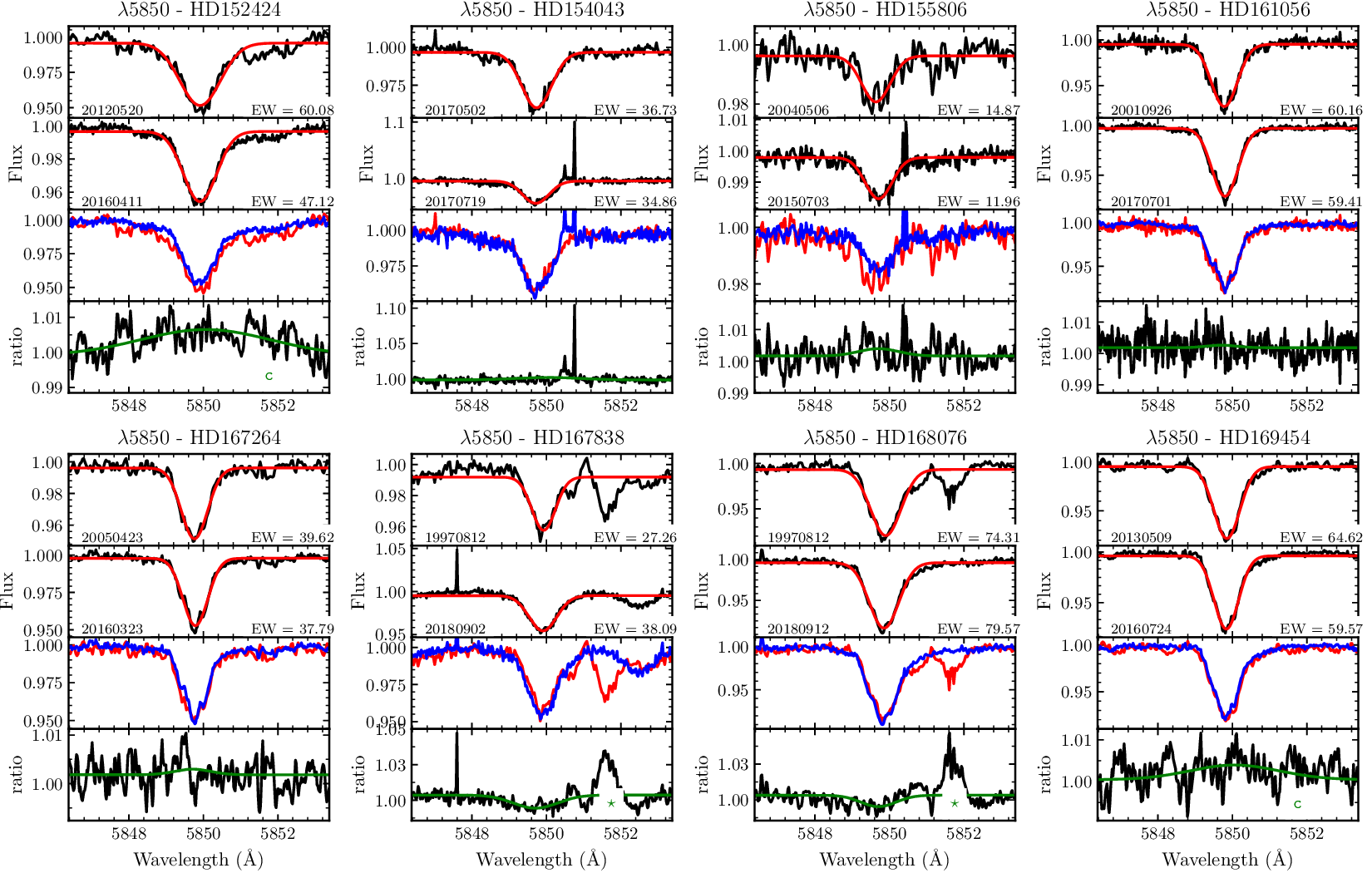}
    \caption{The same as \ref{plt-dib-var1}}
    \label{plt-dib-var68}
\end{figure*}

\begin{figure*}[ht!]
    \centering
    \includegraphics[width=0.99\hsize]{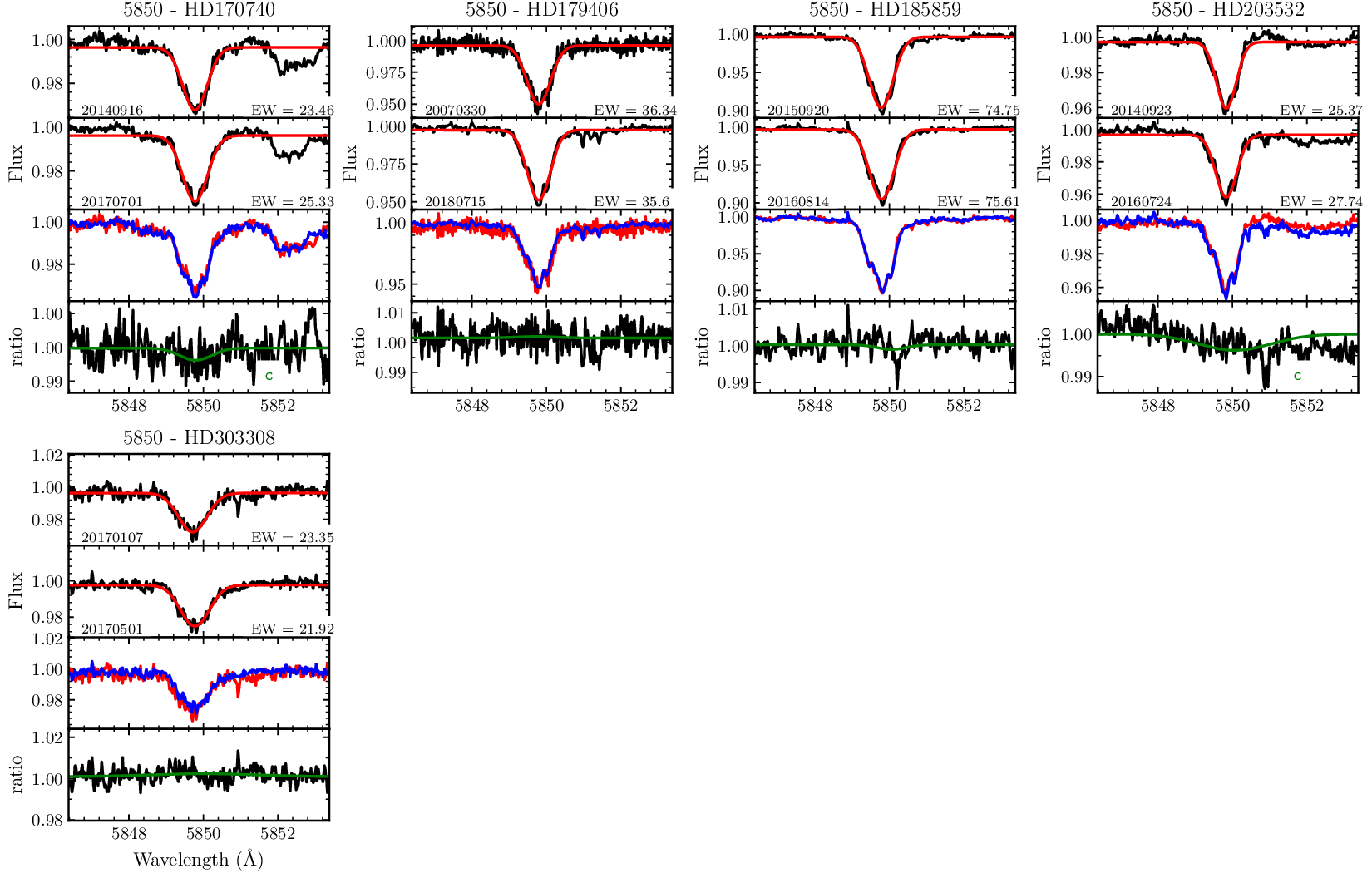}
    \caption{The same as \ref{plt-dib-var1}}
    \label{plt-dib-var69}
\end{figure*}

% %%%%%%%%%%%%%%%%%%
% %%%%%%%%%%%%%%%%%%
\begin{figure*}[ht!]
    \centering
    \includegraphics[width=0.99\hsize]{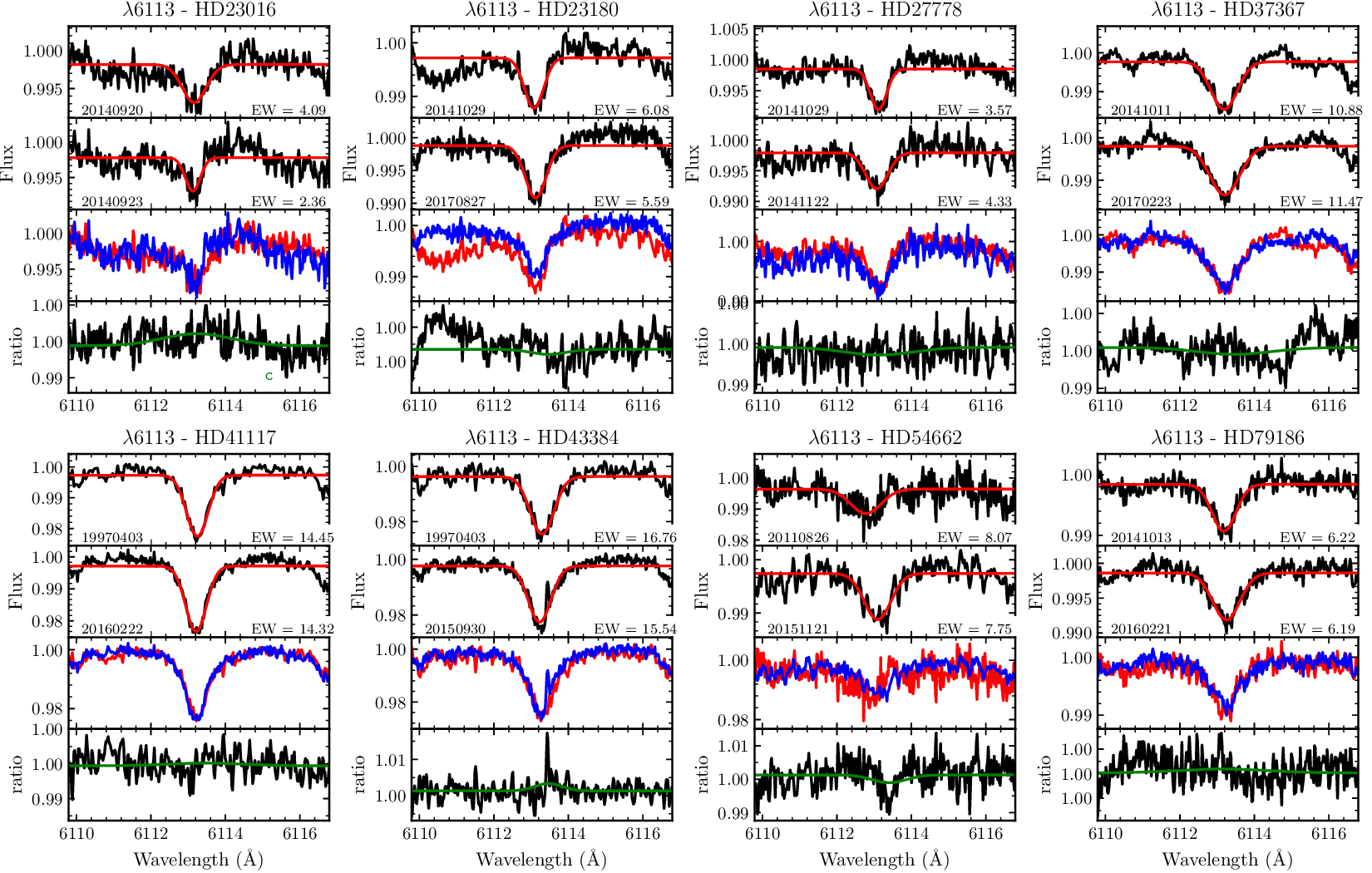}
    \caption{The same as \ref{plt-dib-var1}}
    \label{plt-dib-var70}
\end{figure*}

\begin{figure*}[ht!]
    \centering
    \includegraphics[width=0.99\hsize]{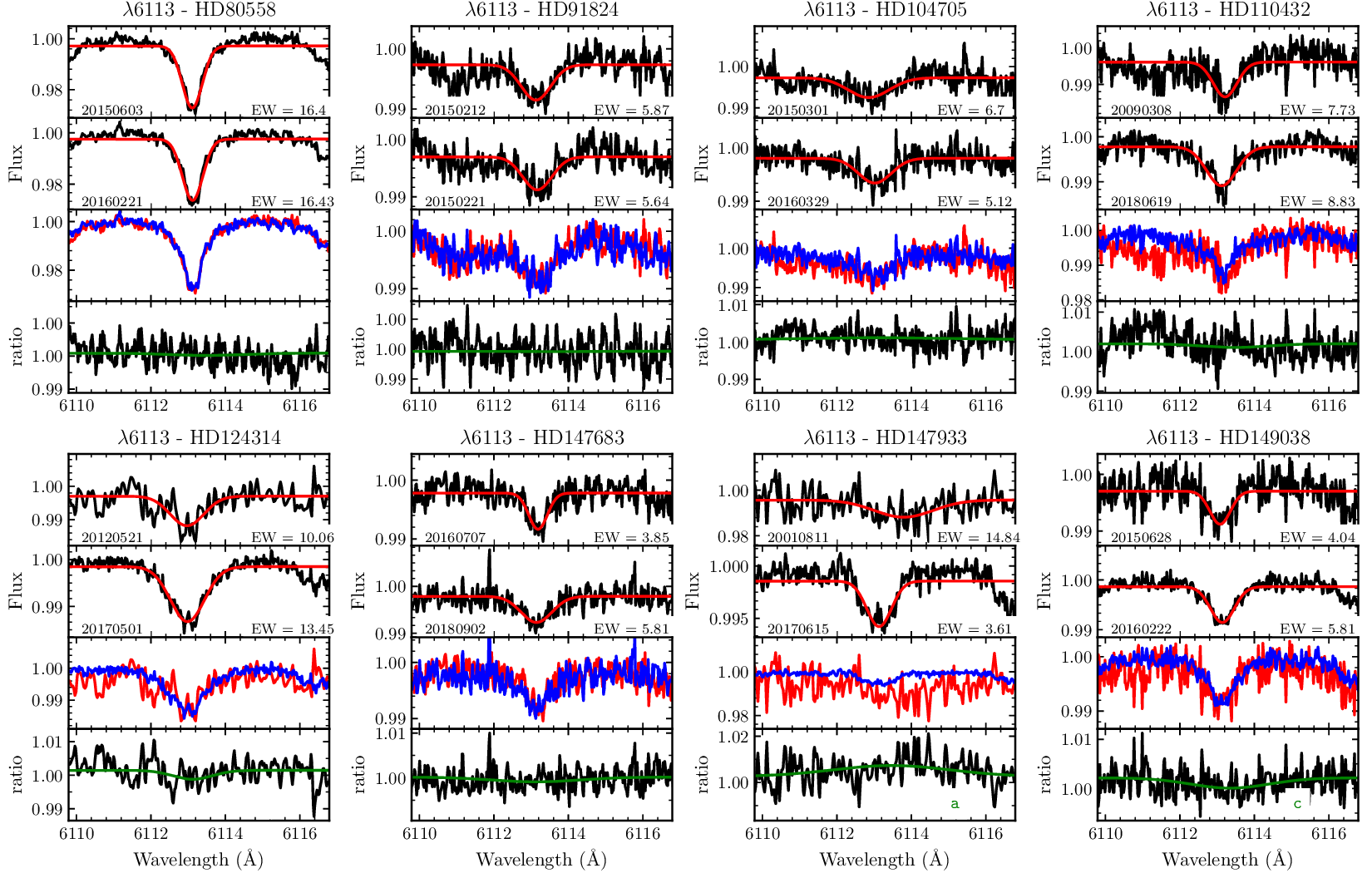}
    \caption{The same as \ref{plt-dib-var1}}
    \label{plt-dib-var71}
\end{figure*}

% %%%%%%%%%%%%%%%%%%
% %%%%%%%%%%%%%%%%%%
\begin{figure*}[ht!]
    \centering
    \includegraphics[width=0.99\hsize]{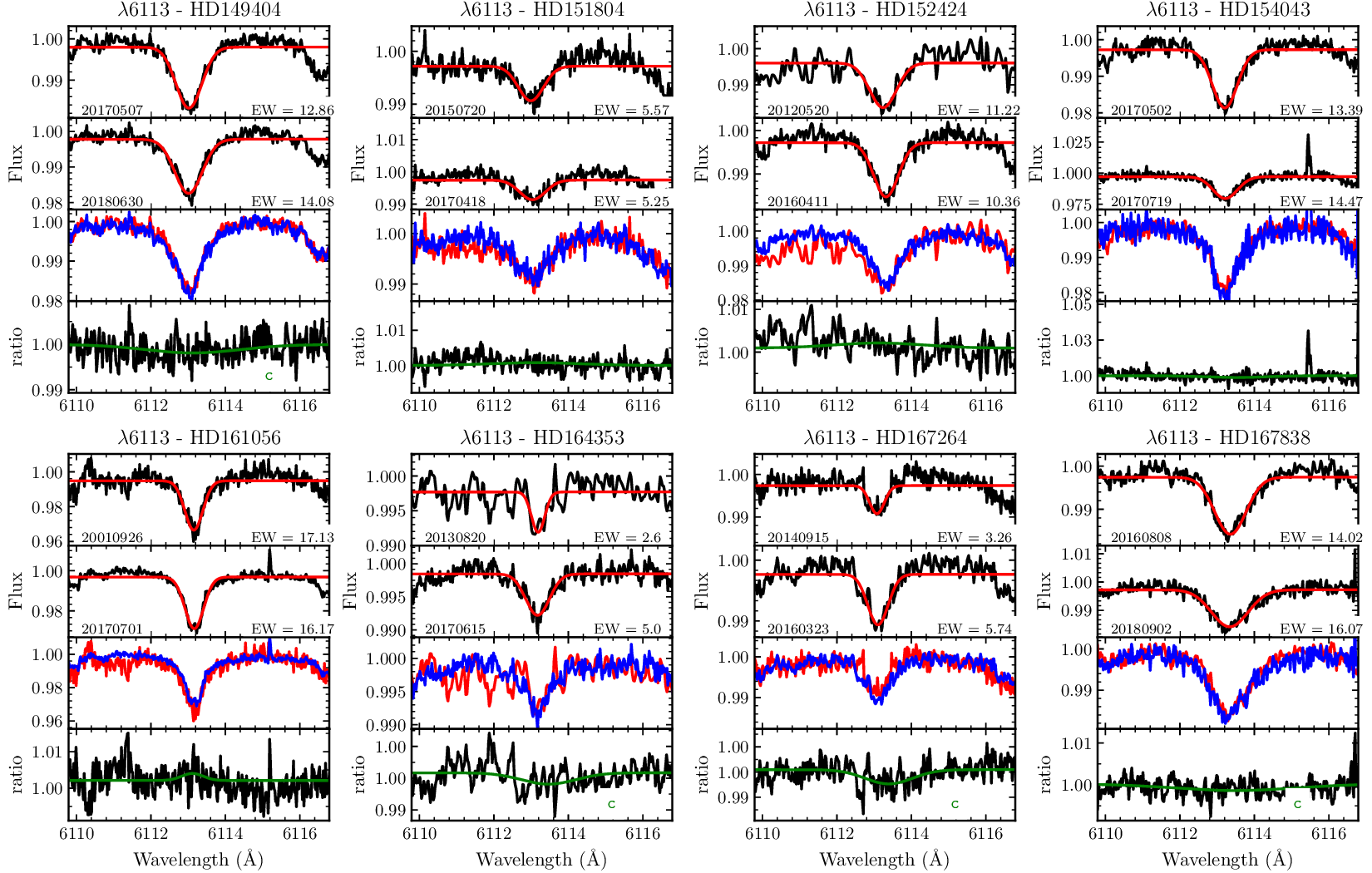}
    \caption{The same as \ref{plt-dib-var1}}
    \label{plt-dib-var72}
\end{figure*}

\begin{figure*}[ht!]
    \centering
    \includegraphics[width=0.99\hsize]{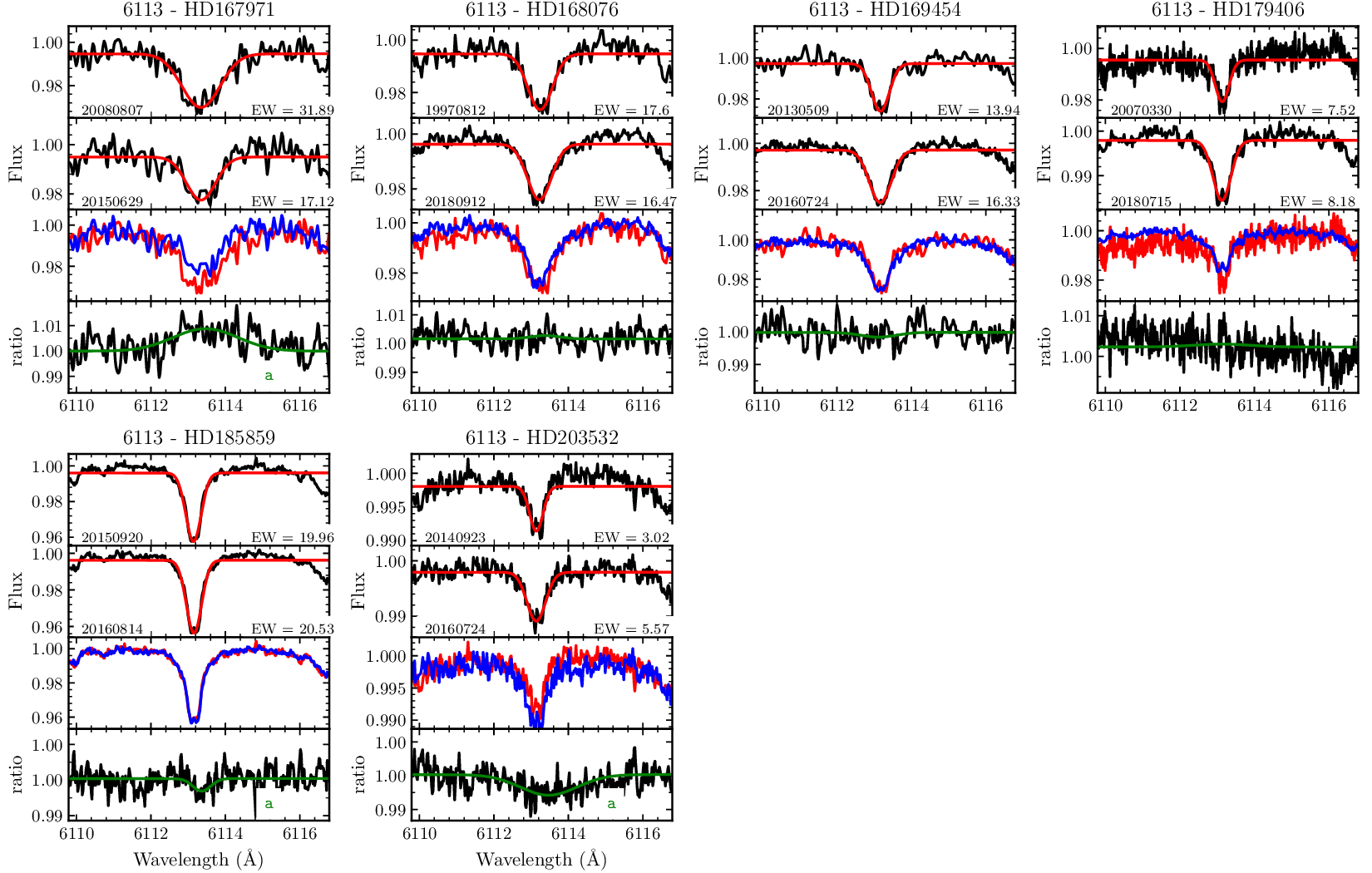}
    \caption{The same as \ref{plt-dib-var1}}
    \label{plt-dib-var73}
\end{figure*}

% %%%%%%%%%%%%%%%%%%
% %%%%%%%%%%%%%%%%%%
\begin{figure*}[ht!]
    \centering
    \includegraphics[width=0.99\hsize]{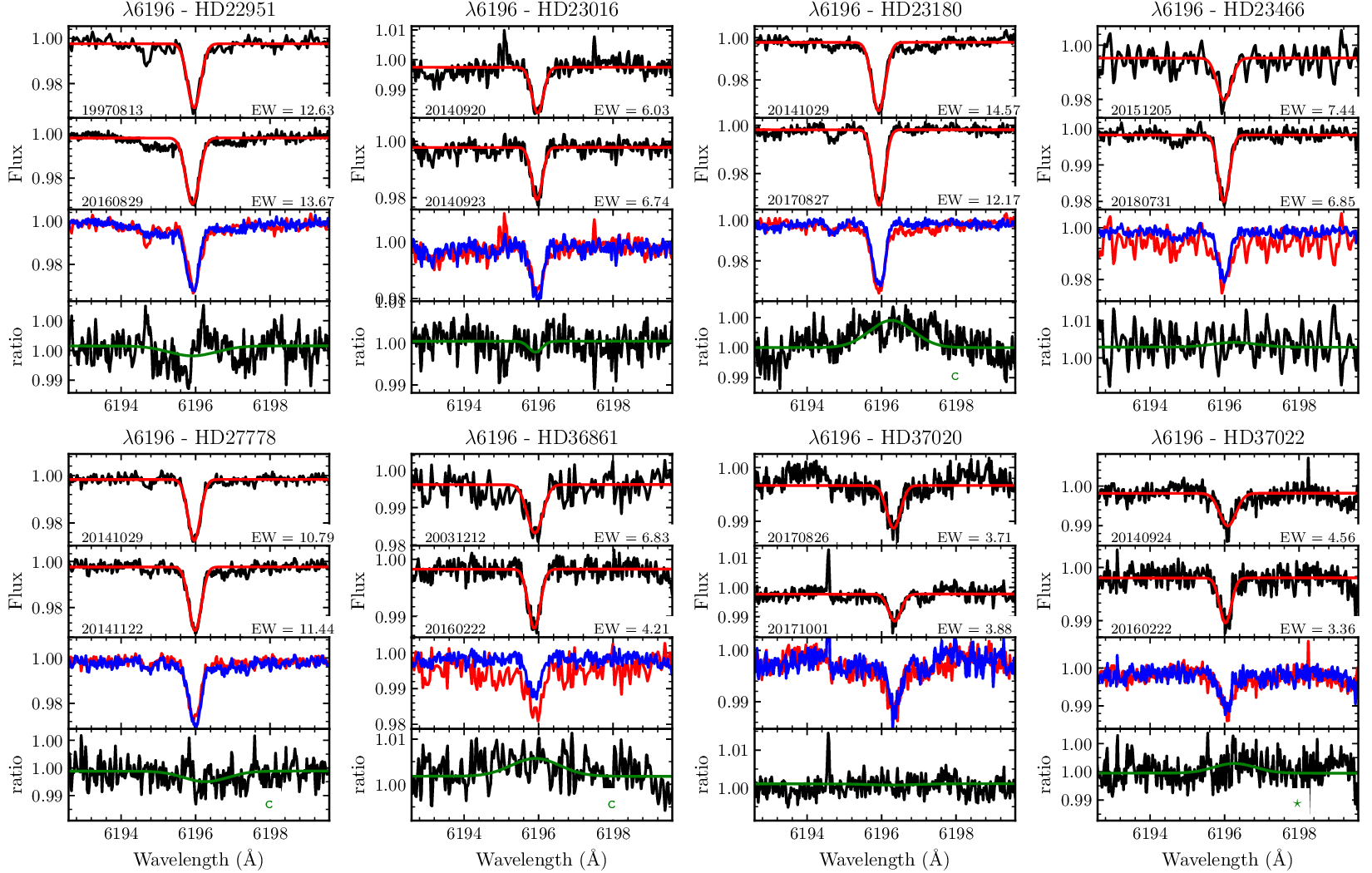}
    \caption{The same as \ref{plt-dib-var1}}
    \label{plt-dib-var74}
\end{figure*}

\begin{figure*}[ht!]
    \centering
    \includegraphics[width=0.99\hsize]{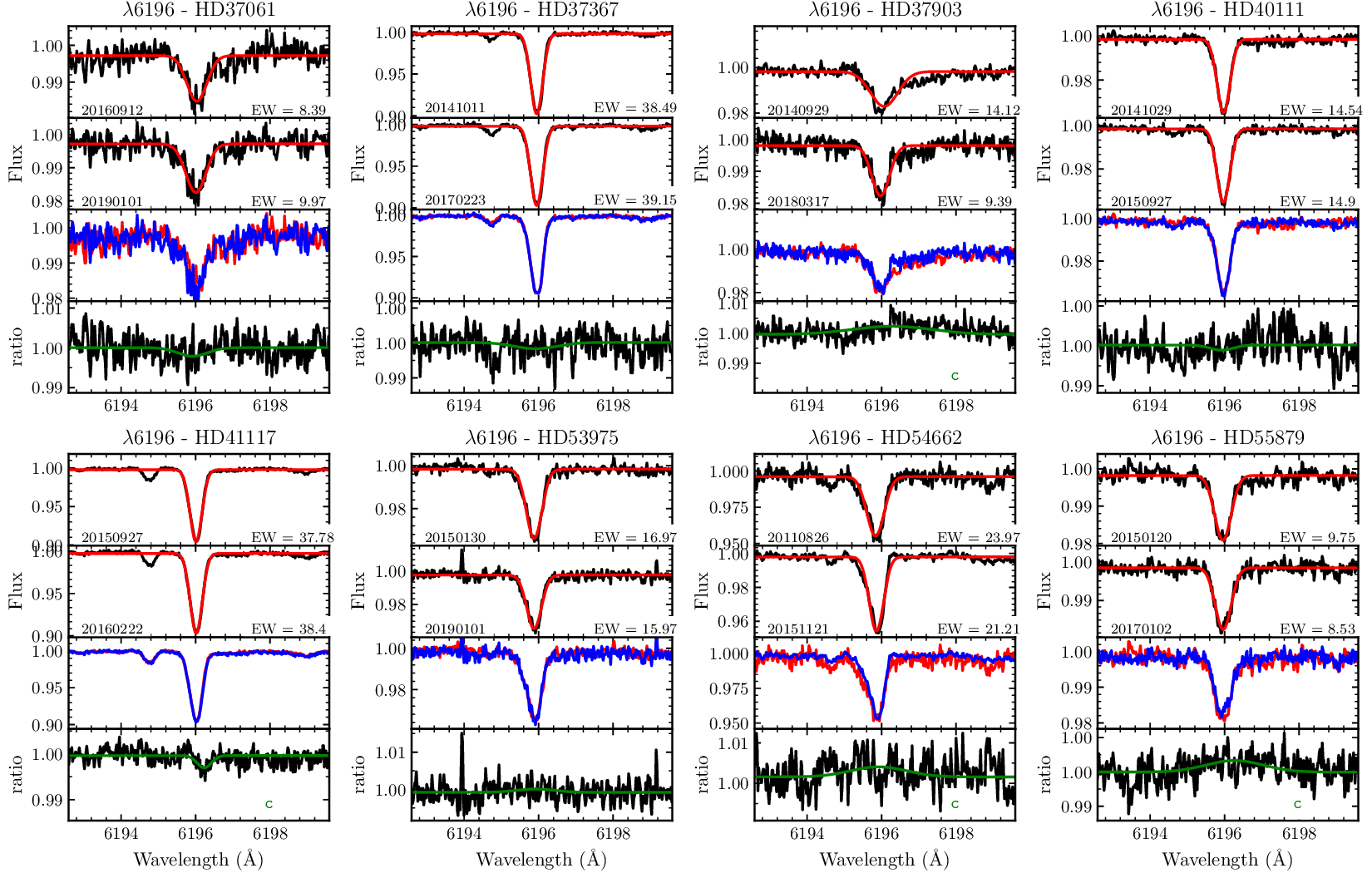}
    \caption{The same as \ref{plt-dib-var1}}
    \label{plt-dib-var75}
\end{figure*}

\begin{figure*}[ht!]
    \centering
    \includegraphics[width=0.99\hsize]{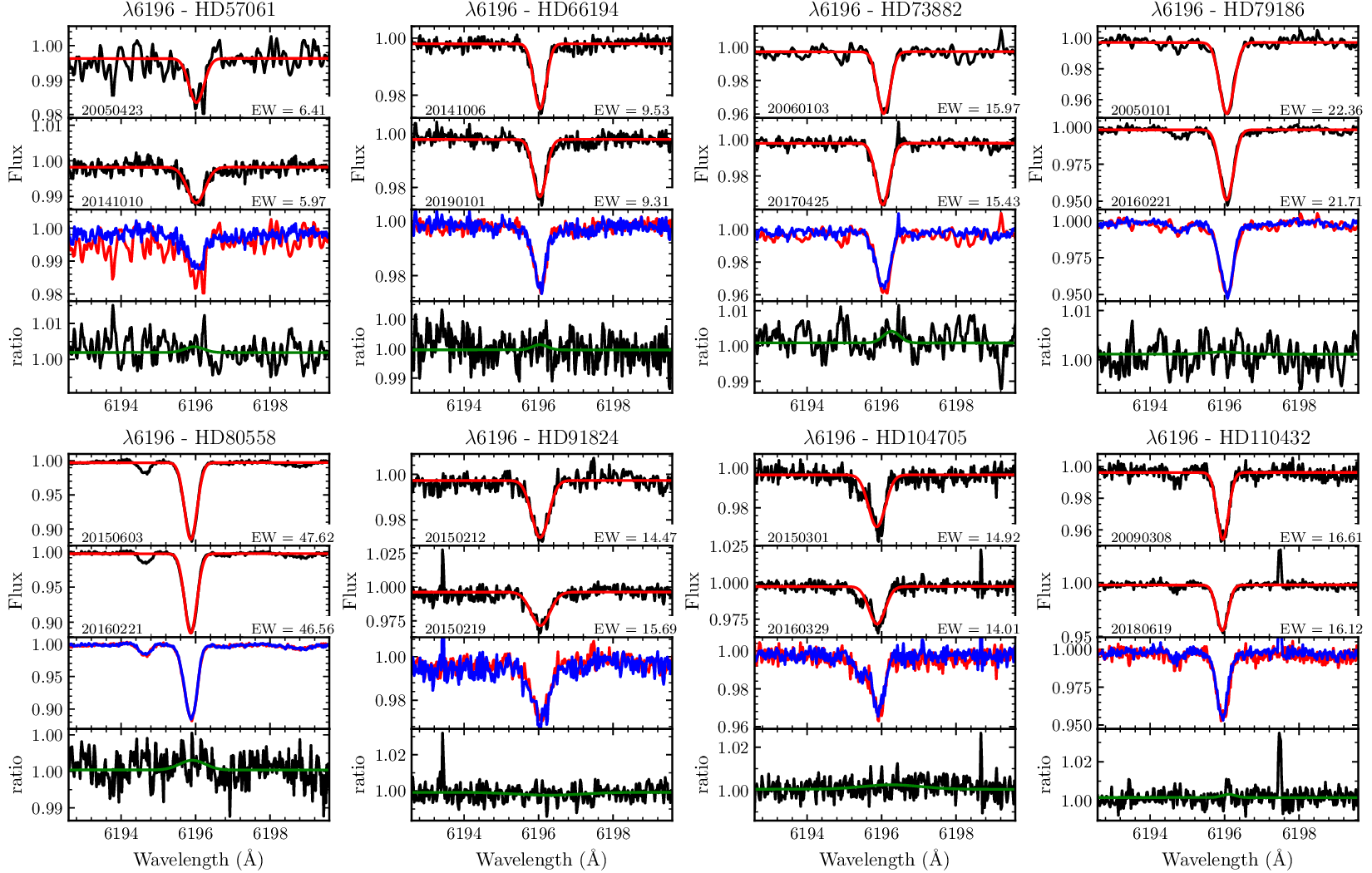}
    \caption{The same as \ref{plt-dib-var1}}
    \label{plt-dib-var76}
\end{figure*}

\begin{figure*}[ht!]
    \centering
    \includegraphics[width=0.99\hsize]{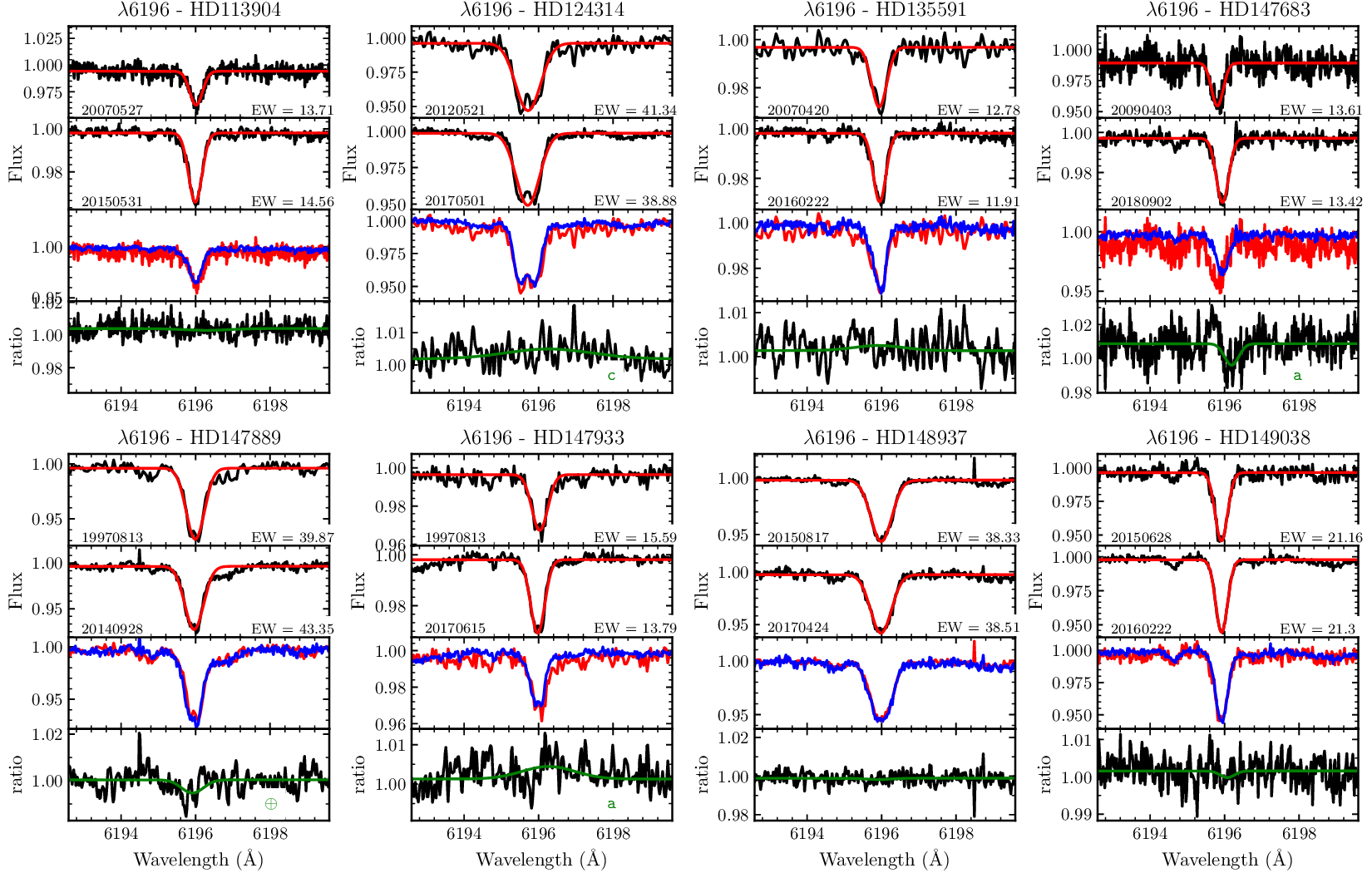}
    \caption{The same as \ref{plt-dib-var1}}
    \label{plt-dib-var77}
\end{figure*}

\begin{figure*}[ht!]
    \centering
    \includegraphics[width=0.99\hsize]{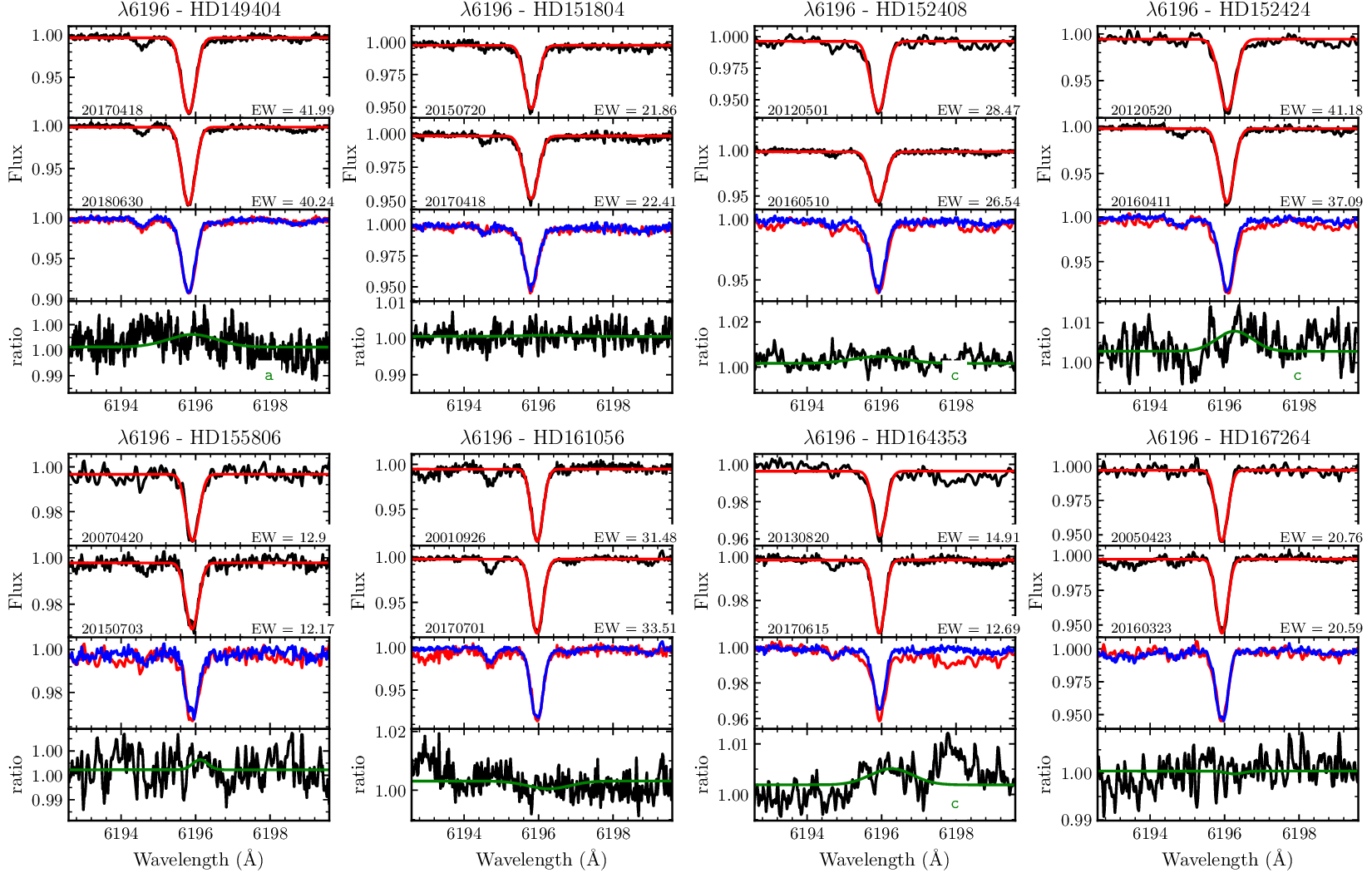}
    \caption{The same as \ref{plt-dib-var1}}
    \label{plt-dib-var78}
\end{figure*}

\begin{figure*}[ht!]
    \centering
    \includegraphics[width=0.99\hsize]{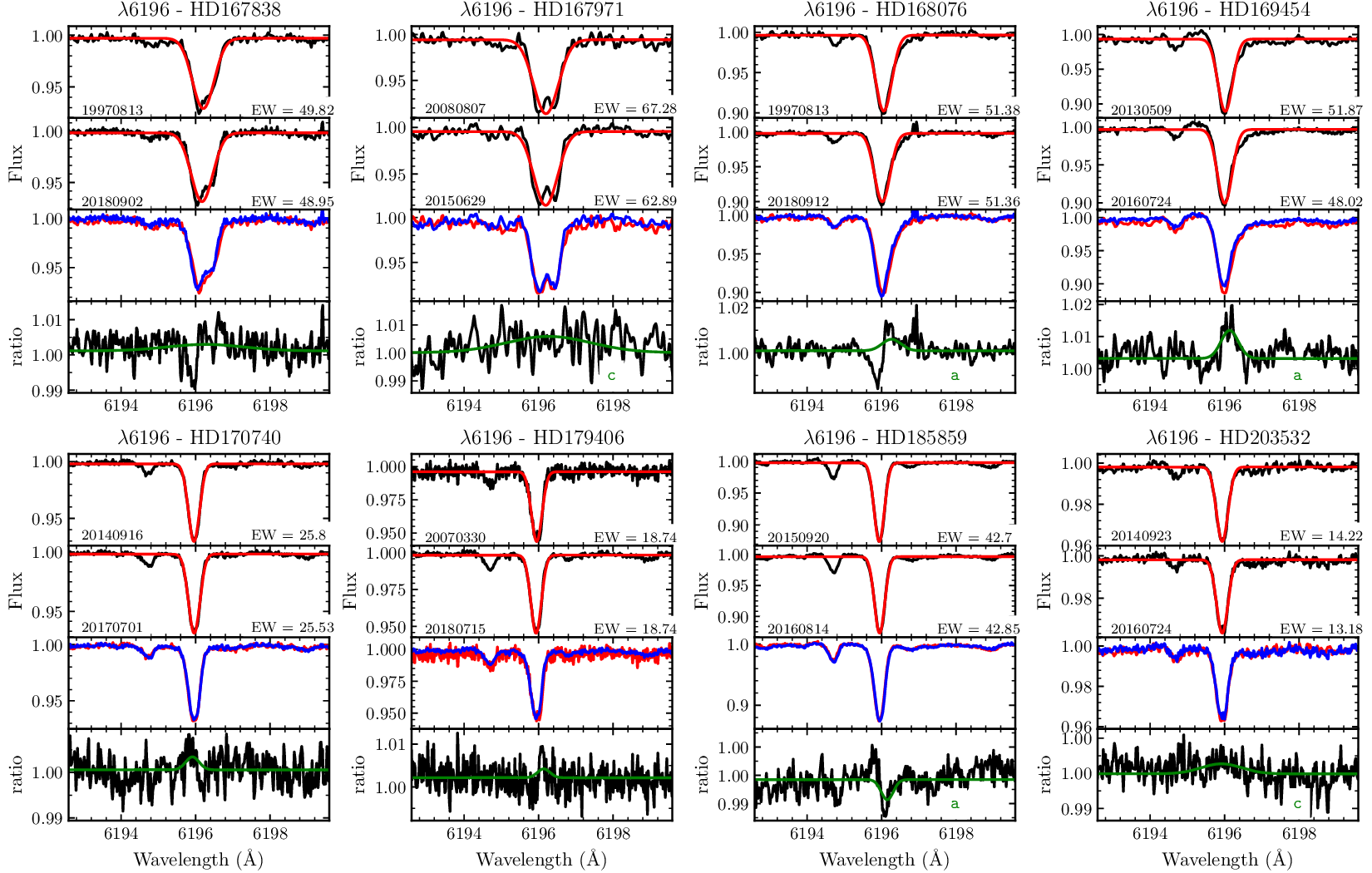}
    \caption{The same as \ref{plt-dib-var1}}
    \label{plt-dib-var79}
\end{figure*}

% %%%%%%%%%%%%%%%%%%
% %%%%%%%%%%%%%%%%%%
\begin{figure*}[ht!]
    \centering
    \includegraphics[width=0.99\hsize]{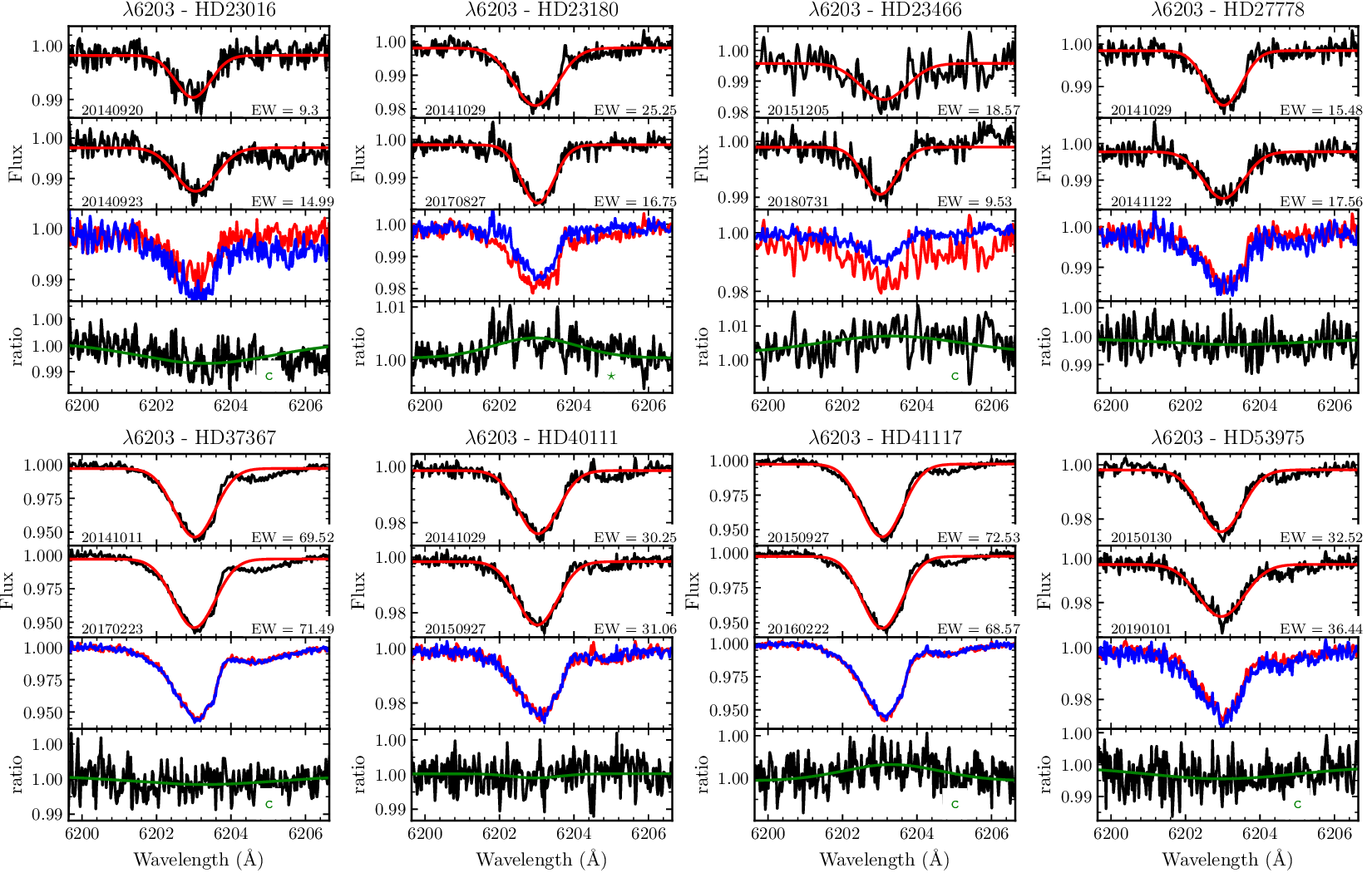}
    \caption{The same as \ref{plt-dib-var1}}
    \label{plt-dib-var80}
\end{figure*}

\begin{figure*}[ht!]
    \centering
    \includegraphics[width=0.99\hsize]{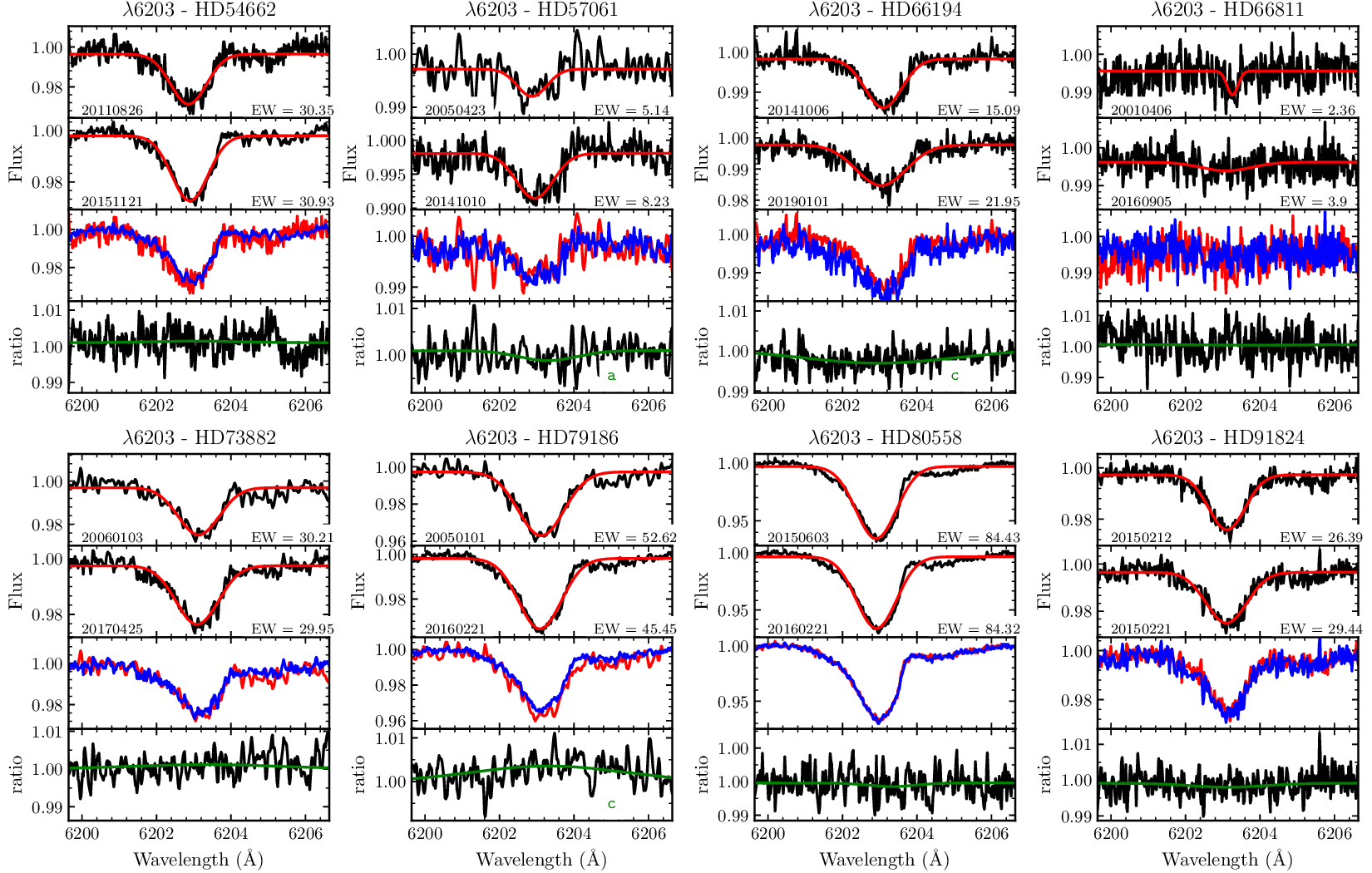}
    \caption{The same as \ref{plt-dib-var1}}
    \label{plt-dib-var81}
\end{figure*}

\begin{figure*}[ht!]
    \centering
    \includegraphics[width=0.99\hsize]{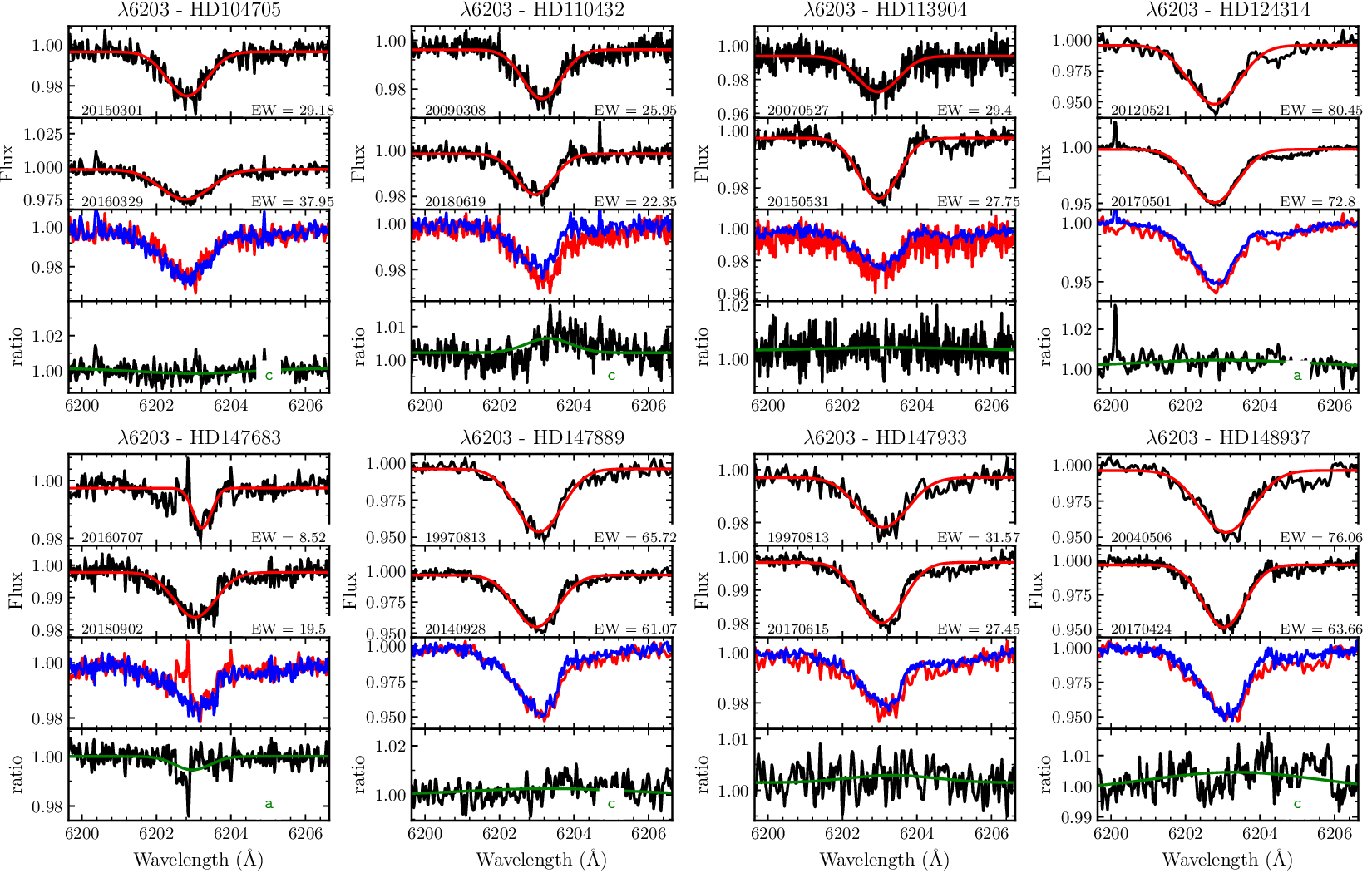}
    \caption{The same as \ref{plt-dib-var1}}
    \label{plt-dib-var82}
\end{figure*}

\begin{figure*}[ht!]
    \centering
    \includegraphics[width=0.99\hsize]{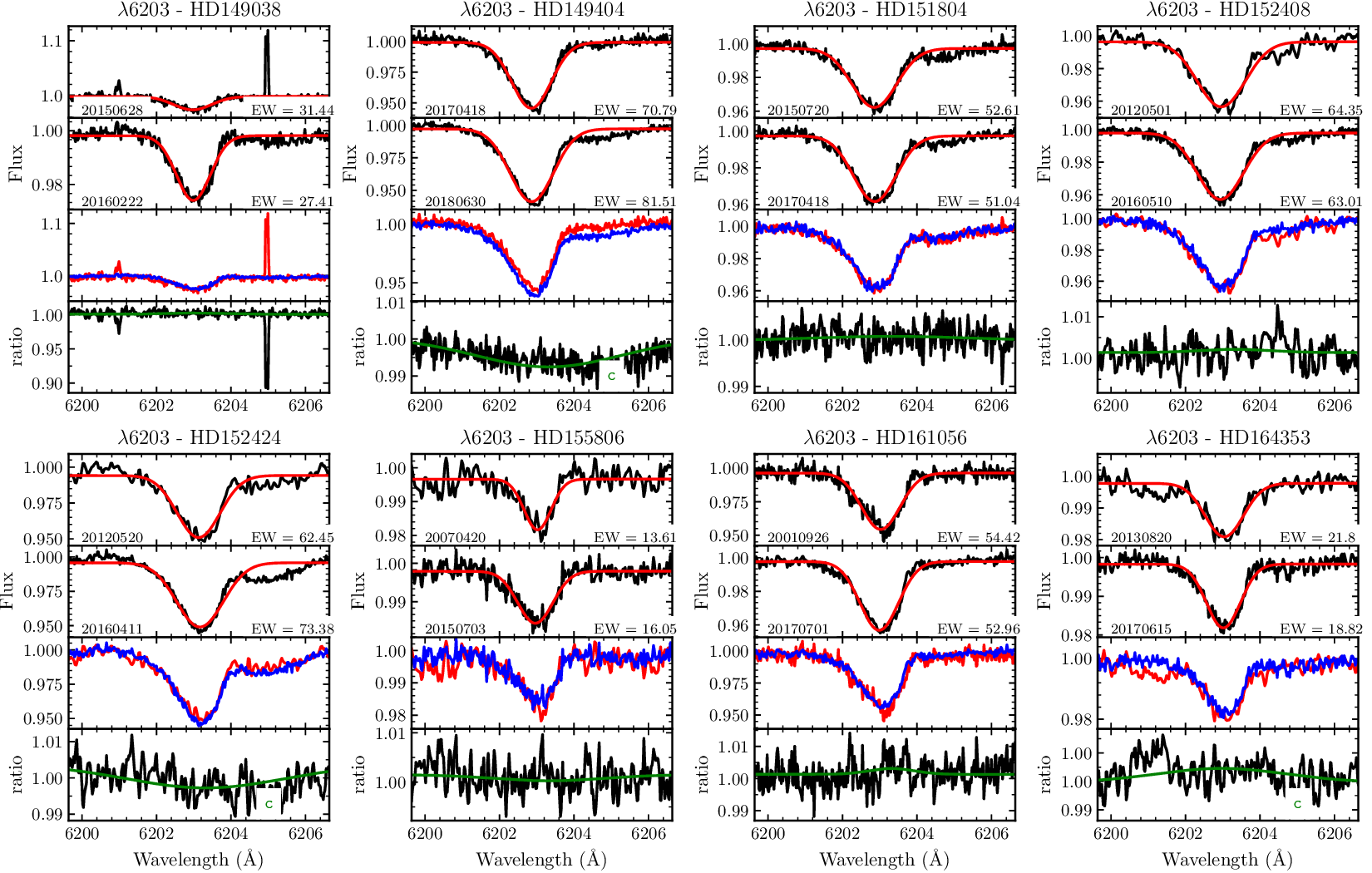}
    \caption{The same as \ref{plt-dib-var1}}
    \label{plt-dib-var83}
\end{figure*}

\begin{figure*}[ht!]
    \centering
    \includegraphics[width=0.99\hsize]{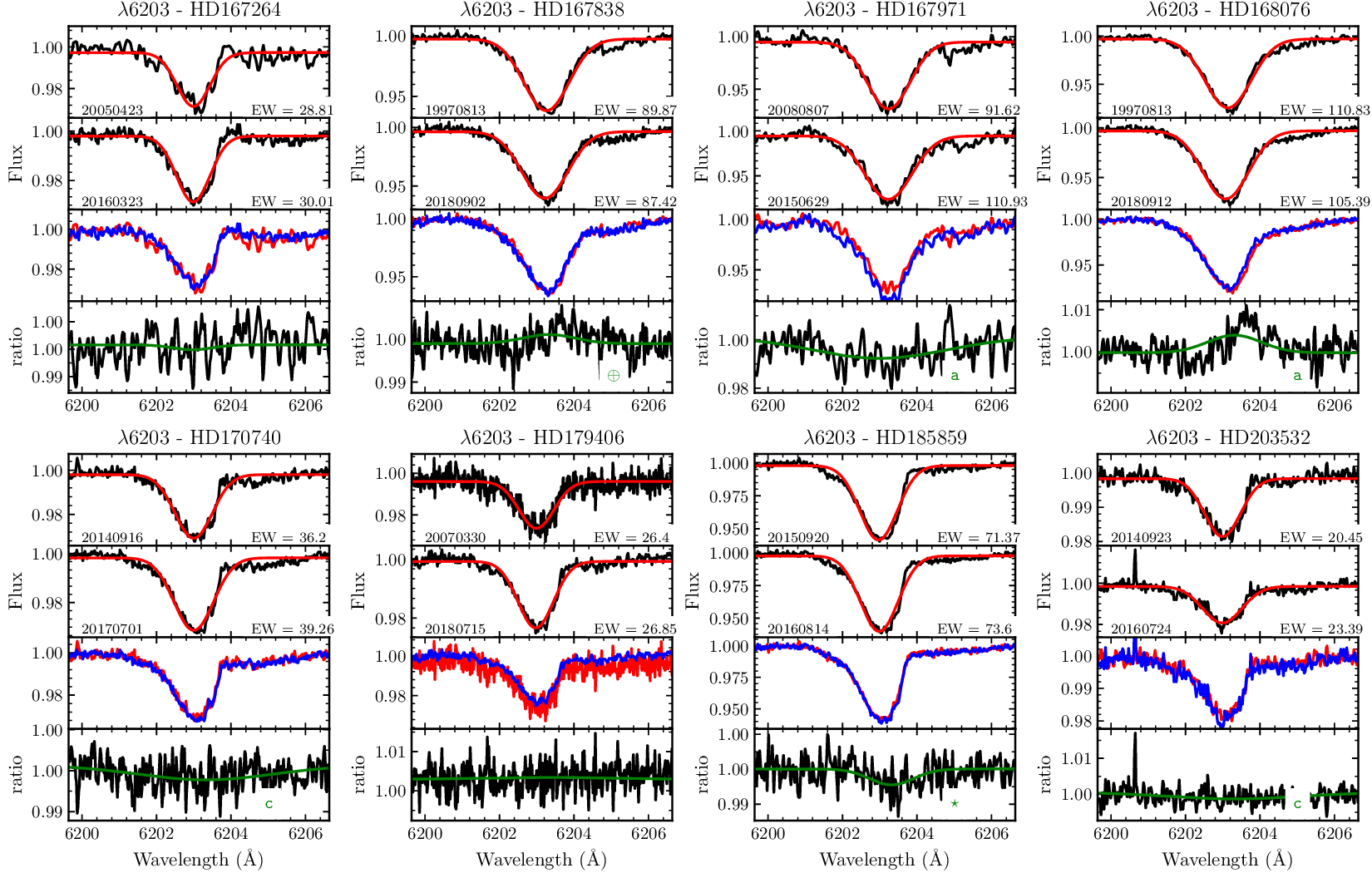}
    \caption{The same as \ref{plt-dib-var1}}
    \label{plt-dib-var84}
\end{figure*}

% %%%%%%%%%%%%%%%%%%
% %%%%%%%%%%%%%%%%%%
\begin{figure*}[ht!]
    \centering
    \includegraphics[width=0.99\hsize]{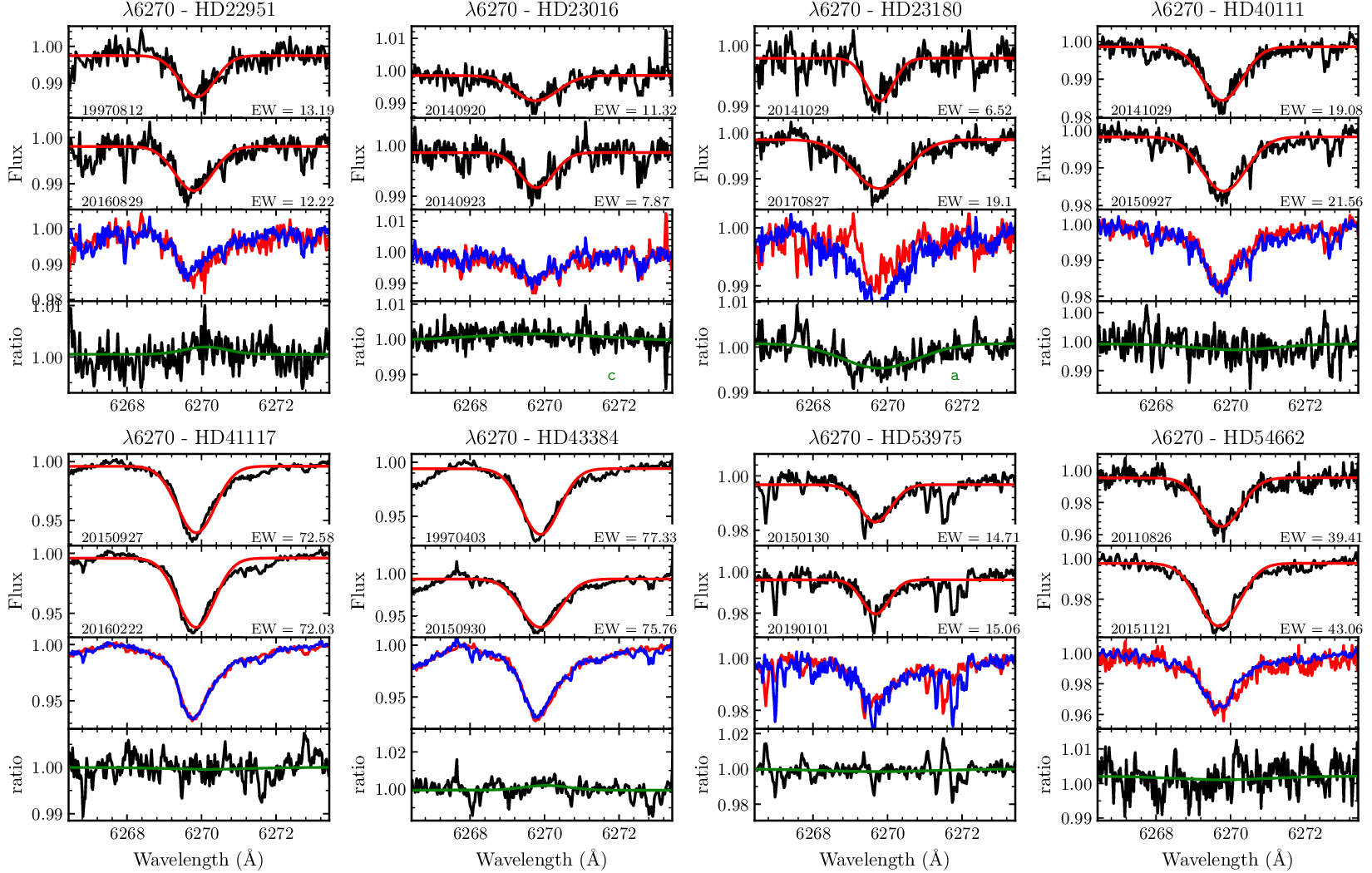}
    \caption{The same as \ref{plt-dib-var1}}
    \label{plt-dib-var85}
\end{figure*}

\begin{figure*}[ht!]
    \centering
    \includegraphics[width=0.99\hsize]{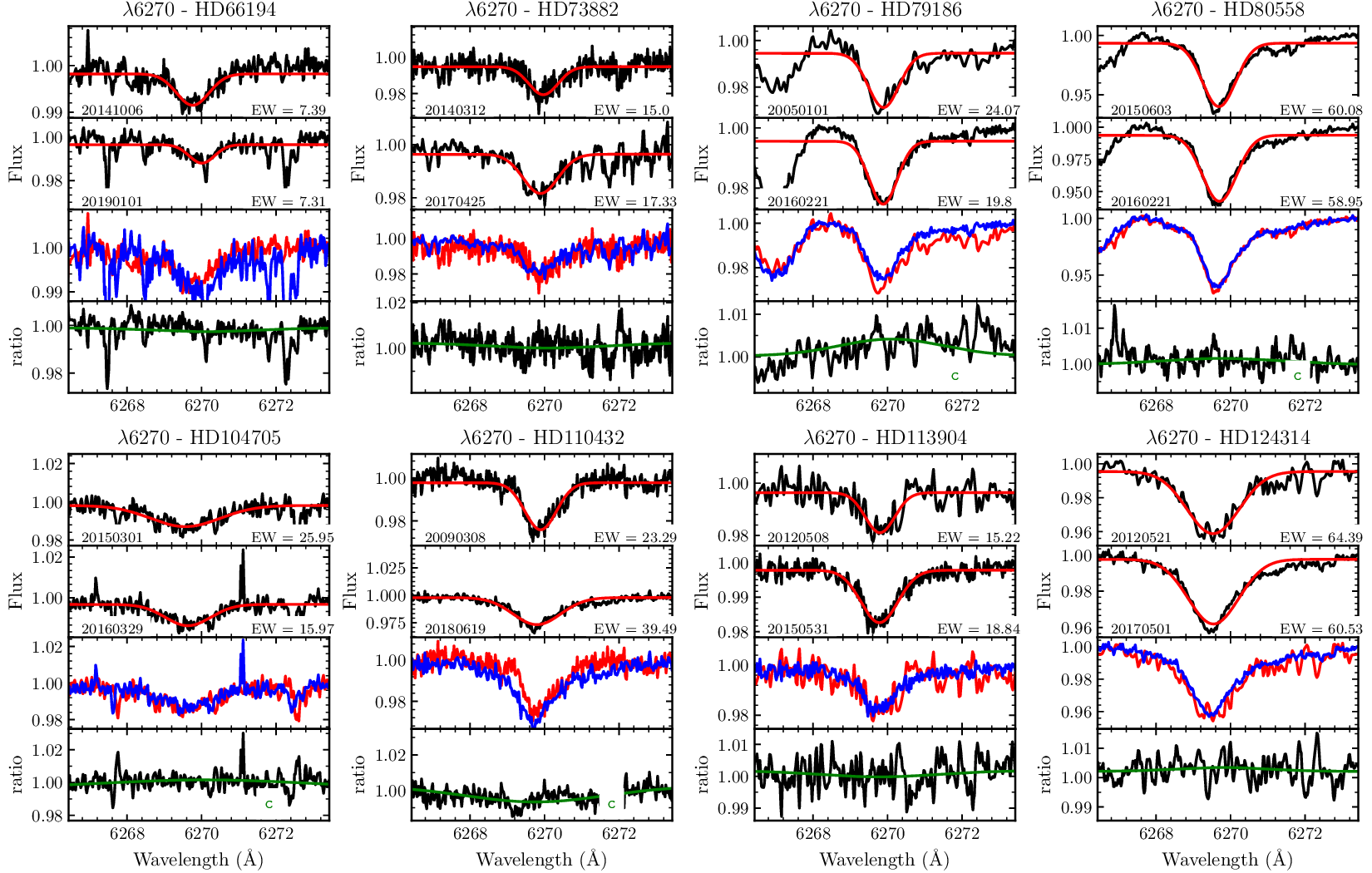}
    \caption{The same as \ref{plt-dib-var1}}
    \label{plt-dib-var86}
\end{figure*}

\begin{figure*}[ht!]
    \centering
    \includegraphics[width=0.99\hsize]{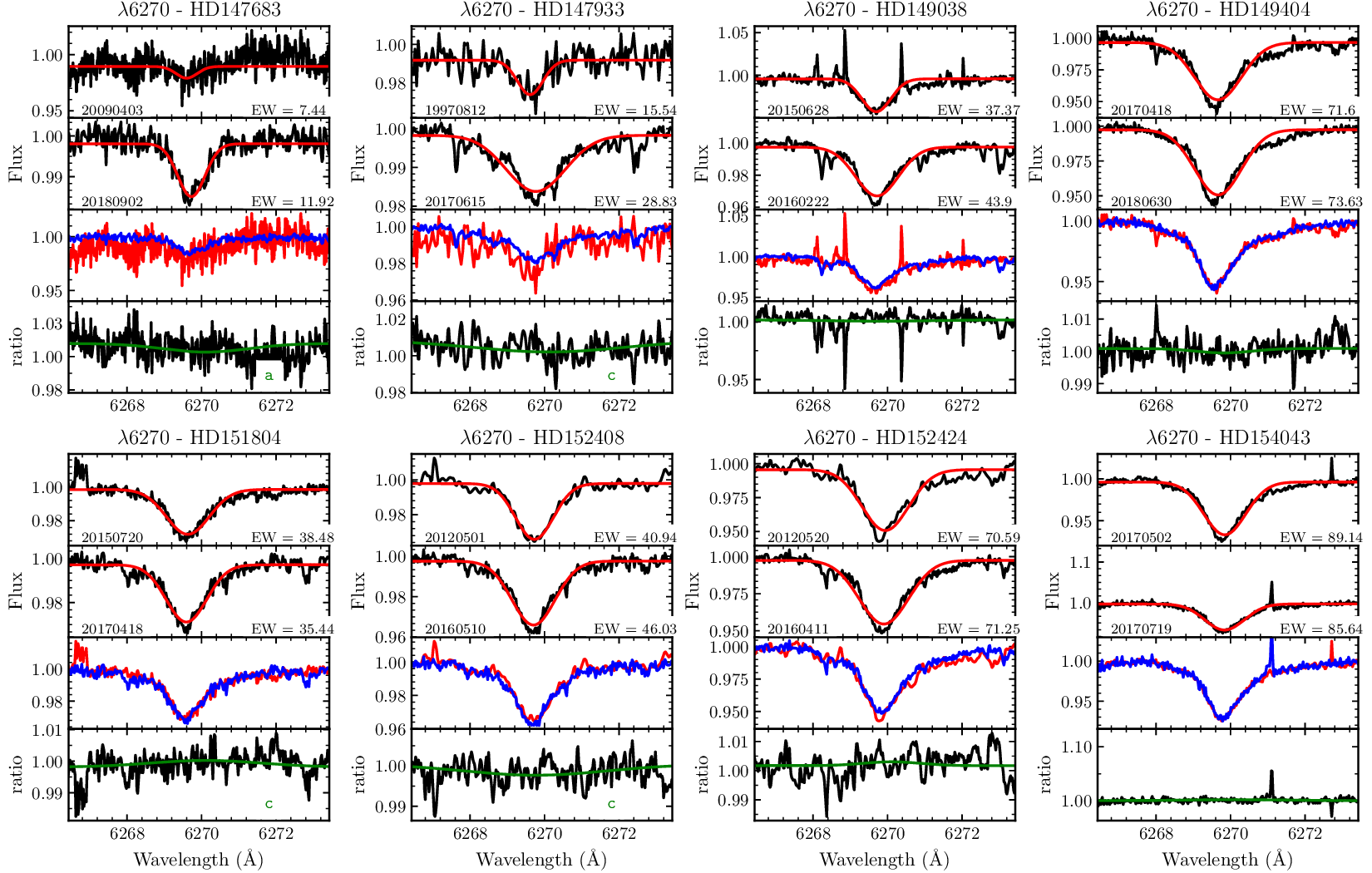}
    \caption{The same as \ref{plt-dib-var1}}
    \label{plt-dib-var87}
\end{figure*}

\begin{figure*}[ht!]
    \centering
    \includegraphics[width=0.99\hsize]{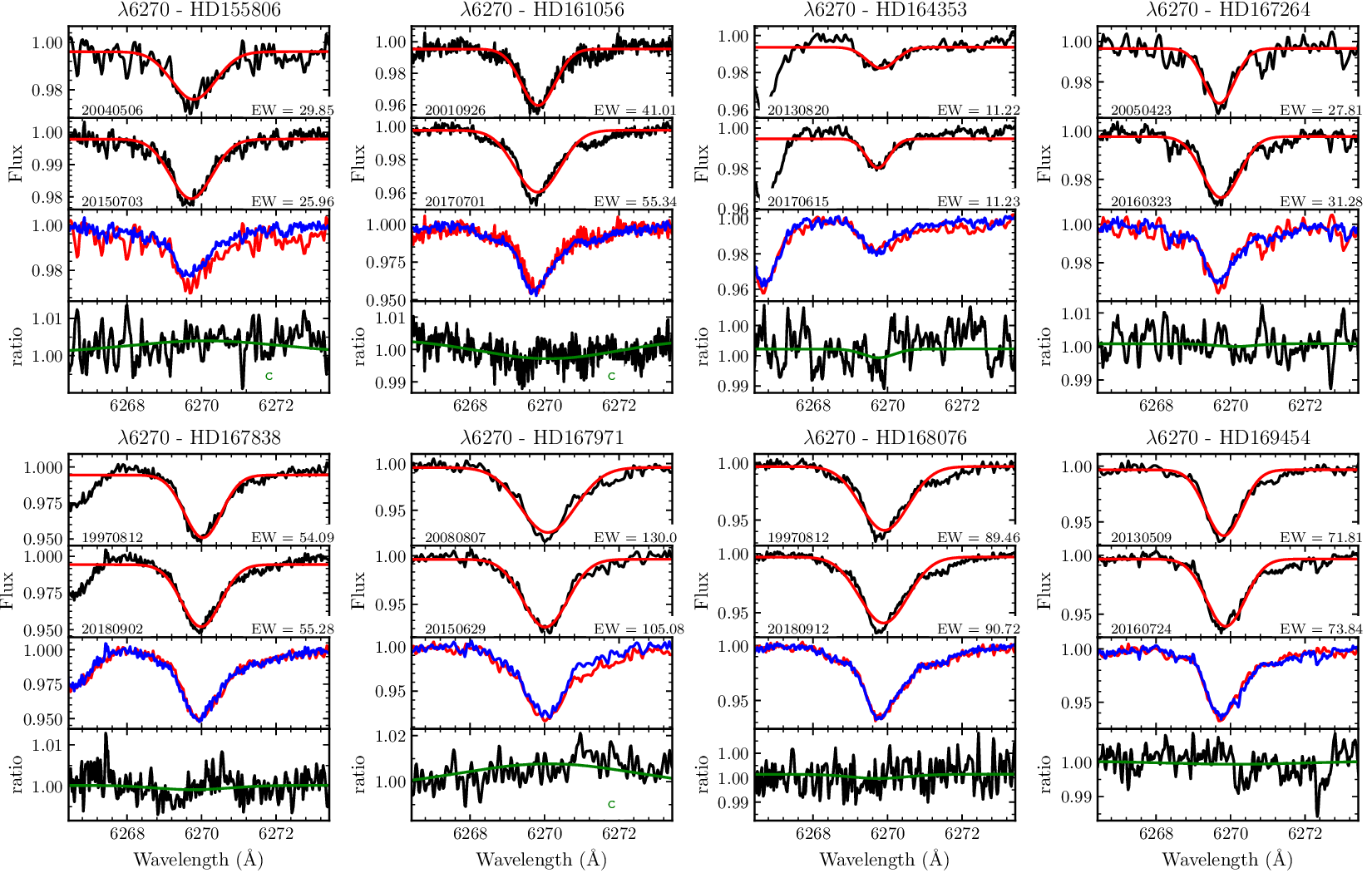}
    \caption{The same as \ref{plt-dib-var1}}
    \label{plt-dib-var88}
\end{figure*}

\begin{figure*}[ht!]
    \centering
    \includegraphics[width=0.99\hsize]{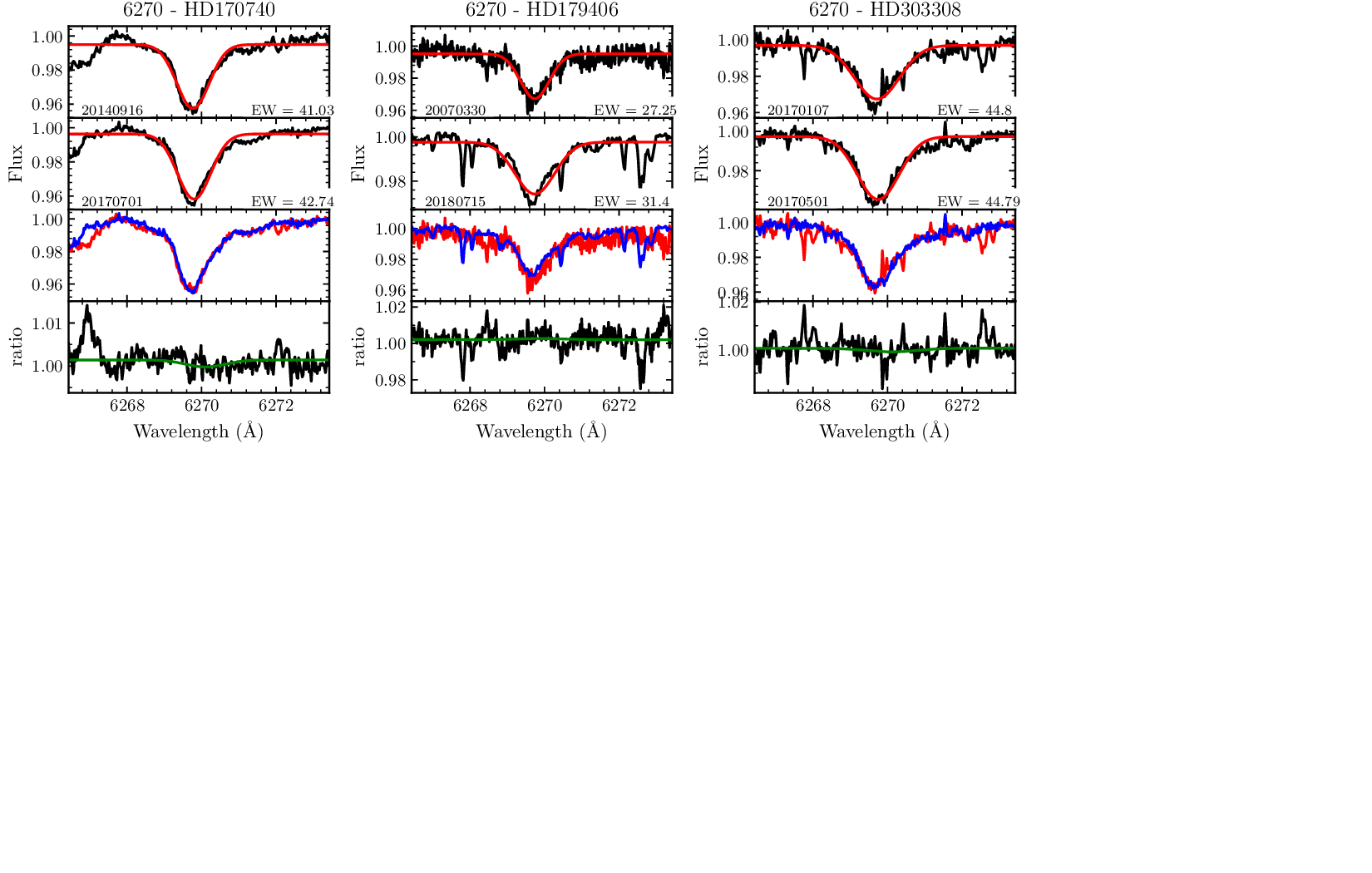}
    \caption{The same as \ref{plt-dib-var1}}
    \label{plt-dib-var89}
\end{figure*}

% %%%%%%%%%%%%%%%%%%
% %%%%%%%%%%%%%%%%%%
\begin{figure*}[ht!]
    \centering
    \includegraphics[width=0.99\hsize]{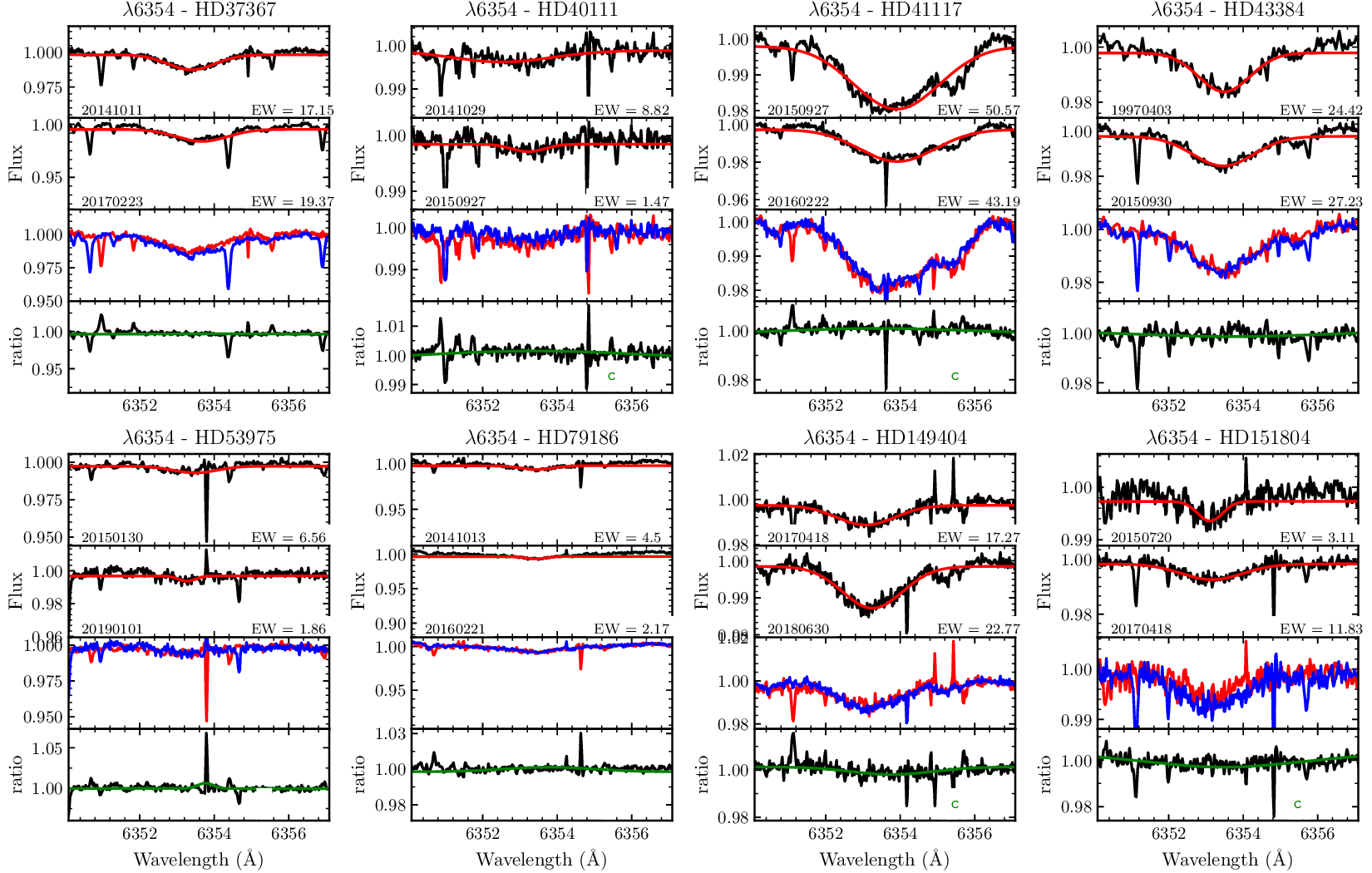}
    \caption{The same as \ref{plt-dib-var1}}
    \label{plt-dib-var90}
\end{figure*}

\begin{figure*}[ht!]
    \centering
    \includegraphics[width=0.99\hsize]{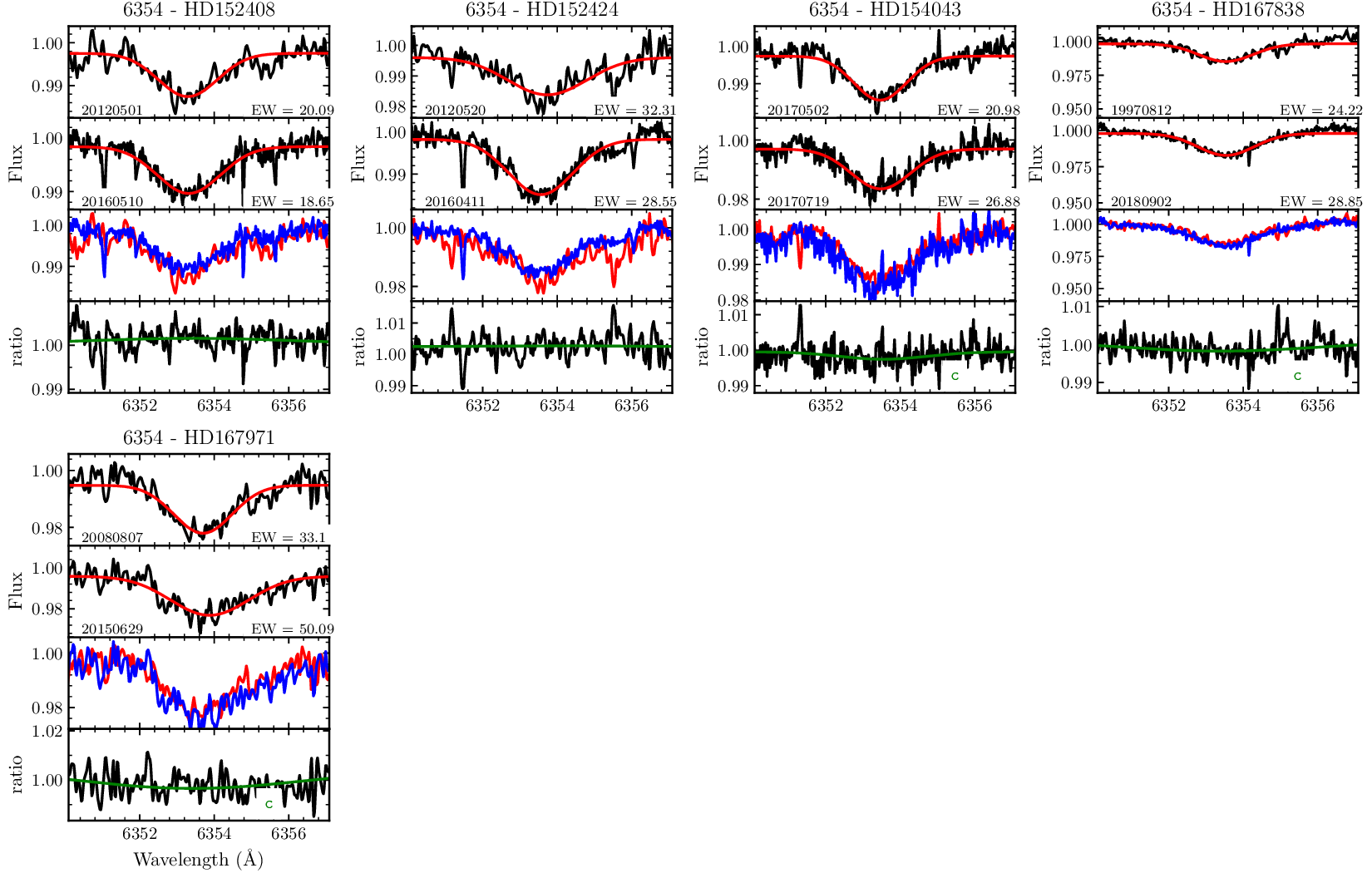}
    \caption{The same as \ref{plt-dib-var1}}
    \label{plt-dib-var91}
\end{figure*}

% %%%%%%%%%%%%%%%%%%
% %%%%%%%%%%%%%%%%%%
\begin{figure*}[ht!]
    \centering
    \includegraphics[width=0.99\hsize]{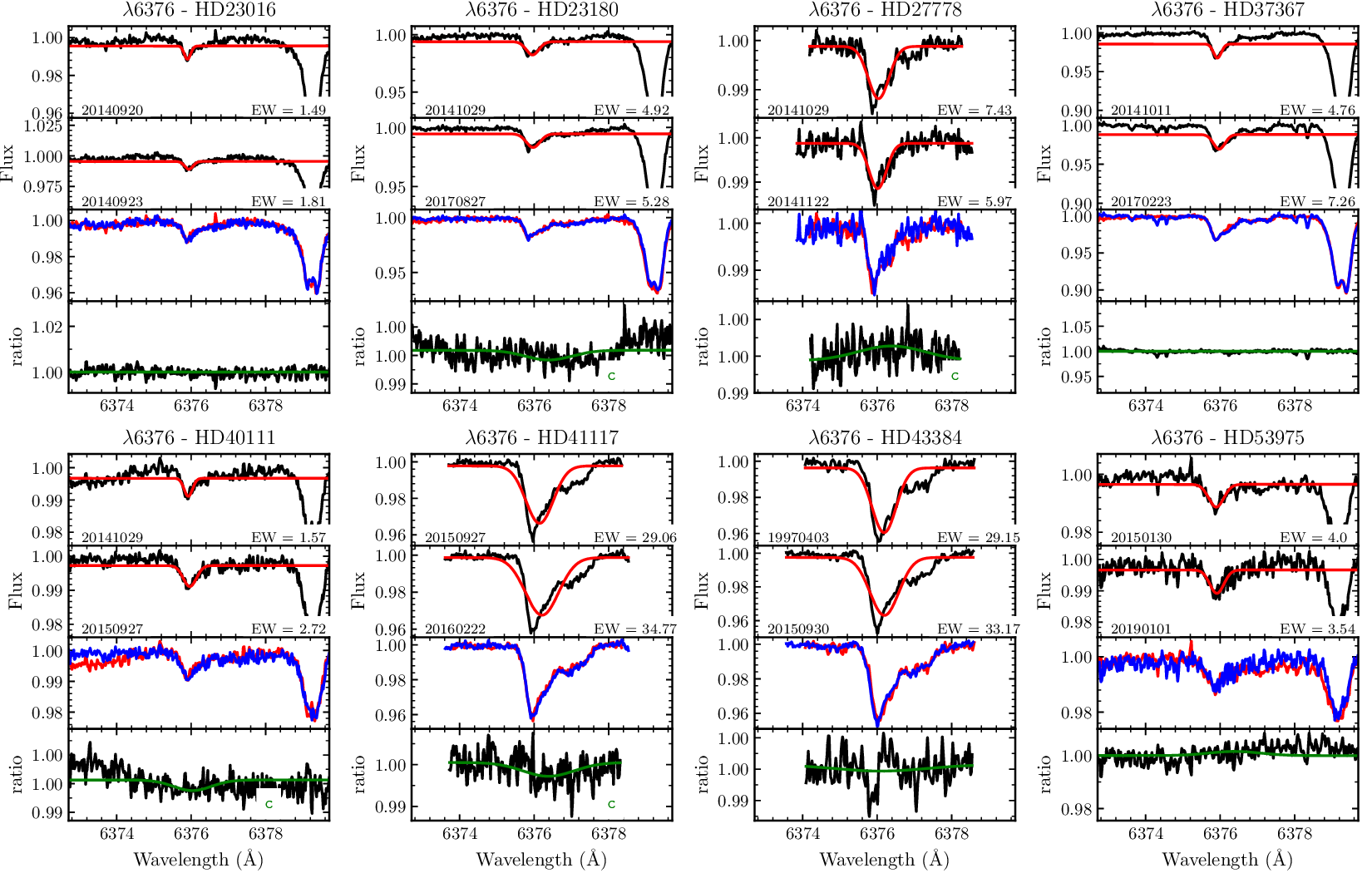}
    \caption{The same as \ref{plt-dib-var1}}
    \label{plt-dib-var92}
\end{figure*}

\begin{figure*}[ht!]
    \centering
    \includegraphics[width=0.99\hsize]{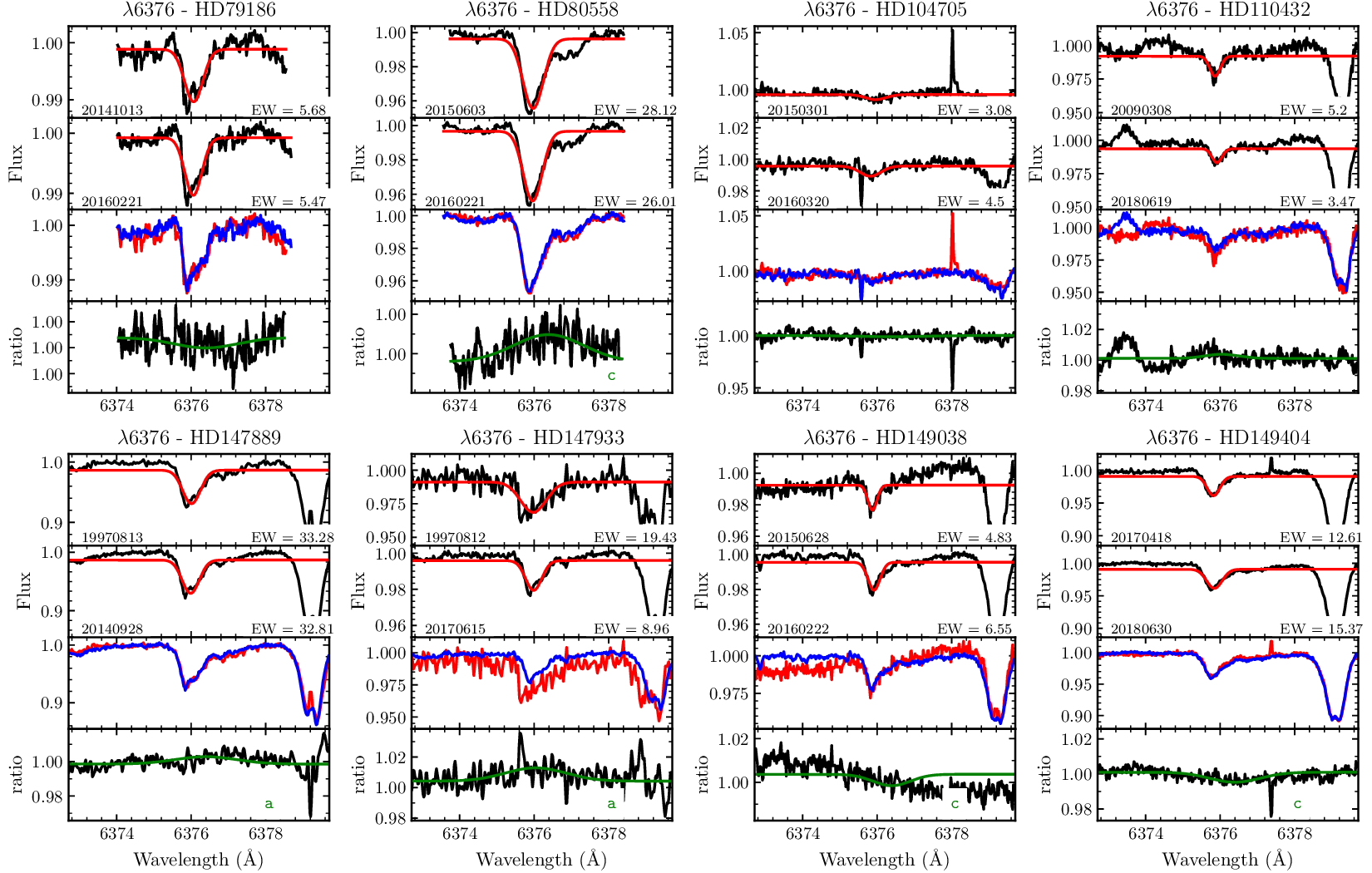}
    \caption{The same as \ref{plt-dib-var1}}
    \label{plt-dib-var93}
\end{figure*}

\begin{figure*}[ht!]
    \centering
    \includegraphics[width=0.99\hsize]{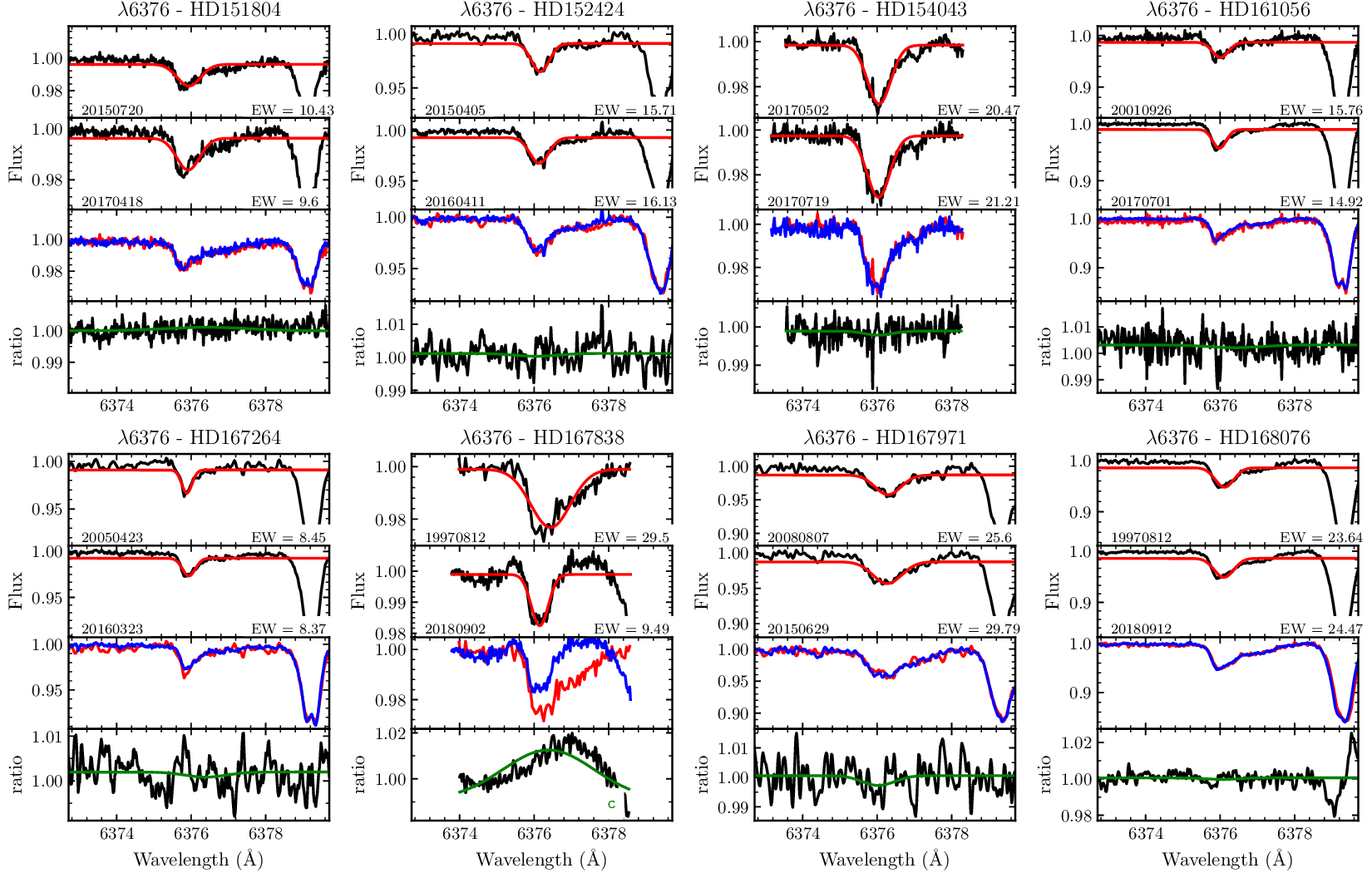}
    \caption{The same as \ref{plt-dib-var1}}
    \label{plt-dib-var94}
\end{figure*}

\begin{figure*}[ht!]
    \centering
    \includegraphics[width=0.99\hsize]{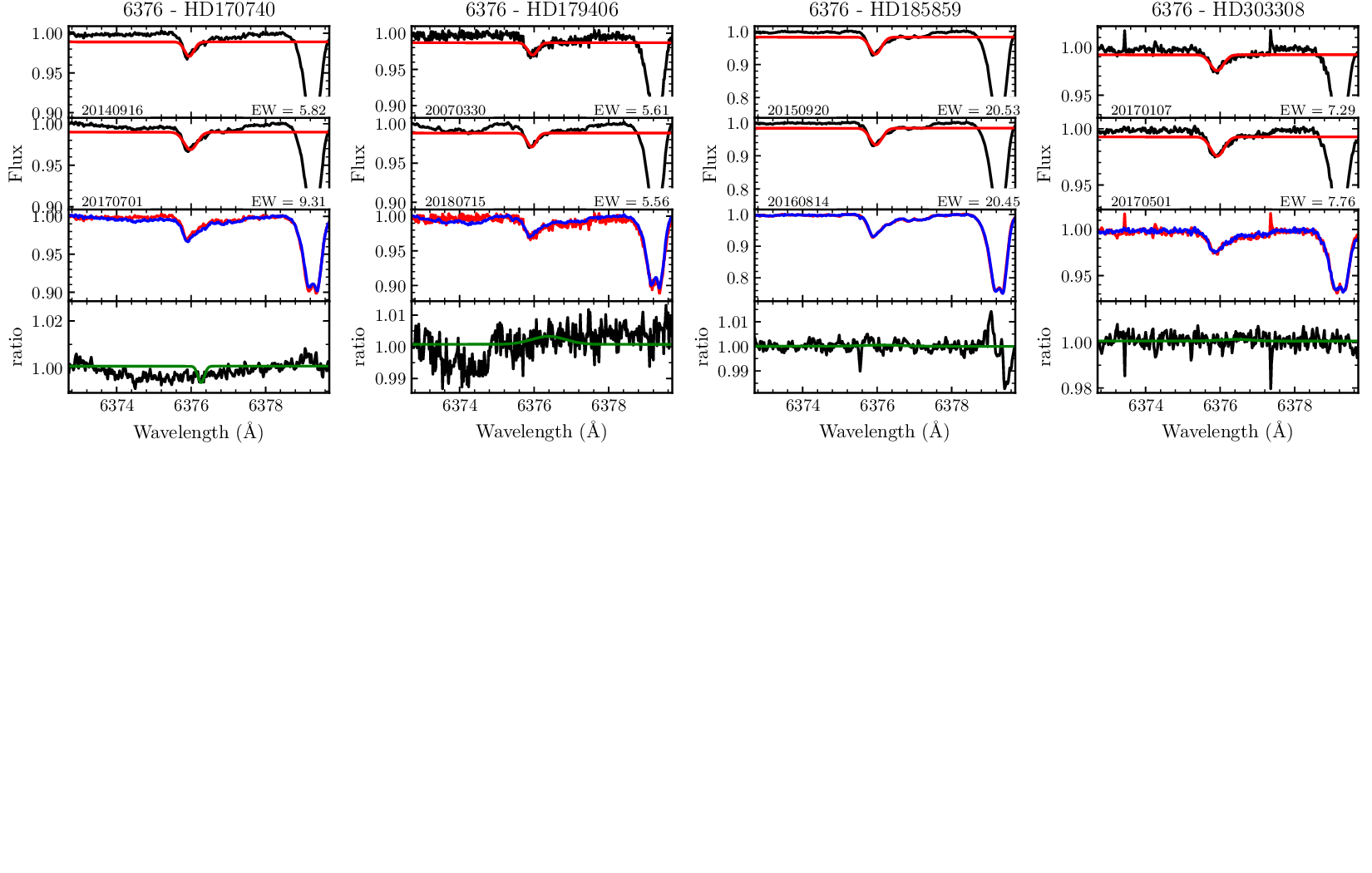}
    \caption{The same as \ref{plt-dib-var1}}
    \label{plt-dib-var95}
\end{figure*}

% %%%%%%%%%%%%%%%%%%
% %%%%%%%%%%%%%%%%%%
\begin{figure*}[ht!]
    \centering
    \includegraphics[width=0.99\hsize]{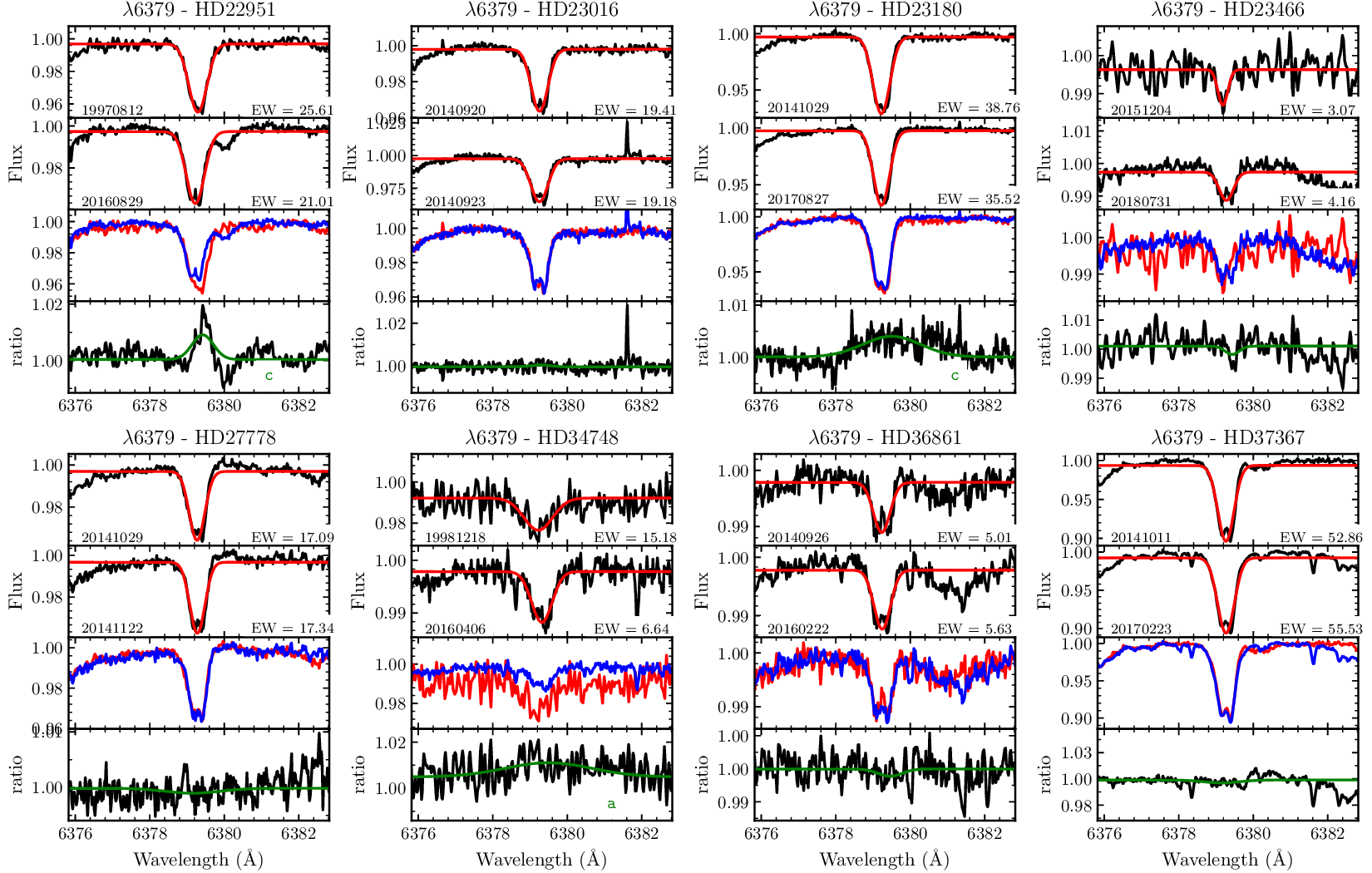}
    \caption{The same as \ref{plt-dib-var1}}
    \label{plt-dib-var96}
\end{figure*}

\begin{figure*}[ht!]
    \centering
    \includegraphics[width=0.99\hsize]{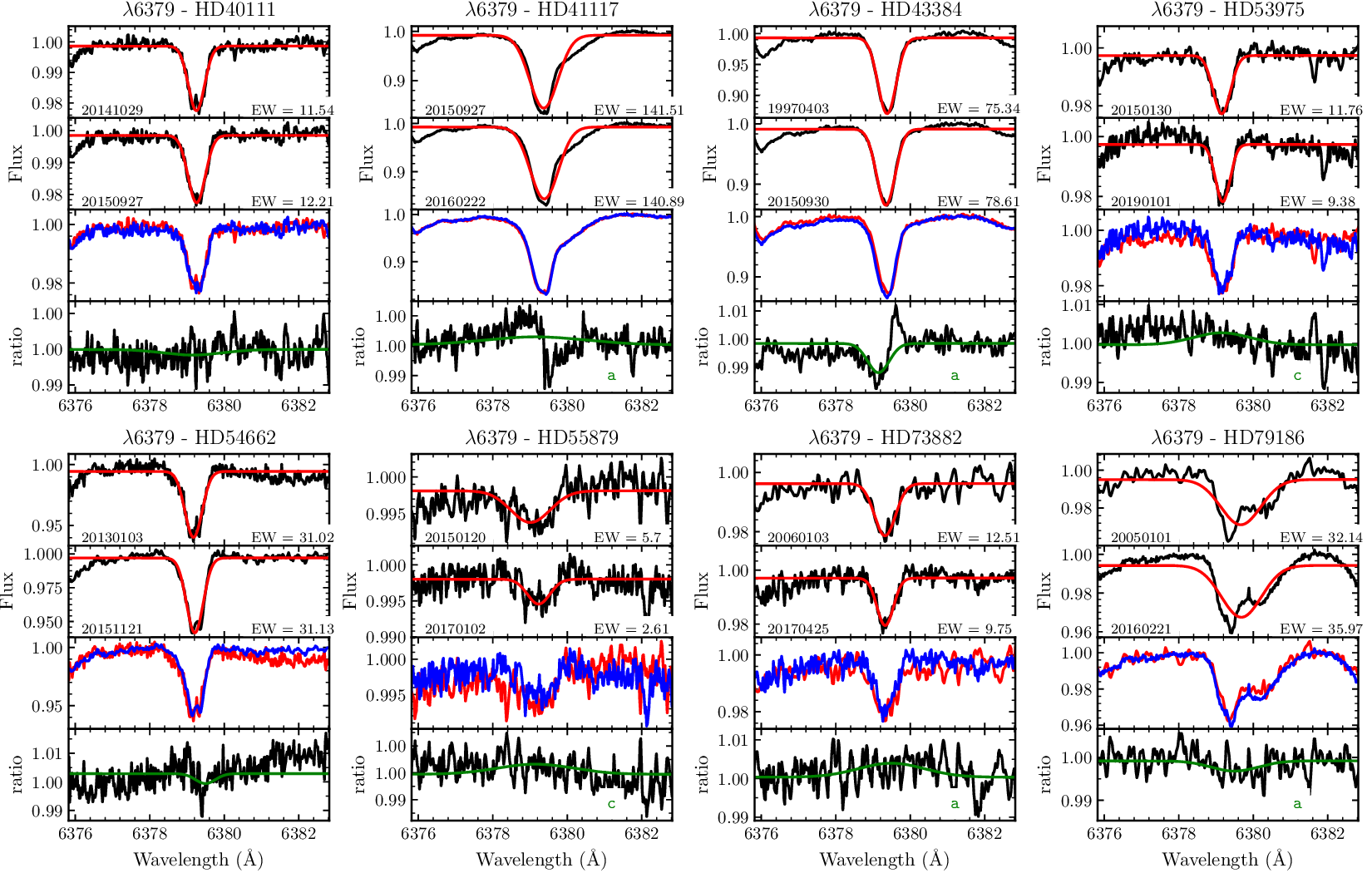}
    \caption{The same as \ref{plt-dib-var1}}
    \label{plt-dib-var97}
\end{figure*}

\begin{figure*}[ht!]
    \centering
    \includegraphics[width=0.99\hsize]{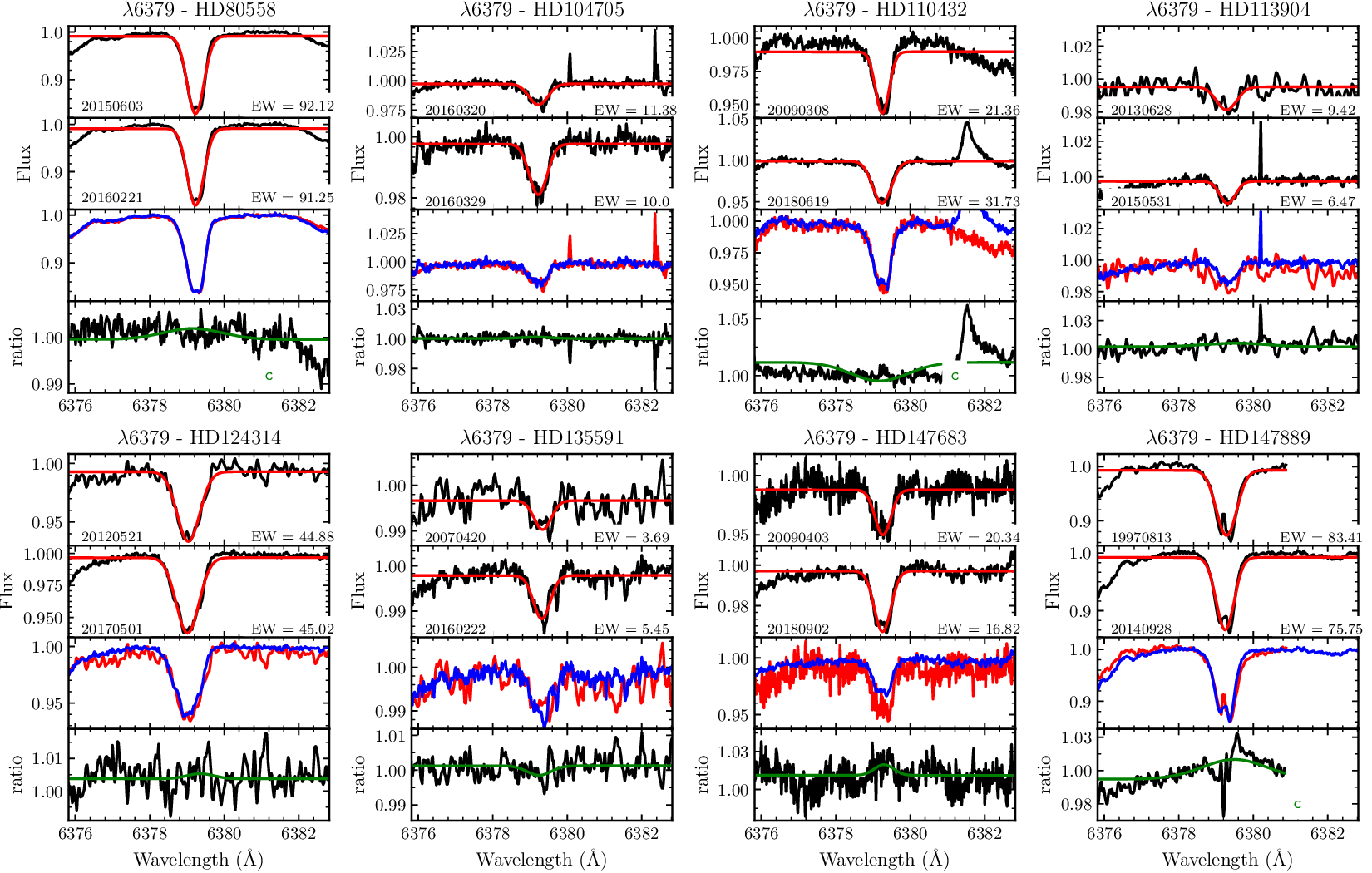}
    \caption{The same as \ref{plt-dib-var1}}
    \label{plt-dib-var98}
\end{figure*}

\begin{figure*}[ht!]
    \centering
    \includegraphics[width=0.99\hsize]{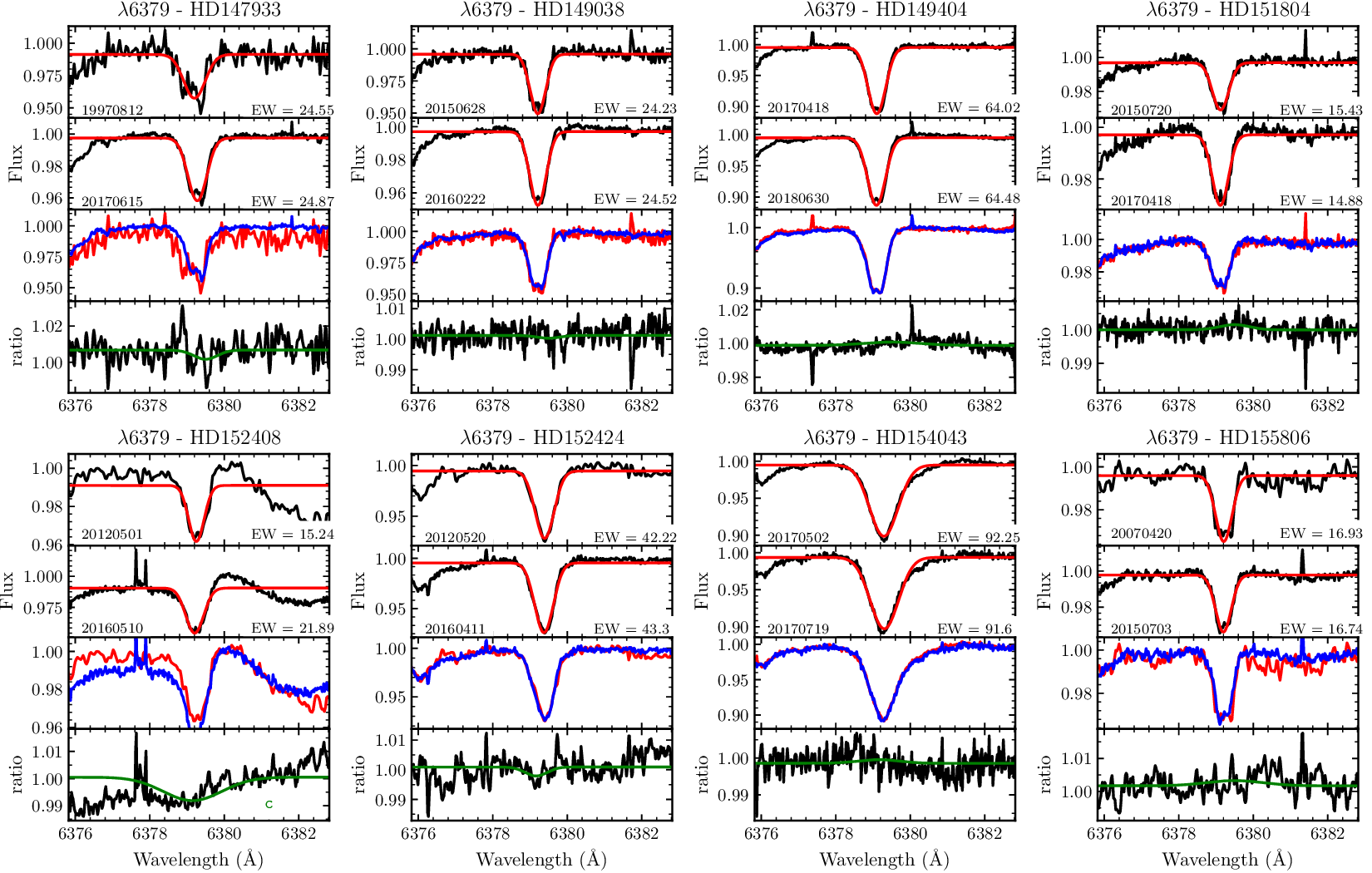}
    \caption{The same as \ref{plt-dib-var1}}
    \label{plt-dib-var99}
\end{figure*}

\begin{figure*}[ht!]
    \centering
    \includegraphics[width=0.99\hsize]{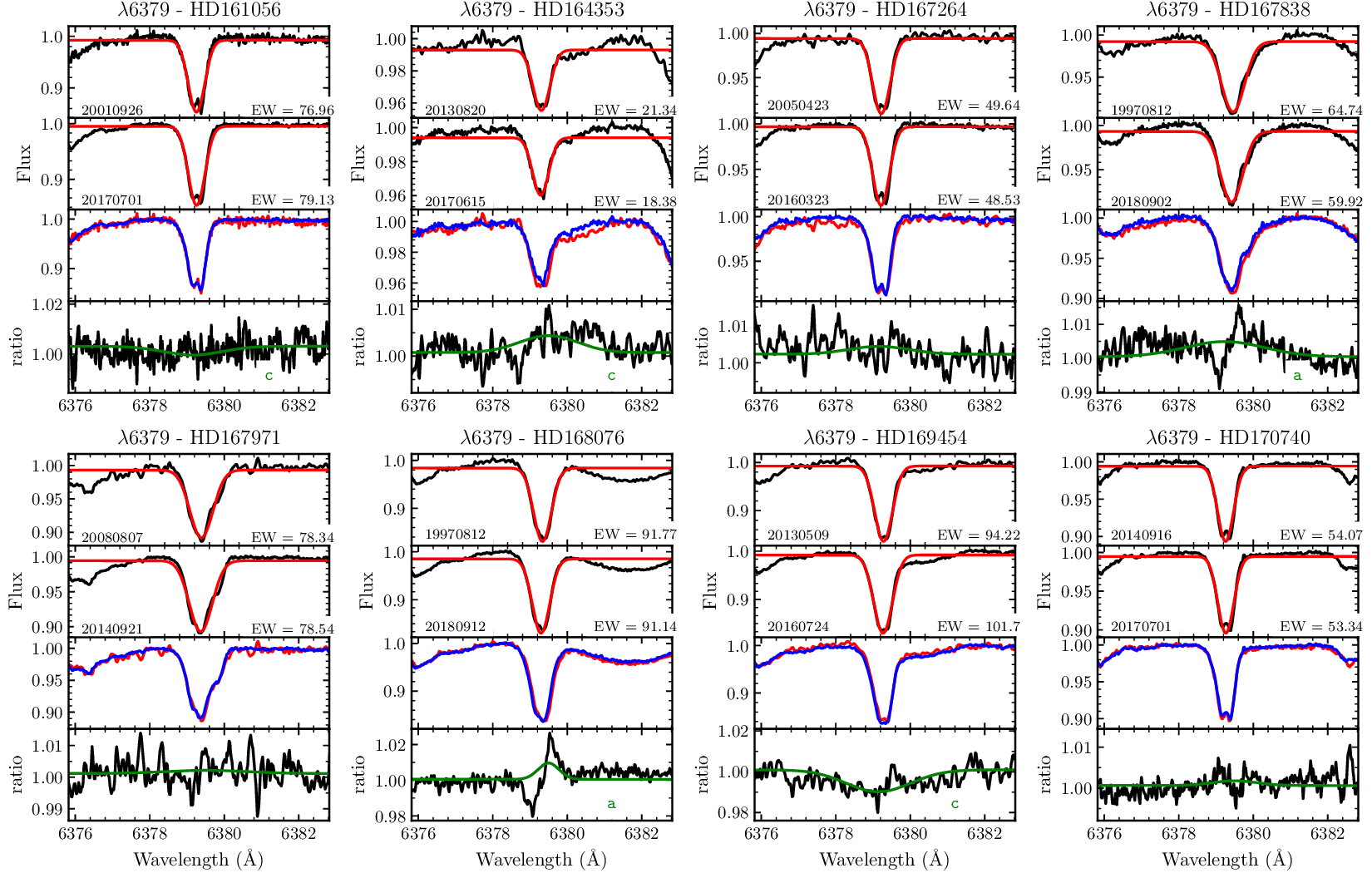}
    \caption{The same as \ref{plt-dib-var1}}
    \label{plt-dib-var100}
\end{figure*}

\begin{figure*}[ht!]
    \centering
    \includegraphics[width=0.99\hsize]{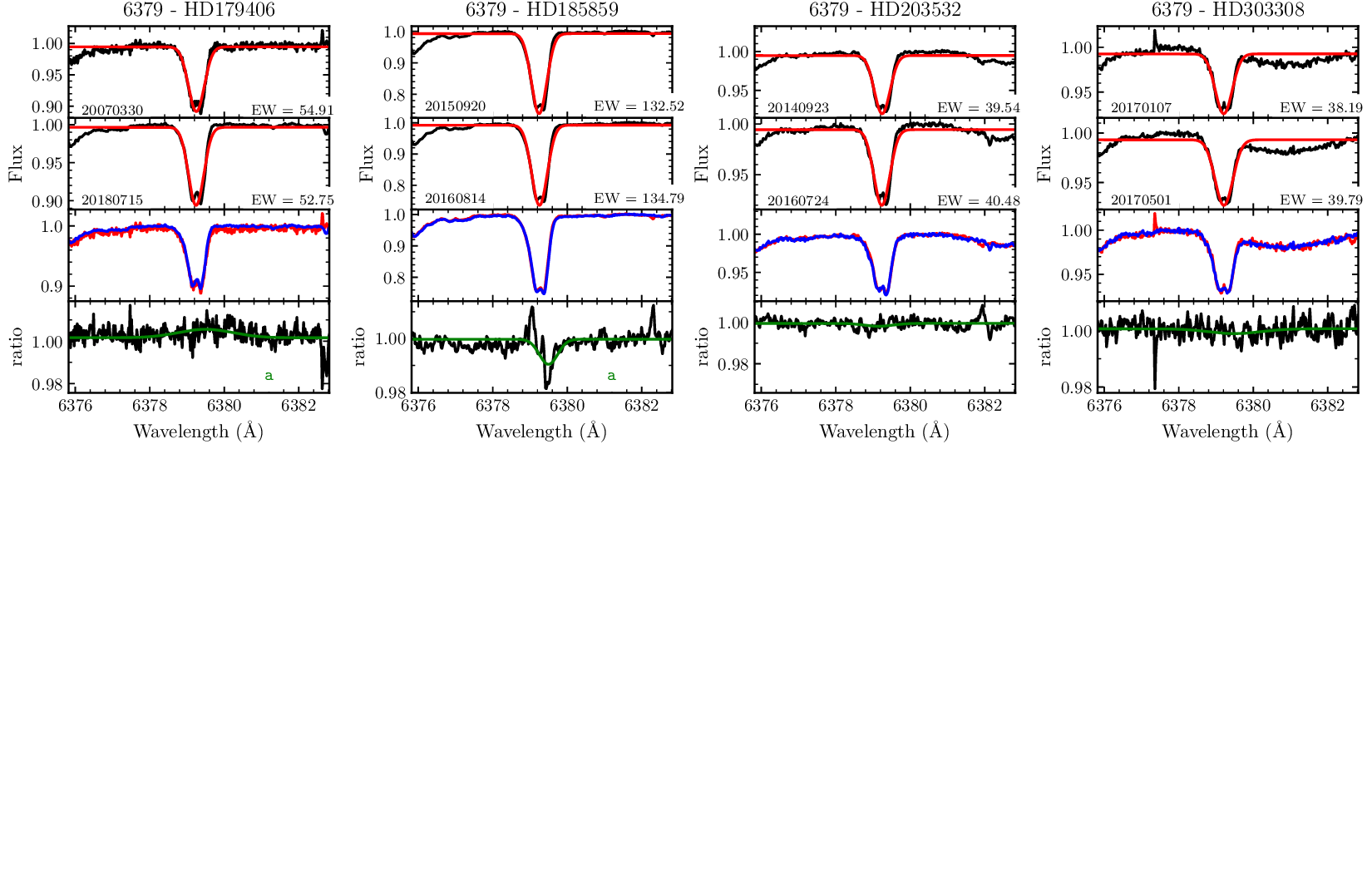}
    \caption{The same as \ref{plt-dib-var1}}
    \label{plt-dib-var101}
\end{figure*}

% %%%%%%%%%%%%%%%%%%
% %%%%%%%%%%%%%%%%%%
\begin{figure*}[ht!]
    \centering
    \includegraphics[width=0.99\hsize]{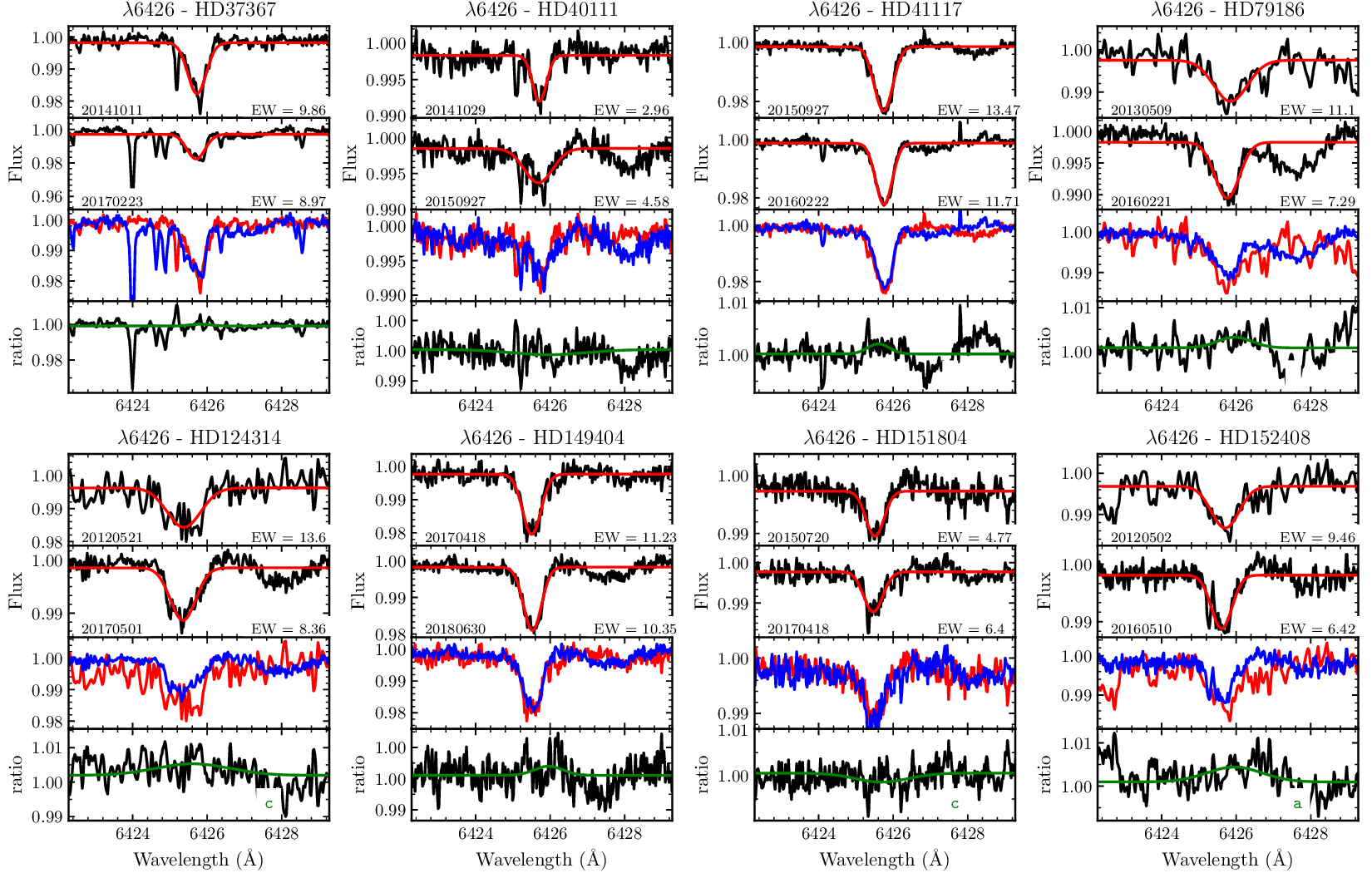}
    \caption{The same as \ref{plt-dib-var1}}
    \label{plt-dib-var102}
\end{figure*}

\begin{figure*}[ht!]
    \centering
    \includegraphics[width=0.99\hsize]{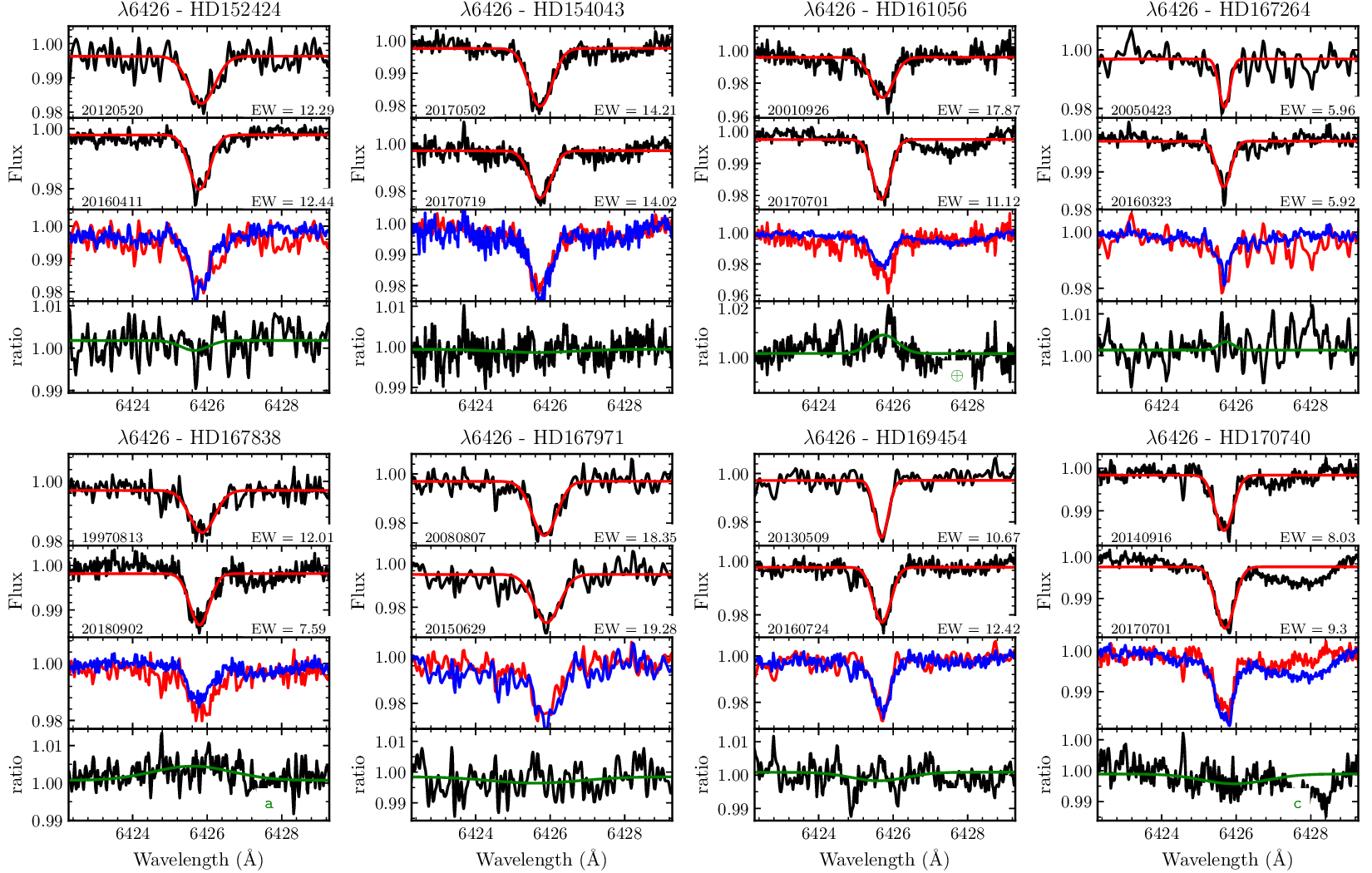}
    \caption{The same as \ref{plt-dib-var1}}
    \label{plt-dib-var103}
\end{figure*}

\begin{figure*}[ht!]
    \centering
    \includegraphics[width=0.99\hsize]{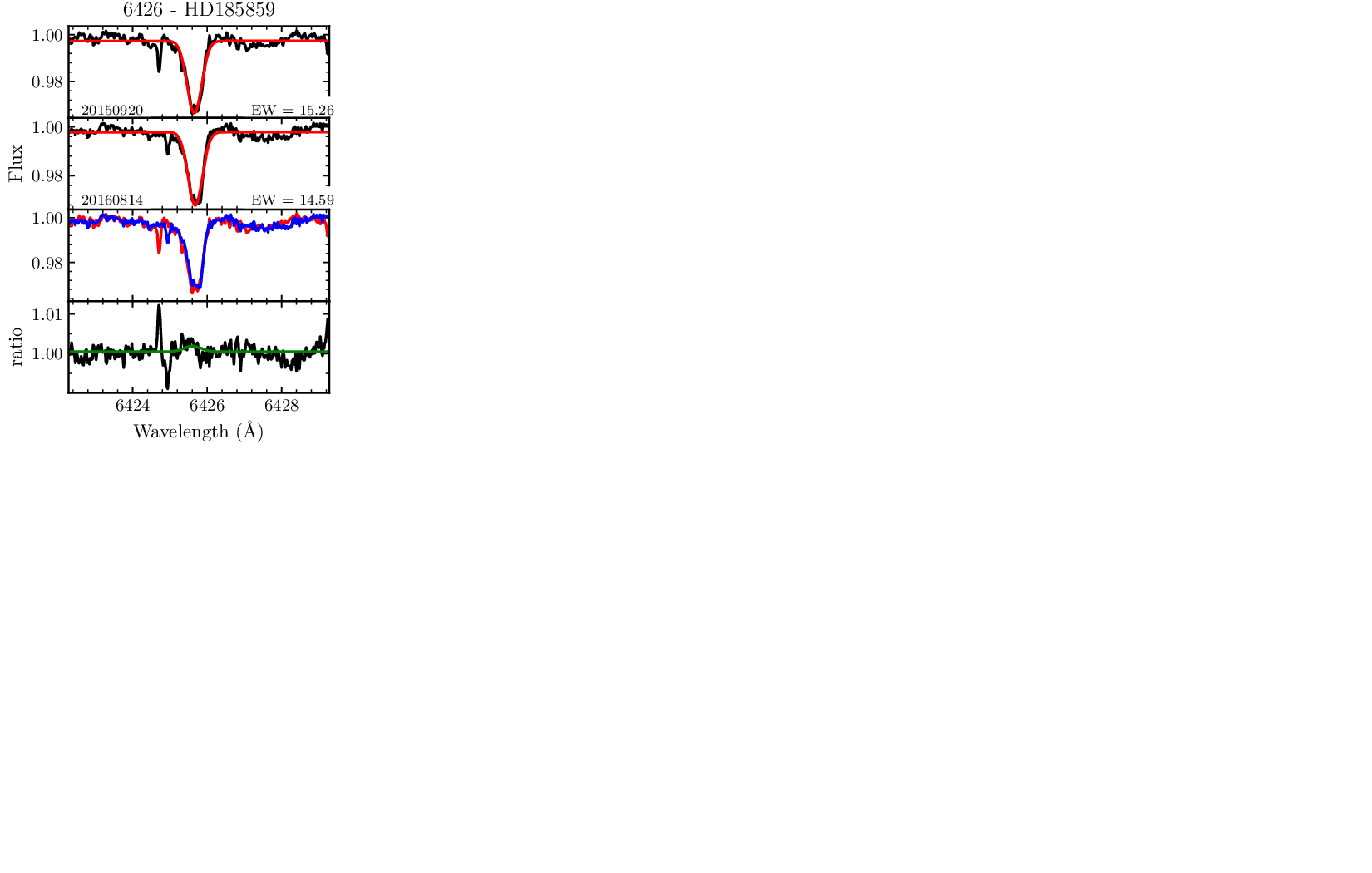}
    \caption{The same as \ref{plt-dib-var1}}
    \label{plt-dib-var104}
\end{figure*}

% %%%%%%%%%%%%%%%%%%
% %%%%%%%%%%%%%%%%%%
\begin{figure*}[ht!]
    \centering
    \includegraphics[width=0.99\hsize]{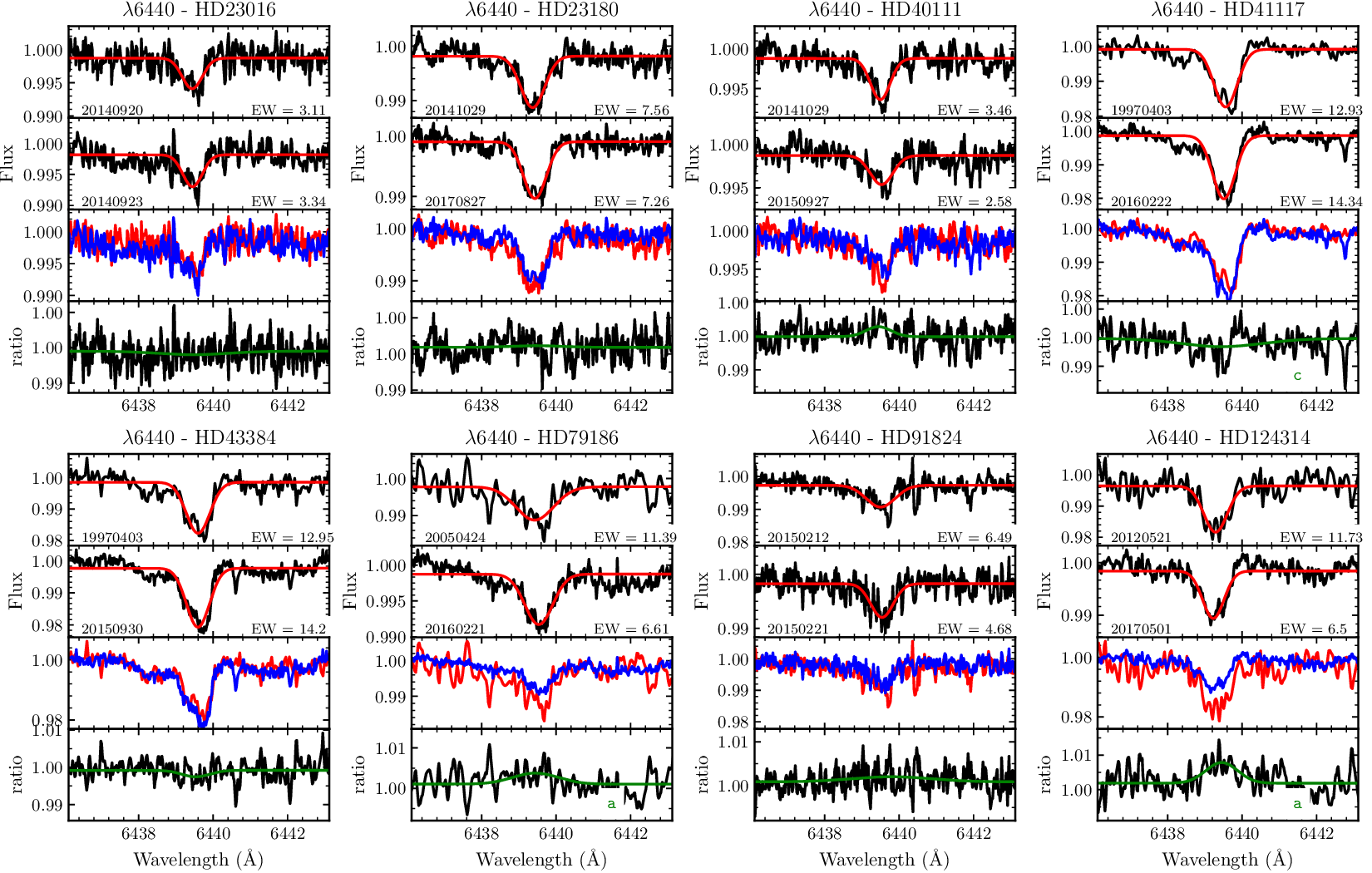}
    \caption{The same as \ref{plt-dib-var1}}
    \label{plt-dib-var105}
\end{figure*}

\clearpage
\begin{figure*}[ht!]
    \centering
    \includegraphics[width=0.99\hsize]{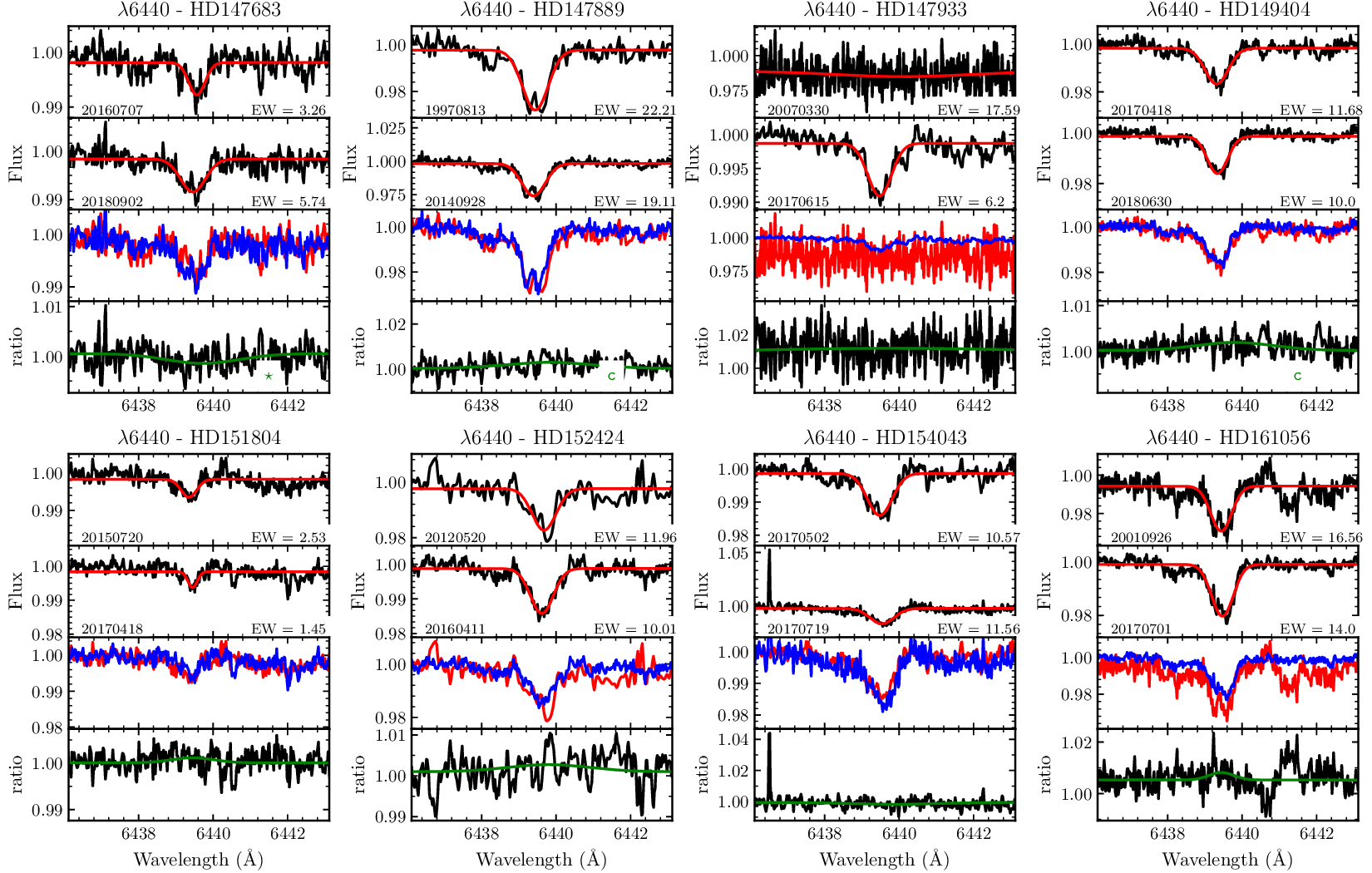}
    \caption{The same as \ref{plt-dib-var1}}
    \label{plt-dib-var106}
\end{figure*}

\begin{figure*}[ht!]
    \centering
    \includegraphics[width=0.99\hsize]{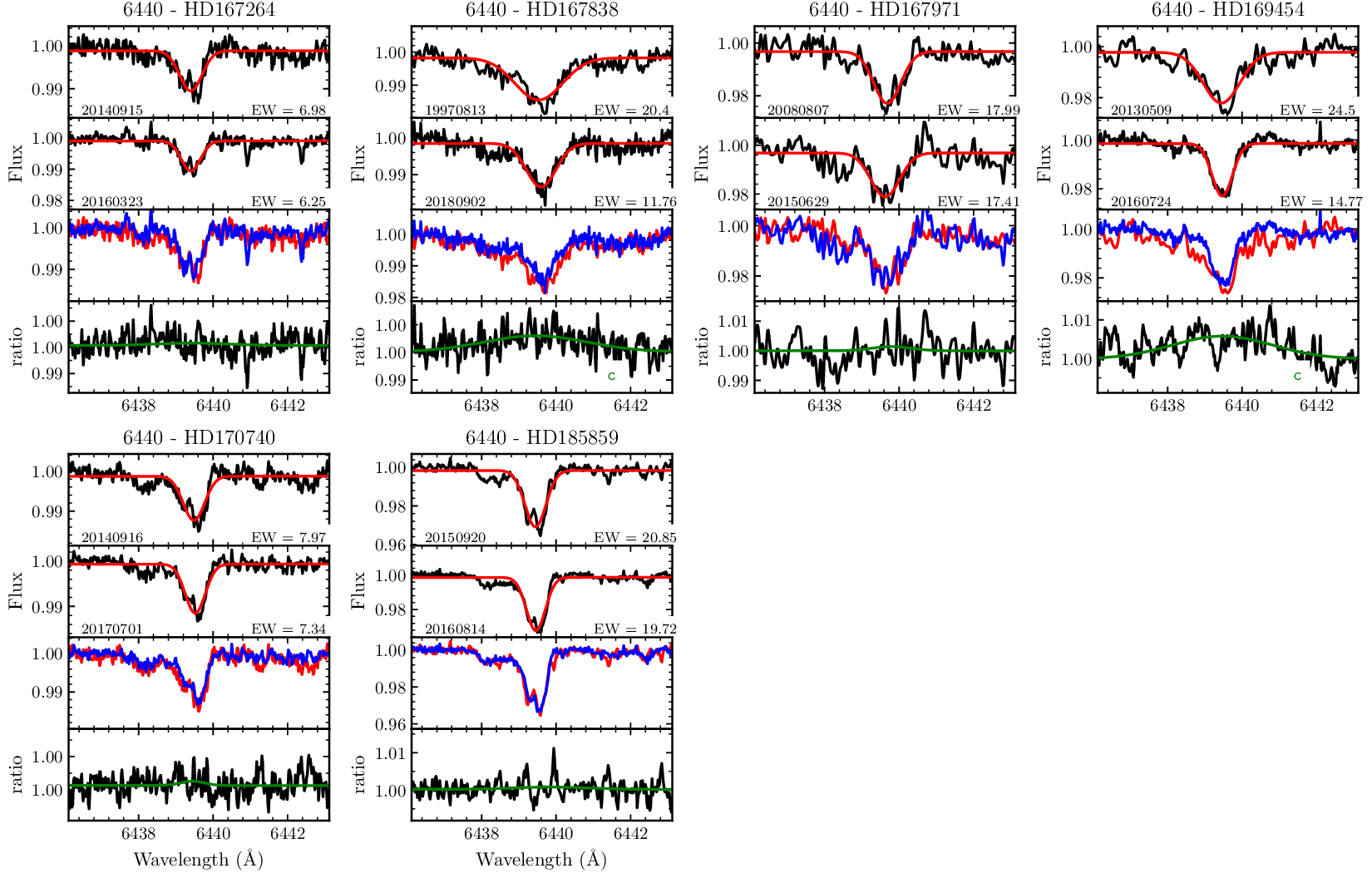}
    \caption{The same as \ref{plt-dib-var1}}
    \label{plt-dib-var107}
\end{figure*}

% %%%%%%%%%%%%%%%%%%
% %%%%%%%%%%%%%%%%%%
\begin{figure*}[ht!]
    \centering
    \includegraphics[width=0.99\hsize]{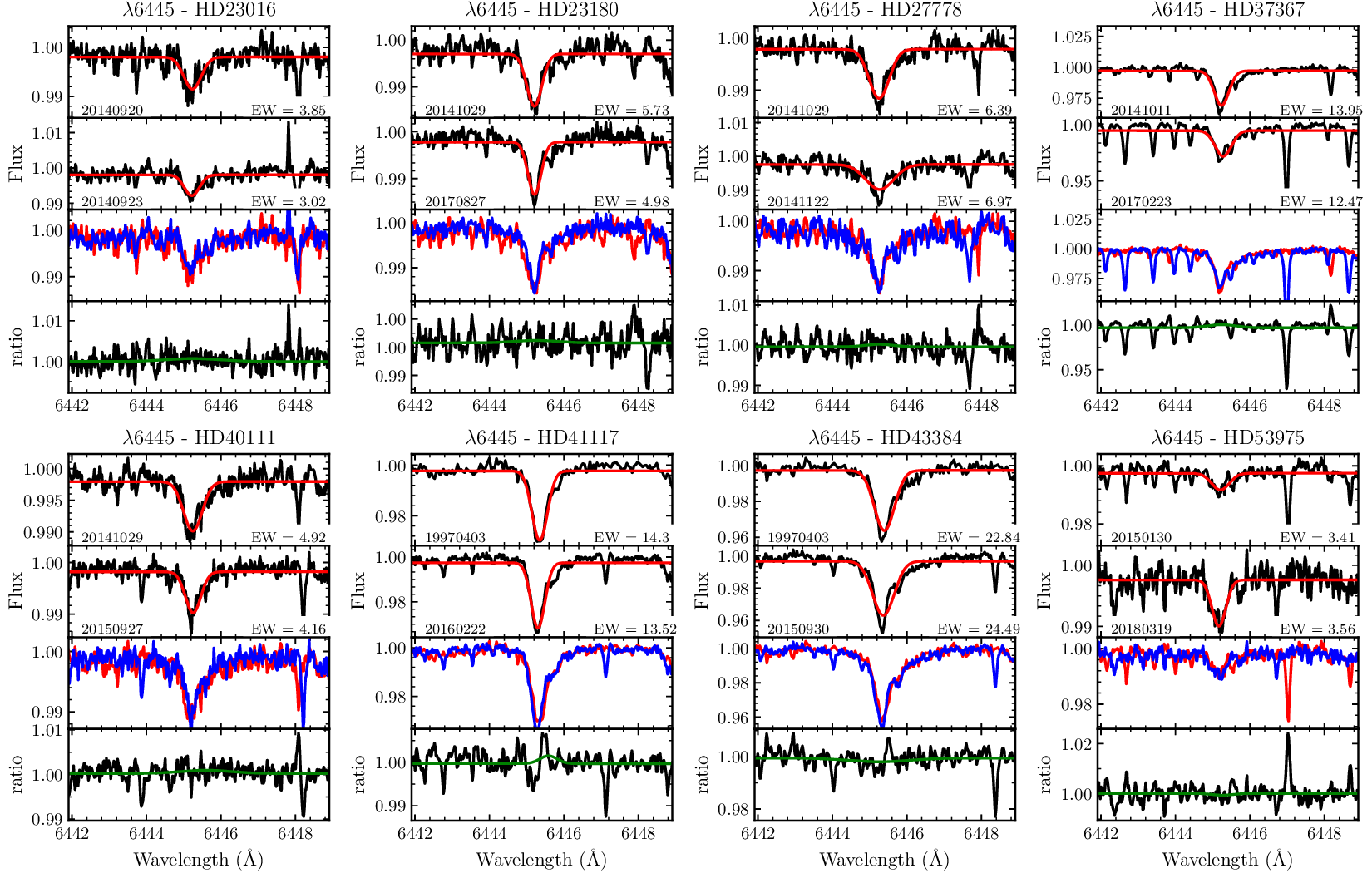}
    \caption{The same as \ref{plt-dib-var1}}
    \label{plt-dib-var108}
\end{figure*}

\begin{figure*}[ht!]
    \centering
    \includegraphics[width=0.99\hsize]{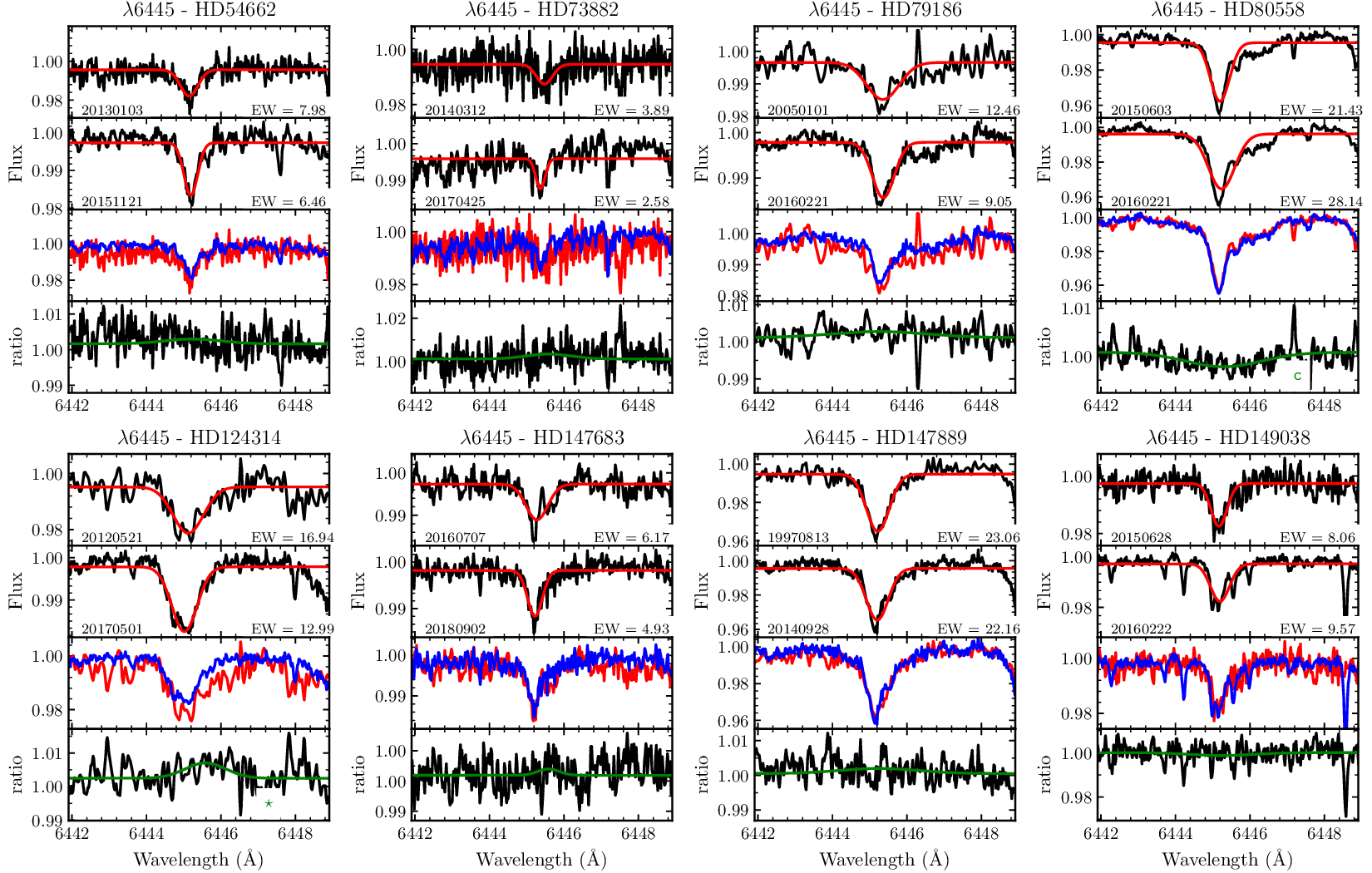}
    \caption{The same as \ref{plt-dib-var1}}
    \label{plt-dib-var109}
\end{figure*}

\begin{figure*}[ht!]
    \centering
    \includegraphics[width=0.99\hsize]{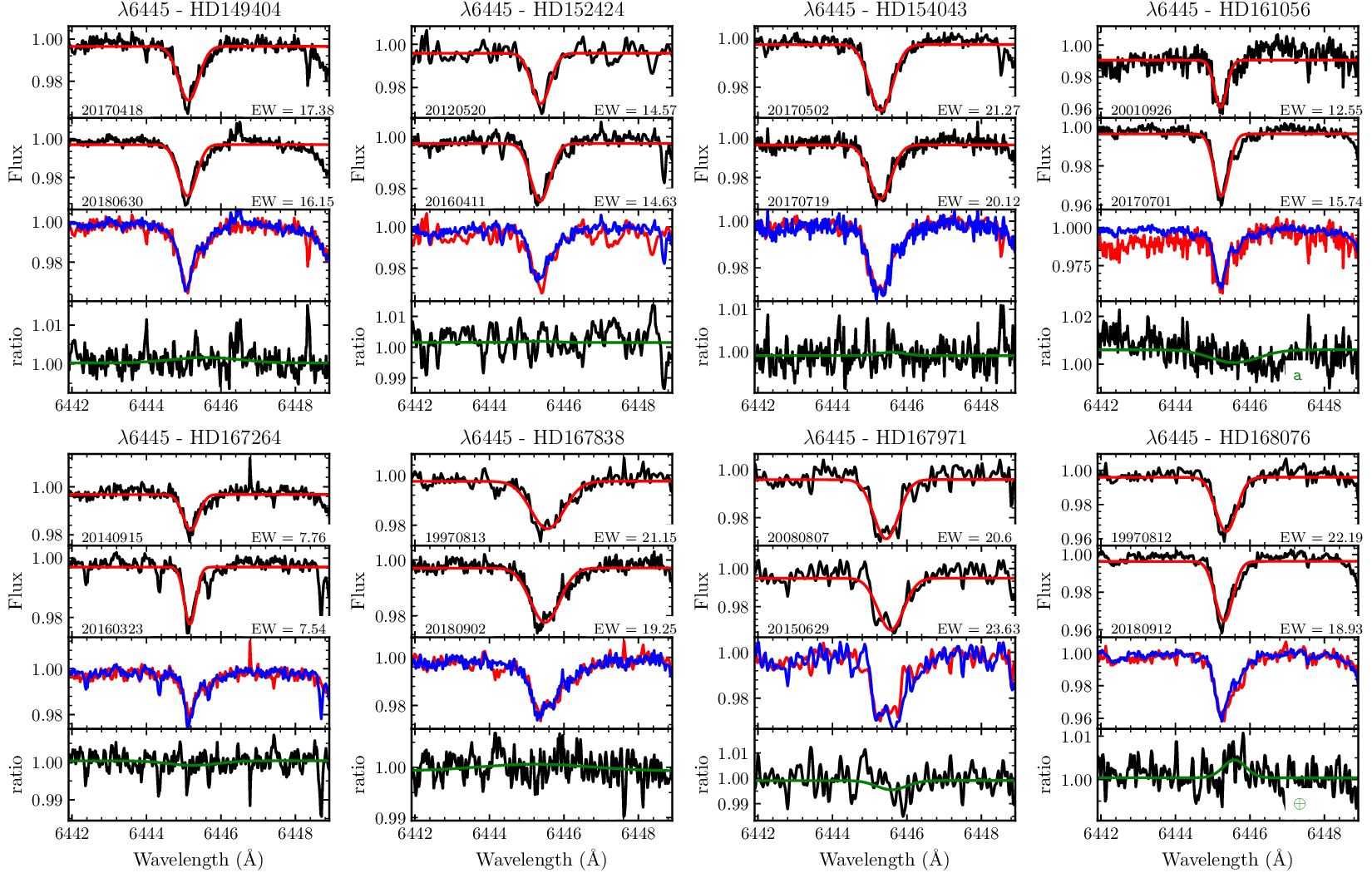}
    \caption{The same as \ref{plt-dib-var1}}
    \label{plt-dib-var110}
\end{figure*}

\begin{figure*}[ht!]
    \centering
    \includegraphics[width=0.99\hsize]{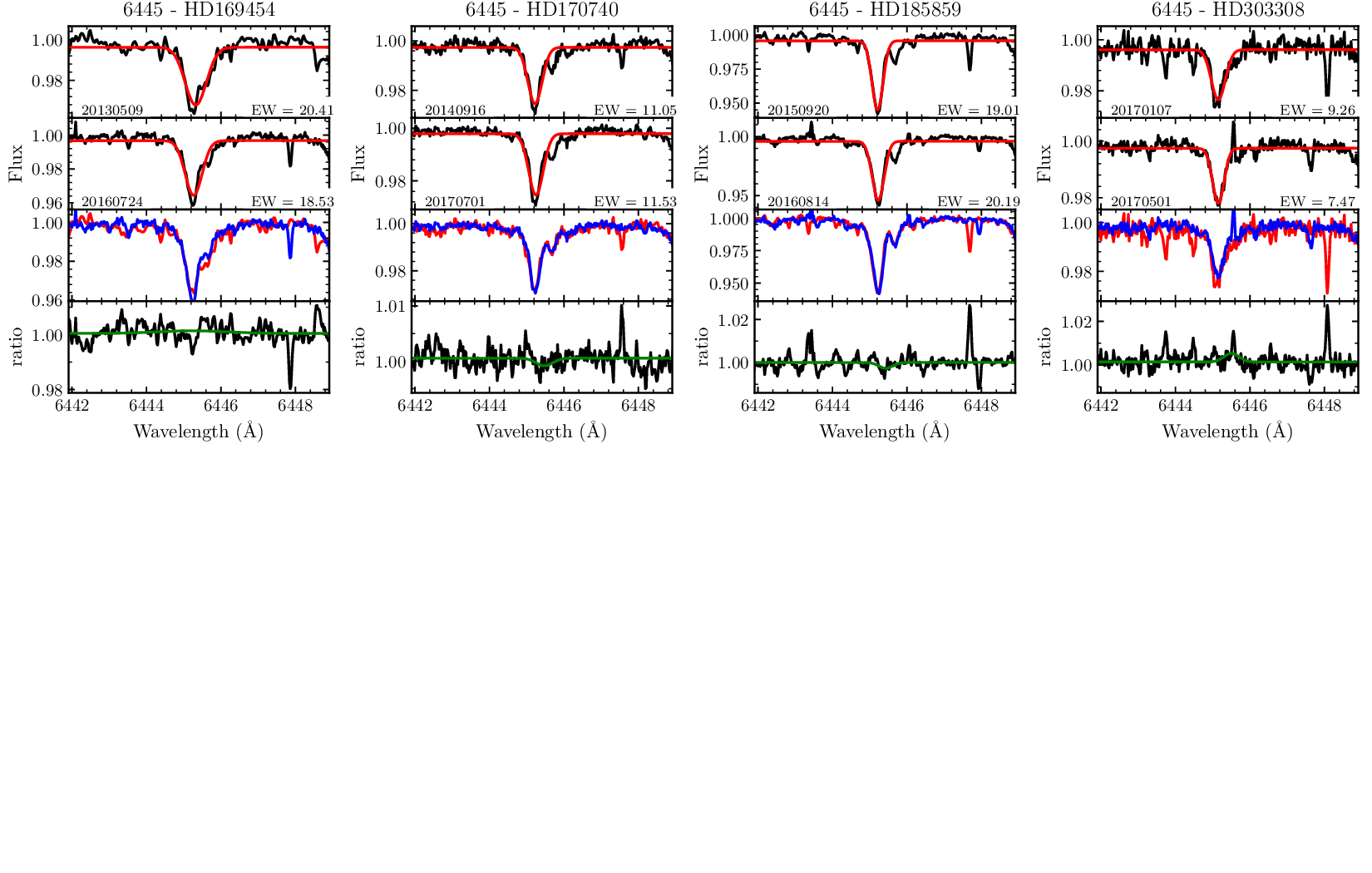}
    \caption{The same as \ref{plt-dib-var1}}
    \label{plt-dib-var111}
\end{figure*}

% %%%%%%%%%%%%%%%%%%
% %%%%%%%%%%%%%%%%%%
\begin{figure*}[ht!]
    \centering
    \includegraphics[width=0.99\hsize]{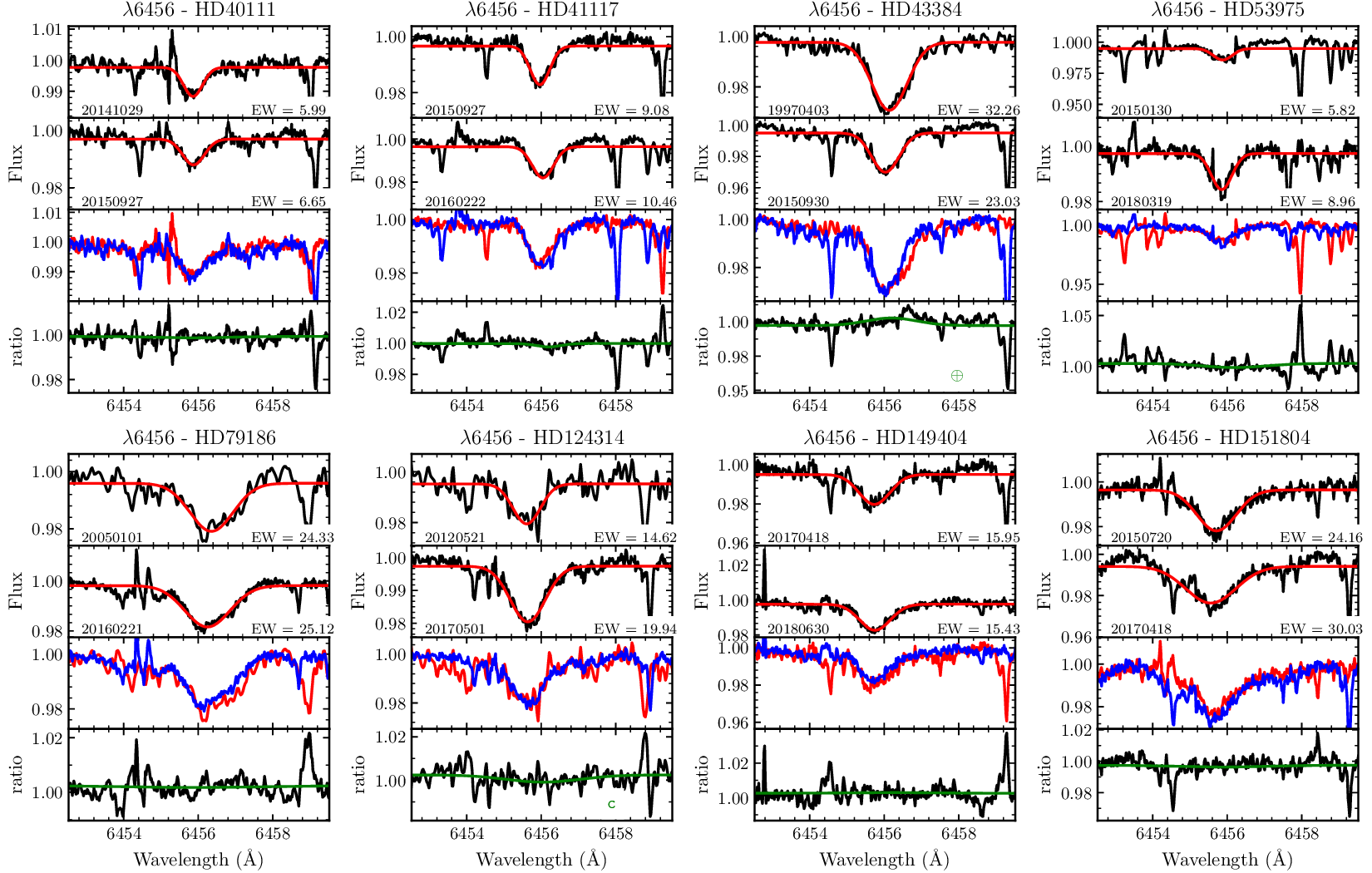}
    \caption{The same as \ref{plt-dib-var1}}
    \label{plt-dib-var112}
\end{figure*}

\begin{figure*}[ht!]
    \centering
    \includegraphics[width=0.99\hsize]{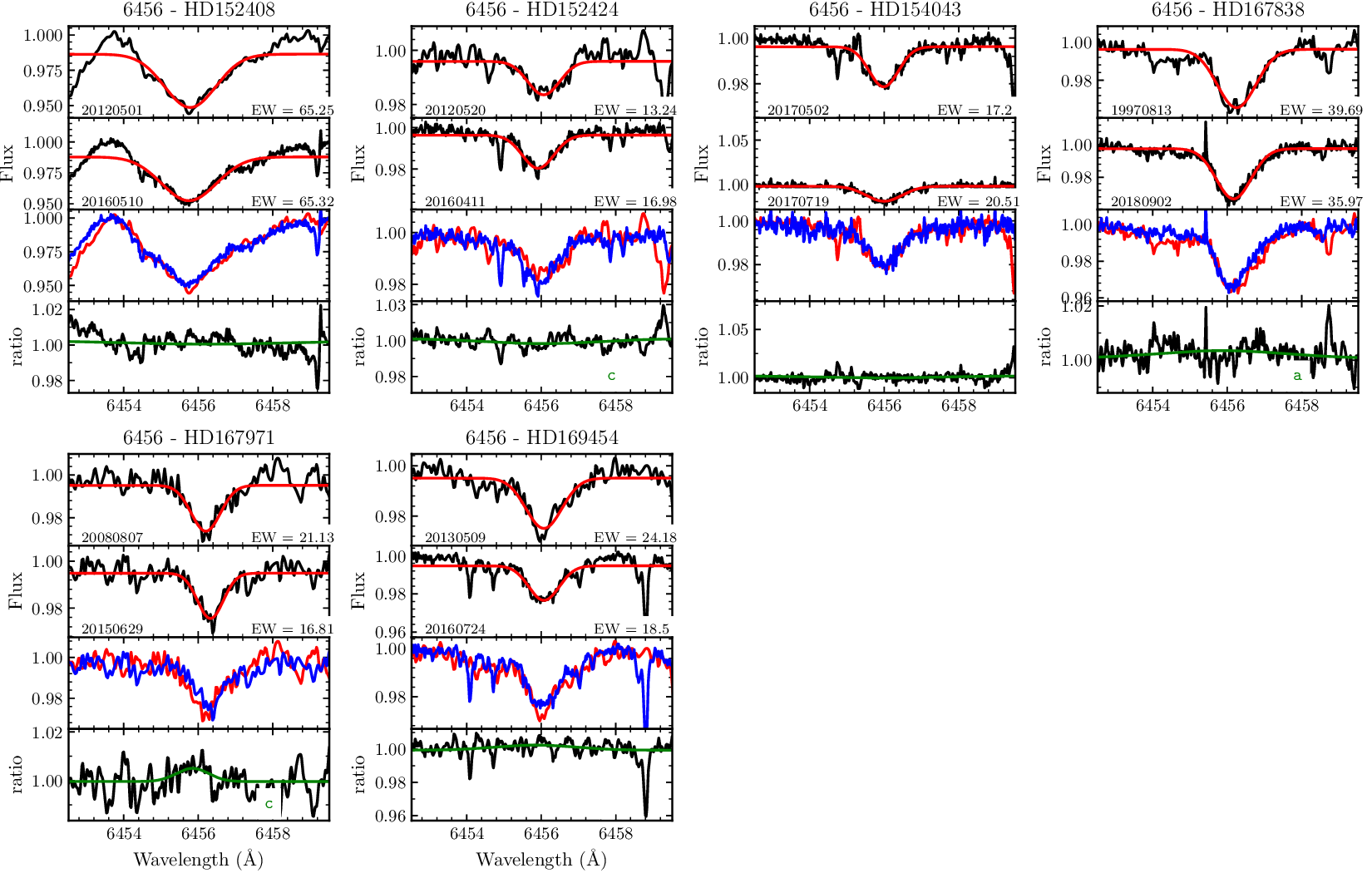}
    \caption{The same as \ref{plt-dib-var1}}
    \label{plt-dib-var113}
\end{figure*}

% %%%%%%%%%%%%%%%%%%
% %%%%%%%%%%%%%%%%%%
\begin{figure*}[ht!]
    \centering
    \includegraphics[width=0.99\hsize]{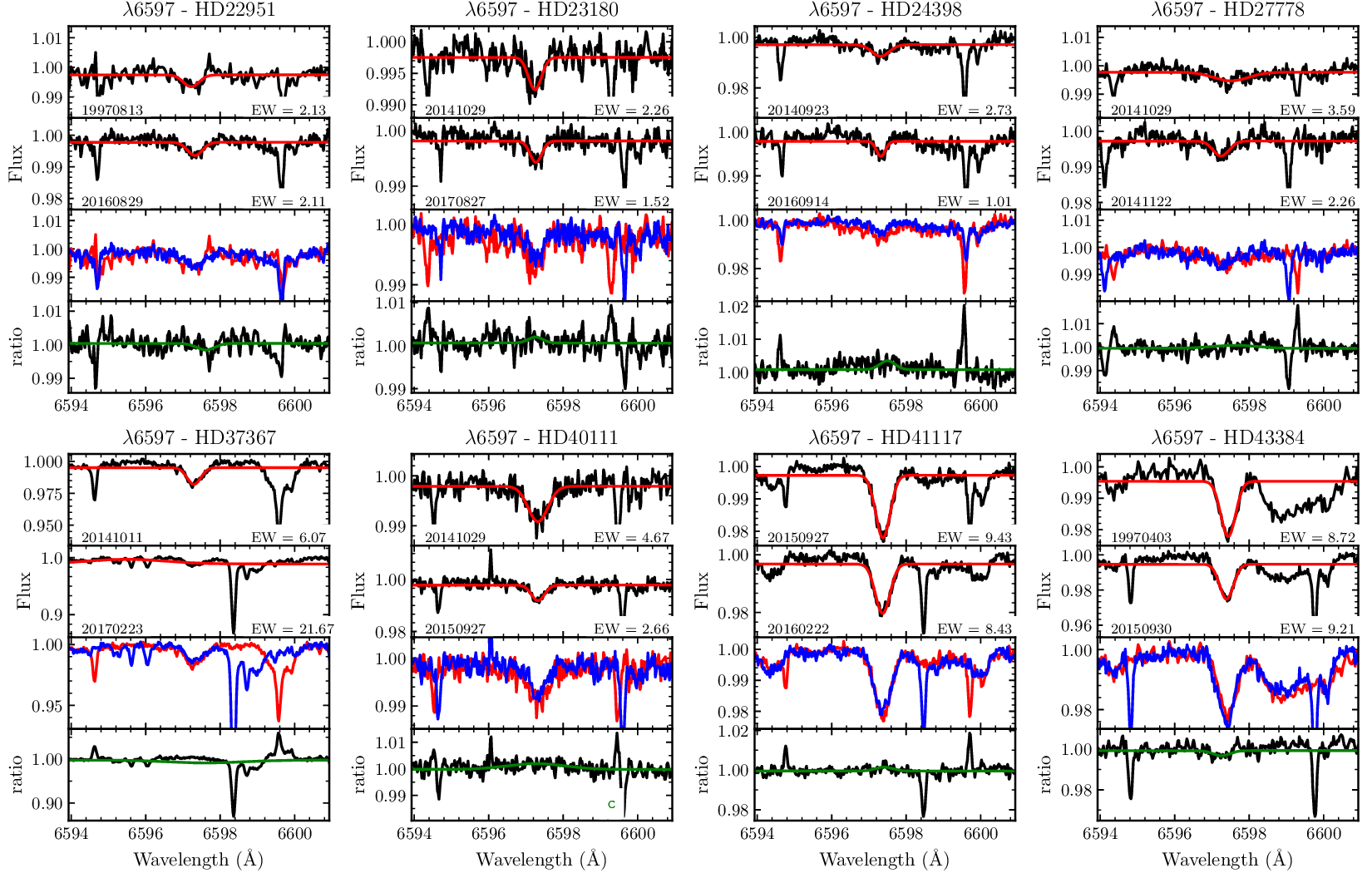}
    \caption{The same as \ref{plt-dib-var1}}
    \label{plt-dib-var114}
\end{figure*}

\begin{figure*}[ht!]
    \centering
    \includegraphics[width=0.99\hsize]{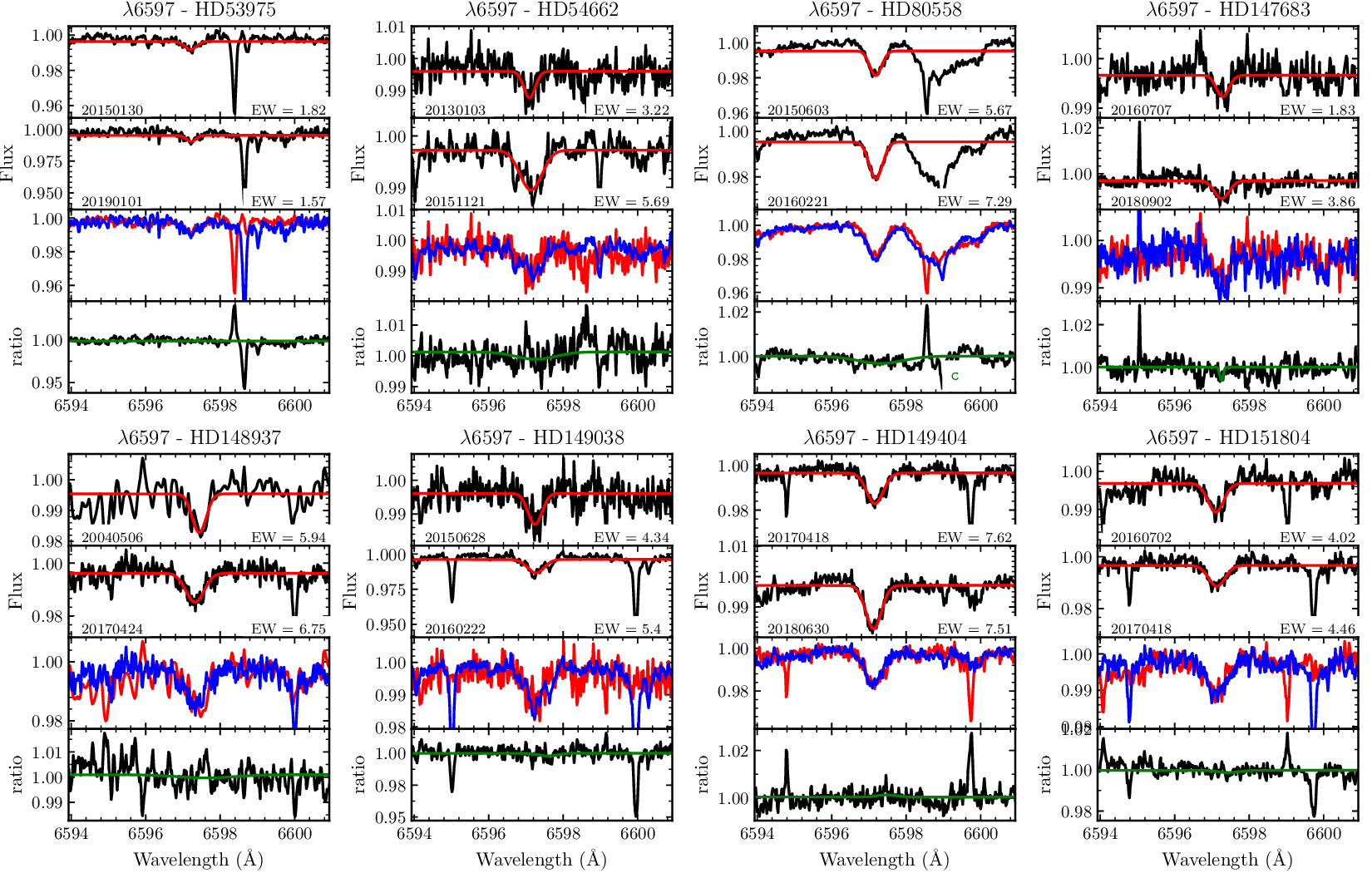}
    \caption{The same as \ref{plt-dib-var1}}
    \label{plt-dib-var115}
\end{figure*}

\begin{figure*}[ht!]
    \centering
    \includegraphics[width=0.99\hsize]{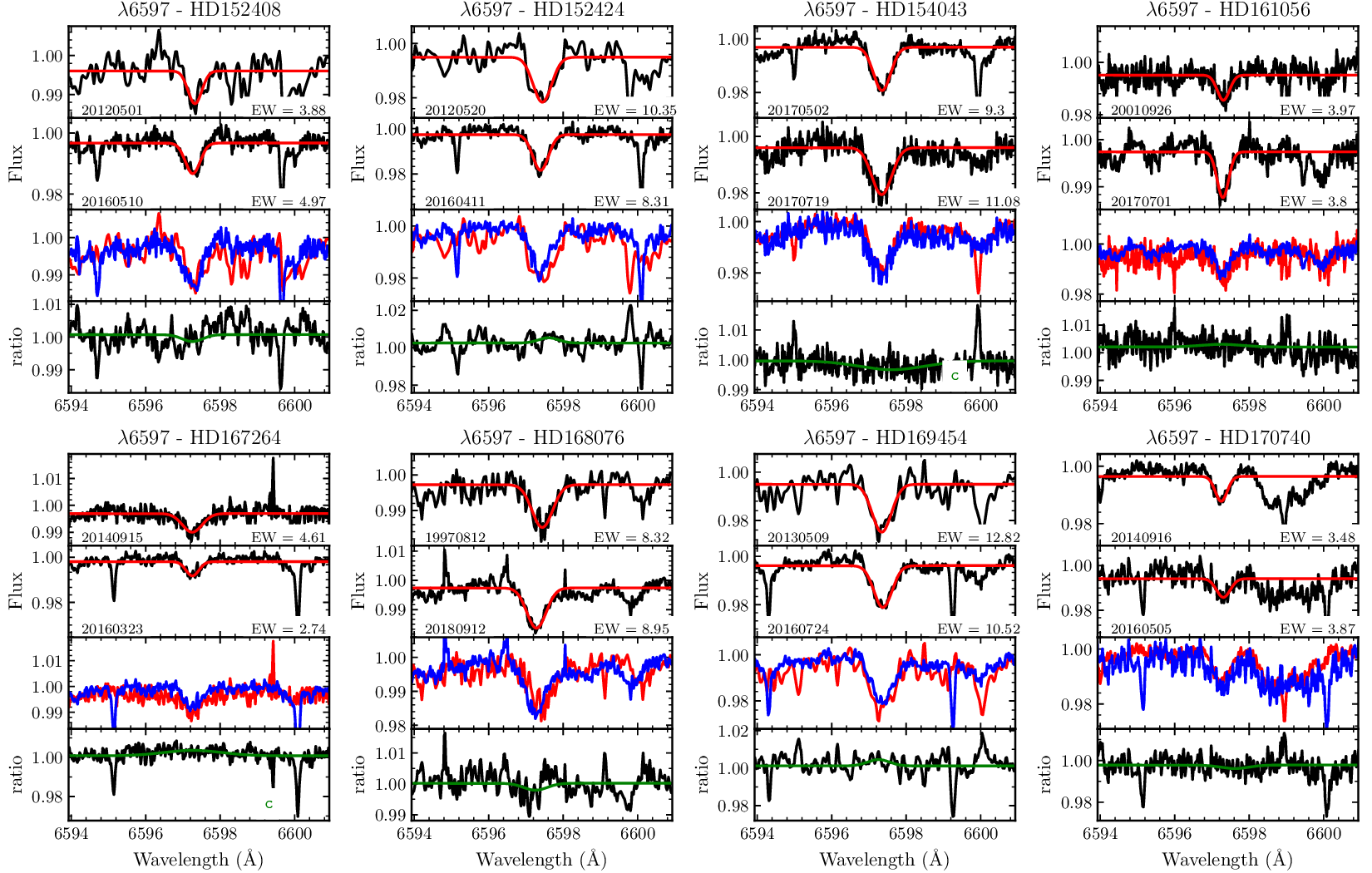}
    \caption{The same as \ref{plt-dib-var1}}
    \label{plt-dib-var116}
\end{figure*}

\begin{figure*}[ht!]
    \centering
    \includegraphics[width=0.99\hsize]{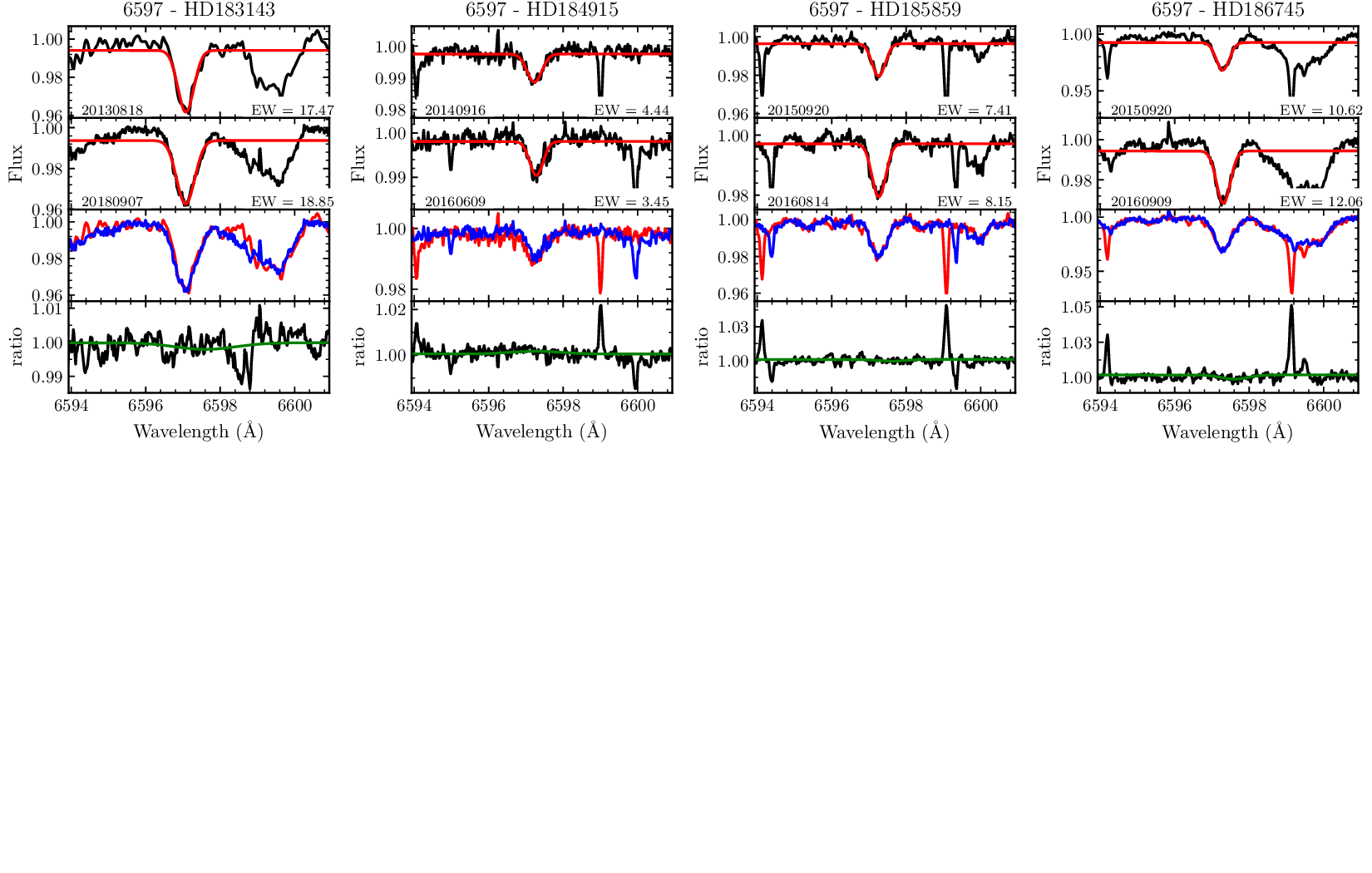}
    \caption{The same as \ref{plt-dib-var1}}
    \label{plt-dib-var117}
\end{figure*}

% %%%%%%%%%%%%%%%%%%
% %%%%%%%%%%%%%%%%%%
\begin{figure*}[ht!]
    \centering
    \includegraphics[width=0.99\hsize]{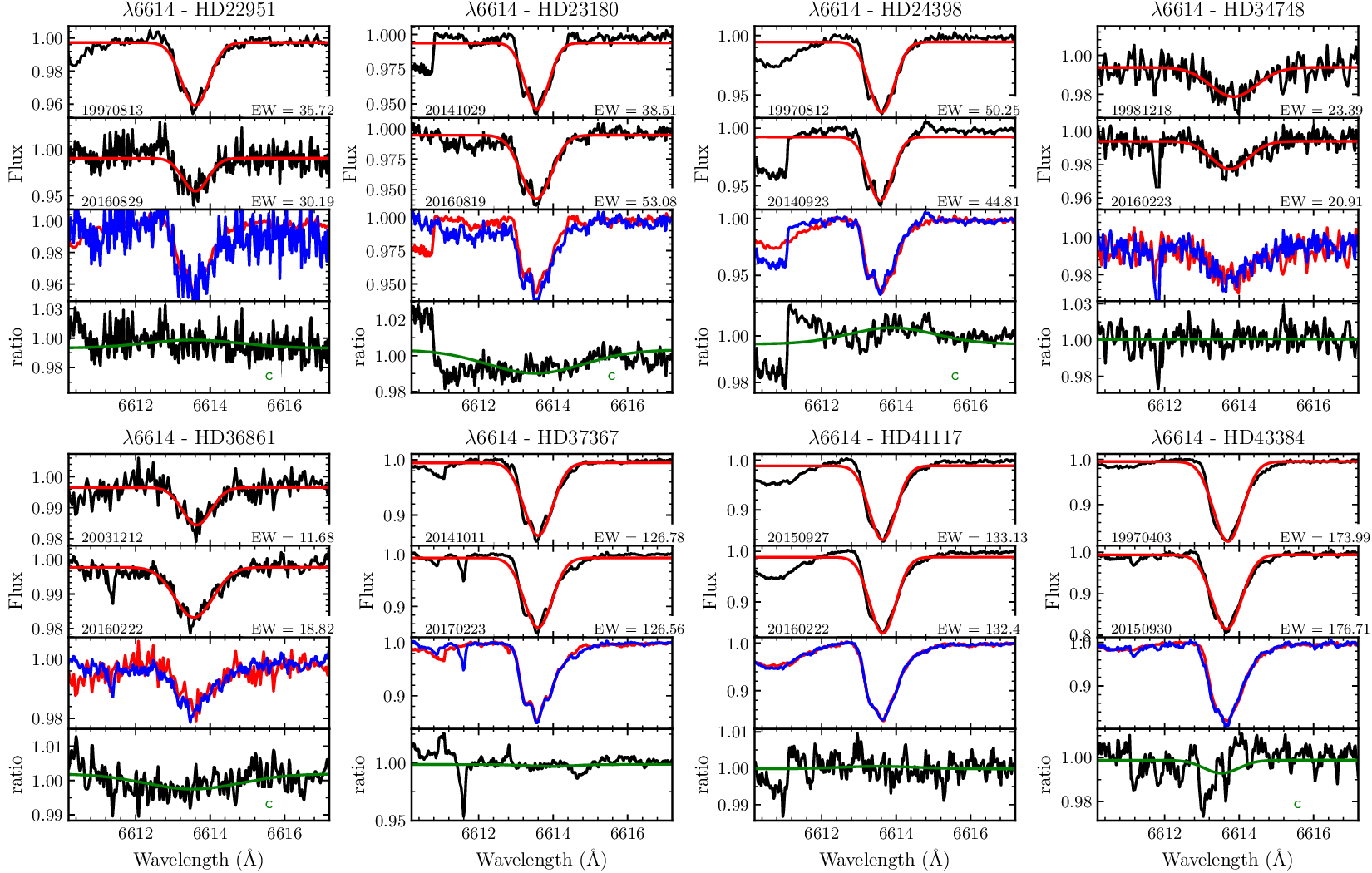}
    \caption{The same as \ref{plt-dib-var1}}
    \label{plt-dib-var118}
\end{figure*}

\begin{figure*}[ht!]
    \centering
    \includegraphics[width=0.99\hsize]{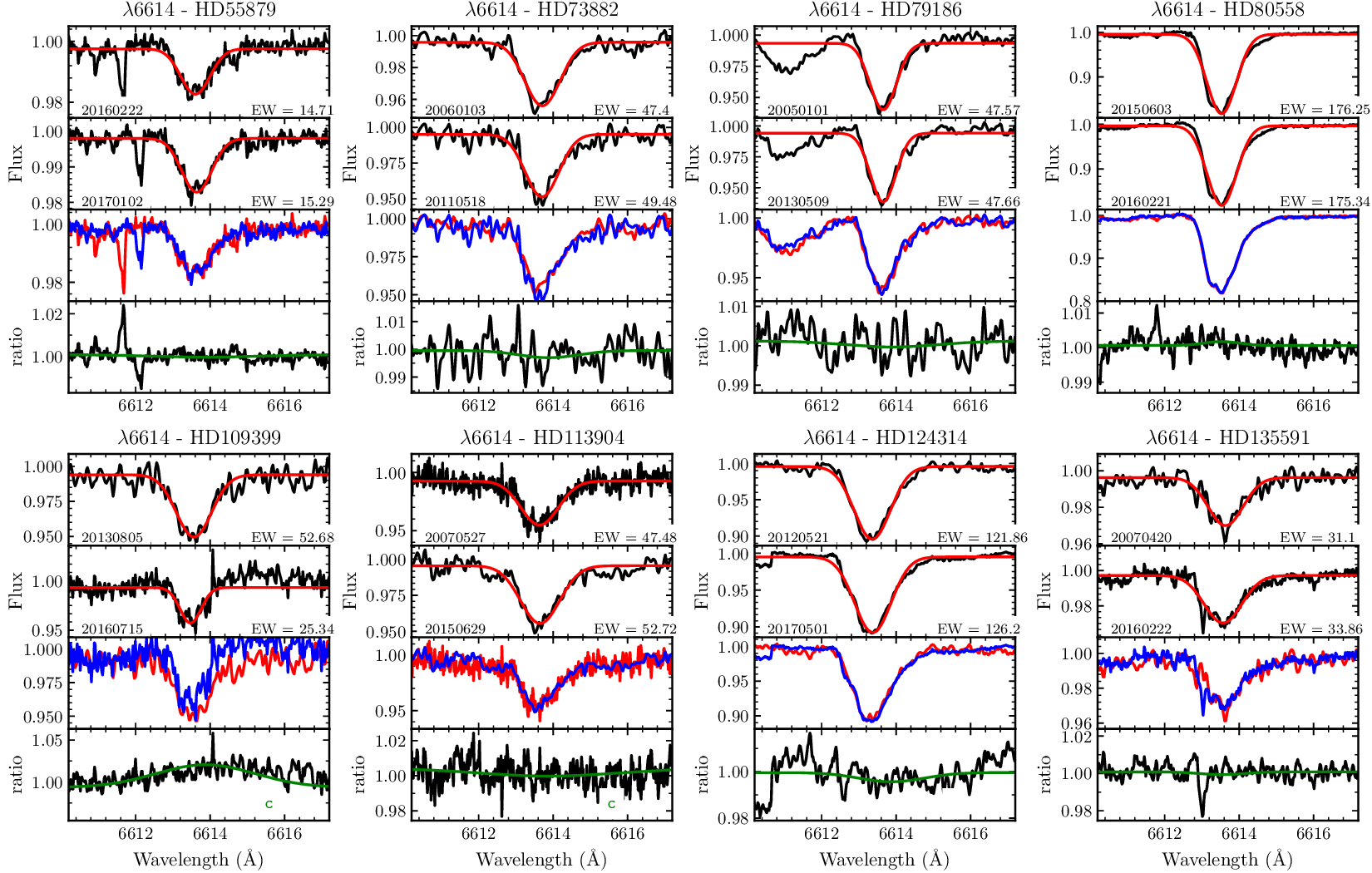}
    \caption{The same as \ref{plt-dib-var1}}
    \label{plt-dib-var119}
\end{figure*}

\begin{figure*}[ht!]
    \centering
    \includegraphics[width=0.99\hsize]{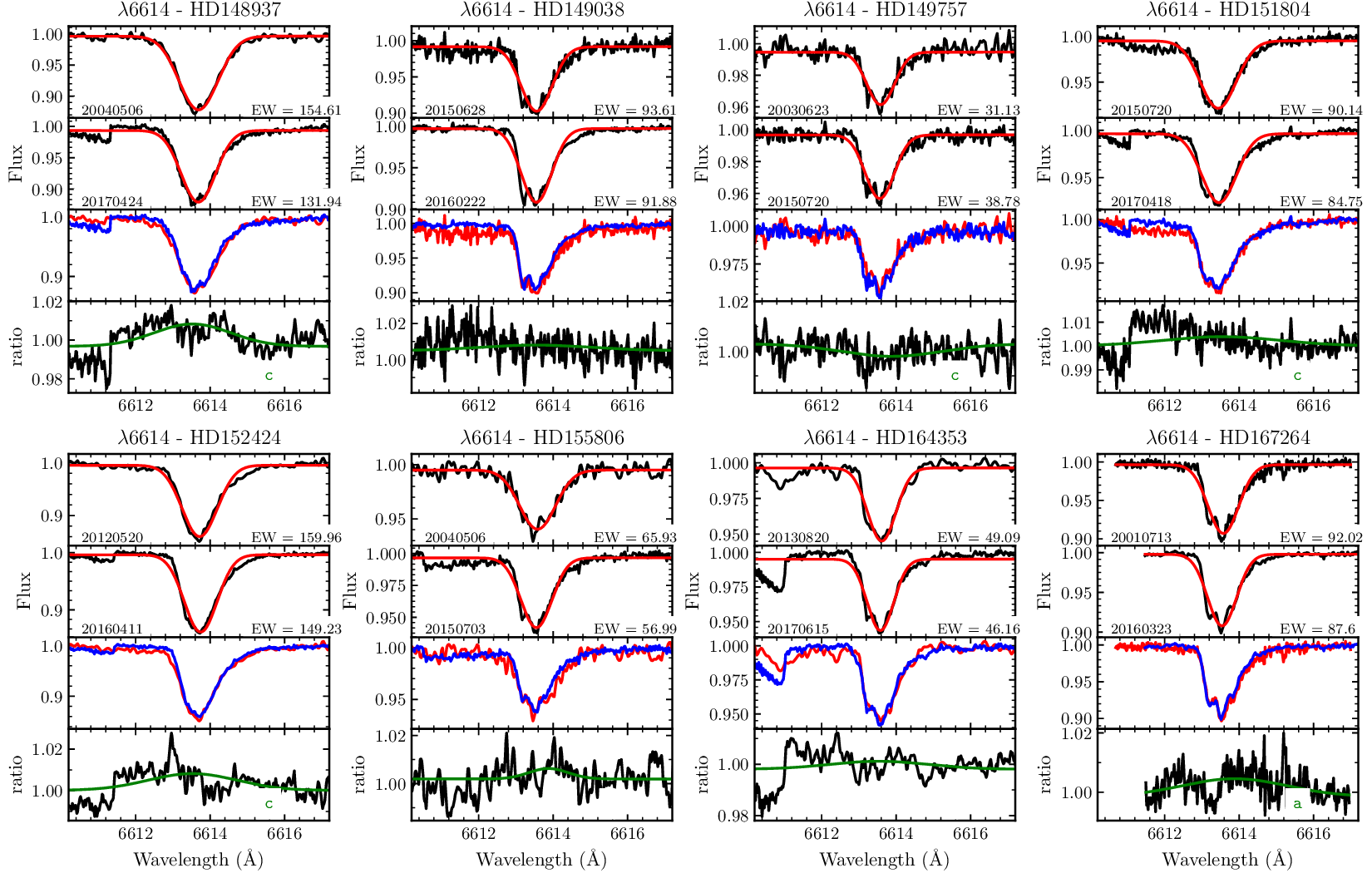}
    \caption{The same as \ref{plt-dib-var1}}
    \label{plt-dib-var120}
\end{figure*}

\begin{figure*}[ht!]
    \centering
    \includegraphics[width=0.99\hsize]{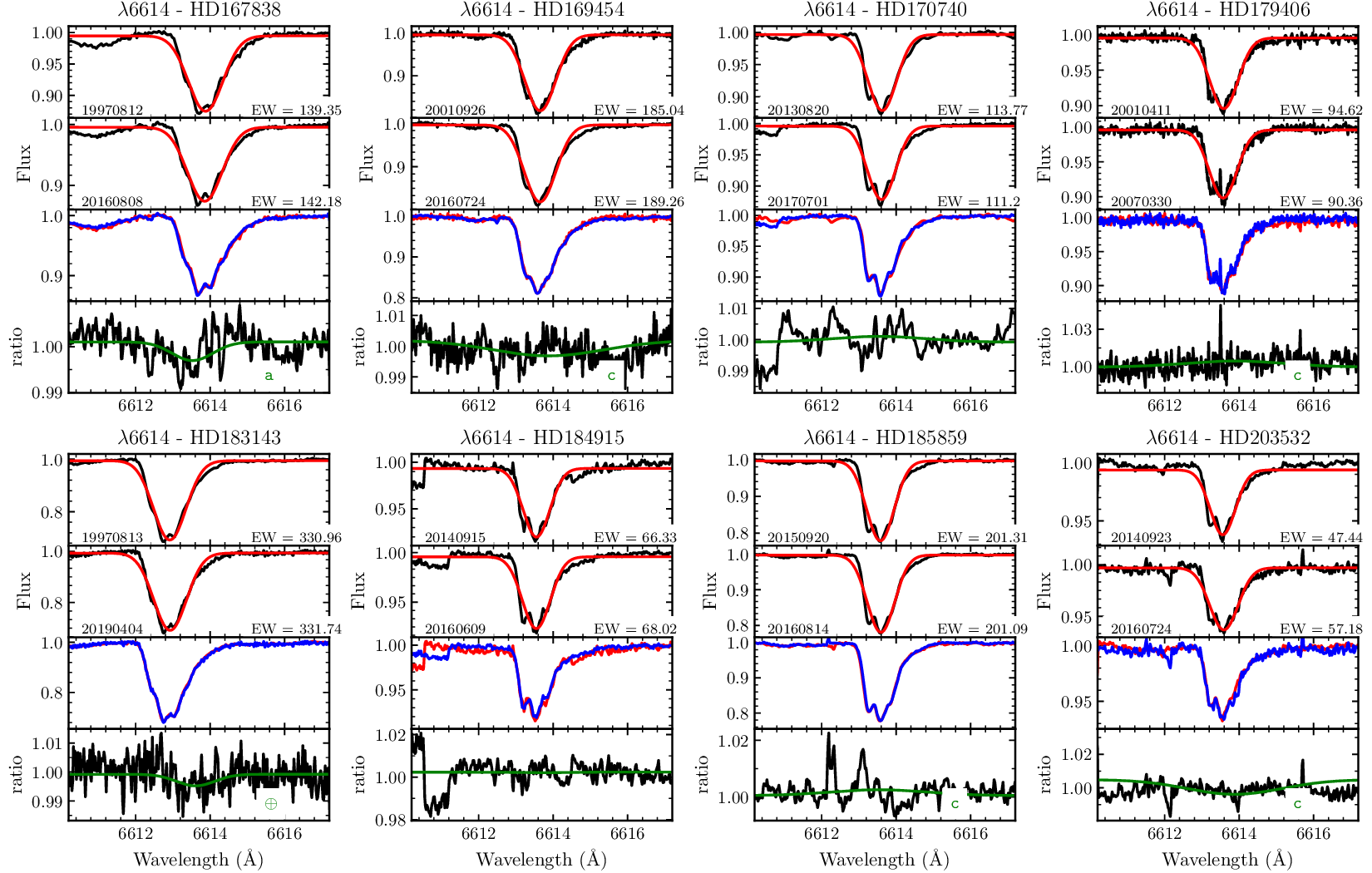}
    \caption{The same as \ref{plt-dib-var1}}
    \label{plt-dib-var121}
\end{figure*}

% %%%%%%%%%%%%%%%%%%
% %%%%%%%%%%%%%%%%%%
\begin{figure*}[ht!]
    \centering
    \includegraphics[width=0.99\hsize]{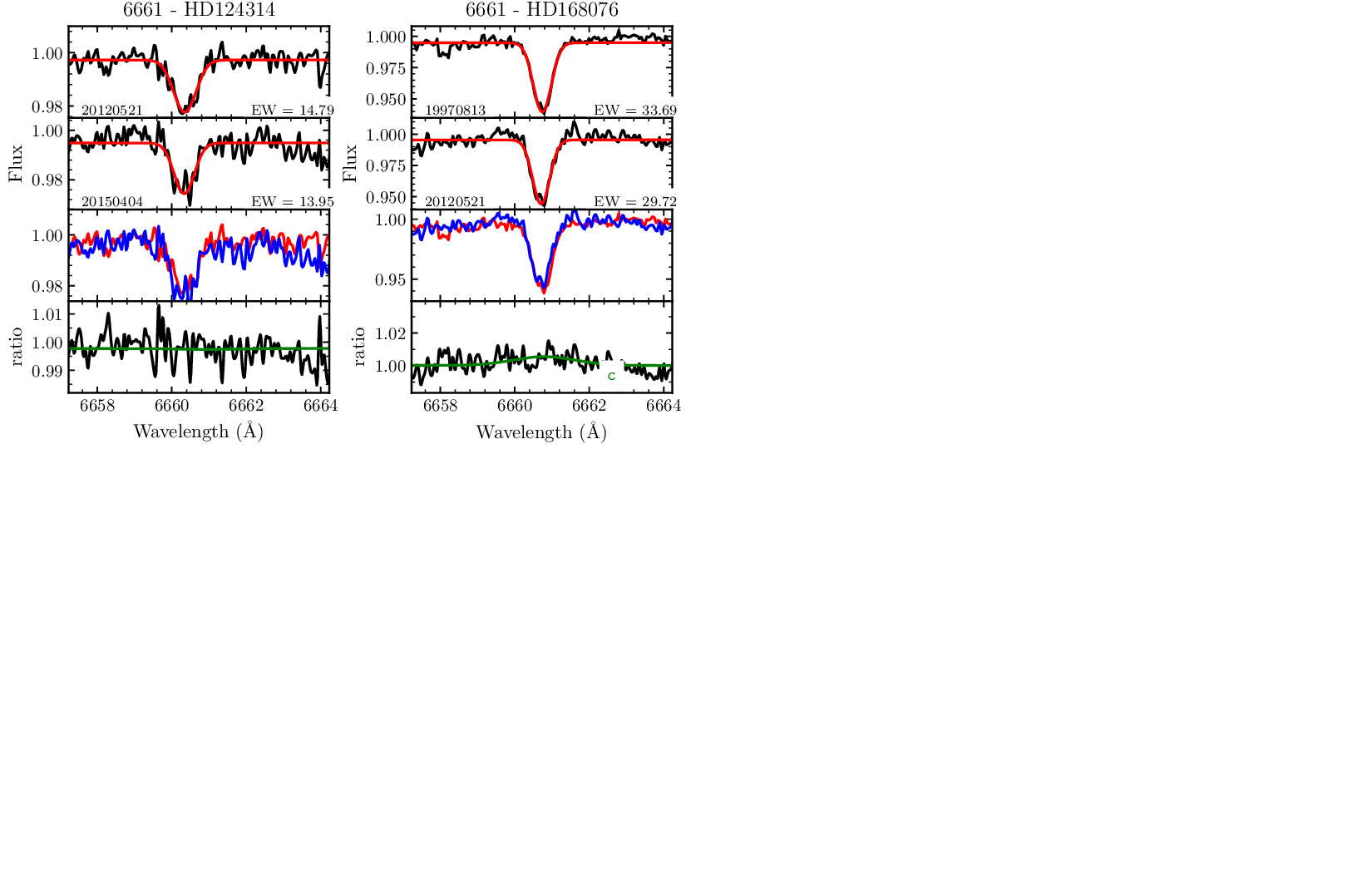}
    \caption{The same as \ref{plt-dib-var1}}
    \label{plt-dib-var122}
\end{figure*}

%% file: table_atomic_fit_results.tex
\onecolumn
\begin{longtable}{p{.7cm}lcccccccc}
\caption{Best-fit Voigt parameters for atomic and molecular lines. Note that all velocities are reported in the heliocentric rest frame.}
\label{tab:ionresults}\\
\toprule

Target & Ion & Epoch 1 &  \multicolumn{3}{c}{Model} & Epoch 2 & \multicolumn{3}{c}{Model} \\
\cline{4-6} \cline{8-10}
~ & ~ & ~ & $\varv$ & $\log N$ & b$_{eff}$ & ~ & $\varv$ & $\log N$ & b$_{eff}$ \\  
~ & ~ & ~ & [km/s] & [cm$^{-2}$] & [km/s] & ~ & [km/s] & [cm$^{-2}$] & [km/s] \\

\midrule
\endfirsthead
\caption{Continued.}\\
\toprule

Target & Ion & Epoch 1 &  \multicolumn{3}{c}{Model} & Epoch 2 & \multicolumn{3}{c}{Model} \\
\cline{4-6} \cline{8-10}
~ & ~ & ~ & $v$ & $\log N$ & b$_{eff}$ & ~ & $v$ & $\log N$ & b$_{eff}$ \\  
~ & ~ & ~ & [km/s] & [cm$^{-2}$] & [km/s] & ~ & [km/s] & [cm$^{-2}$] & [km/s] \\  

\midrule
\endhead

HD~36861 & ~ \\
~ &  CH & 2002-01-20 & 23.8$\pm$0.2 & 12.36$\pm$0.04 & 2.3$\pm$0.5 & 2016-02-22 & 23.5$\pm$0.3 & 12.29$\pm$0.04 & 2.3$\pm$0.5 \\
~ &  CH$^{+}$ & 2002-09-11 & 25.6$\pm$0.3 & 11.92$\pm$0.07 & 1.7$\pm$0.7 & 2016-02-22 & 22.4$\pm$0.6 & 11.65$\pm$0.1 & 1.7$\pm$0.7 \\
~ &  \ion{Ca}{i} & 2002-01-20 & 25.5$\pm$0.1 & 10.05$\pm$0.12 & 3.1$\pm$0.9 & 2016-02-22 & 24.7$\pm$0.2 & 9.95$\pm$0.02 & 3.1$\pm$0.9 \\
~ &  \ion{Ca}{ii} & 2002-01-20 & 6.0$\pm$0.1 & 11.99$\pm$0.1 & 3.2$\pm$0.1 & 2016-02-22 & 5.6$\pm$0.1 & 11.99$\pm$0.1 & 3.2$\pm$0.1 \\
~ &  \ion{Ca}{ii} & 2002-01-20 & 15.3$\pm$0.1 & 11.75$\pm$0.01 & 3.3$\pm$0.1 & 2016-02-22 & 15.2$\pm$0.1 & 11.8$\pm$0.1 & 3.3$\pm$0.1 \\
~ &  \ion{Ca}{ii} & 2002-01-20 & 23.8$\pm$0.1 & 12.11$\pm$0.1 & 3.3$\pm$0.1 & 2016-02-22 & 23.6$\pm$0.1 & 12.0$\pm$0.1 & 3.3$\pm$0.1 \\
~ &  \ion{Fe}{i} & 2002-09-11 & 25.2$\pm$0.5 & 11.49$\pm$0.06 & 4.1$\pm$3.8 & 2016-02-22 & 23.1$\pm$0.7 & 11.29$\pm$2.42 & 4.1$\pm$3.8 \\
\bottomrule
\end{longtable}
\twocolumn